\pdfoutput=1
\documentclass{aastex631}

\usepackage{indentfirst}
\usepackage{xcolor,graphicx,float}
\usepackage{subfigure}
\usepackage{color}
\usepackage{multirow}
\usepackage{booktabs}
\usepackage{appendix}  

\shorttitle{A Study of the Spectral Properties of Gamma-Ray Bursts with the Main and Second Bursts}
\shortauthors{Ze-lian Du et al.}

\graphicspath{{./}{figures/}}

\begin{document}

\title{A Study of the Spectral Properties of Gamma-Ray Bursts with the Main and Second Bursts}

\author{Ze-Lian Du}
\affiliation{College of Physics and Electronics information, Yunnan Normal University, Kunming 650500, People's Republic of China\\}
\author{Zhao-Yang Peng}
\affiliation{College of Physics and Electronics information, Yunnan Normal University, Kunming 650500, People's Republic of China; pengzhaoyang412@163.com\\}
\author{Jia-Ming Chen}
\affiliation{Department of Astronomy, School of Physics and Astronomy, Key Laboratory of Astroparticle Physics of Yunnan Province, Yunnan University, Kunming 650091, People's Republic of China \\}
\author{Yue Yin}
\affiliation{Department of Physics, Liupanshui Normal College, Liupanshui 553004, People's Republic of China\\}
\author{Ting Li}
\affiliation{State-owned Assets and Laboratory Management Office, Yunnan Normal University, Kunming 650500, People's Republic of China}

\begin{abstract}
The origins of the main burst and second burst of gamma-ray bursts (GRBs) and the composition of their jets remain uncertain. To explore this complex subject more thoroughly, we conduct a spectral analysis on 18 GRBs with a main and a second burst observed by Fermi/GBM. First, we employ Bayesian time-resolved spectral analysis to compare the spectral components of the main and the second burst, finding that $83.3\%$ of the main and second bursts contain a thermal component. $67\%$ of the GRBs, the thermal component gradually decreased from the main to the second burst and the number of spectra exceeding the “Synchrotron line-of-death” is significantly higher in the main burst than in the second burst. Subsequently, we ascertain that for both the main and second bursts, $71.4\%$ of the low-energy spectral index $\alpha$ and $77.8\%$ of the peak energy $E_{P}$ evolve in a similar fashion. There are $50.0\%$ and $72.2\%$ of the GRBs exhibit comparable correlations for the $Flux-\alpha$ and $\alpha-E_{P}$, respectively. For $Flux-E_{P}$ both the main and second burst show a positive correlation. Moreover, from the perspective of the temporal evolution of characteristic radii, the transition from the main to the second burst appeared to be seamless. Finally, we find that both the main and the second burst follow the same Amati relation and Yonetoku relation. Our analysis strongly indicates that the second burst is a continuation of the main burst and is highly likely to share a common physical origin.
\end{abstract}


\section{Introduction\label{Sections1}}
Gamma-ray bursts (GRBs) stand as some of the most violent astronomical phenomena in the universe. These are occurrences where the intensity of gamma rays from distant cosmic sources experiences a rapid increase within a brief time span, followed by a rapid decline. Based on the burst duration, denoted by $T_{90}$, GRBs are categorized into two types: long GRBs  ($LGRBs$) with $T_{90} > 2 s$ and short GRBs ($SGRBs$) with $T_{90} < 2 s$. Researchers widely believe that LGRBs originate from black holes or magnetars formed by the collapse of massive stars, while SGRBs are thought to result from the merger of compact binary stars \citep{1986ApJ...308L..43P,1992ApJ...395L..83N,1993ApJ...405..273W,1998ApJ...494L..45P,2006ARA&A..44..507W}. Despite having been discovered over 50 years ago, the origin of the prompt emission (the bursty emission in the hard-X-ray/soft-$\gamma$-ray band)  from GRBs remains poorly comprehended. A key unresolved issue concerns the composition of the jet, which plays a pivotal role in determining the energy dissipation and radiation mechanisms associated with GRBs \citep{1999ApJ...518..356P,1999A&AS..138..511R,2002ApJ...579..706D,2006ApJ...643L..87G,2007ApJ...661.1025L,2017NewAR..79....1L,2011NuPhS.221..324Z,2011ApJ...726...90Z,2013ApJS..207...23X,2015PhR...561....1K,2015ApJ...802..134B,2015AdAst2015E..22P,2019MmSAI..90...57M}. 

Spectral analysis is an essential method for studying GRBs. From the spectrum, valuable information regarding the energy, radiation mechanisms, and jet structure of GRBs can be obtained. Observations suggest that the spectra of GRBs may comprise both thermal and nonthermal components. The thermal radiation is thought to originate from the fireball photosphere, as predicted by the fireball model, while the nonthermal radiation is believed to stem from shocks, magnetic reconnection, and turbulence \citep{2000HEAD....5.3403L,2008MNRAS.384...33K,2011A&A...526A.110D,2011ApJ...726...90Z,2014ApJ...782...92Z,2015ApJ...805..163D,2022ApJ...927..173S,2024ApJ...969...26P}. By analyzing the spectral components, the radiation mechanism of GRBs (synchrotron radiation or photospheric radiation) can be determined, which in turn enables the inference of the jet composition. If the spectrum exhibits a significant thermal component, it indicates that the radiation mechanism is photospheric emission, and the jet is predominantly controlled by a baryon-dominated thermal fireball \citep{1990ApJ...363..218P,1990ApJ...365L..55S,2000A&A...359..855R,2004ApJ...614..827R,2009ApJ...702.1211R,2014ApJ...785..112D,2016LPICo1962.4083R,2017IJMPD..2630018P}. Conversely, if the radiation mechanism is synchrotron emission, it implies that the jet is primarily dominated by Poynting-flux \citep{2006A&A...450..887G}. However, there is an alternative possibility: the central engine of GRBs may involve a hybrid jet model, featuring both thermal fireball components and Poynting-flux components \citep{2018NatAs...2...69Z,2019ApJS..242...16L}. Based on this, \cite{2015ApJ...801..103G} developed a theory of photospheric emission that incorporates a mixture of both thermal fireball and Poynting-flux components. In spectral analysis, commonly used empirical models for fitting GRB spectra include the simple power law (PL), cutoff power law (CPL), Band function (Band), smoothly broken power law (SBPL), and Planck function.

Previous studies have revealed a highly special type of GRB. This type is characterized by two distinct bursts that occur at different times, with an extremely long quiescent period separating them. The phenomenon in which a faint precursor burst precedes the main burst is known as a precursor. This particular aspect has been extensively investigated by numerous researchers in the field of GRBs \citep{1995ApJ...452..145K,2005MNRAS.357..722L,2008ApJ...685L..19B,2009A&A...505..569B,2010ApJ...723.1711T,2019ApJ...884...25Z,2020ApJ...902L..42W}. 

In the research on GRB precursors, \cite{2015MNRAS.448.2624C} gathered data on 2710 LGRBs. They found that the pre - and post - peak emission periods are statistically comparable, potentially suggesting a shared origin. Among the analyzed GRB sample, $24\%$ manifested more than one isolated emission event, $11\%$ had at least one pre - peak emission event, and $15\%$ had at least one post - peak emission event. \cite{2018NatAs...2...69Z} reported on an exceptionally bright GRB, GRB 160625B, which was observed in both gamma - ray and optical wavelengths and has three distinct episodes: a brief precursor, a very bright main burst, and an extended radiation event resembling a second burst. These were separated by two long quiescent intervals of 180 and 300 seconds. Temporal and time - resolved spectral analyses of the precursor and main burst showed that the precursor had a thermal component, while the main burst and the extended radiation events had a non - thermal component. This shift from thermal to non - thermal radiation within a single GRB with clearly separated emission events implies a change in the jet composition from being fireball - dominated to Poynting - flux - dominated. \cite{2019ApJ...884...25Z} selected 18 SGRB samples with precursors from a total of 660 short bursts observed by Swift and Fermi and analyzed their temporal and spectral characteristics. The results showed: (1) the duration of the main bursts was generally longer than that of the precursors; (2) the average flux of the precursors tended to increase as the main burst brightened; (3) based on the distribution of hardness ratios and spectral fitting, precursors were generally softer than the main bursts, with most precursors and main bursts showing non - thermal radiation characteristics; (4) precursors could be important probes for detecting the properties of the progenitor stars of short bursts. \cite{2019ApJS..242...16L} utilized GRB monitoring data from the Fermi satellite to obtain a sample of 43 GRBs with clear multiple pulses. They found that 9 out of 43 ($21\%$) of these bursts may exhibit a transition or shift in the jet composition from a fireball - dominated to a Poynting - flux - dominated regime. A detailed time - resolved spectral analysis of 4 brighter GRBs in these 9 bursts revealed that the precursors had a thermal component. By constraining their outflow properties, it was determined that they were consistent with the typical characteristics of photospheric radiation. Their analysis suggested that a significant portion of the Fermi multiple - pulse GRBs may experience a transition in the jet composition from a matter - dominated fireball to a Poynting - flux - dominated one. 

\cite{2020PhRvD.102j3014C} identified 217 GRBs with precursors by analyzing over 11 years of Fermi/GBM data. Their analysis showed that the duration of the quiescent period between the precursor and the main burst could be well - described by a double Gaussian distribution, indicating that the observed precursors have two distinct physical progenitors. \cite{2021ApJS..252...16L} analyzed short - burst data from Swift/BAT, focusing on events that included precursors, main bursts, and extended radiation. The study explored the similarities in temporal structure between main bursts with a single peak and those with two peaks, revealing no significant anomalies. Observations from Swift and BATSE indicated comparable properties for the main bursts of short bursts. It was observed that the duration of the main burst was slightly longer than that of the precursor but a bit shorter than that of the extended radiation component. Notably, a power - law correlation was found among the peak fluxes of the precursors, main bursts, and extended radiation, suggesting that these three events likely originated from similar central - engine activity.

\cite{2015MNRAS.448.2624C} observed that, following a long period of quiescent time, a weak post-peak emission period can occur after the end of the main burst, similar to the separation found between precursors and main emission periods. The post - main emission investigated in this paper is referred to as the second burst.

Research on GRB precursors and main bursts has yielded some fascinating results. However, at present, there is relatively limited research regarding the main bursts and the second bursts of GRBs. \cite{2022ApJ...940...48D} carried out a statistical study on the spectral characteristics for only two bright bursts that had both main bursts and second bursts (GRB 160509A and GRB 130427A). Their findings indicated that the correlation of spectral parameters displays different behaviors in the main bursts and the second bursts. The evolution trends of spectral parameters in the main bursts and the second bursts also exhibited both congruent and incongruent behaviors. Thus, it remains uncertain whether the main and second bursts share the same origin.

In this study, we employ Bayesian and Markov Chain Monte Carlo methods to conduct detailed time-resolved spectral and time-integrated spectral analyses of the main bursts and second bursts of 18 GRBs. We compare their spectral components and spectral characteristics using the best-fit models. Additionally, based on previous theoretical models, we impose constraints on the outflow properties of the main bursts and second bursts to explore whether they have a common origin and to provide evidence for elucidating their physical origins.

The structure of this paper is organized as follows: Sections \ref{Sections2} and \ref{Sections3} are dedicated to introducing the sample selection and analysis methods, along with the spectral models utilized in this research. Section \ref{Sections4} presents the outcomes of the spectral analysis. Section \ref{Sections5} is focused on deriving the photospheric radiation parameters. Section \ref{Sections6} showcases the Amati and Yonetoku relations for the sample. Finally, Sections \ref{Sections7} and \ref{Sections8} offer the discussion and conclusions, respectively.

\section{Sample Selection\label{Sections2}}
\begin{figure}[htbp]
\centering
\includegraphics [width=7.5cm,height=6cm]{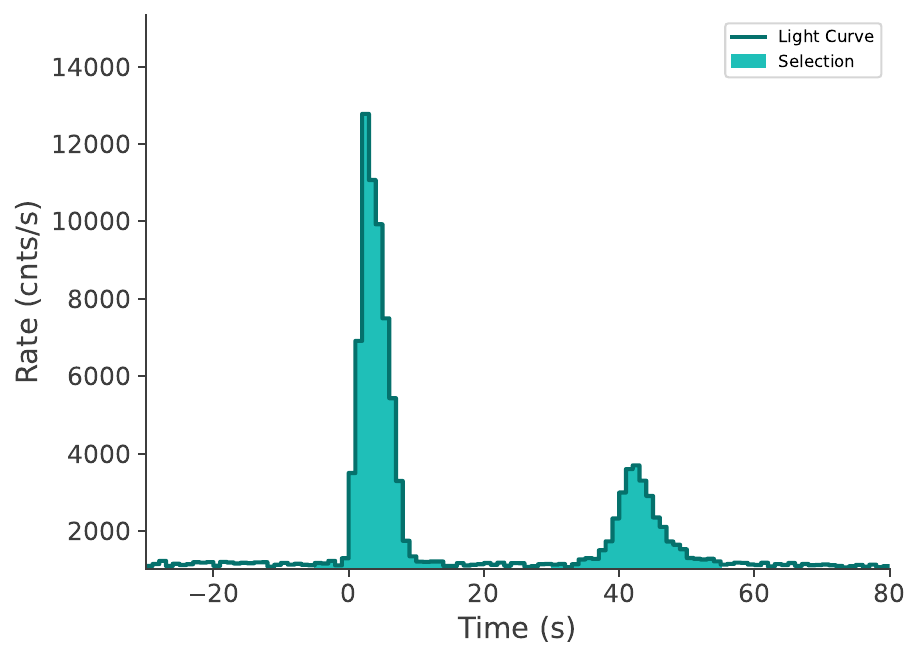}
   \figcaption{GRB with a quiescent period (GRB 081009A). \label{fig 1}}  
\end{figure}

The data used in this paper come from the Fermi satellite. The Fermi Gamma - ray Space Telescope is currently the most effective GRB - detecting satellite in orbit. It has two main instruments: the Large Area Telescope (LAT) and the Gamma - ray Burst Monitor on the Fermi (GBM). The LAT scans $20\%$ of the sky, and the GBM constantly monitors the whole sky not blocked by Earth. On average, the GBM detects 240 GRBs per year, and the LAT detects 18 GRBs per year. The GBM has fourteen detectors, each with 128 energy channels. There are 12 NaI detectors covering an effective energy range from 8 keV to 1 MeV, and two Bismuth Germanate (BGO) detectors with an effective range of 200 keV to 40 MeV. GBM observations are stored in three file formats: CTIME, CSPEC, and TTE. TTE files take up the most memory and usually record data 30 seconds before and 300 seconds after a trigger event. Compared to the other two file types, TTE files have the highest time resolution of  2$\mu s$ and the best energy resolution, which makes them perfect for spectral analysis. In this research, we apply the following two main criteria to choose GRBs with both main and second bursts from the time period between January 2008 and March 2024:

(1) We set the criterion as a flux greater than 1$ \times 10^{ - 4}$ $erg$ $cm^{-2}$ $s^{-1}$.
(2) We require that the two pulses don't overlap. There should be a quiet period between them, and the photon count of the main burst (the first pulse) should be the highest (as depicted in Figure \ref{fig 1}).

Based on these two criteria, we first visually pick out GRBs from the Fermi/GBM catalog that had a quiet period. Then, we verify that the flux of the main burst (the first pulse) is greater than that of the second burst (the second pulse). Finally, we obtain a sample of 18 GRBs that simultaneously have both a main burst and a second burst, along with a relatively long quiet period. We use TTE data and usually choose one to three NaI detectors with the strongest signal and the brightest BGO detector for the spectral analysis. 
\\

\section{Spectral Fitting Method\label{Sections3}}
\subsection{Introduction to Fitting Tools}
We utilize the Bayesian analysis package, the multi - task maximum likelihood framework (3ML; \citealt{2015arXiv150708343V}), to conduct time - integrated and time - resolved spectral analyses of the 18 GRBs. It has been successfully used for data analysis in several previous works (e.g., \citealt{2021ApJ...920...53C,2022ApJ...932...25C,2022ApJ...940...48D,2024ApJ...970...67D}). This fully Bayesian approach is commonly applied in the Fermi/GBM catalog for bright GRB spectra \citep{2019ApJS..242...16L,2019ApJ...886...20Y}.

\subsection{Background Fitting}
The background interval both before and after the radiation event is determined by using the detector with the highest photon count among the NaI detectors. This established background interval is then uniformly applied to all detectors as the standard approach for fitting the background of GRBs. When it comes to fitting each of the 128 energy channels for every detector, we make use of polynomials with orders varying from 0 to 4. After that, we perform integration on these polynomials to derive the background photon count flux as well as the flux error for each individual energy channel.

\subsection{Division of Time Bins in the Light Curve}
Time-resolved spectral analysis of GRBs serves as the primary approach for extracting valuable information from the data. \cite{2014MNRAS.445.2589B} conducted a comparison of four distinct methods for dividing time bins and reached the conclusion that the Bayesian Block (BBlock) method stands out by providing the most refined time bins. Moreover, it effectively minimizes the influence of spectral evolution that could be caused by mixed spectra. The BBlock method is characterized by several notable features: (1) every time bin obtained through this method exhibits a constant Poisson ratio; (2) each time bin has a unique width and signal-to-noise ratio  (S/N); (3) the algorithm is specifically employed to segment the GRB light curve for the purpose of time bin selection; (4) the time bin selection process carried out by the BBlock method accurately reflects the actual variability present in the data. Nevertheless, it should be noted that the BBlock method does not ensure that there are an adequate number of photons in each time bin to enable precise spectral fitting. The research also revealed that both the S/N  method and the BBlock method are among the most effective techniques for time bin division, albeit with their own inherent limitations. Traditional  S/N methods have the advantage of ensuring a sufficient number of photons for spectral fitting, but they may potentially disrupt the physical structure of the data. On the other hand, the BBlock method, while capable of providing refined time bins, cannot guarantee that there is sufficient signal strength in each time bin to support accurate spectral fitting.

Therefore, we initiate the process by utilizing the BBlock method to partition the data into time bins. Following this, we compute the S/N for each individual time bin, thereby obtaining the S/N values for all the time bins generated. Subsequently, we carefully select the time bins that exhibit the optimal S/N values. Then, we implement the BBlock method with a false alarm probability set at $p = 0.01$ \citep{2013ApJ...764..167S} on the  TTE light curve of the brightest NaI detector. After that, the time bin thus determined is uniformly applied to the data from other detectors as well. This ensures consistency in the time binning across different detectors for a more comprehensive and accurate analysis of the GRB data.

\subsection{Spectral Models and Selection of the Best Model}
Spectral analysis is a crucial tool that can offer valuable insights into the radiation mechanisms, energy dissipation processes, and the composition of the outflows associated with GRBs. Observations of GRB prompt emission spectra are often fit using three empirical components: (1) a non-thermal component, often described by the Band function; (2) a quasi-thermal component; and (3) a non-thermal component that extends to higher energies \citep{2011ApJ...730..141Z}.
The majority of GRB spectra have been found to be well-represented by non-thermal empirical models \citep{1993ApJ...413..281B}. In light of this, we begin our analysis by considering two well-established empirical models: the Band function, which is a smooth inflection power-law function, and the cut-off power-law function (CPL) \citep{1993ApJ...413..281B,2011MNRAS.411.1323G}. The specific details of these two empirical models employed in this paper are as follows:

The Band function is:
\begin{equation}
{N_{B{\rm{and}}}}\left( E \right) = A\left\{ \begin{array}{l}
\left( {\frac{E}{{100keV}}} \right)\exp \left( { - \frac{E}{{{E_0}}}} \right),E < \left( {\alpha  - \beta } \right){E_0}\\
{\left( {\frac{{\left( {\alpha  - \beta } \right){E_0}}}{{100keV}}} \right)^{\alpha  - \beta }}\exp \left( {\beta  - \alpha } \right){\left( {\frac{E}{{100keV}}} \right)^\beta },E > \left( {\alpha  - \beta } \right){E_0}
\end{array} \right.
\end{equation}
where A is the normalization constant in units of photons ${s^{ - 1}}c{m^{ - 2}}ke{V^{ - 1}}$, $\alpha$ and $\beta$ are the low-energy and high-energy spectral indices of the photon spectrum, respectively, and $E_{0}$ is the break energy in keV, and $E_{p} = ( 2 + \alpha ) E_{0}$ is the peak energy in units of keV.

The CPL Function is:
\begin{equation}
{N_{CPL}}\left( E \right) = A{\left( {\frac{E}{{100kev}}} \right)^\alpha }\exp \left( { - \frac{\left( {\alpha + 2} \right)E}{{{E_p}}}} \right)
\end{equation}
where A is the normalization factor in units of photons ${s^{ - 1}}c{m^{ - 2}}ke{V^{ - 1}}$, $\alpha$ is the low-energy spectral index and ${E_p}$ is the peak energy in keV.

Some GRBs exhibit additional thermal components, which are typically modelled using the Planck blackbody (BB) function. The BB function is defined as follows:
\begin{equation}
{N_{BB}}\left( E \right) = A\frac{{{E^2}}}{{\exp \left[ {E/kT} \right] - 1}}
\end{equation}
where A is the normalization, k is the Boltzmann constant, and kT is the blackbody temperature.

In this study, we initially apply the Band function and the CPL function to fit the time-resolved spectra of each GRB. First, we compare the fitting results to identify whether the Band or CPL function is the better model for each time bin. Subsequently, we add the BB component on this basis, resulting in Band + BB or CPL + BB models. Then, we re-analyze the data to determine the optimal model among these new combinations. Finally, we compare the best-fit model without the BB component (the best model) and the best-fit model with the BB component (the best model BB) to ascertain the presence of a thermal component.

In this paper, we evaluate the quality of the fitting models using the DIC, defined as follows::
\begin{equation}
{DIC} = -2 \log \left[ {P(\text{data}\mid \vec{\theta})}\right] + 2{P_{DIC}}
\end{equation}
Here, $\vec{\theta}$ represents the posterior mean of the model, and $P_{DIC}$ denotes the effective number of parameters. When we use various models to fit the identical set of data points, a lower DIC value serves as an indication that the corresponding model provides a better fit,  demonstrating its superiority in representing the data compared to models with higher DIC values.

\section{Spectral Analysis\label{Sections4}}
\subsection{Thermal Component Analysis}
As an example, consider GRB 081009A. Using the 3ML fitting tool in combination with the BBlocks method, GRB 081009A is partitioned into 23 time bins. The first pulse (1st) encompasses 15 of these time bins, while the second pulse (2nd) is composed of 8 time bins.

\begin{figure}[htbp]
\centering
\includegraphics [width=7.5cm,height=6cm]{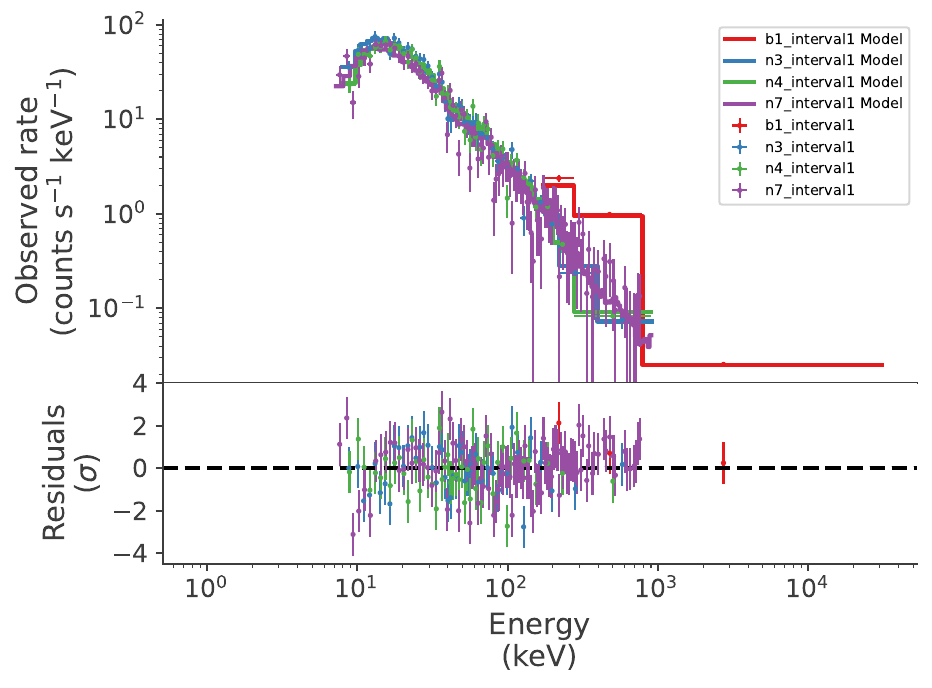}
\includegraphics [width=7.5cm,height=6cm]{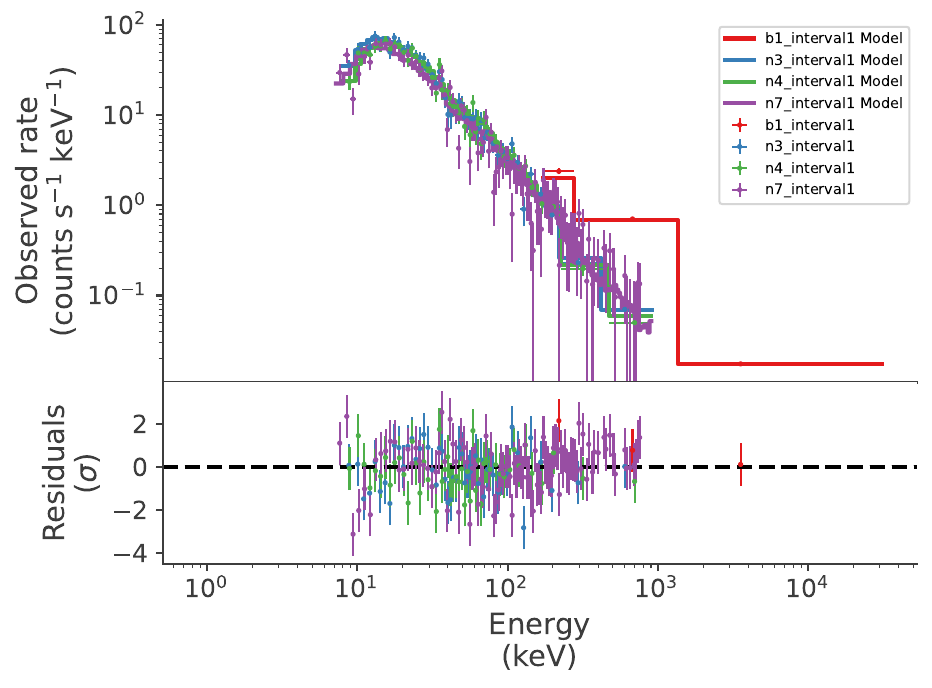}
\includegraphics [width=7.5cm,height=6cm]{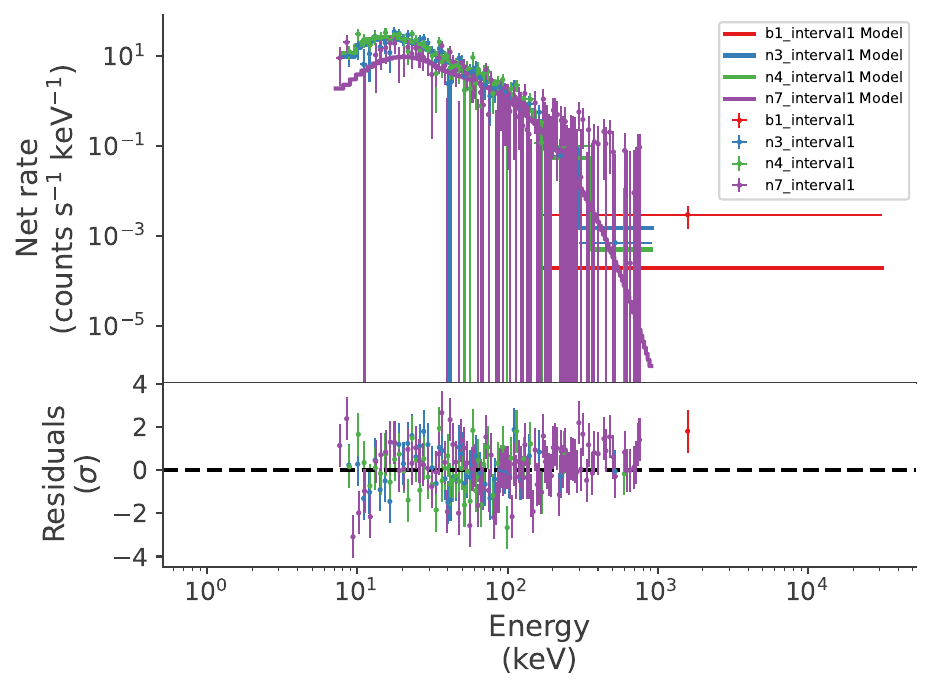}
\includegraphics [width=7.5cm,height=6cm]{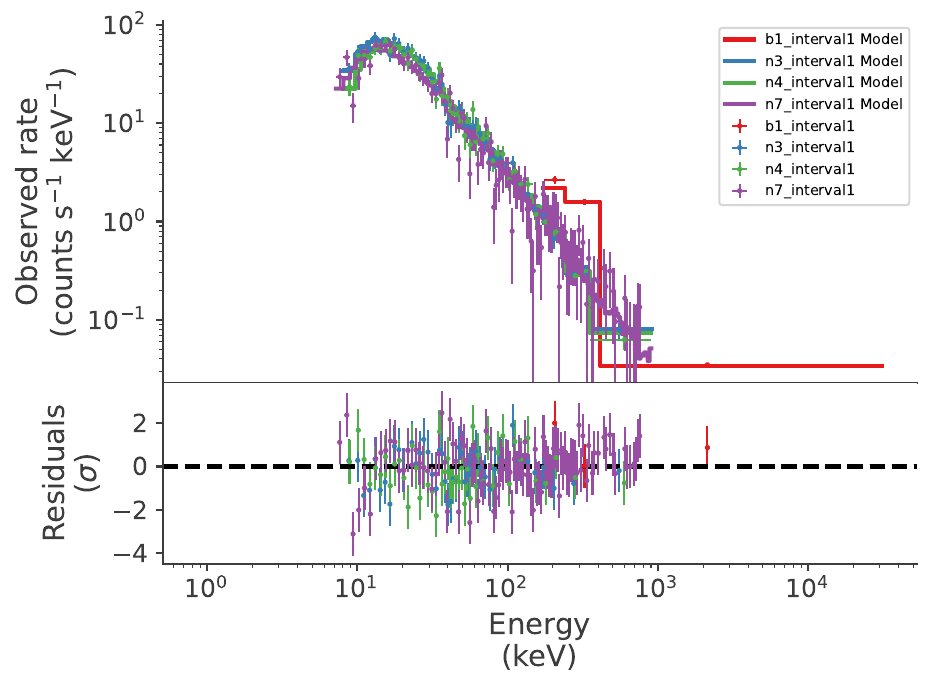}
   \figcaption{The photon count spectrum associated with the initial time bin of GRB 081009A. In the upper-left quadrant of the visual display, the fit of the Band model is presented. The upper-right quadrant showcases the fit of the Band + BB model. In the lower - left area, the fit of the CPL model is depicted, and the lower - right section reveals the fit of the CPL + BB model. \label{fig 2}}

\end{figure}

We initially fit the spectrum with the Band and CPL models. The fitting results are presented in Table \ref{table1}. For the first time bin, spanning from $-0.4-0.3$ s (with its photon  count spectrum shown in Figure \ref{fig 2}), the DIC value for the Band model is 1858.04, and for the CPL model, it is 1899.82. The difference $\Delta DIC = DIC_{Band} - DIC_{CPL} $ is  $-$41.78. This large negative difference provides very strong evidence that the Band model is the optimal choice for the first time slice.

Subsequently, we incorporate thermal components, fitting the spectrum with Band + BB and CPL + BB models. The DIC for the Band + BB model is $1855.10$, and for the CPL + BB model, it is $1853.10$. Here, $\Delta DIC = DIC_{Band+BB} - DIC_{CPL+BB} = 2$. Based on the goodness-of-fit, the performances of Band + BB and CPL + BB are indistinguishable. In this case, we select CPL + BB, which has the smaller DIC, as the better model.

Finally, by comparing the DIC values of the two best models (before and after adding the thermal component), we can determine the presence of a thermal component. Here, $\Delta DIC_{best} = DIC_{best1} - DIC_{best2} $ is 4.94. Since $\Delta DIC_{best} < 10$, it indicates that there is no thermal component in this time bin. (Conversely, if $\Delta DIC_{best} > 10$, it would suggest the presence of a thermal component in the spectrum.)

\begin{center}
\begin{longtable}{ccccccccc}
\caption{Time-resolved spectral fitting result of GRB 081009A}
\label{table1}\\
\hline
\hline
$t_{start}-t_{end}$ & $S$ & Model & $\alpha$ & $\beta$ & $E_p/E_c$ & kT & DIC & pgstat\\
 s& & & & &$keV$ &$keV$&  \\
(1)&(2)&(3)&(4)&(5)&(6)&(7)&(8)&(9)\\
\hline
\endfirsthead
\hline
\hline
$t_{start}-t_{end}$ & $S$ & Model & $\alpha$ & $\beta$ & $E_p/E_c$ & kT & DIC & pgstat\\
 s& & & & &$keV$&$keV$&  \\
(1)&(2)&(3)&(4)&(5)&(6)&(7)&(8)&(9)\\
\hline
\endhead
\hline
\endfoot
\multirow{4}{*}{-0.4-0.3} & \multirow{4}{*}{10} & band & $-1.1^{+0.42}_{-0.13}$ & $-2.24^{+0.11}_{-0.17}$ & $43.5^{+7.31}_{-8.15}$ & ... & 1858.04 & 3.95 \\
 &  & cpl & $-1.48^{+0.17}_{-0.14}$ & ... & $66.27^{+14.42}_{-6.95}$ & ... & 1899.82 & 3.95 \\
 &  & bandbb & $-1.05^{+0.37}_{-0.16}$ & $-2.27^{+0.09}_{-0.28}$ & $41.32^{+6.04}_{-6.66}$ & $42.78^{+21.21}_{-15.97}$ & 1855.1 & 3.97 \\
 &  & cplbb & $-1.09^{+0.24}_{-0.29}$ & ... & $42.95^{+13.17}_{-4.08}$ & $81.48^{+2.67}_{-32.33}$ & 1853.15 & 3.96 \\
\multirow{4}{*}{0.3-1.22} & \multirow{4}{*}{25} & band & $-1.01^{+0.2}_{-0.1}$ & $-2.8^{+0.07}_{-0.07}$ & $25.01^{+0.82}_{-1.09}$ & ... & 2269.58 & 4.74 \\
 &  & cpl & $-1.52^{+0.1}_{-0.07}$ & ... & $25.29^{+2.0}_{-1.57}$ & ... & 2308.1 & 4.8 \\
 &  & bandbb & $-1.01^{+0.09}_{-0.12}$ & $-2.87^{+0.1}_{-0.1}$ & $24.66^{+1.0}_{-0.84}$ & $64.22^{+0.72}_{-39.03}$ & 2266.4 & 4.76 \\
 &  & cplbb & $-1.28^{+0.17}_{-0.04}$ & ... & $25.41^{+1.31}_{-0.95}$ & $90.58^{+2.33}_{-25.87}$ & 2267.19 & 4.77 \\
\multirow{4}{*}{1.22-1.52} & \multirow{4}{*}{24} & band & $-1.16^{+0.12}_{-0.06}$ & $-3.5^{+0.15}_{-0.31}$ & $38.94^{+1.38}_{-1.44}$ & ... & 1137.44 & 2.33 \\
 &  & cpl & $-1.15^{+0.1}_{-0.07}$ & ... & $39.5^{+1.56}_{-1.29}$ & ... & 1122.0 & 2.31 \\
 &  & bandbb & $-1.19^{+0.14}_{-0.05}$ & $-3.58^{+0.24}_{-0.2}$ & $38.53^{+1.66}_{-1.42}$ & $35.23^{+14.37}_{-13.2}$ & 1125.44 & 2.33 \\
 &  & cplbb & $-1.16^{+0.11}_{-0.06}$ & ... & $39.6^{+1.17}_{-1.72}$ & $42.32^{+14.62}_{-18.13}$ & 1109.32 & 2.32 \\
\multirow{4}{*}{1.52-1.92} & \multirow{4}{*}{39} & band & $-0.86^{+0.02}_{-0.09}$ & $-3.8^{+0.22}_{-0.16}$ & $47.75^{+0.82}_{-0.07}$ & ... & 1518.56 & 3.11 \\
 &  & cpl & $-0.83^{+0.08}_{-0.06}$ & ... & $48.43^{+0.93}_{-1.08}$ & ... & 1499.99 & 3.09 \\
 &  & bandbb & $-0.79^{+0.06}_{-0.08}$ & $-3.67^{+0.09}_{-0.34}$ & $47.15^{+1.19}_{-0.9}$ & $39.87^{+6.73}_{-18.59}$ & 1506.77 & 3.13 \\
 &  & cplbb & $-0.83^{+0.04}_{-0.08}$ & ... & $48.34^{+0.81}_{-1.18}$ & $50.89^{+10.06}_{-26.03}$ & 1486.24 & 3.11 \\
\multirow{4}{*}{1.92-2.71} & \multirow{4}{*}{83} & band & $-0.87^{+0.03}_{-0.03}$ & $-3.94^{+0.13}_{-0.26}$ & $68.54^{+0.79}_{-1.04}$ & ... & 2501.0 & 5.17 \\
 &  & cpl & $-0.88^{+0.03}_{-0.03}$ & ... & $68.98^{+1.01}_{-0.81}$ & ... & 2478.94 & 5.14 \\
 &  & bandbb & $-0.87^{+0.03}_{-0.03}$ & $-3.97^{+0.15}_{-0.25}$ & $68.5^{+0.56}_{-1.26}$ & $32.59^{+16.09}_{-9.71}$ & 2488.77 & 5.19 \\
 &  & cplbb & $-0.88^{+0.03}_{-0.03}$ & ... & $69.15^{+0.55}_{-1.36}$ & $37.93^{+16.5}_{-14.58}$ & 2467.0 & 5.17 \\
\multirow{4}{*}{2.71-2.95} & \multirow{4}{*}{57} & band & $-0.85^{+0.04}_{-0.05}$ & $-3.6^{+0.2}_{-0.27}$ & $85.6^{+2.65}_{-1.75}$ & ... & 1236.73 & 2.52 \\
 &  & cpl & $-0.87^{+0.05}_{-0.04}$ & ... & $87.78^{+1.83}_{-2.17}$ & ... & 1219.6 & 2.5 \\
 &  & bandbb & $-0.84^{+0.03}_{-0.06}$ & $-3.57^{+0.17}_{-0.27}$ & $85.96^{+1.94}_{-2.68}$ & $47.69^{+12.1}_{-21.46}$ & 1224.19 & 2.53 \\
 &  & cplbb & $-0.87^{+0.05}_{-0.04}$ & ... & $87.69^{+1.46}_{-3.12}$ & $55.44^{+12.65}_{-27.47}$ & 1207.53 & 2.52 \\
\multirow{4}{*}{2.95-3.27} & \multirow{4}{*}{55} & band & $-0.8^{+0.01}_{-0.08}$ & $-3.7^{+0.14}_{-0.28}$ & $57.52^{+1.77}_{-0.03}$ & ... & 1386.39 & 2.83 \\
 &  & cpl & $-0.83^{+0.05}_{-0.05}$ & ... & $58.82^{+1.19}_{-0.86}$ & ... & 1370.73 & 2.82 \\
 &  & bandbb & $-0.79^{+0.05}_{-0.05}$ & $-3.66^{+0.14}_{-0.26}$ & $57.36^{+0.93}_{-1.36}$ & $53.44^{+2.48}_{-26.73}$ & 1373.91 & 2.84 \\
 &  & cplbb & $-0.8^{+0.03}_{-0.07}$ & ... & $58.16^{+1.47}_{-1.04}$ & $79.73^{+2.65}_{-49.16}$ & 1357.09 & 2.83 \\
\multirow{4}{*}{3.27-4.68} & \multirow{4}{*}{93} & band & $-0.77^{+0.04}_{-0.02}$ & $-4.08^{+0.13}_{-0.19}$ & $42.65^{+0.42}_{-0.37}$ & ... & 3298.3 & 6.83 \\
 &  & cpl & $-0.82^{+0.02}_{-0.03}$ & ... & $43.33^{+0.39}_{-0.34}$ & ... & 3276.07 & 6.81 \\
 &  & bandbb & $-0.76^{+0.02}_{-0.04}$ & $-4.08^{+0.09}_{-0.26}$ & $42.57^{+0.39}_{-0.43}$ & $44.14^{+0.59}_{-20.97}$ & 3285.64 & 6.86 \\
 &  & cplbb & $-0.79^{+0.02}_{-0.05}$ & ... & $42.75^{+0.59}_{-0.45}$ & $55.06^{+6.92}_{-17.98}$ & 3260.71 & 6.83 \\
\multirow{4}{*}{4.68-5.5} & \multirow{4}{*}{62} & band & $-0.95^{+0.04}_{-0.04}$ & $-3.72^{+0.11}_{-0.24}$ & $55.17^{+1.03}_{-0.78}$ & ... & 2432.94 & 5.03 \\
 &  & cpl & $-0.98^{+0.05}_{-0.03}$ & ... & $56.21^{+0.75}_{-0.89}$ & ... & 2418.06 & 5.02 \\
 &  & bandbb & $-0.96^{+0.05}_{-0.03}$ & $-3.75^{+0.13}_{-0.23}$ & $55.03^{+0.96}_{-0.84}$ & $40.13^{+6.25}_{-19.06}$ & 2421.06 & 5.06 \\
 &  & cplbb & $-0.97^{+0.05}_{-0.03}$ & ... & $55.57^{+0.92}_{-1.06}$ & $52.83^{+13.3}_{-27.94}$ & 2405.36 & 5.04 \\
\multirow{4}{*}{5.5-6.64} & \multirow{4}{*}{49} & band & $-0.6^{+0.06}_{-0.05}$ & $-4.27^{+0.19}_{-0.22}$ & $39.45^{+0.52}_{-0.6}$ & ... & 2765.91 & 5.71 \\
 &  & cpl & $-0.66^{+0.0}_{-0.09}$ & ... & $39.86^{+0.38}_{-0.49}$ & ... & 2738.36 & 5.68 \\
 &  & bandbb & $-0.61^{+0.04}_{-0.06}$ & $-4.19^{+0.13}_{-0.28}$ & $39.41^{+0.48}_{-0.6}$ & $30.27^{+7.99}_{-11.89}$ & 2753.35 & 5.73 \\
 &  & cplbb & $-0.61^{+0.07}_{-0.04}$ & ... & $39.69^{+0.47}_{-0.62}$ & $46.81^{+6.89}_{-25.02}$ & 2723.55 & 5.7 \\
\multirow{4}{*}{6.64-7.24} & \multirow{4}{*}{26} & band & $-0.97^{+0.02}_{-0.14}$ & $-4.02^{+0.28}_{-0.14}$ & $23.69^{+0.0}_{-1.37}$ & ... & 1792.88 & 3.68 \\
 &  & cpl & $-0.95^{+0.07}_{-0.09}$ & ... & $24.16^{+0.7}_{-0.83}$ & ... & 1767.02 & 3.65 \\
 &  & bandbb & $-1.0^{+0.05}_{-0.11}$ & $-3.96^{+0.2}_{-0.19}$ & $23.12^{+0.63}_{-0.82}$ & $33.75^{+5.6}_{-14.21}$ & 1781.03 & 3.7 \\
 &  & cplbb & $-0.94^{+0.03}_{-0.12}$ & ... & $24.31^{+0.39}_{-1.26}$ & $40.58^{+1.7}_{-20.79}$ & 1755.32 & 3.67 \\
\multirow{4}{*}{7.24-7.79} & \multirow{4}{*}{12} & band & $-1.12^{+0.03}_{-0.17}$ & $-3.82^{+0.23}_{-0.17}$ & $13.63^{+0.01}_{-1.59}$ & ... & 1629.35 & 3.34 \\
 &  & cpl & $-1.11^{+0.16}_{-0.12}$ & ... & $14.01^{+1.12}_{-1.0}$ & ... & 1606.64 & 3.32 \\
 &  & bandbb & $-1.03^{+0.05}_{-0.18}$ & $-3.86^{+0.26}_{-0.15}$ & $13.96^{+0.54}_{-1.22}$ & $30.97^{+10.5}_{-12.36}$ & 1615.24 & 3.35 \\
 &  & cplbb & $-1.08^{+0.02}_{-0.25}$ & ... & $14.09^{+0.27}_{-2.02}$ & $44.27^{+32.76}_{-14.85}$ & 1595.4 & 3.34 \\
\multirow{4}{*}{7.79-8.82} & \multirow{4}{*}{6} & band & $-1.4^{+0.15}_{-0.01}$ & $-3.72^{+0.23}_{-0.22}$ & $7.22^{+1.37}_{-0.27}$ & ... & 2379.71 & 4.92 \\
 &  & cpl & $-1.15^{+0.18}_{-0.11}$ & ... & $9.28^{+1.92}_{-0.24}$ & ... & 2366.24 & 4.9 \\
 &  & bandbb & $-1.14^{+0.14}_{-0.03}$ & $-3.65^{+0.06}_{-0.32}$ & $8.69^{+1.4}_{-0.18}$ & $29.12^{+12.41}_{-4.99}$ & 2370.63 & 4.95 \\
 &  & cplbb & $-1.06^{+0.2}_{-0.01}$ & ... & $9.43^{+1.29}_{-0.14}$ & $30.35^{+10.39}_{-7.98}$ & 2350.31 & 4.92 \\
\hline
\hline
\multirow{4}{*}{38.08-39.78} & \multirow{4}{*}{11} & band & $-1.09^{+0.24}_{-0.12}$ & $-3.73^{+0.15}_{-0.28}$ & $15.09^{+1.46}_{-0.93}$ & ... & 3196.36 & 6.67 \\
 &  & cpl & $-1.12^{+0.25}_{-0.05}$ & ... & $15.42^{+1.83}_{-0.4}$ & ... & 3181.59 & 6.64 \\
 &  & bandbb & $-1.11^{+0.22}_{-0.09}$ & $-3.79^{+0.23}_{-0.18}$ & $14.98^{+1.42}_{-0.7}$ & $23.03^{+6.58}_{-7.78}$ & 3189.67 & 6.7 \\
 &  & cplbb & $-1.14^{+0.32}_{-0.05}$ & ... & $15.1^{+2.14}_{-0.1}$ & $28.34^{+11.5}_{-8.28}$ & 3171.7 & 6.67 \\
\multirow{4}{*}{39.78-40.48} & \multirow{4}{*}{15} & band & $-1.1^{+0.22}_{-0.13}$ & $-3.47^{+0.18}_{-0.24}$ & $17.69^{+1.47}_{-1.23}$ & ... & 1892.06 & 3.92 \\
 &  & cpl & $-1.16^{+0.12}_{-0.1}$ & ... & $17.95^{+1.57}_{-1.14}$ & ... & 1885.28 & 3.91 \\
 &  & bandbb & $-1.04^{+0.12}_{-0.15}$ & $-3.49^{+0.18}_{-0.23}$ & $17.63^{+1.1}_{-1.04}$ & $30.39^{+18.52}_{-10.2}$ & 1883.84 & 3.94 \\
 &  & cplbb & $-1.13^{+0.26}_{-0.06}$ & ... & $17.87^{+1.94}_{-0.54}$ & $59.9^{+0.78}_{-32.13}$ & 1867.97 & 3.93 \\
\multirow{4}{*}{40.48-43.21} & \multirow{4}{*}{39} & band & $-0.97^{+0.02}_{-0.09}$ & $-3.99^{+0.08}_{-0.25}$ & $16.87^{+0.31}_{-0.57}$ & ... & 3941.37 & 8.19 \\
 &  & cpl & $-1.13^{+0.1}_{-0.04}$ & ... & $16.38^{+0.78}_{-0.36}$ & ... & 3917.82 & 8.17 \\
 &  & bandbb & $-0.98^{+0.09}_{-0.08}$ & $-4.03^{+0.13}_{-0.2}$ & $16.79^{+0.5}_{-0.5}$ & $22.43^{+9.5}_{-6.77}$ & 3927.51 & 8.23 \\
 &  & cplbb & $-1.12^{+0.08}_{-0.05}$ & ... & $16.33^{+0.72}_{-0.38}$ & $34.97^{+5.31}_{-15.95}$ & 3904.86 & 8.2 \\
\multirow{4}{*}{43.21-45.06} & \multirow{4}{*}{26} & band & $-1.25^{+0.01}_{-0.0}$ & $-2.93^{+0.02}_{-0.01}$ & $13.23^{+0.07}_{-0.02}$ & ... & 3508.9 & 7.31 \\
 &  & cpl & $-1.35^{+0.17}_{-0.07}$ & ... & $10.9^{+1.4}_{-0.73}$ & ... & 3314.55 & 6.91 \\
 &  & bandbb & $-1.23^{+0.03}_{-0.16}$ & $-3.8^{+0.08}_{-0.21}$ & $11.36^{+0.37}_{-1.35}$ & $29.41^{+0.01}_{-13.24}$ & 3320.74 & 6.95 \\
 &  & cplbb & $-1.29^{+0.17}_{-0.05}$ & ... & $11.17^{+1.28}_{-0.36}$ & $40.69^{+2.19}_{-16.82}$ & 3304.21 & 6.94 \\
\multirow{4}{*}{45.06-46.87} & \multirow{4}{*}{15} & band & $-0.94^{+0.0}_{-0.0}$ & $-3.46^{+0.08}_{-0.03}$ & $11.75^{+0.0}_{-0.08}$ & ... & 3319.27 & 6.9 \\
 &  & cpl & $-1.6^{+0.14}_{-0.0}$ & ... & $6.01^{+1.54}_{-0.14}$ & ... & 3187.13 & 6.62 \\
 &  & bandbb & $-1.0^{+0.01}_{-0.02}$ & $-3.72^{+0.08}_{-0.15}$ & $8.71^{+0.13}_{-0.01}$ & $66.02^{+2.0}_{-0.2}$ & 3176.71 & 6.64 \\
 &  & cplbb & $-1.52^{+0.27}_{-0.06}$ & ... & $5.97^{+2.14}_{-0.52}$ & $45.58^{+6.81}_{-14.27}$ & 3160.73 & 6.63 \\
\multirow{4}{*}{46.87-51.75} & \multirow{4}{*}{7} & band & $-0.65^{+0.07}_{-0.02}$ & $-3.05^{+0.01}_{-0.05}$ & $9.29^{+0.91}_{-0.21}$ & ... & 4638.12 & 9.67 \\
 &  & cpl & $-0.93^{+0.25}_{-0.01}$ & ... & $9.65^{+0.74}_{-0.15}$ & ... & 4610.15 & 9.62 \\
 &  & bandbb & $-0.75^{+0.15}_{-0.02}$ & $-3.49^{+0.06}_{-0.14}$ & $8.35^{+0.97}_{-0.03}$ & $39.84^{+11.6}_{-9.26}$ & 4599.84 & 9.65 \\
 &  & cplbb & $-1.29^{+0.34}_{-0.1}$ & ... & $6.9^{+1.58}_{-0.38}$ & $36.47^{+3.58}_{-13.29}$ & 4587.13 & 9.63 \\
\hline
\hline
\end{longtable}
\end{center}

\begin{figure}[htbp]
\centering
\includegraphics [width=7.5cm,height=6cm]{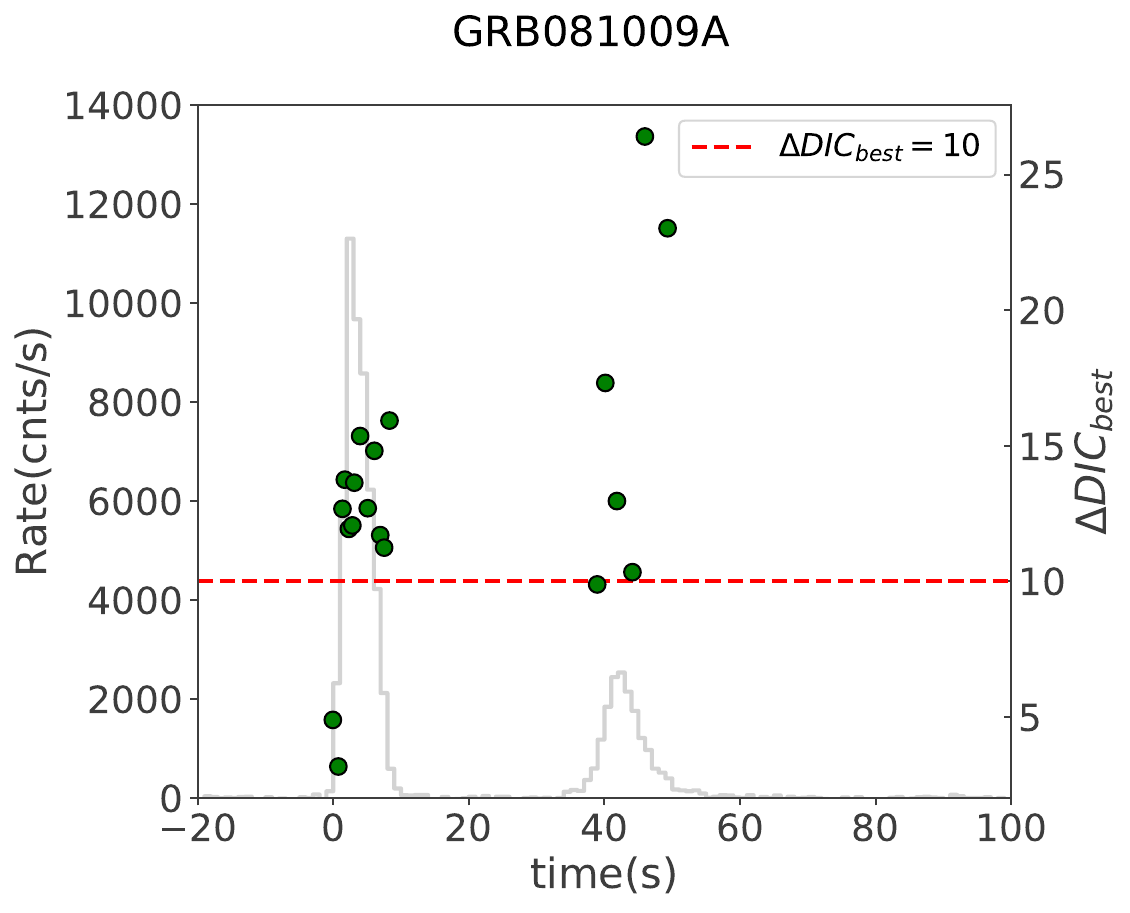}
   \figcaption{The evolution of $\Delta DIC_{best}$ over time. The red dotted line indicates $\Delta DIC_{best} = 10$. \label{fig 3}}

\end{figure}

We also perform the same analysis of the thermal component for other time-resolved spectra adopt the similar criterion. It can be clearly observed in Figure \ref{fig 3} that within the main burst, 11 time bins exhibit $\Delta$DIC values greater than 10. While in the second burst, 5 time bins have $\Delta$DIC values exceeding 10. This indicates the presence of thermal components in both the main burst and the second burst of GRB 081009A. Employing the same approach, we carry out a thermal-component analysis on the other 17 GRBs. The results are presented in Table \ref{table2}. From Table \ref{table2}, it is evident that thermal components are detected in both the main bursts and second bursts of all 18 GRBs. Among which three bursts (GRB 170510A, GRB 210202A, and GRB 221119A) have a thermal radiation component in all selected time bins of both the main burst and the second burst. When considering the proportion change of thermal components from the main burst to the second burst, $67\%$ (12 out of 18) of the GRBs show a gradual decrease. $28\%$ (5 out of 18) show no change, and $5\%$ (1 out of 18) of the GRBs (GRB 131108A) show an increase.
\\

\begin{center}
\begin{longtable}{cccccccc}
\caption{Thermal Component Analysis Results}
\label{table2}\\
\hline
\hline
\multirow{2}*{GRB}&\multicolumn{2}{c}{the time period}&\multirow{2}*{Selected probe}&Fluence&\multirow{2}*{$T_{90}$}&\multicolumn{2}{c}{Proportion of thermal components}\\ 
\cline{2-3}
\cline{7-8}
 &main bursts&second bursts& &$(\times10^{ - 5}erg^{-1}s^{-1}cm^{-2})$& &main bursts&second bursts \\
\hline
\endhead

GRB081009A & \(-5 \sim 15s\) & \(30 \sim 55s\) & $n4,n3,n7,b1$ & 3.83 & 41.35s & $84.62\%(11/13)$ & $83.33\%(5/6)$ \\
GRB100719C & \(0 \sim 10s\) & \(15 \sim 30s\) & n5,n4,b0 & 5.19 & 21.83s & 100.00\%(7/7) & 50.00\%(2/4) \\
GRB111228B & \(-1 \sim 15s\) & \(40 \sim 60s\) & $n7,n8,nb,b1$ & 1.81 & 99.84s & $71.43\%(5/7)$ & $60.00\%(3/5)$ \\
GRB120412B & \(-2 \sim 10s\) & \(68 \sim 98s\) & n5,n1,n2,b0 & 0.703 & 101.18s & 100.00\%(5/5) & 75.00\%(3/4) \\
GRB130404B & \(-1 \sim 10s\) & \(-21 \sim 38s\) & n6,n7,b1 & 0.836 & 34.56s & 71.43\%(5/7) & 42.86\%(3/7) \\
GRB131108A & \(-2 \sim 15s\) & \(228 \sim 275s\) & n8,n7,nb,b1 & 0.282 & 14.59s & 83.33\%(5/6) & 100.00\%(3/3) \\
GRB140108A & \(-2 \sim 13s\) & \(71 \sim 101s\) & n2,n1,n9,b1 & 1.97 & 91.39s & 85.71\%(6/7) & 85.71\%(6/7) \\
GRB150220A & \(-2 \sim 25s\) & \(120 \sim 150s\) & na,nb,n9,b1 & 2.41 & 144.64s & 90.00\%(9/10) & 75.00\%(3/4) \\
GRB160802A & \(-5 \sim 10s\) & \(13 \sim 21s\) & n2,n1,na,b0 & 6.84 & 16.38s & 100.00\%(7/7) & 75.00\%(3/4) \\
GRB170510A & \(0 \sim 50s\) & \(100 \sim 140s\) & n9,na,b1 & 4.39 & 127.75s & 100.00\%(10/10) & 100.00\%(6/6) \\
GRB171120A & \(-1 \sim 5s\) & \(35 \sim 50s\) & n3,n1,n0,b0 & 1.61 & 44.06s & 86.67\%(13/15) & 75.00\%(6/8) \\
GRB180612A & \(-1 \sim 23s\) & \(80 \sim 120s\) & n2,n5,n1,b1 & 1.55 & 101.12s & 90.00\%(9/10) & 75.00\%(3/4) \\
GRB190901A & \(0 \sim 30s\) & \(150 \sim 225s\) & n4,n5,n1,b0 & 6.30 & 473.10s & 88.89\%(8/9) & 75.00\%(3/4) \\
GRB210202A & \(-1 \sim 10s\) & \(10 \sim 20s\) & n1,n2,n0,b0 & 1.55 & 17.66s & 100.00\%(6/6) & 100.00\%(3/3) \\
GRB220927A & \(-1 \sim 9s\) &\(32 \sim 48s\) & n9,n1,na,b1 & 0.848 & 76.80s & 100.00\%(6/6) & 66.67\%(2/3) \\
GRB221119A &\(0 \sim 11s\) & \(39 \sim 57s\) & n4,n3,n8,b0 & 3.52 & 66.05s & 100.00\%(9/9) & 100.00\%(7/7) \\
GRB231104A & \(0 \sim 22s\) & \(28 \sim 40s\) & n9,n6,na,b1 & 5.12 & 37.12s & 100.00\%(12/12) & 50.00\%(2/4) \\
GRB240229A & \(0 \sim 17s\) & \(23 \sim 31s\) & n9,na,nb,b1 & 2.09 & 28.67s & 100.00\%(6/6) & 83.33\%(5/6) \\
\hline
\hline
\end{longtable}
\end{center}

\subsection{The Evolution of Spectral Parameters.}
Previous research has indicated that the evolution patterns of spectral parameters ($E_{p}$ and $\alpha$) in GRBs display two prominent characteristics: the hard-to-soft (h.t.s.) and flux-tracking (f.t.) behaviors. \cite{2019ApJS..242...16L} and \cite{2016A&A...588A.135Y} identified a "tracking" trend for the parameter $E_{p}$. This implies that the variations in spectral parameters are correlated with the increase and decrease of the flux, rather than being directly associated with the rise and decay of the pulses. Additionally, there are other evolution patterns such as the soft-to-hard (s.t.h) and hybrid types. In this study, we conduct a comprehensive statistical analysis of the spectral parameters for 18 GRBs that possess a second burst, as tabulated in Table \ref{table3}. From an examination of Table \ref{table3}, it is evident that, in terms of the evolution of $\alpha$ and $E_{p}$, both the main bursts and the second bursts demonstrate comparable behaviors. We illustrate the temporal evolution of the spectral parameters ($\alpha$, $E_{p}$) for these 18 GRBs in Figures \ref{fig 10} and \ref{fig 11} of the Appendix.

Regarding the evolution of the $\alpha$ (refer to Table \ref{table3}, columns 2 and 4): During the main burst phase, six GRBs display a "hard-to-soft-to-hard" (h.t.s.t.h) mode of evolution. Seven GRBs show a f.t. mode, where the changes in $\alpha$ are closely related to the variations in the flux. One GRB exhibits a "hard-to-soft-to-flux-tracking" (h.t.s.t.f.t) mode, and four GRBs do not show any distinct or obvious evolution pattern. In the second burst phase, nine GRBs follow the "h.t.s.t.h" mode. Five GRBs display the "f.t." mode, and four GRBs lack an identifiable or clear evolution mode for the $\alpha$.

Concerning the evolution of the $E_{p}$ (refer to Table \ref{table3}, columns 3 and 5): In the main burst phase, sixteen GRBs demonstrate a f.t. mode, indicating that the changes in $E_{p}$ are in tandem with the flux variations. One GRB shows a "h.t.s.t.f.t" mode, and another GRB exhibits an "h.t.s.t.h" mode. During the second burst phase, sixteen GRBs exhibit the "f.t." mode. One GRB shows a "flux-tracking-to-soft-to-hard" (f.t.t.s.t.h) mode, and one GRB displays an "h.t.s.t.h" mode. This detailed analysis of the evolution of $\alpha$ and $E_{p}$ provides valuable insights into the behavior of the spectral parameters in the main and second bursts of these 18 GRBs.

\begin{center}
\begin{longtable}{ccc|cc|ccc|ccc}
\caption{Evolution Pattern of Spectral Parameters and Correlation Coefficient Statistics}
\label{table3}\\
\hline
\hline
\multirow{3}*{GRB}&\multicolumn{4}{c|}{Evolution Pattern of  Spectral Parameters}&\multicolumn{6}{c}{Correlation Coefficient Between Spectral Parameters}\\ 
\cline{2-5}
\cline{6-11}
 &\multicolumn{2}{c|}{main bursts}&\multicolumn{2}{c|}{second bursts}&\multicolumn{3}{c|}{main bursts}&\multicolumn{3}{c}{second bursts}\\
\cline{2-3}
\cline{4-5}
\cline{6-8}
\cline{9-11}
 &$\alpha$&$E_{p}$&$\alpha$ &$E_{p}$&$\alpha-F$ &$\alpha-E_{p}$&$E_{p}-F$&$\alpha-F$ &$\alpha-E_{p}$&$E_{p}-F$\\
\hline
\endhead
GRB081009A & f.t & f.t & h.t.s.t.h & f.t.t.s.t.h & 0.59 & 0.45 & 0.85 & 0.37 & 0.89 & 0.49 \\
GRB100719C & h.t.s.t.h & f.t & h.t.s.t.h & f.t & -0.68 & -0.71 & 0.46 & -0.80 & -0.80 & 1.00 \\
GRB111228B & h.t.s.t.h & f.t & h.t.s.t.h & f.t & 0.11 & -0.25 & 0.86 & 0.10 & 0.00 & 0.90 \\
GRB120412B & f.t & f.t & h.t.s.t.h & f.t & 0.70 & 0.70 & 1.00 & -0.80 & -0.80 & 1.00 \\
GRB130404B & h.t.s.t.h & f.t & h.t.s.t.h & h.t.s.t.h & -0.54 & -0.25 & 0.32 & -0.68 & 0.14 & 0.25 \\
GRB131108A & h.t.s.t.h & f.t & h.t.s.t.h & f.t & -0.83 & -0.60 & 0.77 & -1.00 & -0.50 & 0.50 \\
GRB140108A & ... & f.t & ... & f.t & -0.57 & -0.50 & 0.82 & -0.54 & -0.71 & 0.89 \\
GRB150220A & h.t.s.t.f.t & f.t & h.t.s.t.h & f.t & 0.21 & 0.31 & 0.94 & -1.00 & -1.00 & 1.00 \\
GRB160802A & f.t & h.t.s.t.h & f.t & f.t & 0.50 & -0.36 & 0.54 & 0.40 & 0.40 & 1.00 \\
GRB170510A & ... & f.t & ... & f.t & 0.48 & 0.28 & 0.72 & -0.83 & -0.26 & 0.43 \\
GRB171120A & f.t & f.t & f.t & f.t & 0.51 & 0.54 & 0.82 & 0.55 & 0.67 & 0.88 \\
GRB180612A & h.t.s.t.h & f.t & h.t.s.t.h & f.t & 0.15 & 0.12 & 0.95 & -0.90 & -0.90 & 0.70 \\
GRB190901A & ... & f.t & ... & f.t & 0.42 & 0.02 & 0.67 & -0.40 & -0.40 & 1.00 \\
GRB210202A & f.t & f.t & f.t & f.t & 0.43 & -0.09 & 0.71 & 1.00 & 1.00 & 1.00 \\
GRB220927A & f.t & h.t.s.t.f.t & h.t.s.t.h & f.t & 0.26 & -0.37 & 0.66 & -1.00 & -0.50 & 0.50 \\
GRB221119A & f.t & f.t & f.t & f.t & 0.63 & 0.55 & 0.97 & 0.79 & 0.79 & 1.00 \\
GRB231104A & ... & f.t & ... & f.t & 0.52 & 0.41 & 0.94 & -0.20 & -0.20 & 1.00 \\
GRB240229A & f.t & f.t & f.t & f.t & 0.09 & -0.09 & 0.94 & 0.77 & 0.89 & 0.77 \\
\hline
\hline
\end{longtable}
\end{center}

\subsection{Correlations of Spectral Parameters}
In this section, we conduct a systematic examination and comparison of the correlations among the spectral parameters ($\alpha$, $E_{p}$, F (the flux)) of the main and second bursts of the 18 GRBs. The aim is to provide evidence to determine whether the main bursts and second bursts stem from the same source. The results of the correlation analysis for $\alpha-F$, $E_{p}-\alpha$, and $E_{p}-F$ for each GRB are presented in Table \ref{table3}.

\begin{figure}[htbp]
\centering
\includegraphics [width=14cm,height=6cm]{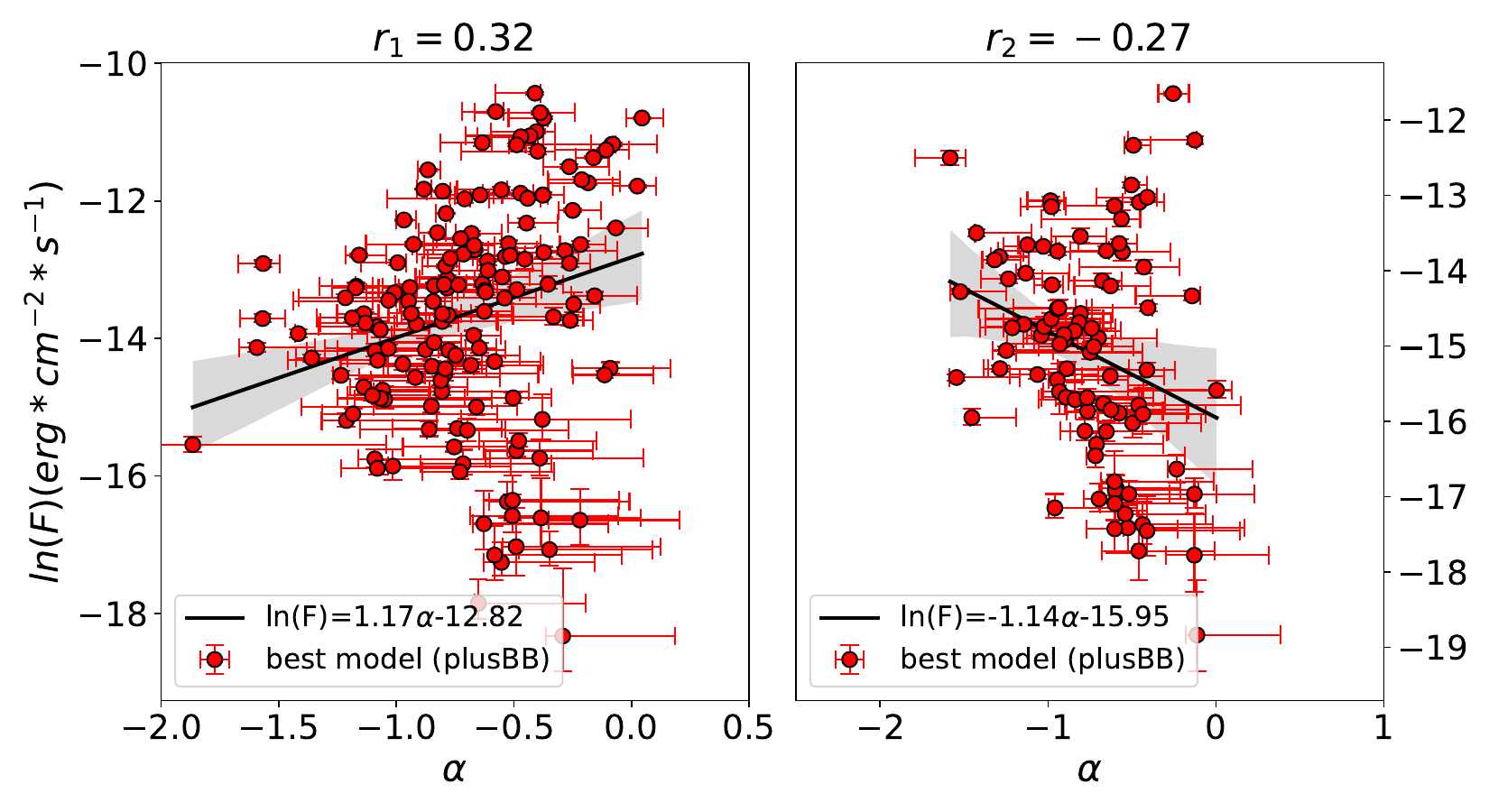}
   \figcaption{The Correlations between $F$ and $\alpha$ for all GRBs.\label{fig 4}}
\end{figure}

\begin{figure}[htbp]
\centering
\includegraphics [width=14cm,height=6cm]{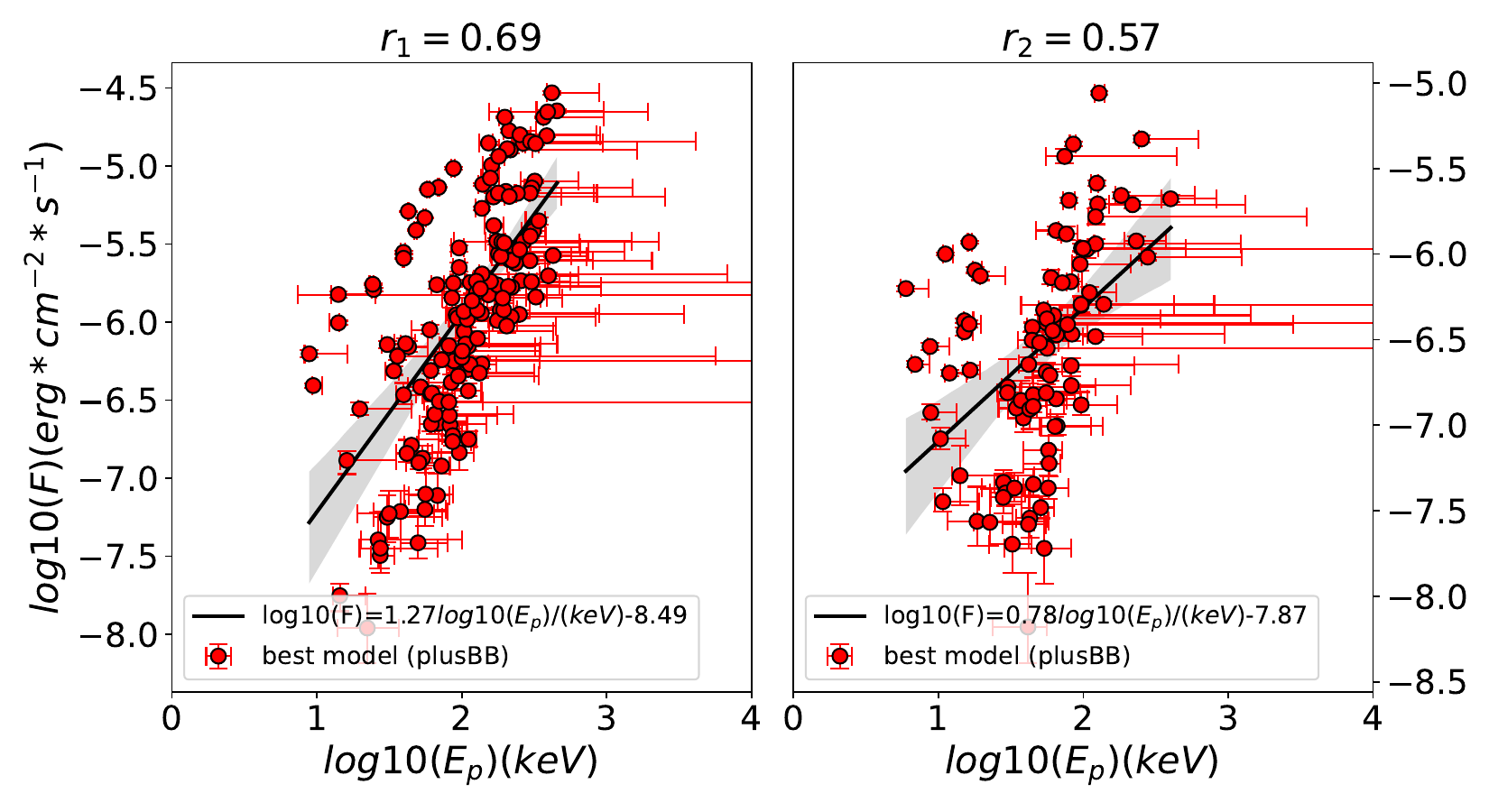}
   \figcaption{The Correlations between $F$ and $E_{p}$ for all GRBs.\label{fig 5}} 
\end{figure}

\begin{figure}[htbp]
\centering
\includegraphics [width=14cm,height=6cm]{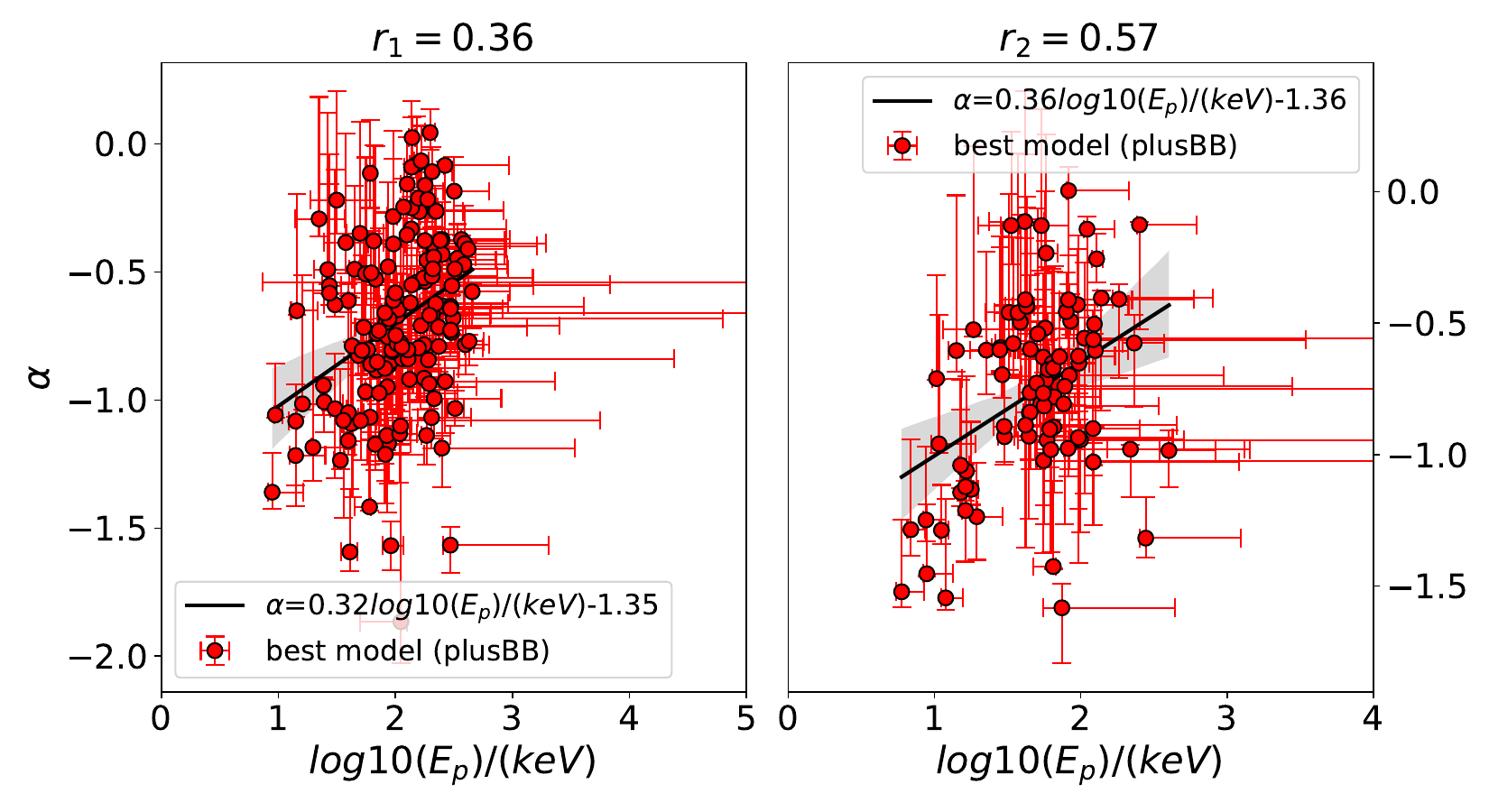}
   \figcaption{The Correlations between $\alpha$ and $E_{p}$ for all GRBs.\label{fig 6}}
\end{figure}

Figure \ref{fig 4} depicts the correlation between $F$ and $\alpha$ for all the GRBs. For the main bursts, the relationship is given by $\ln F$ = $1.17$ $\alpha - 12.82$ with a correlation coefficient $r = 0.32$. For the second bursts, it is $\ln F$ = $- 1.14$ $\alpha - 15.95$ and $r = - 0.27$. Similarly, Figure \ref{fig 5} shows the correlation between $F$ and $E_{p}$ for all GRBs. Through function fitting, we derive the expressions for these two relationships. In the main bursts, $\log F$ = $1.27$ $\log E_{p} - 8.49$ with $r = 0.69$, while in the second bursts, $\log F$ = $0.78$ $\log E_{p} - 7.87$ and $r = 0.57$. Additionally, we analyze the correlation between $E_{p}$ and $\alpha$ for all GRBs, as illustrated in Figure \ref{fig 6}. In the main bursts, $\alpha$ = $0.32$ $\log E_{p} - 1.35$ and $r = 0.36$. In the second bursts, $\alpha$ = $0.36$ $\log E_{p} - 1.36$  and $r = 0.57$.

\subsection{Distributions of Spectral Parameters}
\begin{figure}[H]
\centering
\includegraphics [width=14cm,height=6cm]{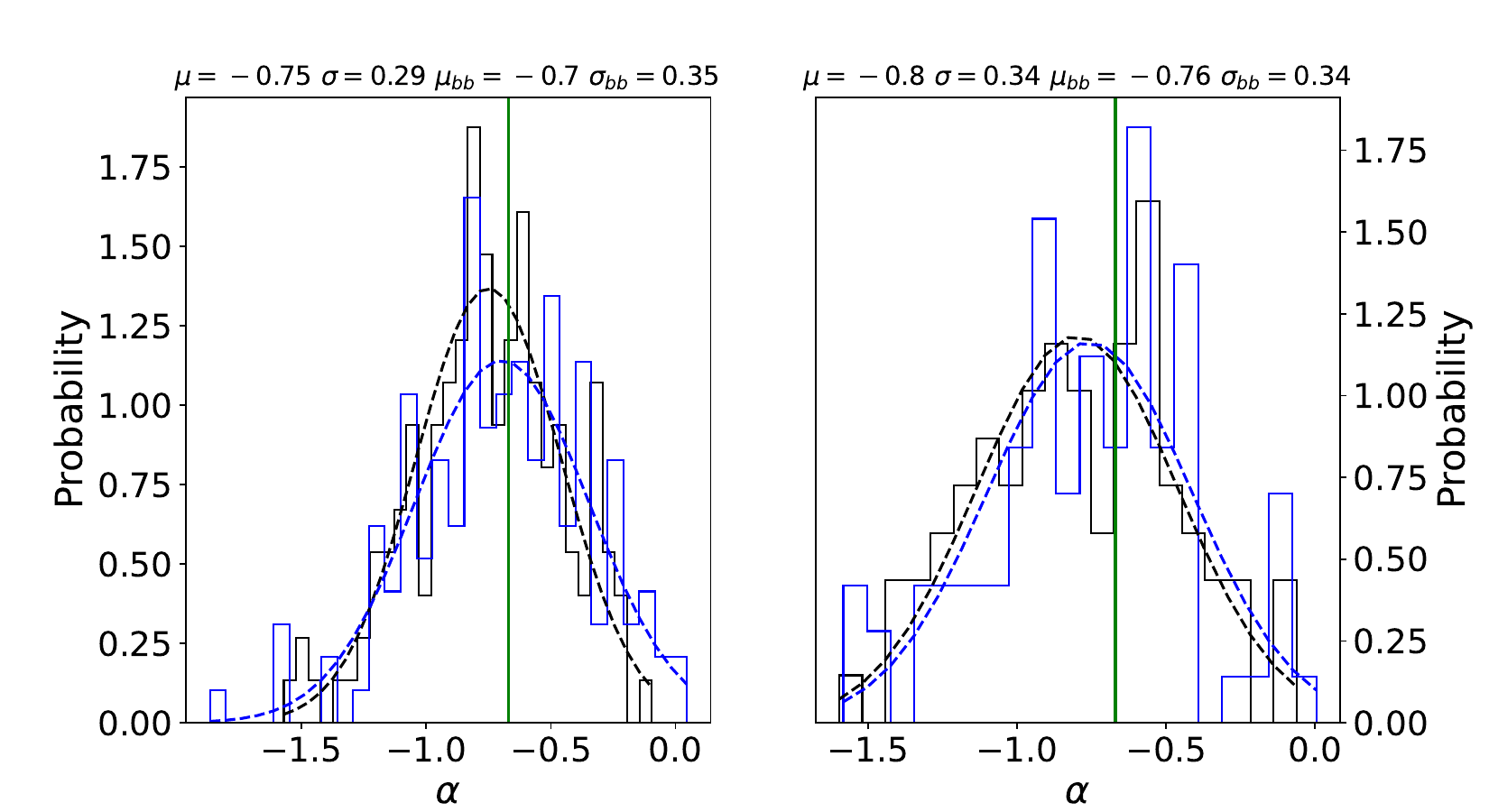}
\includegraphics [width=14cm,height=6cm]{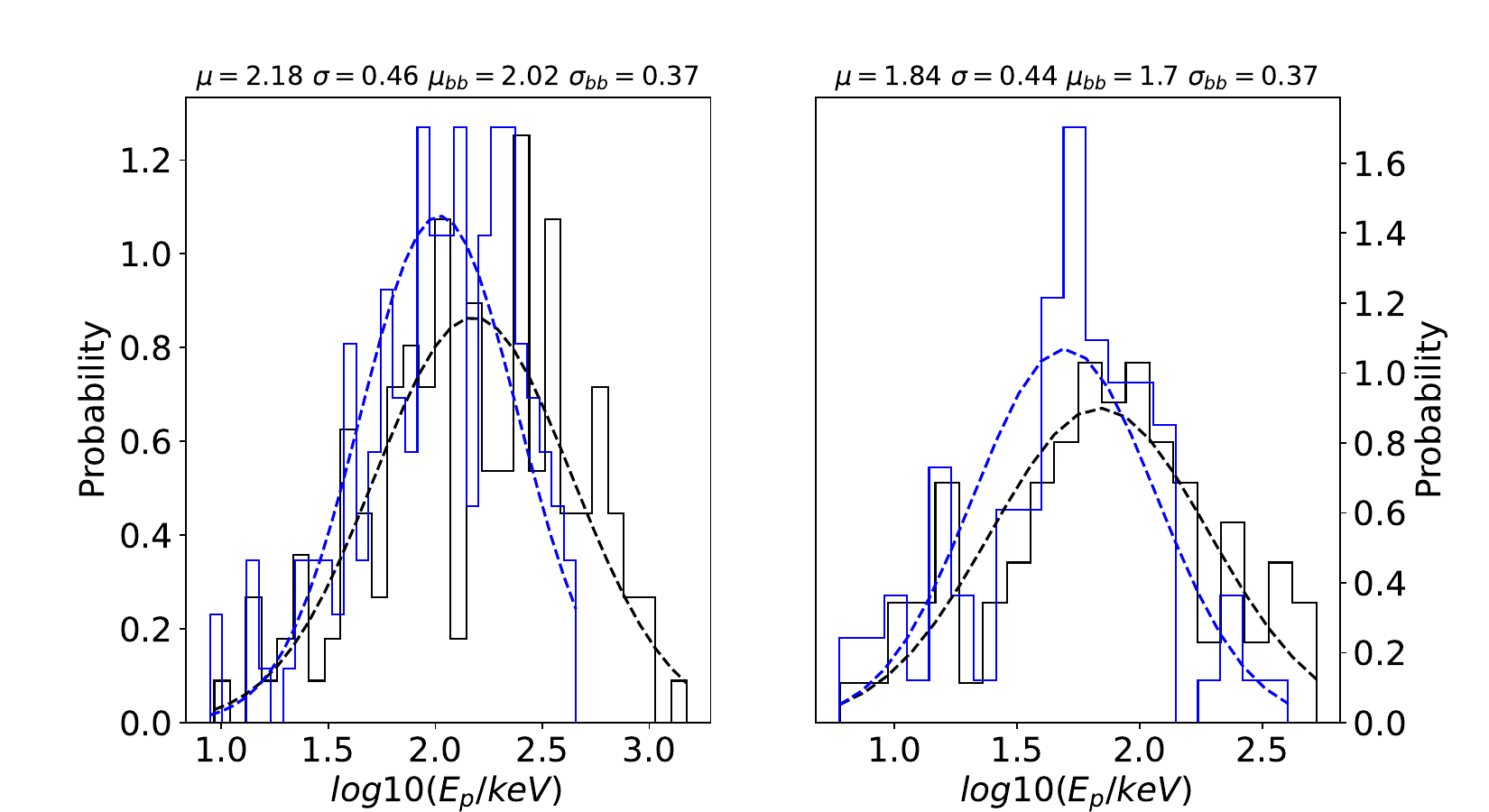}
   \figcaption{The upper panel illustrates the distribution of the $\alpha$, while the lower panel depicts the distribution of the $E_{p}$. The left panel corresponds to the time-resolved spectra of the main burst, and the right panel represents those of the second burst. The green solid line in the panels denotes $\alpha = - 2/3$. The blue dashed line signifies the best-fit models for the time-resolved spectra, and the black dashed line indicates the best-fit models after incorporating the BB component.\label{fig 7}}
      
\end{figure}

As depicted in the upper panel of Figure \ref{fig 7}, we display the distributions of the low - energy spectral indices $\alpha$ for the main bursts and second bursts of 18 GRBs. These distributions are fitted with Gaussian functions to determine their respective mean values and standard deviations. For the main bursts of GRBs, the value of $\alpha$ is $\alpha = - 0.75 \pm 0.29$, and when the BB component is added, it becomes $\alpha = - 0.70 \pm 0.35$. For the second bursts, the value of \(\alpha\) is $\alpha = - 0.80 \pm 0.34$, and after incorporating the BB component, it is $\alpha = - 0.76 \pm 0.34$. From these statistical results, it is evident that the $\alpha$ values in the main bursts and second bursts are similar, and the $\alpha$ values after adding the BB component also exhibit similarity. Additionally, compared to the second bursts, there are more time - resolved spectra in the main bursts that exceed the 'synchronization death line'. Moreover, for both the main and second bursts, the $\alpha$ values do not change significantly after adding the BB component, yet these values are closer to the typical value $\alpha = - 1$, \cite{2000ApJS..126...19P}).

Subsequently, in the lower panel of Figure \ref{fig 7}, we present a distribution map of the  $E_{p}$. Similarly, the distributions of the $E_{p}$ for the main bursts and second bursts of GRBs are fitted with Gaussian functions. For the main bursts, we obtain $\log E_{p} = 2.18 \pm 0.46$, and after adding the BB component, it is $\log E_{p} = 2.02 \pm 0.37$. For the second bursts, we obtain $\log E_{p} = 1.84 \pm 0.44$, and after adding the BB component, it is $\log E_{p} = 1.70 \pm 0.37$. We observe that the $E_{p}$ value of the main burst is larger than that of the second burst. Furthermore, within the margin of error, the $E_{p}$ values of the main burst and the second burst remain nearly unchanged before and after adding the BB component.

\section{Photosphere radiation parameters\label{Sections5}}
The acceleration mechanisms of GRB jets can be classified into two principal types: thermally driven and magnetically driven. Thermally driven jets are linked to hot fireballs and develop rapidly. In contrast, magnetically driven jets are associated with Poynting - flux - dominated outflows and evolve relatively slowly \citep{2015ApJ...801..103G}. In this section, we analyze the radiation sources within the photospheres of both the main bursts and the second bursts. We will achieve this by employing specific empirical relationships to constrain the properties of the outflow generated by the thermal pulse.

For each time - resolved spectrum, we initially apply the method described in \cite{2007RSPTA.365.1171P} to estimate the outflow parameters $\Gamma$ and $r_{0}$. Subsequently, following the approach of \cite{2009ApJ...702.1211R}, we calculate the effective transverse size of the radiative region ($\Re$), the photospheric radius ($r_{ph}$), and the saturation radius ($r_{s}$).

\subsection{ Parameter $\Re $}
In the framework of spherical symmetry \citep{2007RSPTA.365.1171P}, the ratio of the observed quantities $F_{BB}$ and $T$ is defined as $\Re $. This ratio can be calculated using the following equation, which holds when ${r_{ph}} > {r_s}$:
\begin{equation}
\Re  = {\left( {\frac{{F_{BB}^{{\rm{ob}}}}}{{\sigma {T^{o{b^4}}}}}} \right)^{1/2}} = \left( {1.06} \right)\frac{{{{\left( {1 + z} \right)}^2}}}{{{d_L}}}\frac{{{r_{ph}}}}{\Gamma }
\end{equation}
Here, $\sigma$ is the Stefan - Boltzmann constant, $z$ represents the redshift, and $d_{L}$ is the luminosity distance. For bursts with a known redshift, the parameter $\Re$ can be regarded as the effective lateral size of the radiating region \citep{2009ApJ...702.1211R}. Thus, a constant $\Re$ indicates that the effective radiative area of the photosphere remains invariant over time. For GRBs with an unknown redshift, we assume $z = 1$ \citep{2015ApJ...813..127P}.

The temporal evolution of $\Re$ for the main burst and second burst is presented in Appendix Figure \ref{fig 15}, and the average values are tabulated in Table \ref{table 4}, being $(5.87 \pm 0.98) \times 10^{-20}$ and $(5.69 \pm 0.98) \times 10^{-20}$, respectively. Within the margin of error, these two average values are approximately equal, and the effective transverse sizes $\Re$ of the main burst and second burst approximately lie on a straight line. This implies that the $\Re$ values of most ($72.2\%= 13/18$) main bursts and second bursts are approximately the same.

\subsection{Parameter $\Gamma $}
The Lorentz factor for the gliding phase (${r_{ph}} > {r_s}$) can be expressed in the following form:
\begin{equation}
\Gamma  \propto {\left( {F/\Re } \right)^{1/4}}{Y^{1/4}}
\end{equation}
Here, Y is related to the radiative efficiency of the burst and is defined by the equation:
\begin{equation}
Y = \frac{{{L_0}}}{{{L_{obs,\gamma }}}}
\end{equation}
In this equation, $L_{0}$ represents the total kinetic luminosity, and $L_{obs,\gamma}$ is the observed gamma - ray luminosity.

In Appendix Figure \ref{fig 16}, we display the temporal evolution of the Lorentz factor $\Gamma $ for 18 GRBs, and the average value of the time-resolved $\Gamma$ is presented in Table \ref{table 4}. The average values of $\Gamma $ for the main bursts and second bursts within the sample are $(582 \pm 65.9)Y^{1/4}$ and $(395 \pm 65.6)Y^{1/4}$, respectively. In the sample, $88.9\%$ of the GRBs exhibit a higher average $\Gamma $ for the main burst compared to the second burst. Furthermore, as can be seen from Figure \ref{fig 16}, it is evident that the $\Gamma $ generally decreases as the burst progresses from the main burst to the second burst.

\subsection{Parameter ${{\rm{r}}_{\rm{0}}}$, ${{\rm{r}}_{\rm{s}}}$, ${{\rm{r}}_{{\rm{ph}}}}$}
The nozzle radius $r_{0}$ represents the radius at which the jet commences its acceleration. Once $\Re$ is determined, the formula for $r_{0}$ applicable when ${r_{ph}} > {r_s}$ is given as \citep{2007RSPTA.365.1171P}:
\begin{equation}
{r_0} \propto {\left( {{F_{BB}}/FY} \right)^{3/2}}\Re
\end{equation}

The saturation radius ${{\rm{r}}_{\rm{s}}}$ is the radius at which the Lorentz factor reaches its maximum value. Using ${{\rm{r}}_0}$, we can estimate the saturation radius ${{\rm{r}}_{\rm{s}}}$ through the following equation:
\begin{equation}
{r_s} = \Gamma {r_0}
\end{equation}

In the context of relativistic holonomic motion, for photons travelling a distance $ds$, the optical depth $\tau$ is defined as:
\begin{equation}
\tau  = \int\limits_{{r_{ph}}}^\infty  {\frac{{n{\sigma _T}}}{{2{\Gamma ^2}}}dr}
\end{equation}
where ${\sigma _T}$ is the Thomson cross - section and n is the electron number density. The expression for $ds$ is $ds = \left( {1 - \beta \cos \theta } \right)dr/\cos \theta $, with $\theta  = 0$. Assuming a constant Lorentz factor, the radius of the photosphere can be calculated using the following equation (when $\tau = 1$):
\begin{equation}
{r_{ph}} = \frac{{{L_0}{\sigma _T}}}{{8\pi {m_p}{c^3}\Gamma _{ph}^3}}
\end{equation}
where ${L_0}$ is the total kinetic luminosity, expressed as ${L_0} = 4\pi d_L^2Y{F_{tot}}$. Here, ${d_L}$ is the luminosity distance and $F_{tot}$ is the observed $\gamma$-ray flux.

In Appendix Figure \ref{fig 17}, we present the temporal evolution of the initial radius $r_{0}$, the saturated radius $r_{s}$, and the photospheric radius $r_{ph}$. The average values of $r_{0}$, $r_{s}$, and $r_{ph}$ for the main and second bursts of 18 GRBs are tabulated in Table \ref{table 4}. From Table \ref{table 4}, we find that for the main burst and second burst, the average values of $r_{0}$ are $(2.77 \pm 1.07) \times 10^7 Y^{-3/2}$ and $(3.39 \pm 1.50) \times 10^7 Y^{-3/2}$, respectively; the average values of $r_{s}$ are $(1.50 \pm 0.55) \times 10^{10} Y^{-5/4}$ and $(1.12 \pm 0.49) \times 10^{10} Y^{-5/4}$, respectively; the average values of $r_{ph}$ are $(1.58 \pm 0.26) \times 10^{11} Y^{1/4}$ and $(1.06 \pm 0.19) \times 10^{11} Y^{1/4}$, respectively. From Figure \ref{fig 17}, it is evident that the values of $r_{0}$, $r_{s}$, and $r_{ph}$ at the end of the main burst and the start of the second burst are remarkably similar. The transition from the main burst to the second burst seems to be smooth. These observed statistical characteristics suggest a connection between these two periods.

\begin{table}[htbp]
\centering
\caption{Average values of parameters $\Re$ , $\Gamma$ , $r_{0}$ , $r_{s}$ , $r_{ph}$\label{tab 4}}
\label{table 4}
\resizebox{\textwidth}{!}{
\begin{tabular}{cccccc|ccccc}
  \toprule
  \multirow{2}*{GRB}&\multicolumn{5}{c|}{main burst}&\multicolumn{5}{c}{second burst}\\ 
  \cline{2-6}
  \cline{7-11}
&$\Re\times10^{-20}$&$\Gamma\times10^{2}$&$r_{0}\times10^{7}$&$r_{s}\times10^{9}$&$r_{ph}\times10^{11}$&$\Re\times10^{-20}$&$\Gamma\times10^{2}$&$r_{0}\times10^{7}$&$r_{s}\times10^{9}$&$r_{ph}\times10^{11}$\\
\hline
  \midrule
GRB081009A & 6.23±0.52 & 6.42±0.42 & 0.27±0.26 & 1.22±1.11 & 2.02±0.20 & 7.32±0.82 & 4.23±0.33 & 0.19±0.10 & 0.68±0.31 & 1.55±0.13 \\
GRB100719C & 7.61±1.70 & 9.75±1.42 & 6.05±1.56 & 48.71±11.41 & 3.21±0.57 & 4.13±0.84 & 4.83±0.92 & 1.52±1.38 & 4.95±4.11 & 0.97±0.26 \\
GRB111228B & 6.00±0.64 & 4.98±0.37 & 0.43±0.24 & 2.38±1.48 & 1.48±0.09 & 9.91±1.84 & 3.65±0.38 & 0.61±0.40 & 1.92±1.12 & 1.81±0.31 \\
GRB120412B & 5.53±0.79 & 3.09±0.52 & 1.26±0.67 & 2.74±0,98 & 0.90±0.21 & 6.77±0.61 & 2.74±0.38 & 5.58±3.18 & 12.65±5.75 & 0.96±0.17 \\
GRB130404B & 7.82±1.93 & 4.47±0.88 & 8.50±3.42 & 34.89±14.11 & 1.56±0.41 & 6.20±1.18 & 3.97±1.17 & 1.69±0.55 & 6.95±2.91 & 0.98±0.19 \\
GRB131108A & 5.72±0.85 & 3.10±0.40 & 3.10±0.98 & 9.12±2.53 & 0.92±0.17 & 5.81±0.71 & 2.54±0.46 & 7.48±3.82 & 22.30±14.08 & 0.79±0.23 \\
GRB140108A & 6.56±1.21 & 3.52±0.48 & 3.51±1.52 & 10.33±4.68 & 1.13±0.20 & 6.23±1.24 & 4.40±0.66 & 5.19±2.39 & 15.63±5.64 & 1.23±0.20 \\
GRB150220A & 5.13±0.36 & 4.65±0.39 & 1.69±0.58 & 6.92±2.12 & 1.20±0.11 & 5.40±0.52 & 2.80±0.47 & 1.54±0.92 & 4.23±2.63 & 0.74±0.05 \\
GRB160802A & 7.46±1.44 & 9.34±1.13 & 7.15±2.06 & 60.05±18.38 & 3.21±0.57 & 6.17±1.06 & 5.52±1.04 & 6.67±2.55 & 35.19±13.75 & 1.92±0.64 \\
GRB170510A & 4.10±0.81 & 6.68±0.52 & 1.36±0.64 & 8.61±4.45 & 1.32±0.24 & 5.91±0.63 & 2.86±0.33 & 4.21±0.49 & 12.14±2.23 & 0.88±0.16 \\
GRB171120A & 6.68±0.82 & 5.49±0.60 & 2.21±0.72 & 14.10±5.30 & 1.75±0.20 & 5.37±1.01 & 4.69±0.53 & 3.70±1.53 & 14.83±5.55 & 1.26±0.27 \\
GRB180612A & 4.87±0.84 & 4.89±0.69 & 0.92±0.47 & 3.55±1.91 & 1.07±0.19 & 3.92±1.13 & 3.46±0.92 & 2.76±0.94 & 8.17±3.79 & 0.50±0.07 \\
GRB190901A & 6.44±1.36 & 5.64±0.68 & 3.36±1.11 & 16.10±5.11 & 1.57±0.20 & 3.55±0.99 & 3.75±0.94 & 6.24±2.77 & 16.39±6.59 & 0.56±0.10 \\
GRB210202A & 4.61±0.98 & 5.26±0.72 & 4.97±2.39 & 25.59±14.54 & 1.29±0.37 & 4.80±0.30 & 3.78±0.47 & 3.19±0.46 & 11.65±0.86 & 0.92±0.05 \\
GRB220927A & 6.86±0.41 & 4.54±0.24 & 2.25±1.12 & 8.99±4.31 & 1.60±0.12 & 6.94±1.64 & 2.32±0.30 & 2.92±0.94 & 7.11±3.05 & 0.81±0.17 \\
GRB221119A & 6.08±1.06 & 5.61±0.51 & 0.59±0.18 & 3.05±0.97 & 1.69±0.32 & 3.76±0.65 & 5.41±0.54 & 0.98±0.67 & 6.17±4.68 & 0.97±0.11 \\
GRB231104A & 2.34±0.43 & 11.74±1.21 & 1.10±0.44 & 9.31±3.31 & 1.18±0.16 & 5.93±1.61 & 4.44±0.80 & 4.96±2.76 & 16.10±8.53 & 1.16±0.17 \\
GRB240229A & 5.57±1.46 & 5.51±0.69 & 1.09±0.81 & 4.37±3.12 & 1.40±0.36 & 4.33±0.92 & 5.62±1.17 & 1.64±1.12 & 5.25±3.22 & 1.01±0.13 \\
\hline
\bottomrule
\end{tabular}
}
\noindent
\footnotesize \textbf{Note.} The mean values of the photosphere radiation parameters presented in this table are computed based on the assumption that $ Y = 1$. For a more in - depth and detailed understanding, kindly refer to the relevant text.

\end{table}

\section{ Amati Relation and Yonetoku Relation\label{Sections6}}
To gain a more comprehensive comparison of the spectral properties of the main burst and the second burst, we investigate whether they adhere to the same Amati ($E_p - E_{iso}$) relation and Yonetoku ($E_p - L_{iso}$) relation. In our sample, the redshift is known only for GRB 131108A. For the GRBs with unknown redshifts, we assume $z = 1$.

\begin{figure}[htbp]
\centering
\includegraphics [width=7.5cm,height=6cm]{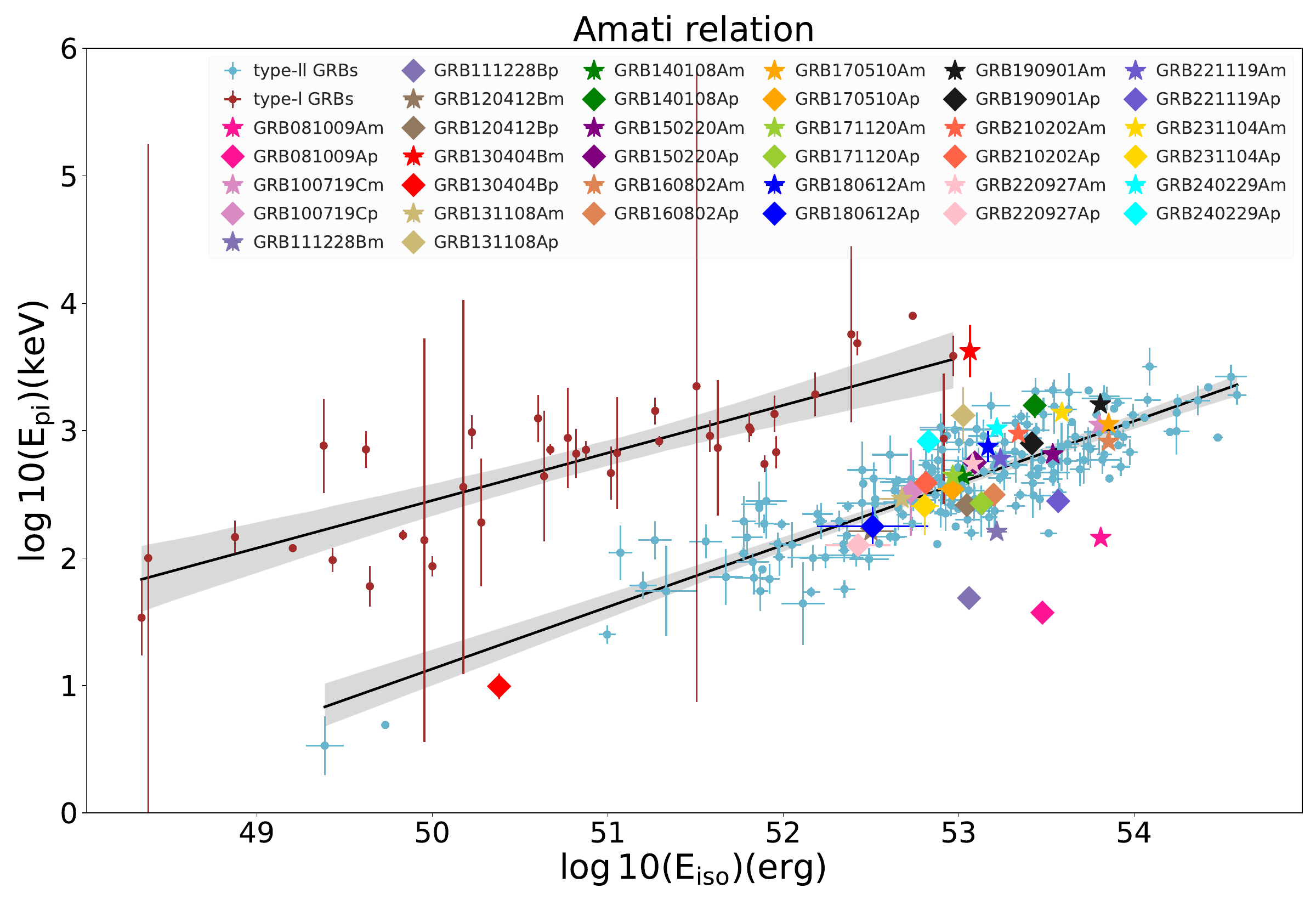}
\includegraphics [width=7.5cm,height=6cm]{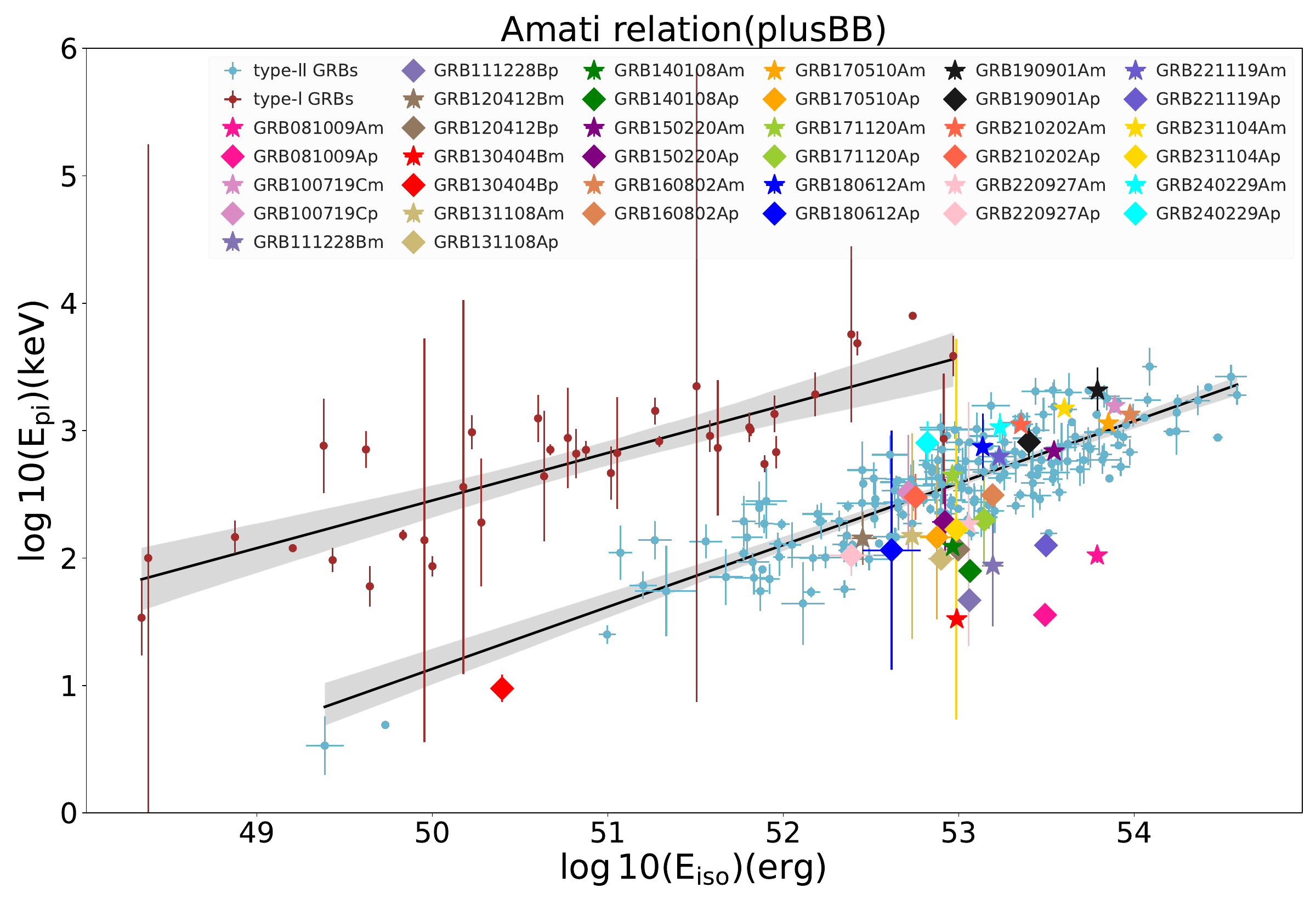}
\includegraphics [width=7.5cm,height=6cm]{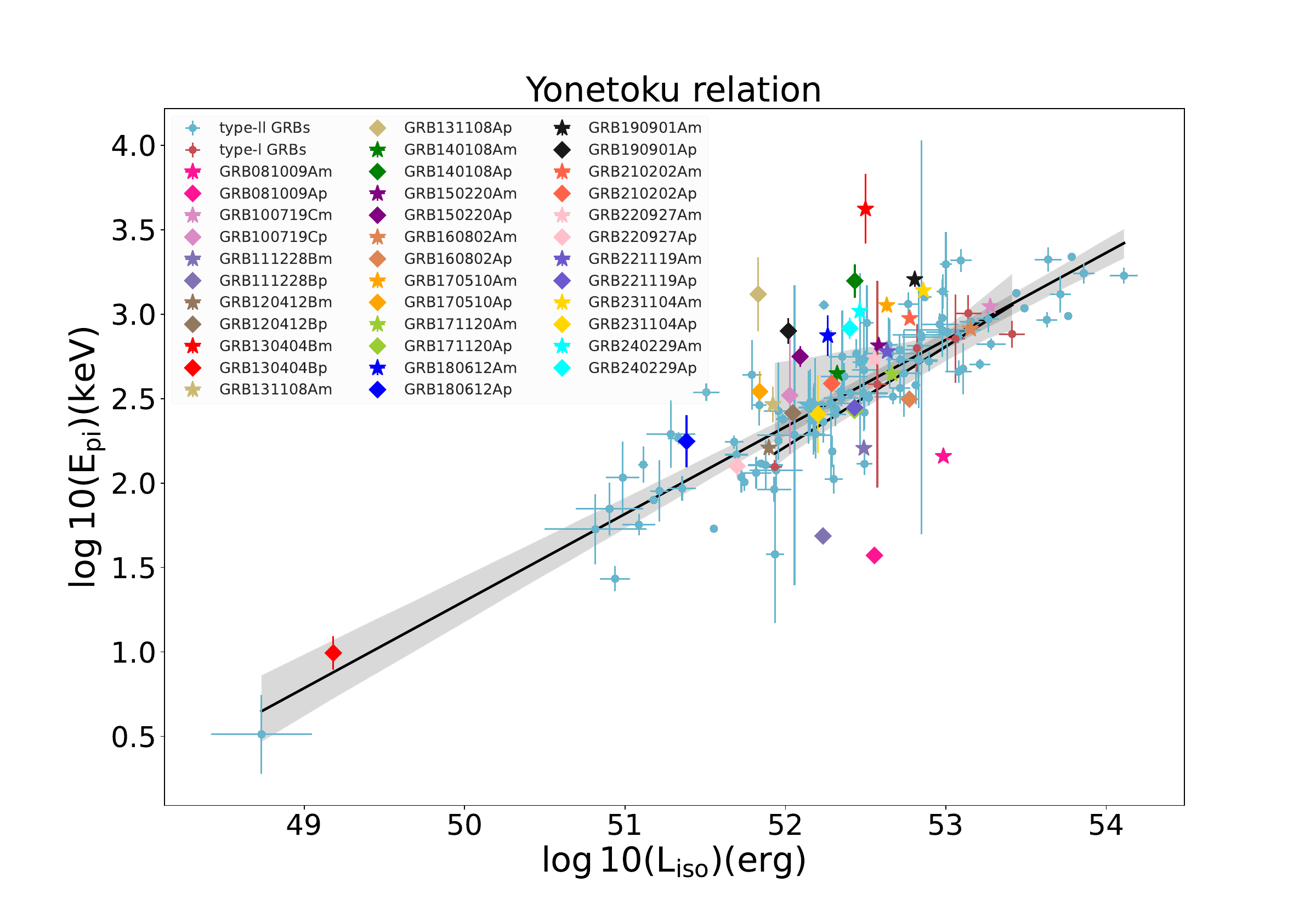}
\includegraphics [width=7.5cm,height=6cm]{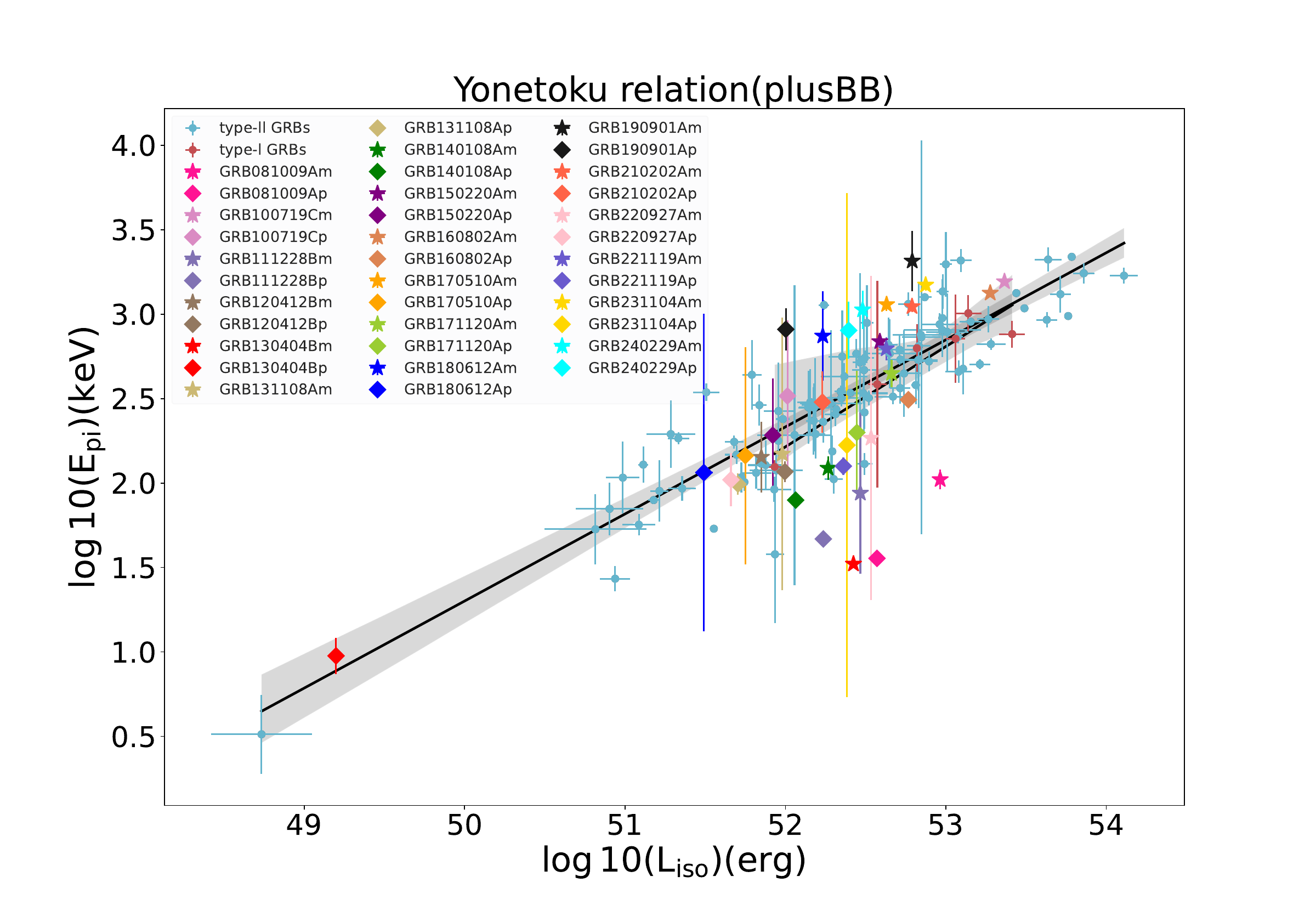}
   \figcaption{In the Amati and Yonetoku relationships, the maroon circles denote type I GRBs, and the blue circles represent type II GRBs. The pentagrams stand for the main burst, and the diamonds symbolize the second burst. The left panel displays the best - fit model, while the right panel shows the best - fit model incorporating BB components. \label{fig 8}}

\end{figure}

Figure \ref{fig 8} depicts the time - integrated spectral relations of Amati and Yonetoku for 18 GRBs. Regarding the Amati relation, it is evident from the figure that the majority of both the main bursts and second bursts follow this relation. Although a few bursts deviate from the Amati relation after adding the BB component, they still lie within the same general region. This indicates a consistent behavior pattern between the main and second bursts. Similarly, for the Yonetoku relation, most of the main and second bursts conform to it. Only very few bursts show deviation from the Yonetoku relation after the addition of the BB component, yet they remain within the same area. In other words, the changes in the behavior of the main and second bursts are consistent.

The observation that both the main bursts and second bursts in the sample follow the same Amati and Yonetoku relations strongly suggests that they might share the same progenitor origin. Thus, based on these correlations, we are inclined to conclude that the main burst and second burst have a common origin.

\section{Discussion\label{Sections7}}
We have carried out time - resolved and time - integrated spectral analyses on the main bursts and second bursts of 18 GRBs. The primary objective is to determine whether these bursts share the same origin and to offer insights into their physical origins. \cite{2014ApJ...789..145H} found that the spectral indices of precursors and main bursts are similar, suggesting that these different radiation events might have a common physical origin. In contrast, for the two GRBs studied by \cite{2022ApJ...940...48D}, there is no correlation between the spectral indices, leading their research to suggest that the main and second bursts may have different origins.

For our entire analyzed sample, from the time - integrated spectra, upon adding a BB component, $100\%$ (18 out of 18) of the main bursts and $88.9\%$ (16 out of 18) of the second bursts have the best - fitting model as Band + BB/CPL + BB. Moreover, the goodness of fit improves for both, indicating that most main bursts and second bursts contain a thermal component. Based on the time - resolved spectra, the variation in the proportion of thermal components between the main bursts and second bursts is as follows: $67\%$ (12 out of 18) of the GRBs show a gradual decrease, $28\% $ (5 out of 18) show no change, and $5\%$ (1 out of 18) show an increase.

We have observed that the $\alpha$ values are comparable in both the main bursts and second bursts. Additionally, there are more time-resolved spectra in the main bursts that exceed the “Synchrotron line-of-death” compared to the second bursts. Furthermore, after adding a BB component, the value of $\alpha$ does not change substantially in either the main bursts or second bursts, yet it gets closer to the typical value of $\alpha$. Concerning the $E_{p}$, after adding the BB component, the $E_{p}$ values in both the main bursts and second bursts decrease slightly, with the $E_{p}$ values of the main bursts being higher than those of the second bursts. This implies that the jet composition changes between the main bursts and second bursts. Our results lend support to the idea of a transition in the GRB jet composition from being fireball - dominated to Poynting - flux - dominated. We also hypothesize that the main bursts and second bursts may have the same physical origin.

Significant alterations in the jet components can disclose the nature of the GRB central engine. Drawing on the analysis by \cite{2007ApJ...670.1247W} regarding the jets related to GRB precursors and main bursts, we infer that during the main bursts and second bursts, the superaccretion of the central engine causes the formation of a matter - dominated fireball in the initial burst, a consequence of neutrino - antineutrino annihilations. The jets exhibit intermittent activity. After the main burst (the rapid accretion phase), the central engine enters a quiescent phase. \cite{2006MNRAS.370L..61P} hypothesized that energy release might take place through the repeated accretion of the accumulated flux around the central engine. After the main burst, the central engine restarts and becomes highly magnetized, giving rise to synchrotron radiation. This process, which is analogous to the prominent late X - ray flares observed in GRBs, supplies the energy for the second burst event (the second burst).

\cite{2021ApJS..252...16L} studied short - burst events with concurrent precursors, main bursts, and extended radiation. They discovered a correlation among the peak fluxes of these three events, which supports the idea that the three events originate from similar central engines. In the sample of this paper, a relationship is observed: the higher the peak flux of the main burst, the higher the peak flux of the second burst. This indicates a certain correlation between the peak fluxes of the main and second bursts, suggesting that the main and second bursts may share the same origin.

The spectral parameters $\alpha$ and $E_{p}$ derived in this paper exhibit a hard - to - soft variation with time and demonstrate flux - tracking behavior, which is in line with the results reported by \cite{2019ApJS..242...16L}. Regarding the temporal evolution patterns of the $\alpha$ values for the main burst and second burst, 4 GRBs did not show a clear evolution pattern. Among the remaining 14 GRBs, $71.4\%$ (10 out of 14) exhibited the same evolution pattern. Specifically, 5 of them followed the hard - to - soft - to - hard (h.t.s.t.h) pattern, and 5 showed the flux - tracking (f.t) pattern. For the evolution pattern of $E_{p}$, $77.8\%$ (14 out of 18) evolved in the same f.t pattern.The evolution of $E_{p}$ from a hard - to - soft (h.t.s) state might be accounted for by the ICMART model \citep{2011ApJ...726...90Z}. On the other hand, the f.t behavior could be due to internal shocks or the photosphere \citep{2012ApJ...756..112L}. If both the main burst and second burst are predominantly governed by thermal components, it implies that this flux tracking could be associated with the photosphere. Conversely, if thermal components are negligible or non - existent, it may be induced by internal shocks.

\cite{2019MNRAS.484.1912R} employed the function $F = F_{0}e^{ k\alpha(t)}$ to depict the relationship between the two parameters and found that the median value of $K$ was approximately 3. Similarly, we utilize this function to fit the relationship between these two parameters in our study. For the main burst, we obtained $\ln F$ = $1.17$ $\alpha - 12.82$ ($r = 0.32$), and for the second burst, $\ln F$ = $1.77$ $\alpha - 16.49$ ($r = - 0.38$). Through analysis, we determine that the median value of $K$ in the main burst of our sample is 1.60, while the median value of $K$ in the second burst is 2.68. Both of these values fall within the range of $K$ where $1 \leq K \leq 5$. Additionally, we note that after adding the BB component, the median values of $K$ decrease.

The radiation mechanism during the prompt emission phase of GRBs can be explored through the correlations between spectral parameters \citep{2019MNRAS.484.1912R}. For the entire sample, both the $F-E_{p}$ correlation and the $\alpha-E_{p}$ correlation display a weak positive correlation in both the main burst and the second burst. Notably, in the $F$ and $\alpha$ correlation, the main burst and second burst have opposite correlation, though neither correlation is highly significant. This may suggest that the radiation mechanisms of the main burst and second burst could be different. However, based on the statistical results for the $F - \alpha$ correlation, $50.0\%$ (9 out of 18) of the GRBs exhibit similar correlations, while $38.9\%$ (7 out of 18) show opposite correlations. Regarding the $\alpha - E_{p}$ correlation, \cite{2019ApJS..242...16L} concluded that the $\alpha - E_{p}$ correlation could be either positive or negative. In our sample, $72.2\%$ (13 out of 18) of the GRBs show similar correlations, while $22.2\%$ (4 out of 18) show different correlations. For the $F - E_{p}$ correlation, both the main burst and second burst show a positive correlation, with $83.3\%$ (15 out of 18) showing a correlation of moderate strength. Therefore, we are inclined to think that the main burst and second burst have the same origin. Nevertheless, whether their radiation mechanisms are identical requires further verification using more GRBs and methods.

\cite{2009ApJ...702.1211R} demonstrated that thermal radiation can serve as a valuable tool for studying the properties of the photosphere and the physical parameters of the GRB fireball. In the case of the 18 GRBs under our study, the average effective lateral size $\Re$ of the main burst and the second burst are approximately equivalent. As is evident from Appendix Figure \ref{fig 15}, $72.2\%$ (13 out of 18) of the GRBs have $\Re$ values that approximately follow a straight line, indicating a stable trend as time progresses.The Lorentz factors and initial radius values of all bursts fall within the ranges ${10^2} \le \Gamma {Y^{-1/4}} \le {10^3}$ and ${10^{6}} cm \le {r_0} {Y^{3/2}} \le {10^{9}} cm$, respectively. These findings are in accordance with the results reported by \cite{2015ApJ...813..127P}, \cite{2013MNRAS.433.2739I}, and \cite{2016MNRAS.456.2157I}. From Appendix Figure \ref{fig 17}, it can be observed that for the 18 GRBs, the photospheric radius $r_{ph}$ of both the main burst and the second burst lies within the range of 
${10^{10}} cm$ and ${10^{12}} cm$, with values tending to be close to ${10^{12}} cm$. This observation suggests that both the main burst and the second burst display typical characteristics of photospheric radiation, which are roughly in line with the photospheric radius predicted by the conventional fireball model. Moreover, as shown in Appendix Figure \ref{fig 17}, the three characteristic radius values ($r_{0}$, $r_{s}$, $r_{ph}$) at the conclusion of the main burst and the onset of the second burst are remarkably close to one another. This indicates that the transition from the main burst to the second burst is smooth. Based on these observed statistical characteristics, we deduce that the main burst and the second burst may share the same origin. Nevertheless, considering the presence of a certain quiet period between the main burst and the second burst, the underlying cause of this smooth transition between the two bursts warrants further in-depth investigation.

In addition, we analyze the time - integrated spectra of the main burst and the second burst. We compare the Amati relation and Yonetoku relation, and also conduct a hardness ratio (HR) analysis for the main and second bursts within the sample. We find that most of the main and second bursts adhere to the same Amati and Yonetoku relationships. They are located in the same region and exhibited consistent change behaviors. This suggests that the main burst and second burst may originate from the same progenitor star and share the same physical origin. However, in our sample, only the redshift of GRB 131108A is known. For GRBs with unknown redshifts, we assume a typical value of z = 1. Identifying additional samples with known redshifts is of great significance for further exploring the properties of the main burst and second burst. Through statistical analysis of the spectral HR  (defined as the ratio of the number of photons received in the energy range of $8-50$ keV to that in the $50-300$ keV range) of the main and second bursts, we discover a hard - to - soft trend from the main burst to the second burst. The average HR in the main burst is 0.96, with a median of 0.90. In the second burst, the average HR is 0.64, with a median of  0.54. Moreover, there is a correlation between the HRs of the main burst and second burst, as evidenced by a strong correlation in the $\log HR_{m}-\log HR_{s}$ plot for the 18 GRBs.

\section{Conclusion\label{Sections8}}
This paper utilizes data from the Fermi observatory to identify 18 GRBs that exhibit both a main burst and a second burst. A detailed time - resolved spectral and time - integrated spectral analysis of the sample is carried out using models such as Band, CPL, and BB. By comparing and analyzing the spectral properties of the main bursts and second bursts, the following interesting conclusions are drawn:
1.Through time - resolved spectral analysis of these 18 GRBs, it is found that the majority ($83.3\%$) of the main bursts and second bursts contain thermal components. For $67\%$ of the GRBs, the thermal components gradually decrease from the main burst to the second burst.

2.Regarding the evolution of the $\alpha$ in the main bursts and second bursts, 4 GRBs show no distinct evolutionary pattern. Among the remaining 14 GRBs, $71.4\%$ (10 out of 14) display the same evolutionary pattern. Specifically, 5 follow the h.t.s.t.h pattern, and 5 exhibit the f.t pattern. For the evolution of the $E_{p}$, $77.8\%$ (14 out of 18) mainly show an evolution pattern dominated by f.t pattern.

3.Based on statistical results, the $\alpha$ values are similar in both the main burst and the second burst. However, there are more time slices in the main burst that exceed the “Synchrotron line - of - death”, while there are fewer in the second burst. Moreover, after adding the BB component, the value of $\alpha$ changes little but gets closer to the typical value of $\alpha$. For the $E_{p}$, within the error range, the value of $E_{p}$ remains nearly unchanged after adding the BB component, with the $E_{p}$ value of the main burst being larger than that of the second burst.

4.After constraining the photospheric radiation parameters ($\Re$, $\Gamma$, $r_{0}$, $r_{s}$, and $r_{ph}$) for the main burst and second burst in the sample, it is discovered that $72.2\%$ (13 out of 18) of the GRBs have $\Re$ values that approximately lie on a straight line, showing a stable evolution trend from the main burst to the second burst. The Lorentz factor $\Gamma$ shows a trend of first increasing and then decreasing over time, with a range of ${10^2} \le \Gamma \le {10^3}$. The values of the initial radius $r_{0}$ and the saturation radius $r_{s}$ vary significantly. The range of $r_{0}$ is ${10^6} \le r_{0} \le {10^8}$, and the range of $r_{s}$ is ${10^8} \le r_{s} \le {10^{11}}$. The photospheric radius $r_{ph}$ ranges from ${10^{11}} \leq r_{ph} \leq {10^{12}}$ and is close to ${10^{12}}$ cm. From the time - evolution plot of the characteristic radius, it can be seen that the transition from the main burst to the second burst is smooth.

5.By performing time - integrated spectral analysis of the main bursts and second burst in our sample and comparing the Amati and Yonetoku relations, it is found that most of the main bursts and second bursts follow the same Amati and Yonetoku relations. Both the main and second burst events are located in the same region and exhibit consistent behavior.

In conclusion, we support the hypothesis that the main burst and second burst share the same physical origin.

\section*{acknowledgements}
We acknowledge the use of the public data from the Fermi data archives. This work is supported by the National Natural Science Foundation of China(grant 12163007, 11763009), the Key Laboratory of Colleges and Universities in Yunnan Province for High-energy Astrophysics, National Astronomical Observatory Yunnan Normal University Astronomy Science Education Base.


\section*{appendix}
\begin{figure}[H]
\centering
\includegraphics [width=5cm,height=4cm]{D1.pdf}
\includegraphics [width=5cm,height=4cm]{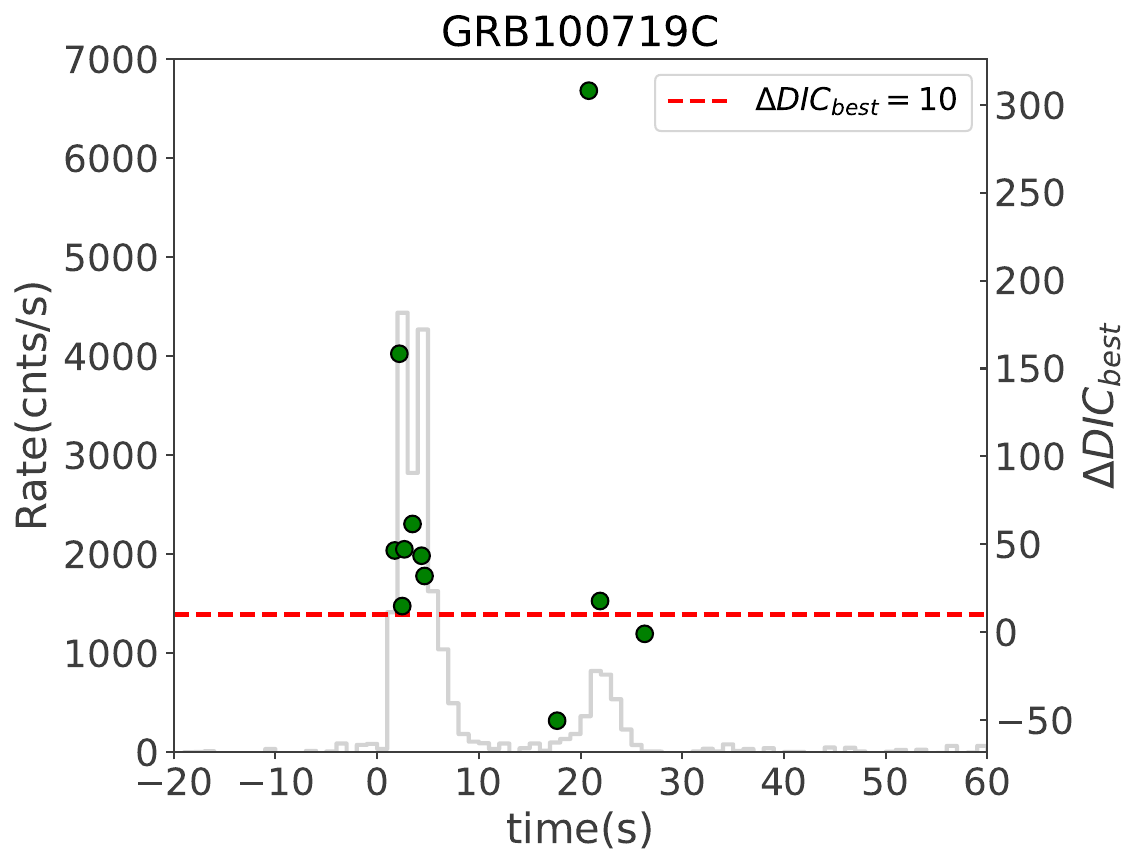}
\includegraphics [width=5cm,height=4cm]{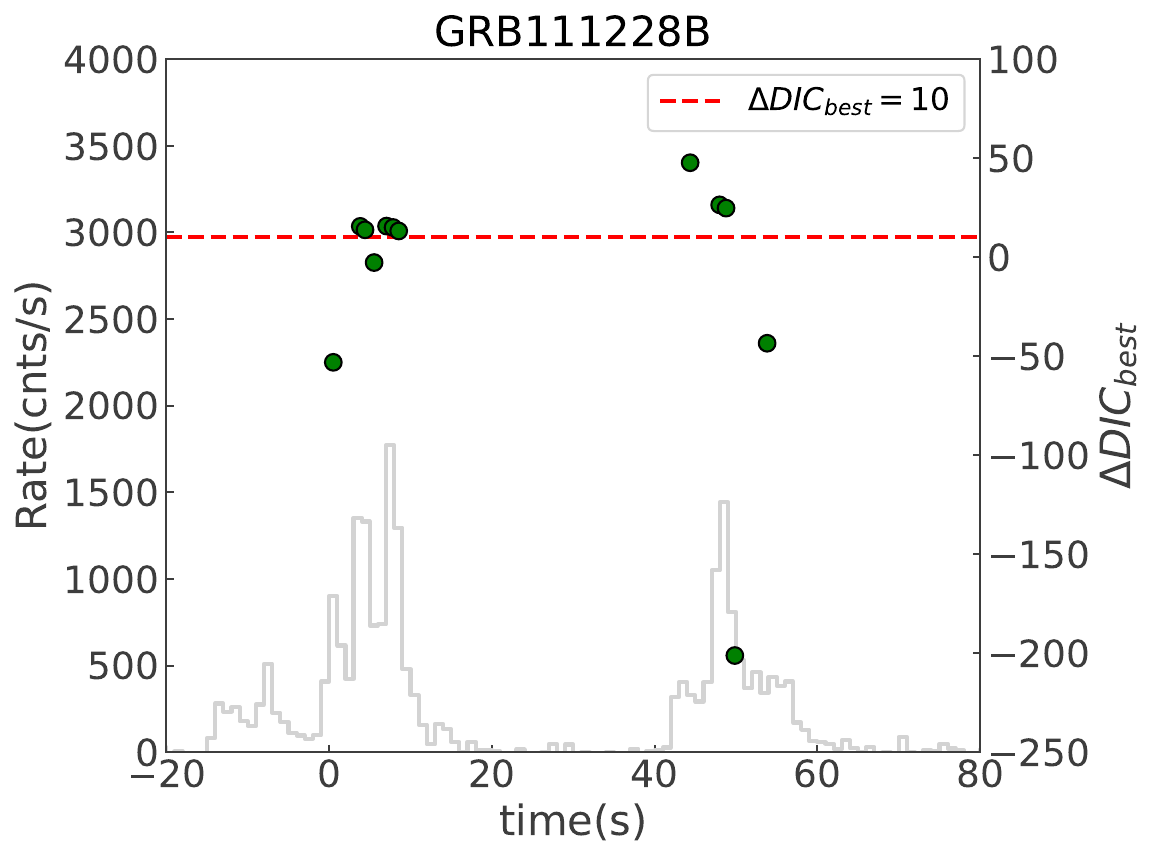}
\includegraphics [width=5cm,height=4cm]{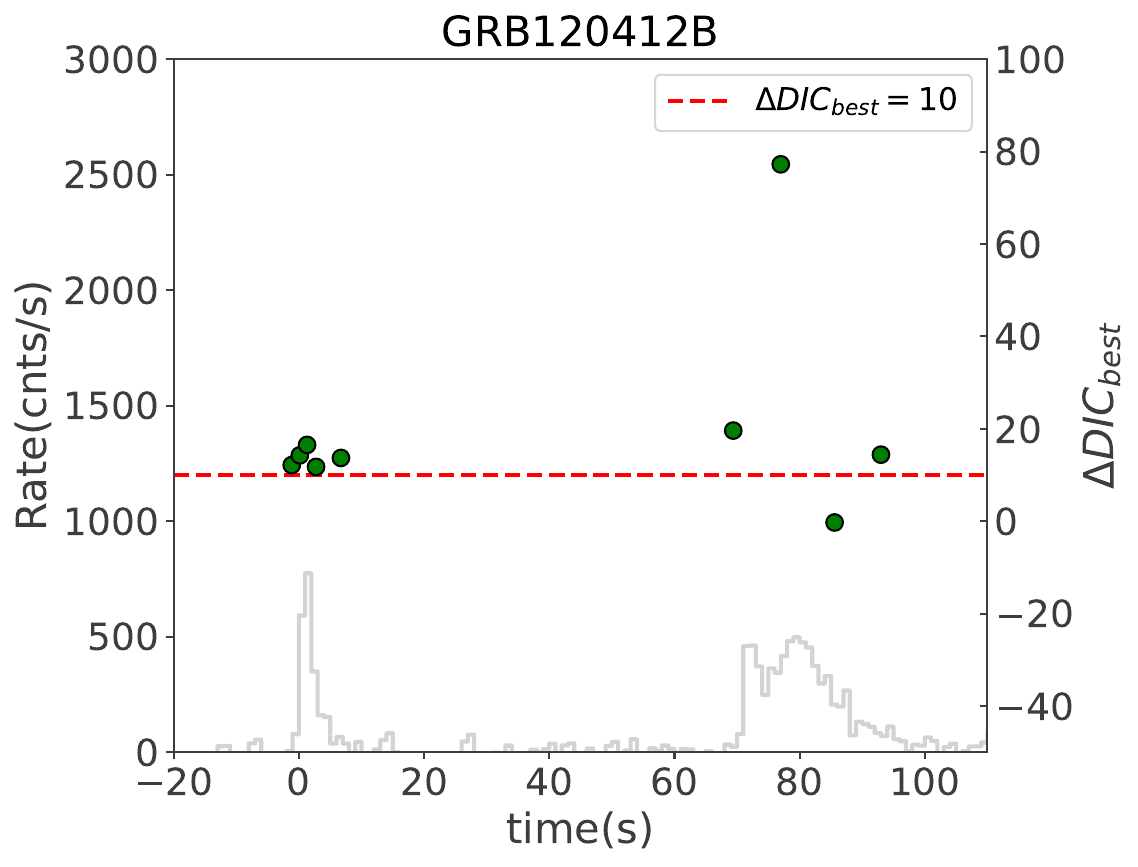}
\includegraphics [width=5cm,height=4cm]{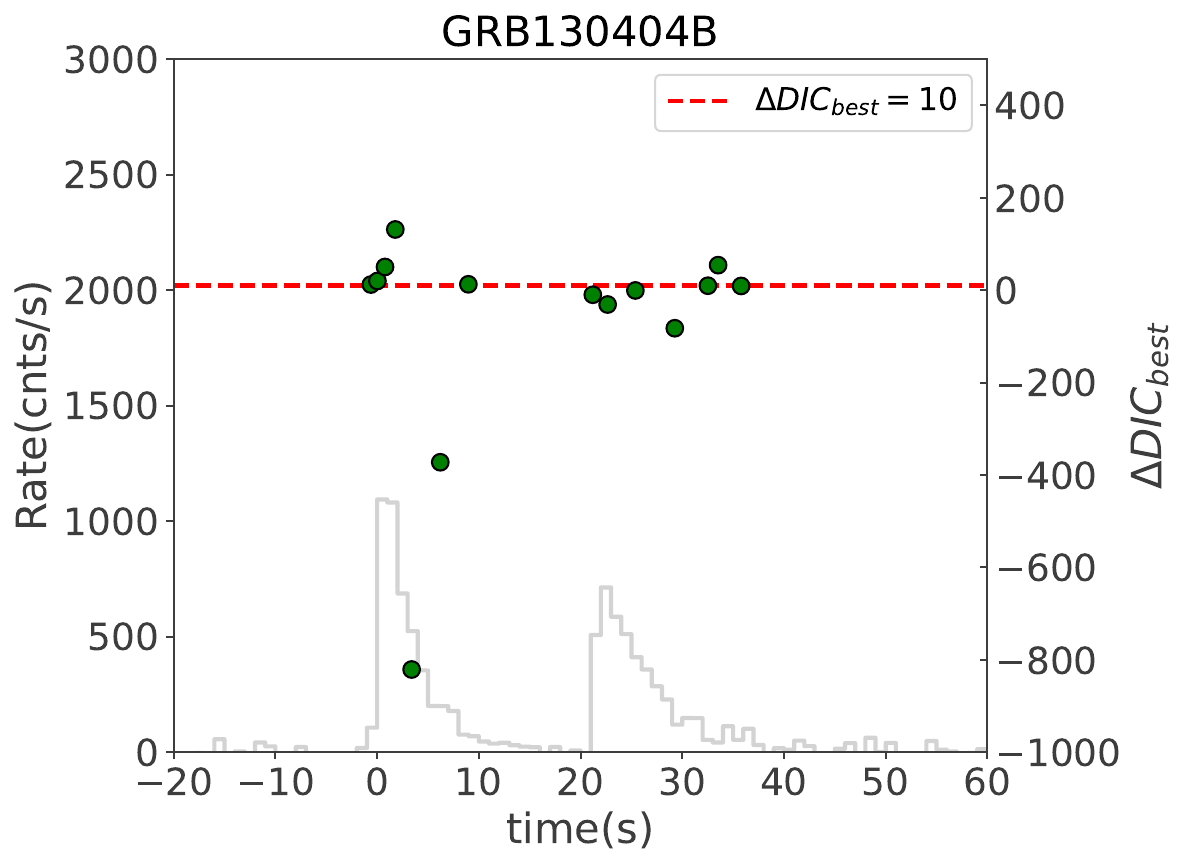}
\includegraphics [width=5cm,height=4cm]{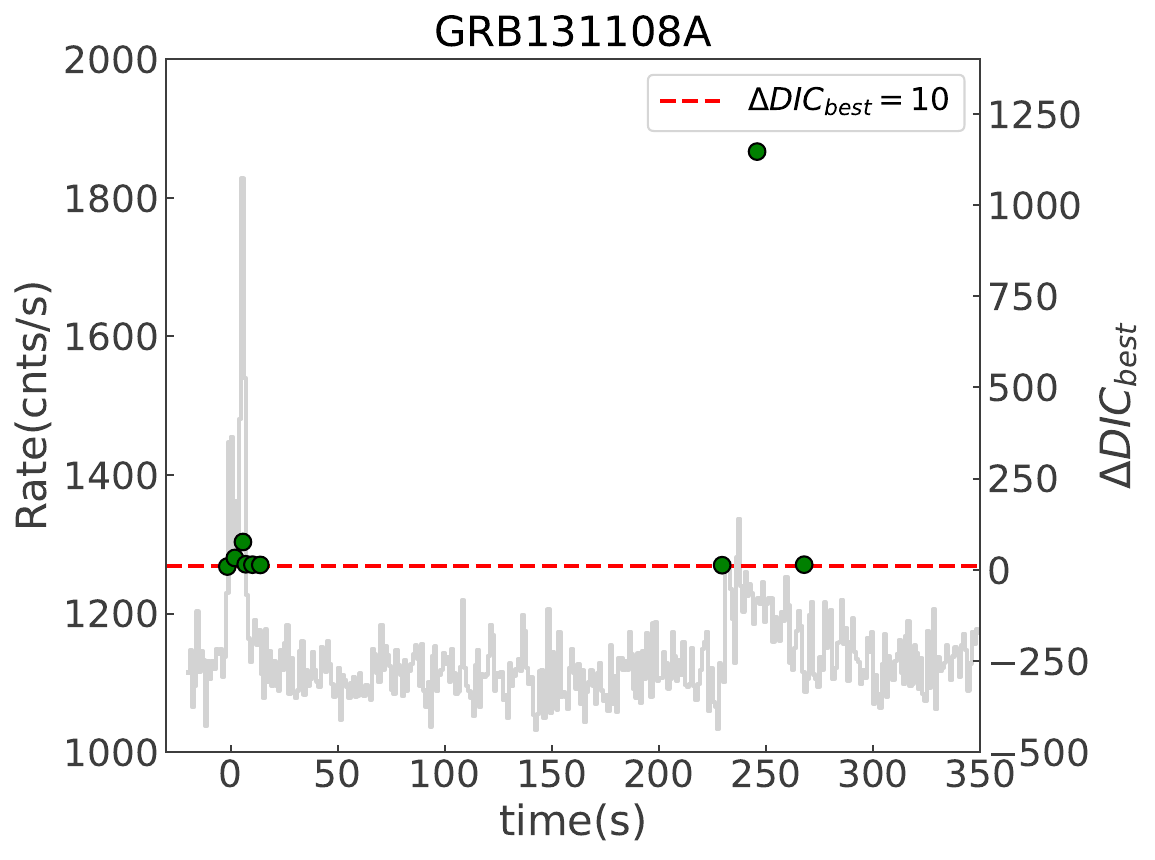}
\includegraphics [width=5cm,height=4cm]{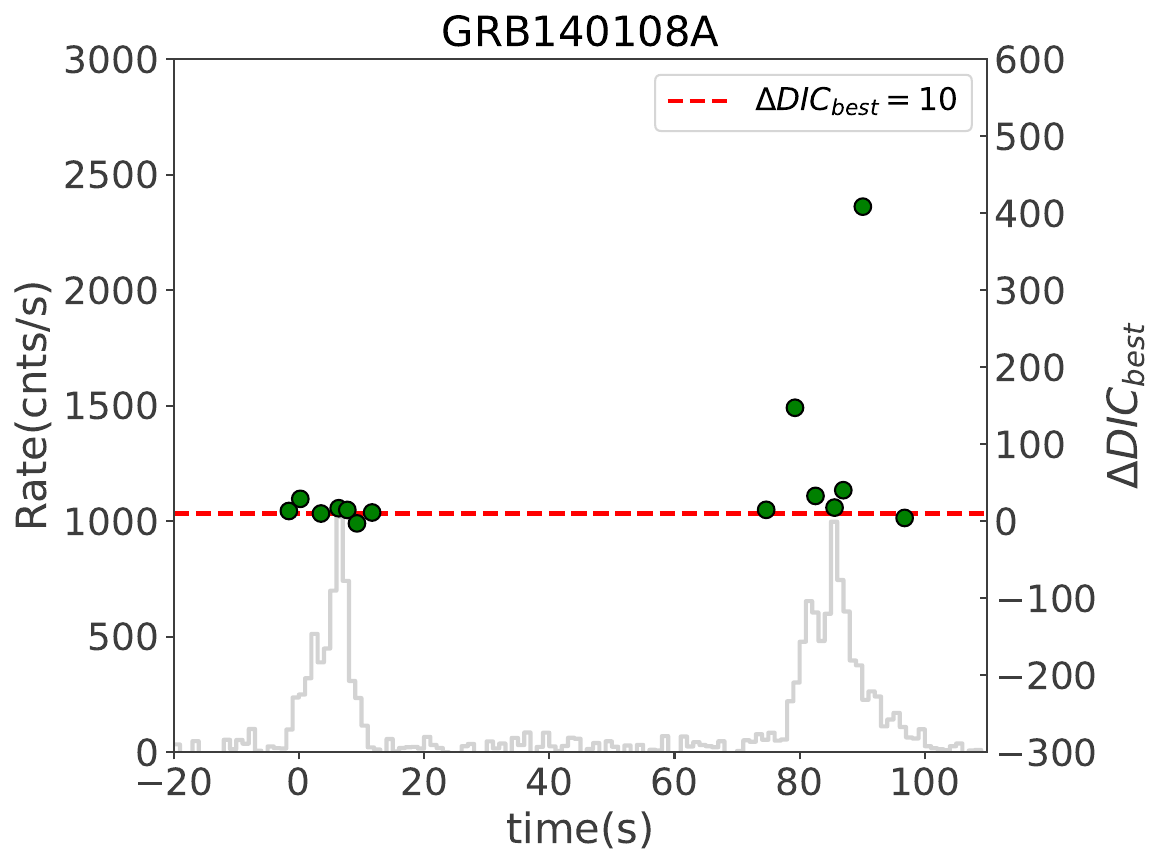}
\includegraphics [width=5cm,height=4cm]{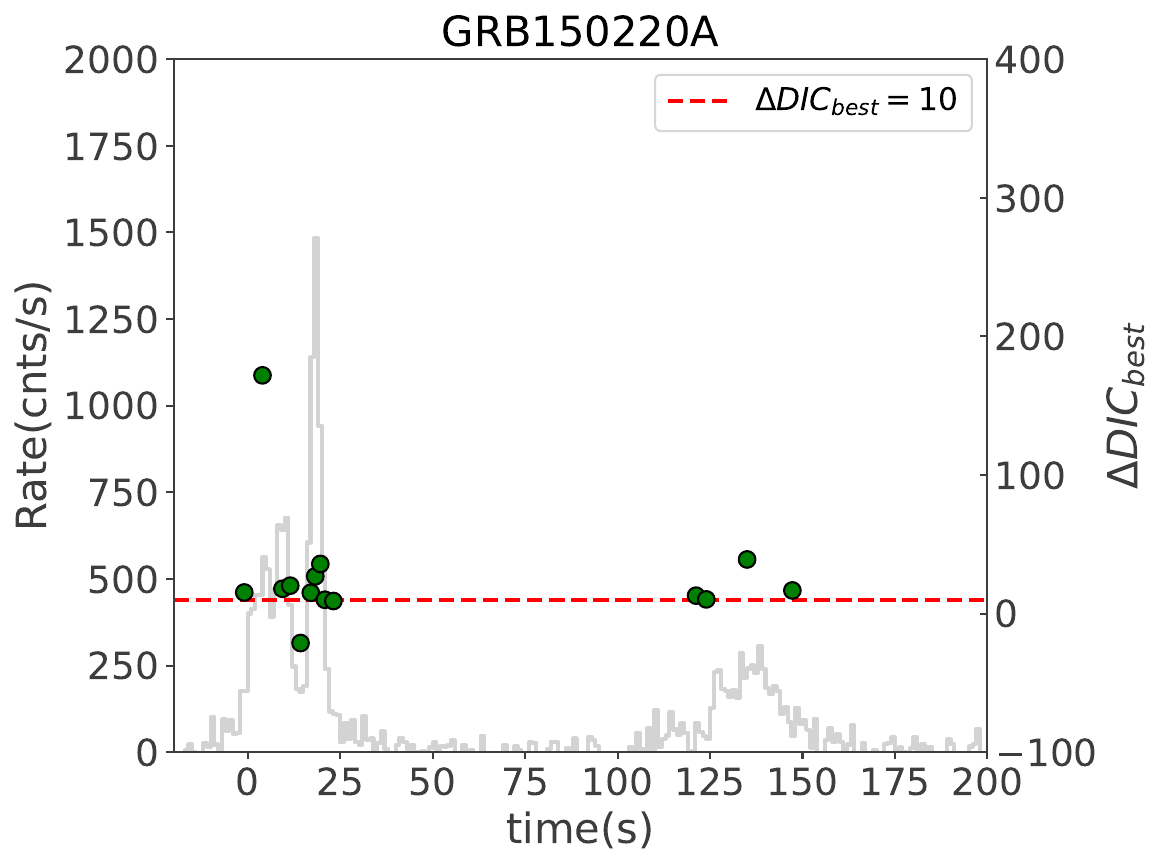}
\includegraphics [width=5cm,height=4cm]{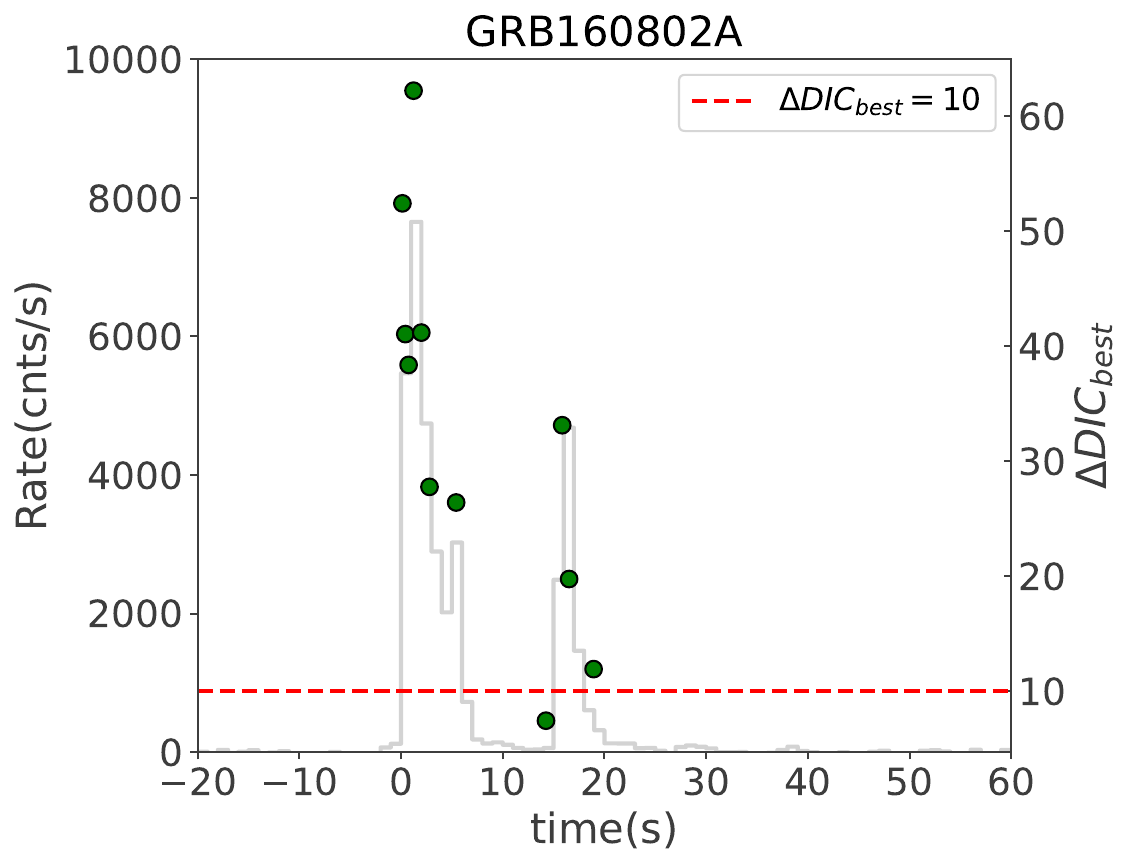}
\includegraphics [width=5cm,height=4cm]{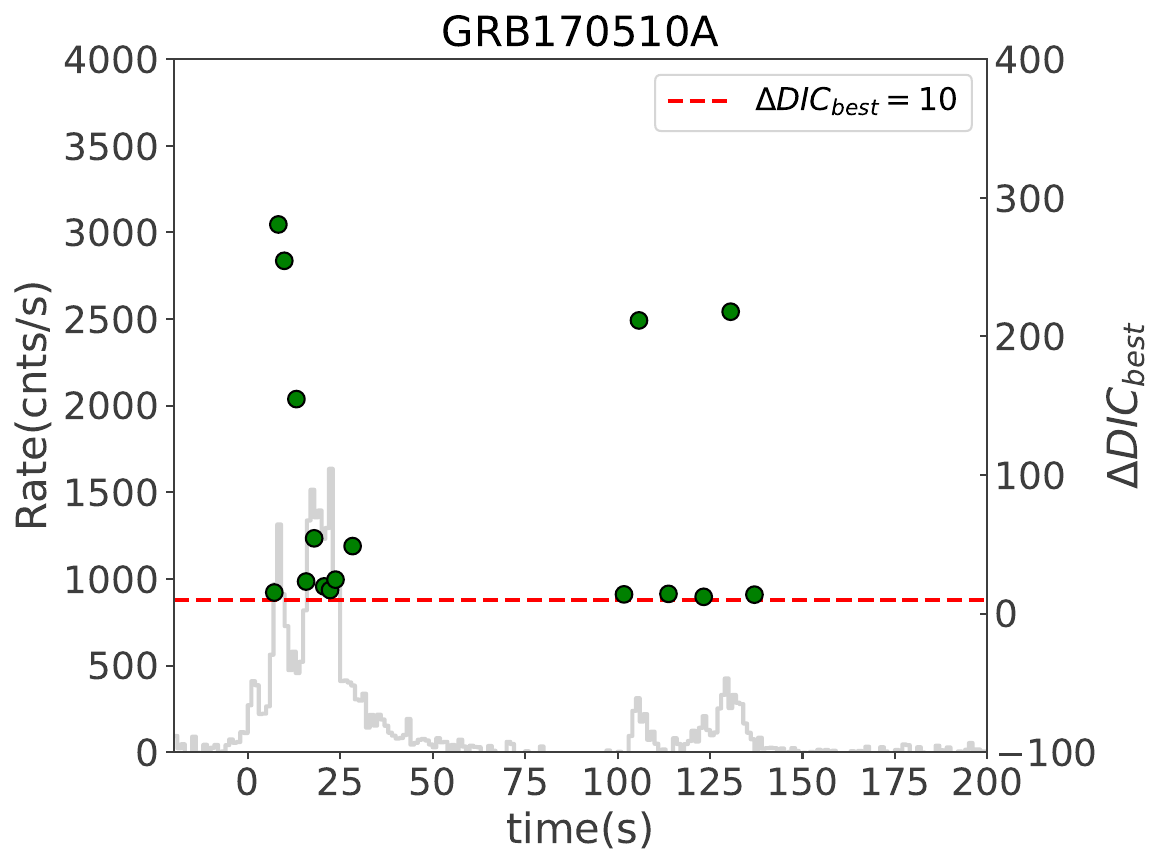}
\includegraphics [width=5cm,height=4cm]{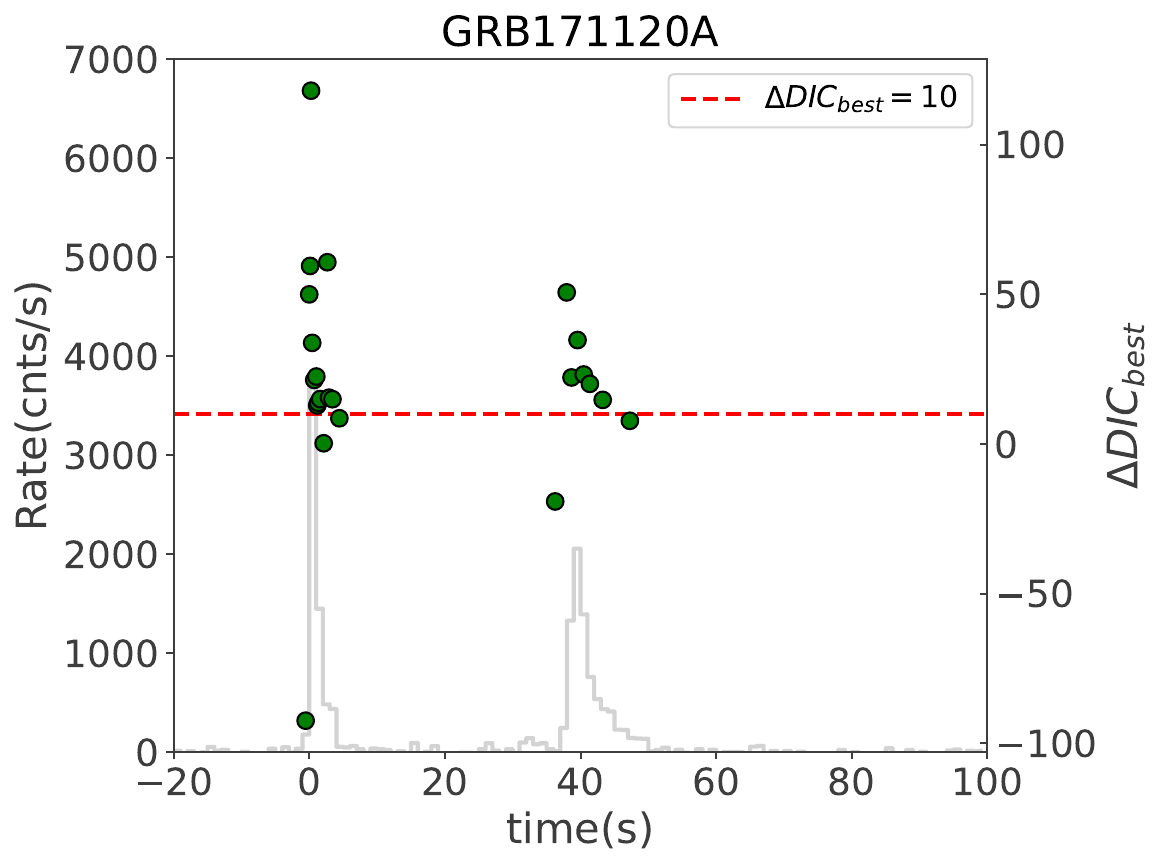}
\includegraphics [width=5cm,height=4cm]{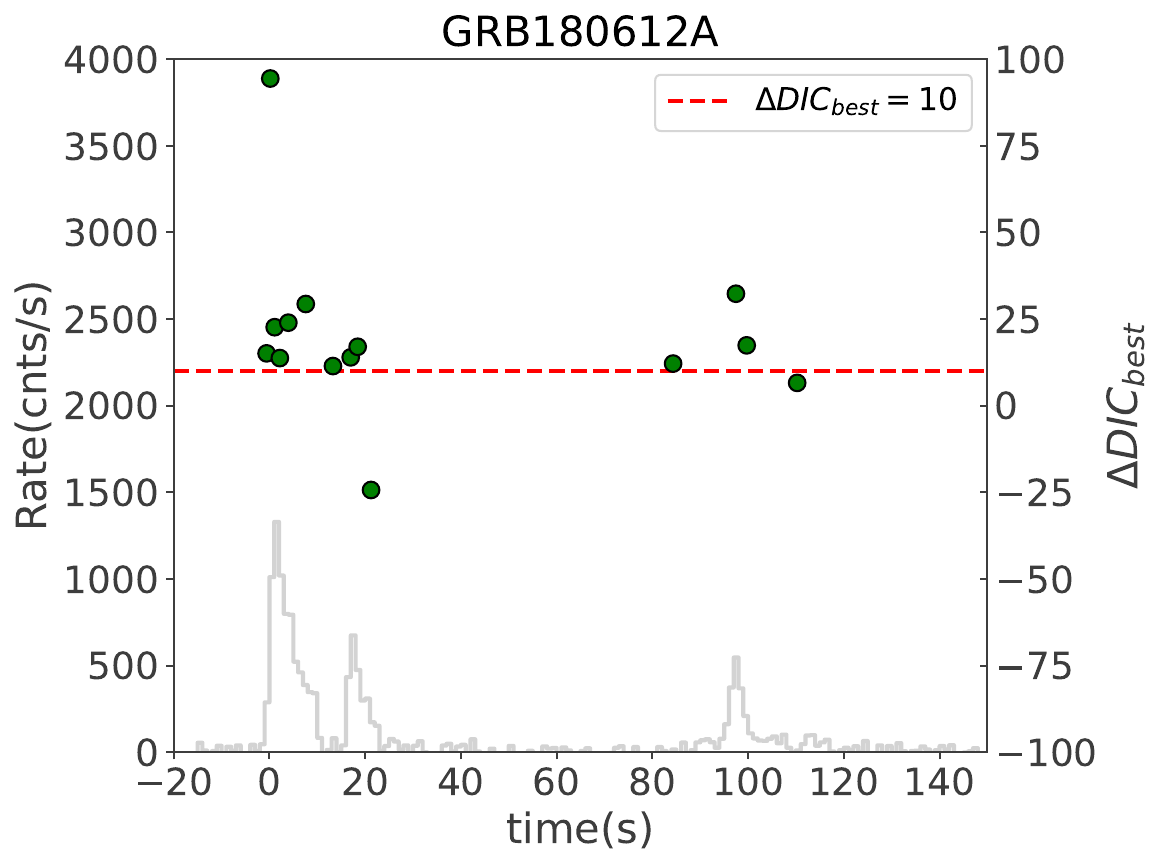}
\includegraphics [width=5cm,height=4cm]{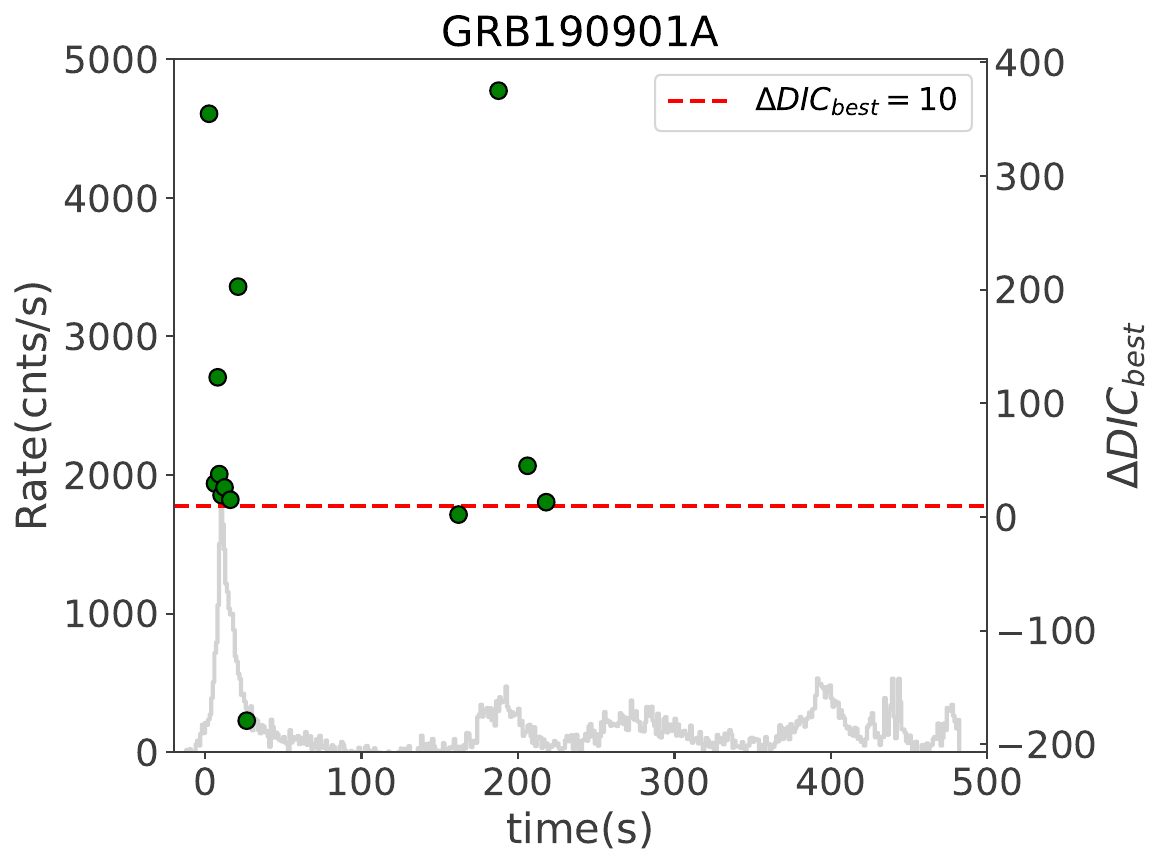}
\includegraphics [width=5cm,height=4cm]{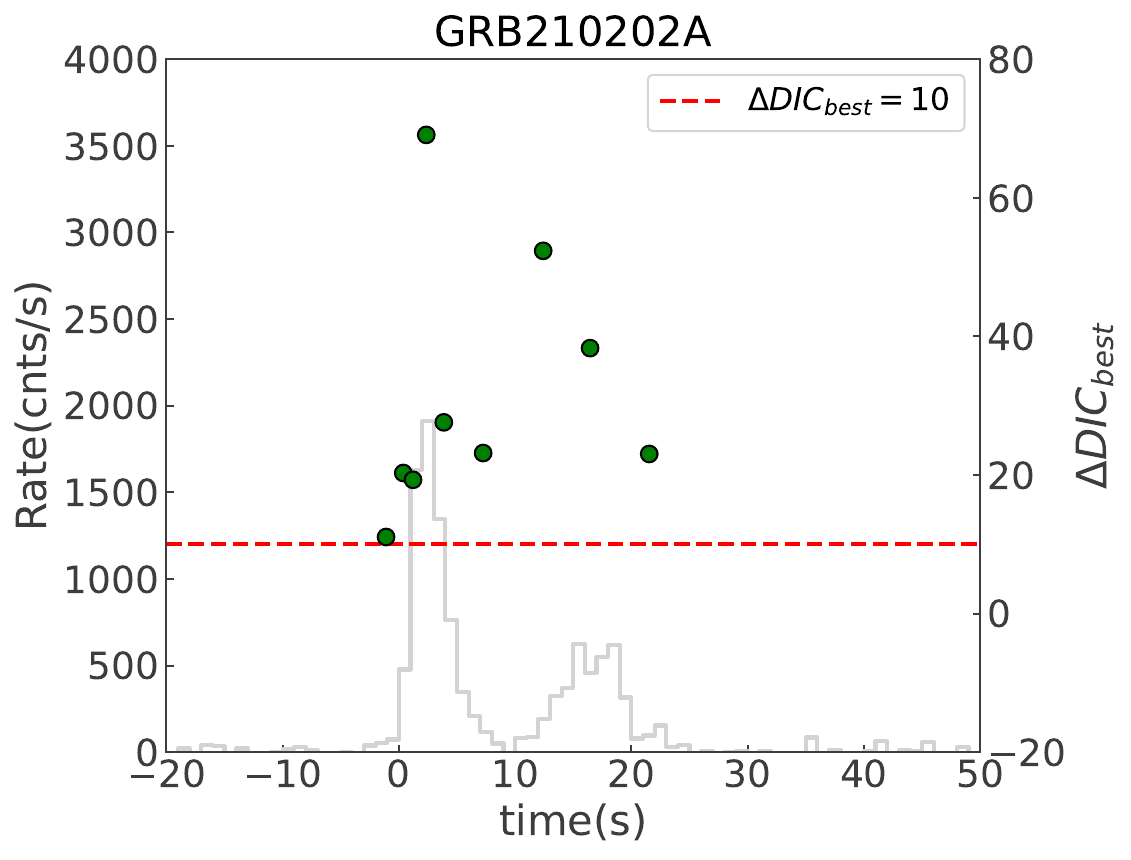}
\includegraphics [width=5cm,height=4cm]{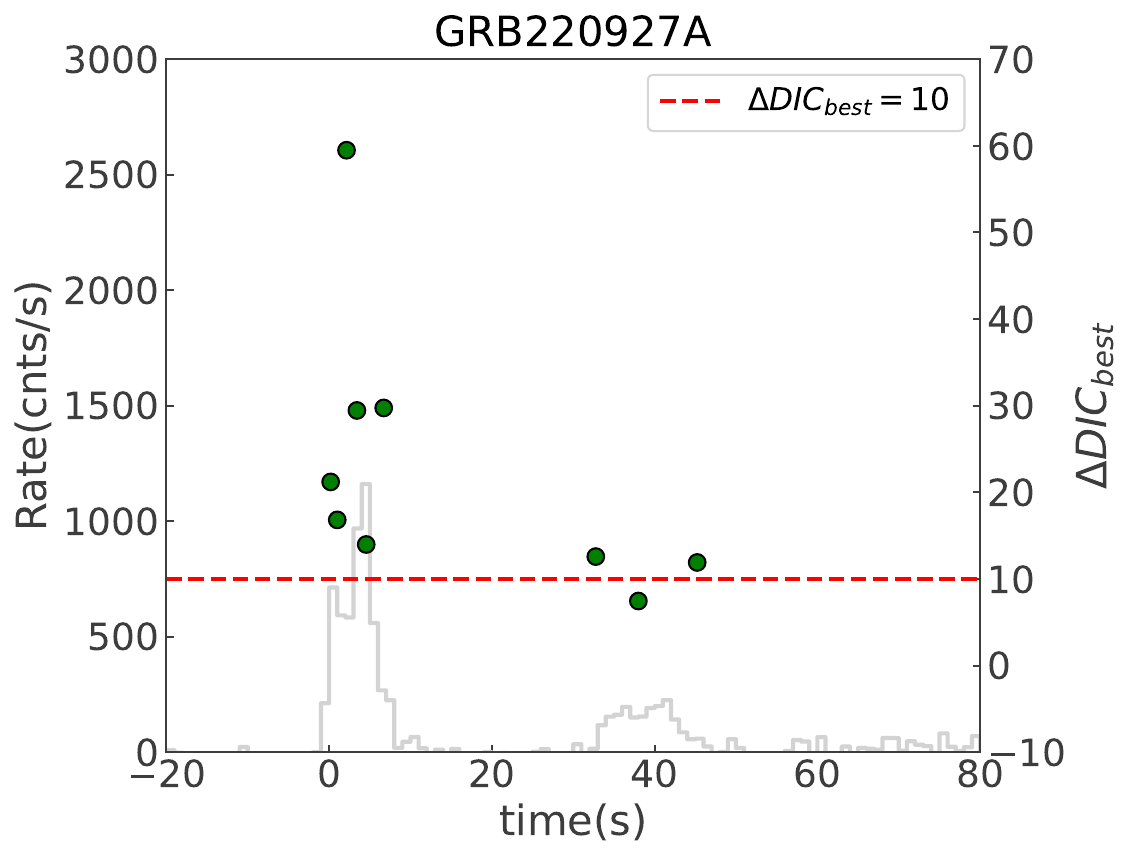}
\includegraphics [width=5cm,height=4cm]{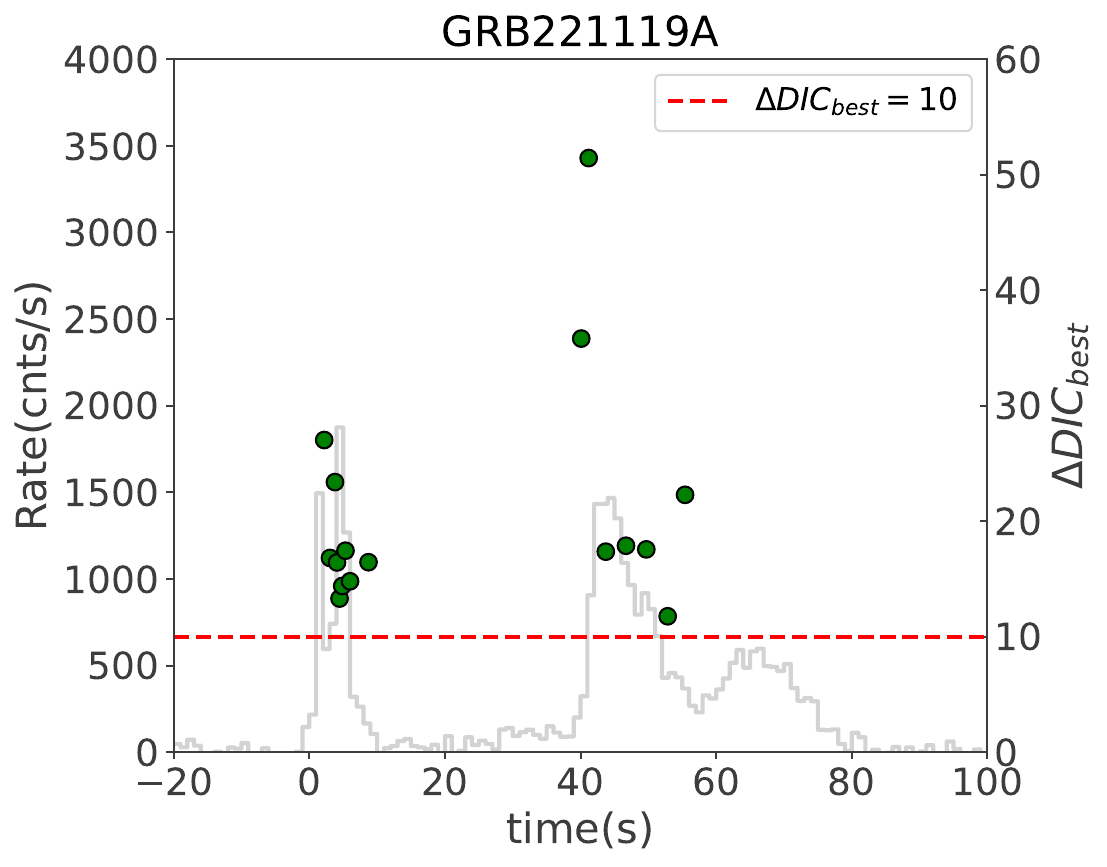}
\includegraphics [width=5cm,height=4cm]{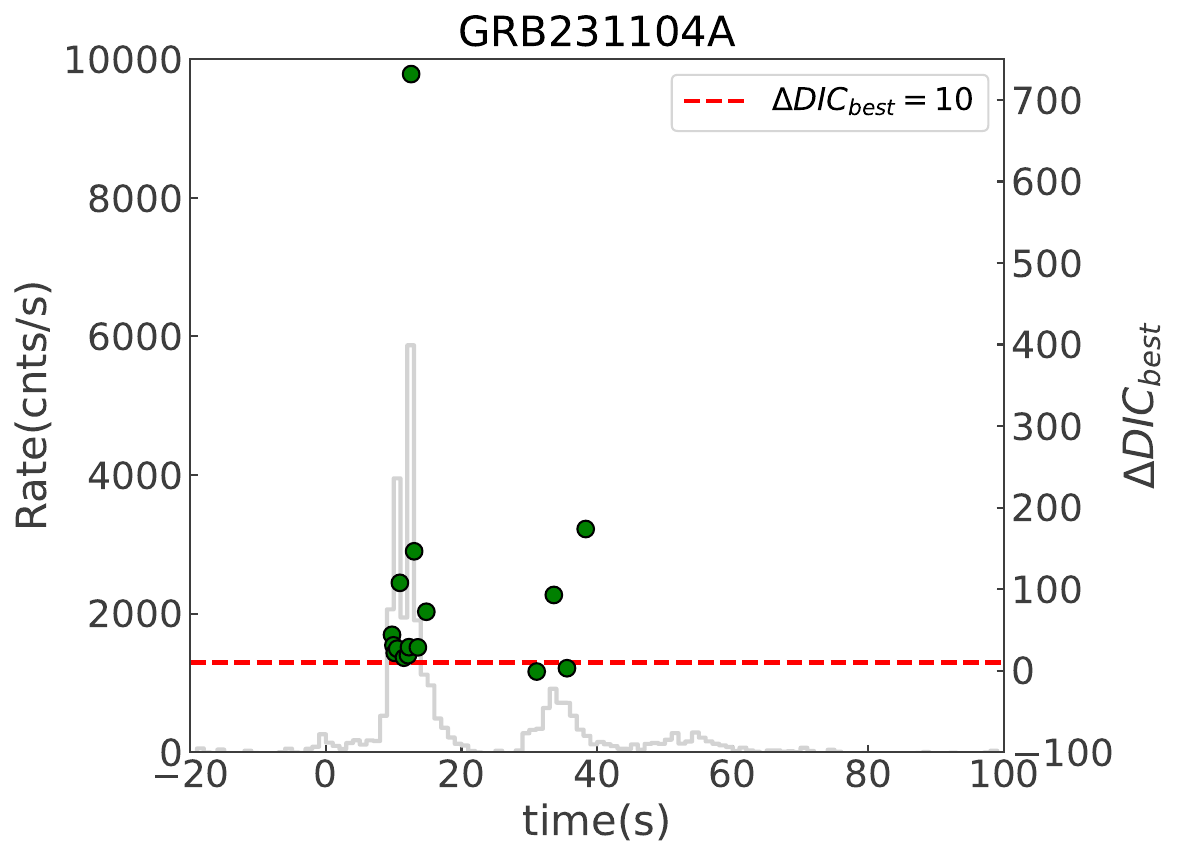}
\includegraphics [width=5cm,height=4cm]{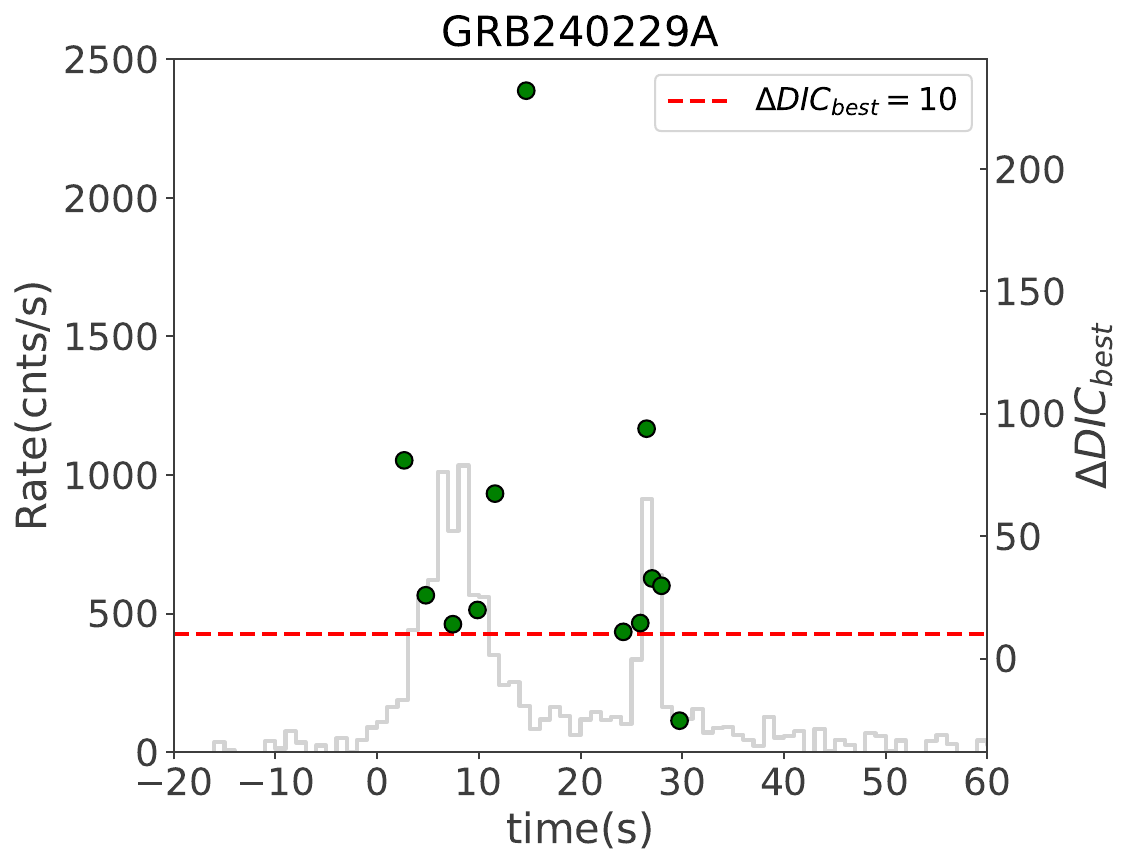}
   \figcaption{The evolution of $\Delta DIC_{best}$ over time. The red dotted line indicates $\Delta DIC_{best} = 10$. If $\Delta DIC_{best} > 10$, it indicates strong evidence that the time bin contains thermal components. \label{fig 9}}

\end{figure}

In this Appendix, we provide additional figures, including the time evolution of $\Delta DIC_{best}$ (Figure \ref{fig 9}), $\alpha$ (Figure \ref{fig 10}), and $E_{p}$ (Figure \ref{fig 11}); the correlations between the spectral parameters of the main burst and the second burst ($\alpha$, $E_{p}$, F) (Figures \ref{fig 12}, \ref{fig 13}, and \ref{fig 14}); as well as the time evolution of the photospheric radiation parameters (Figures \ref{fig 15}, \ref{fig 16}, and \ref{fig 17}).

\setcounter{figure}{9}  
\begin{figure}[H]
  \centering
  \includegraphics [width=8cm,height=4cm]{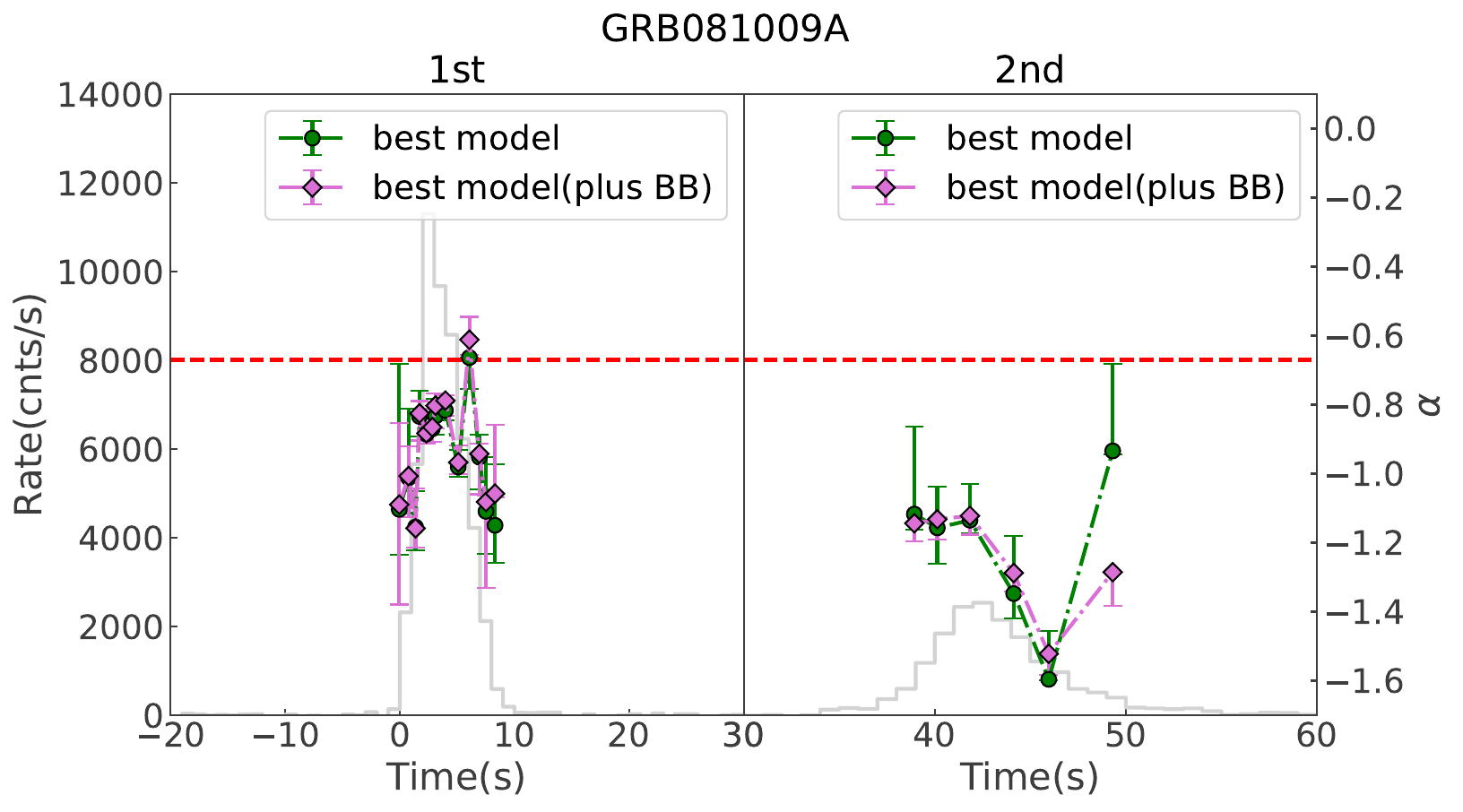} 
  \includegraphics [width=8cm,height=4cm]{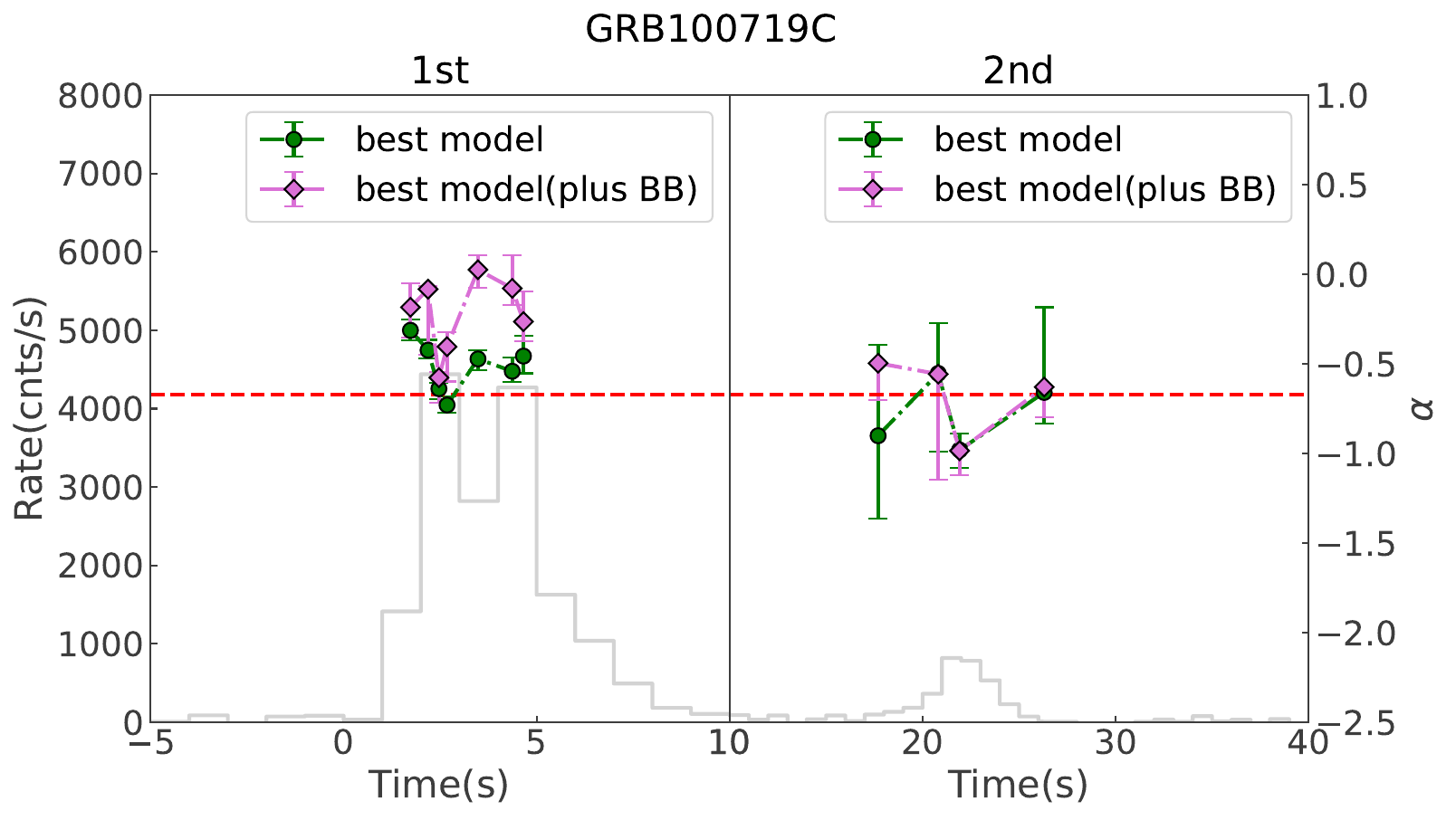} 
  \includegraphics [width=8cm,height=4cm]{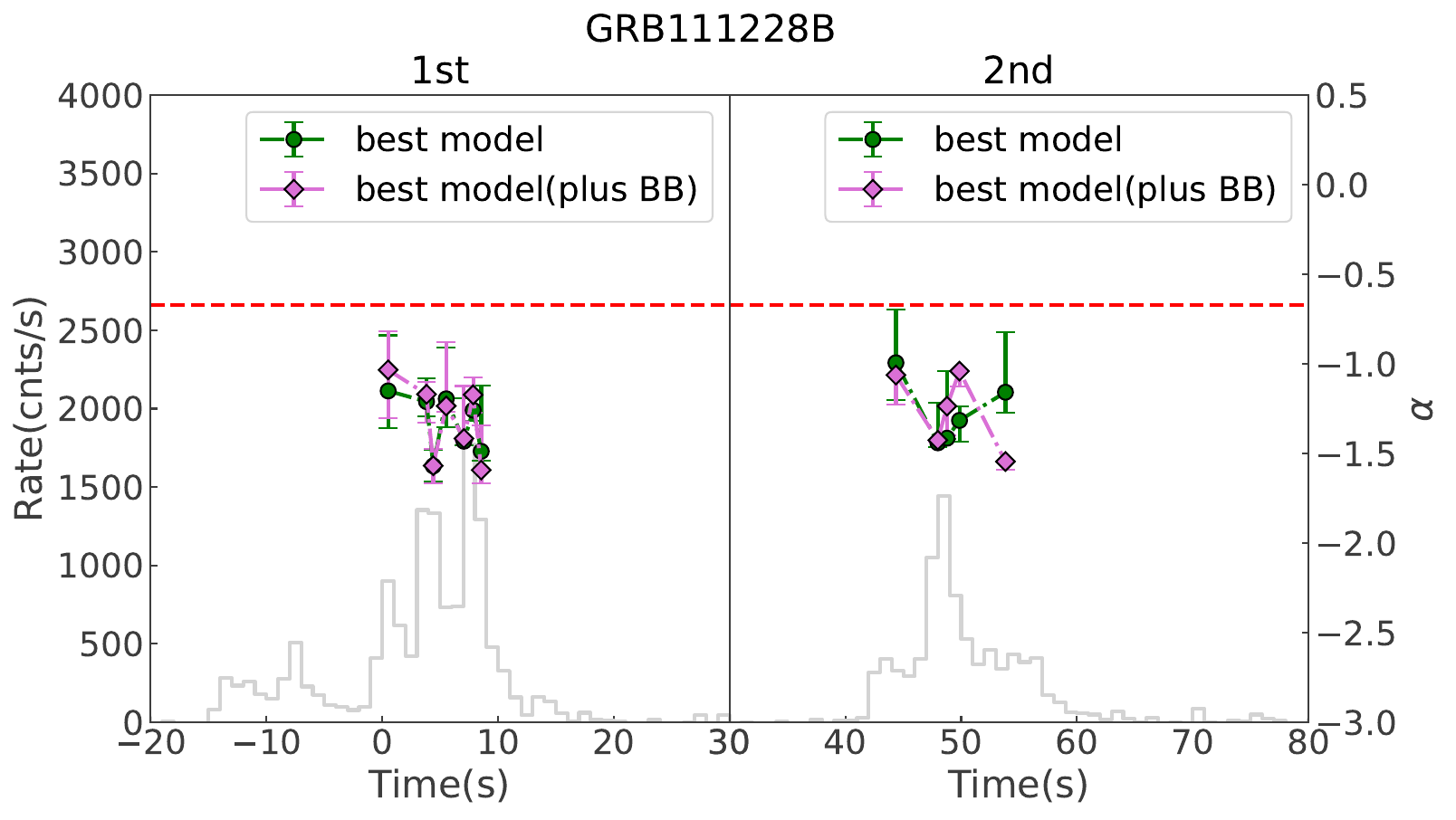} 
  \includegraphics [width=8cm,height=4cm]{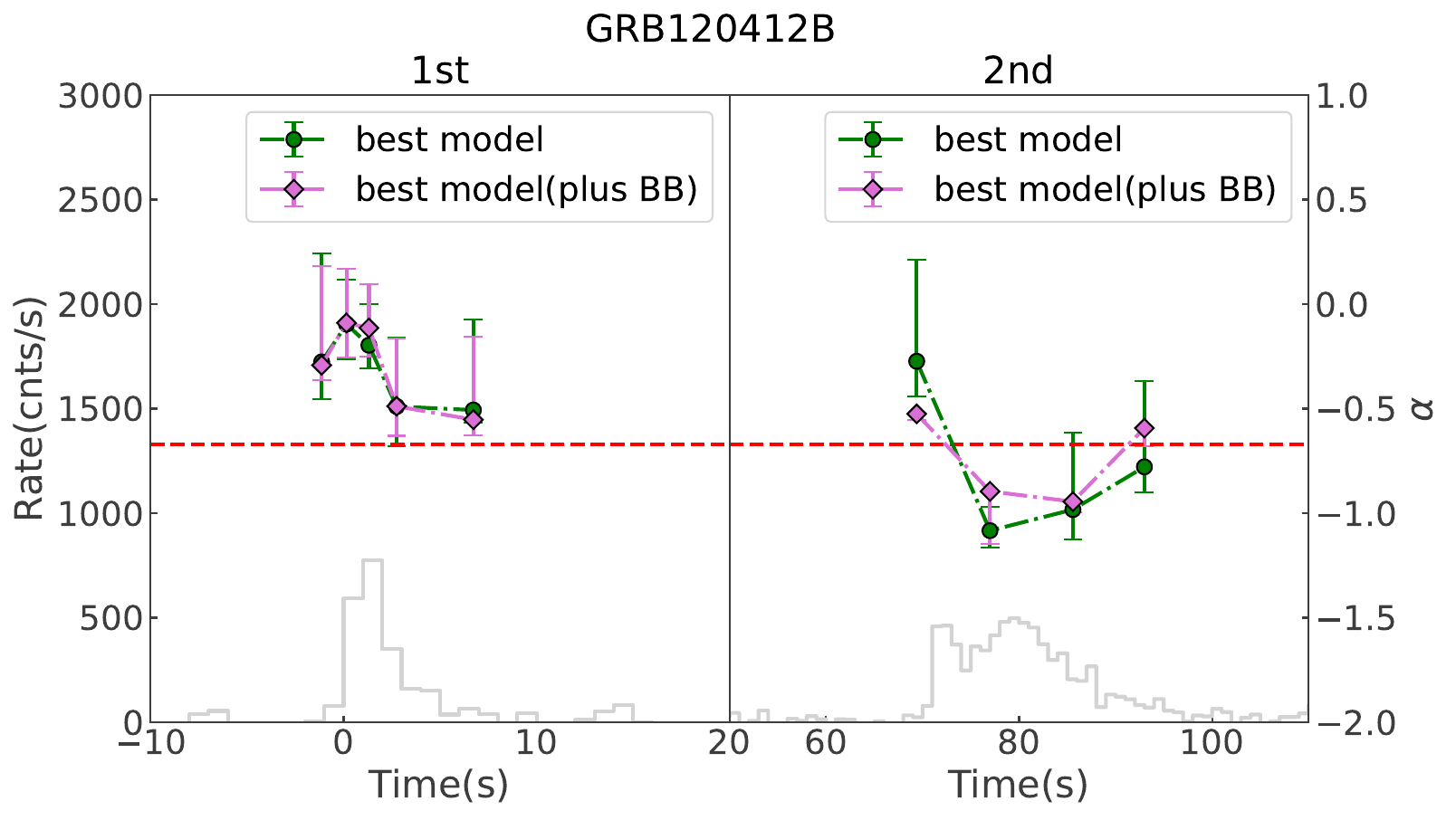}
  \includegraphics [width=8cm,height=4cm]{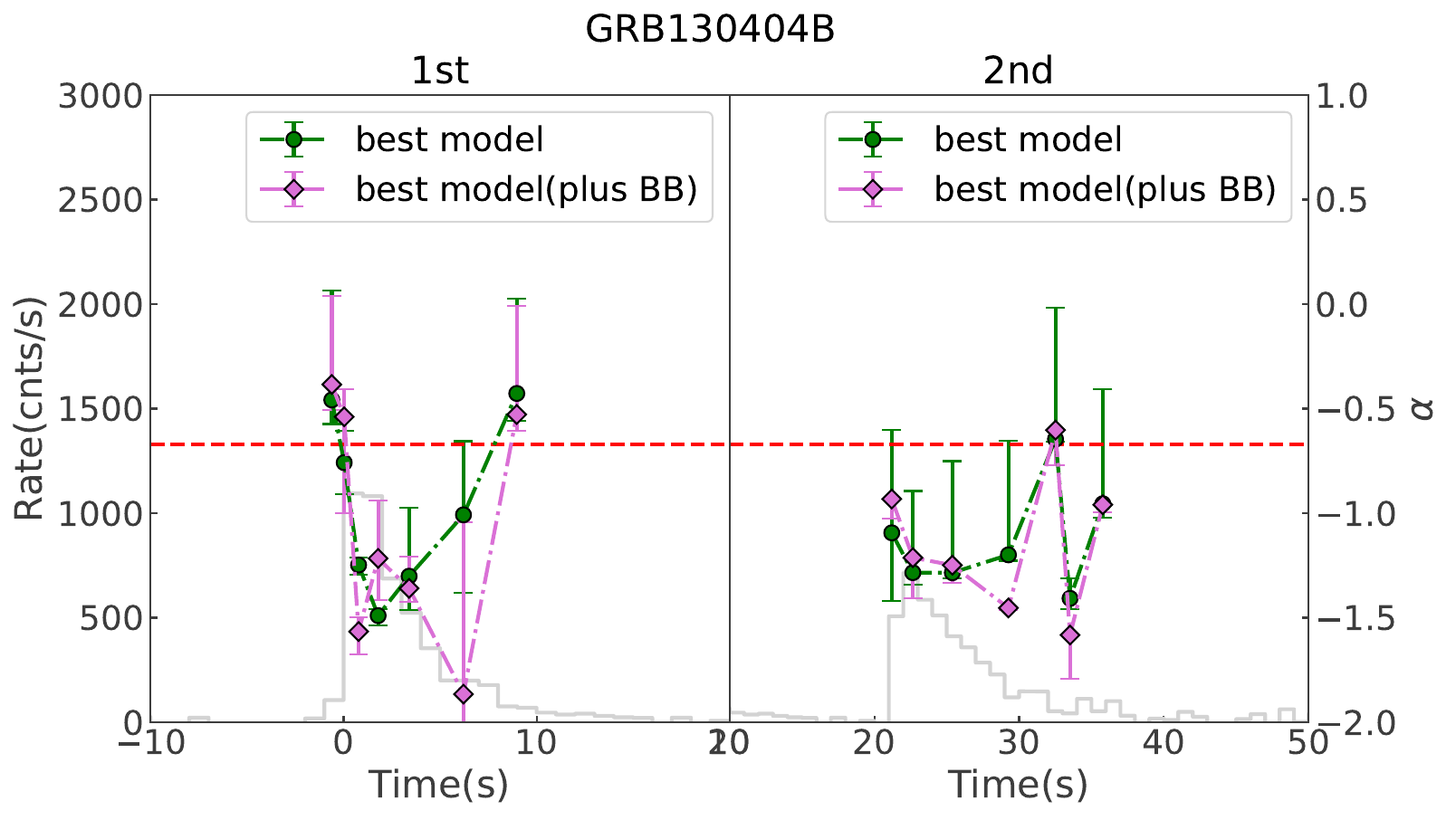}
  \includegraphics [width=8cm,height=4cm]{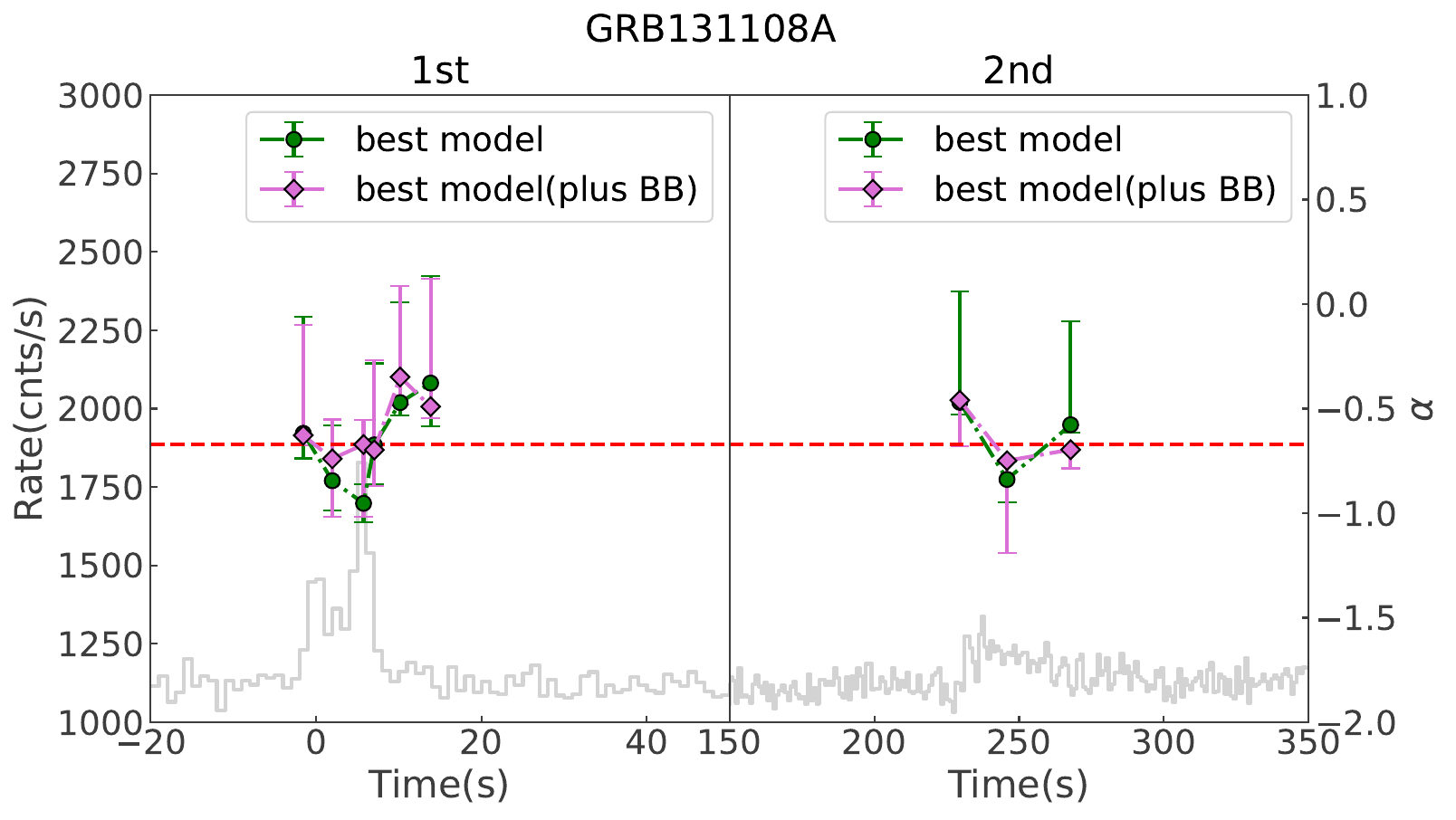}
  \includegraphics [width=8cm,height=4cm]{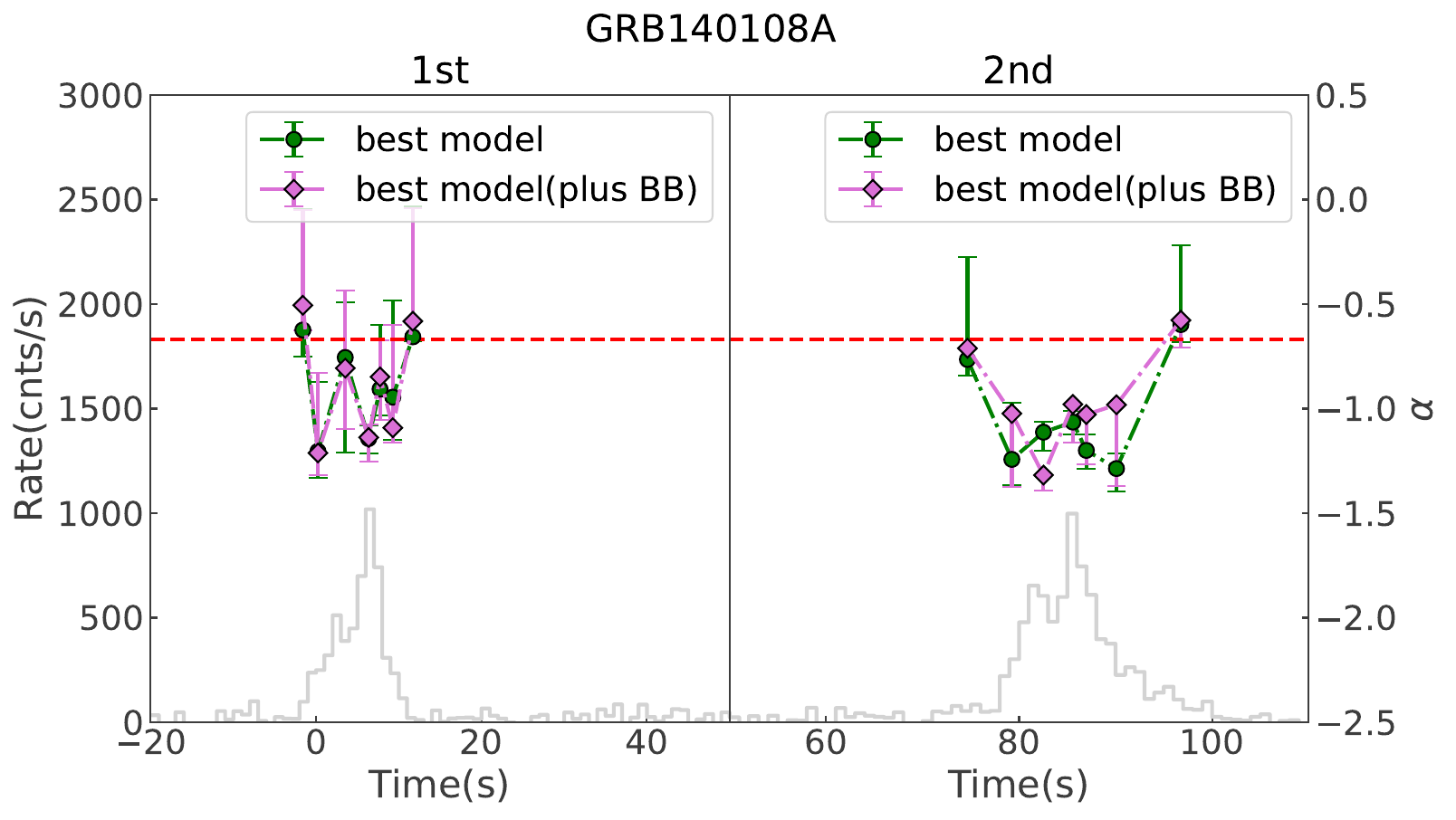}
  \includegraphics [width=8cm,height=4cm]{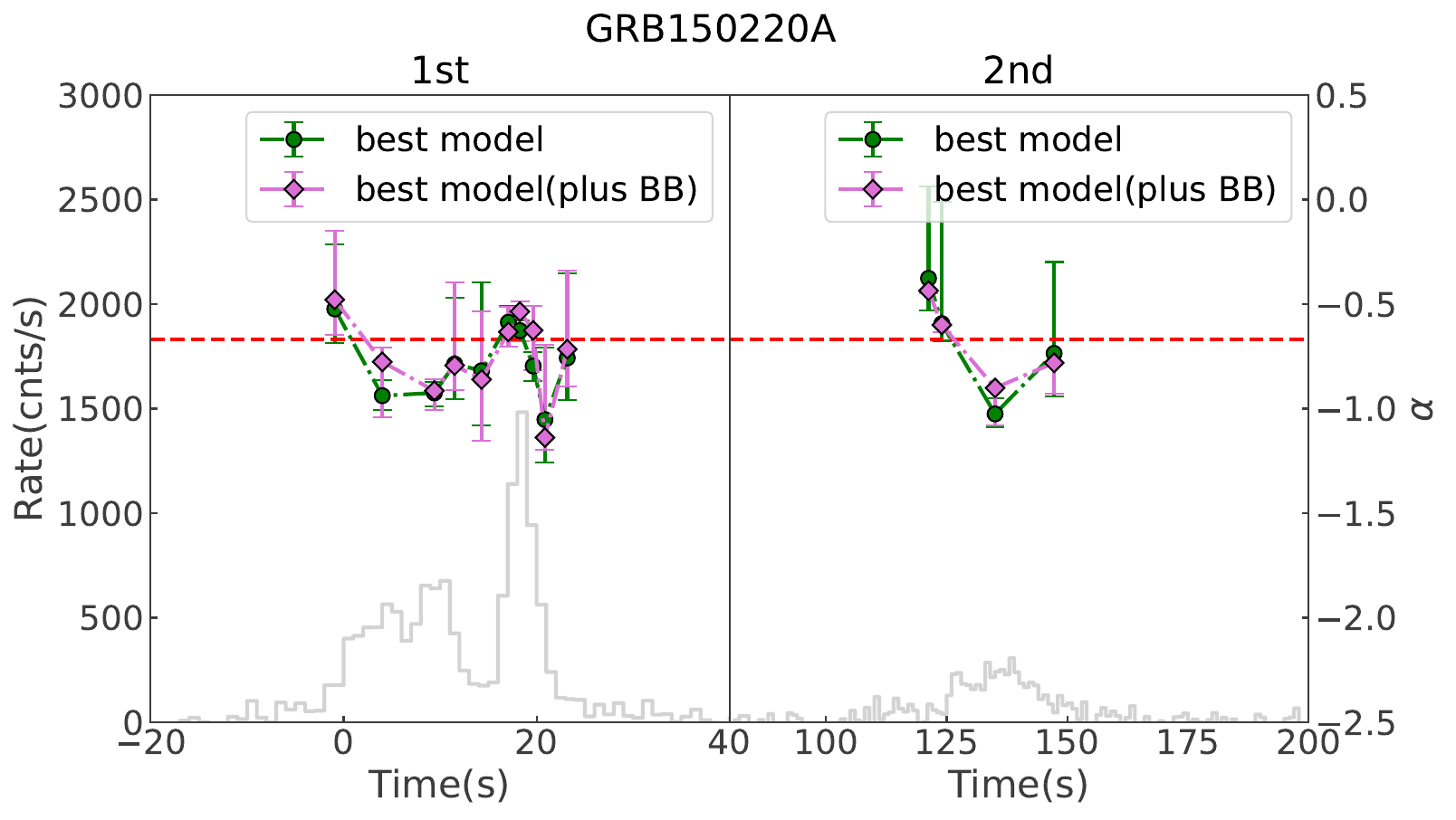}
  \includegraphics [width=8cm,height=4cm]{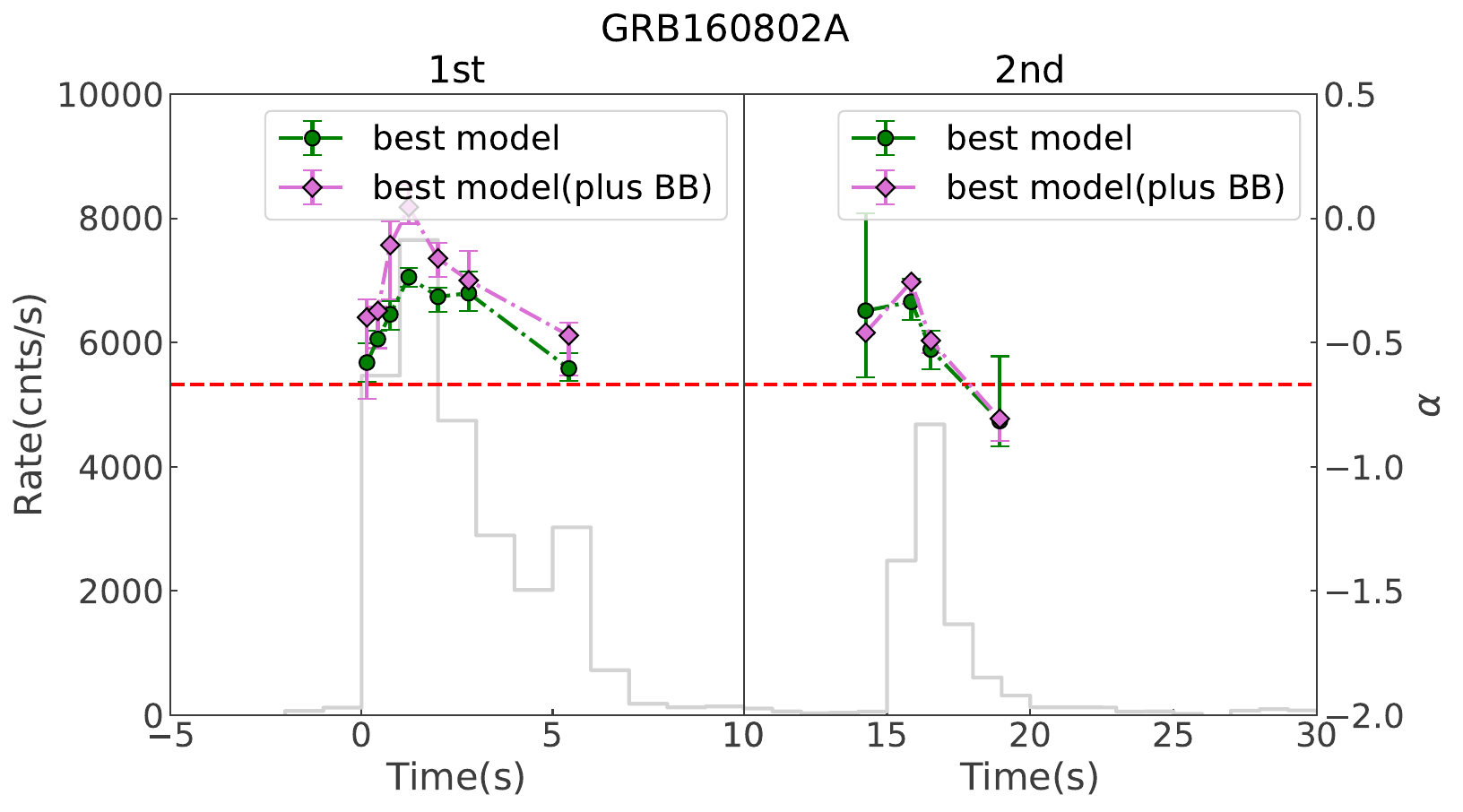}
  \includegraphics [width=8cm,height=4cm]{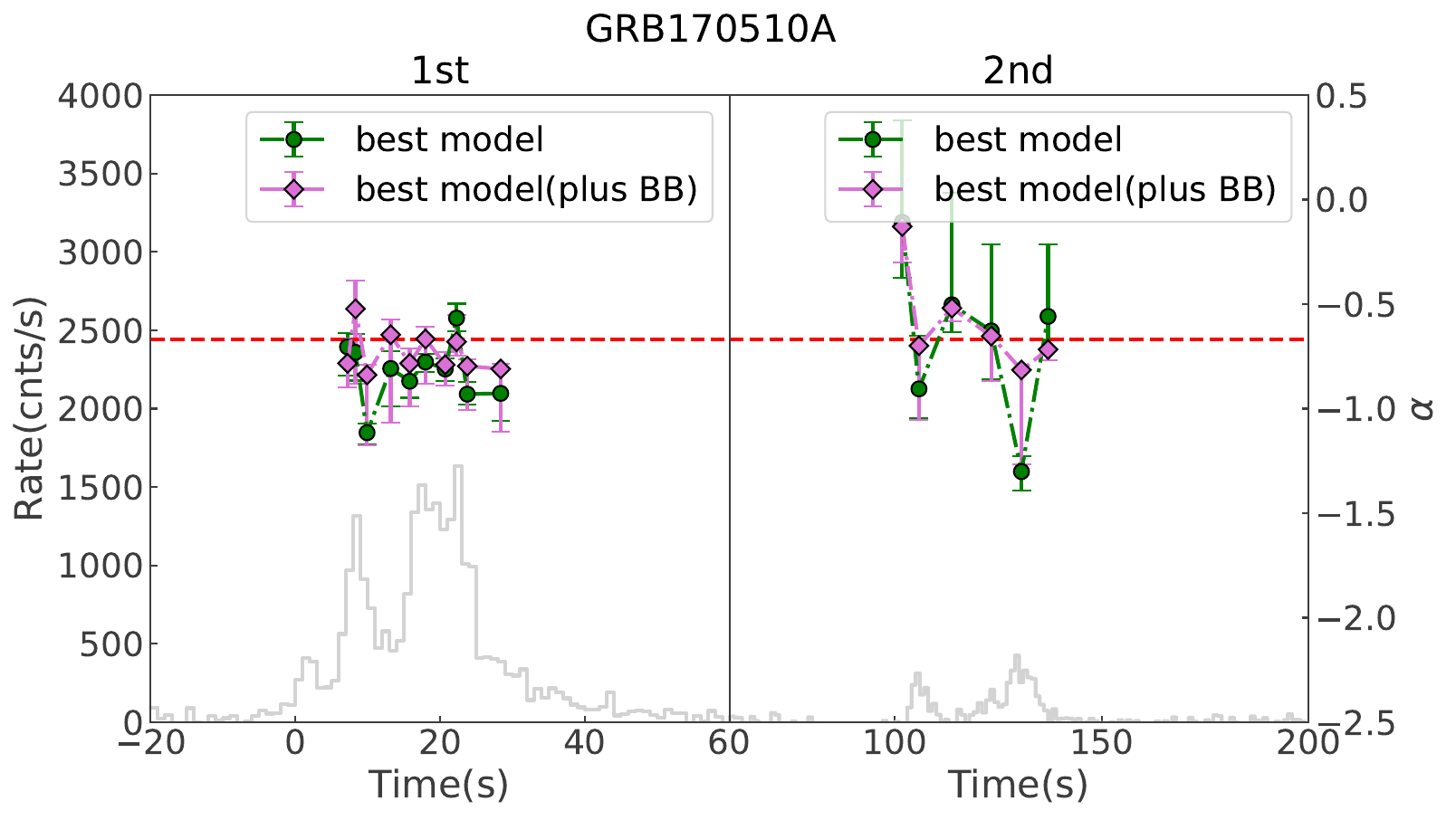}
  \caption{The evolution of the spectral parameter $\alpha$ over time, fitted with the best model. The green and purple data points represent the best model and the best model $+$ BB. The red dashed line indicates $\alpha = -0.67$. ``1st'' represents the main burst, and ``2nd'' represents the second burst. \label{fig 10}}

\end{figure}

\setcounter{figure}{9}  
\begin{figure}[H]

  \centering
  \includegraphics [width=8cm,height=4cm]{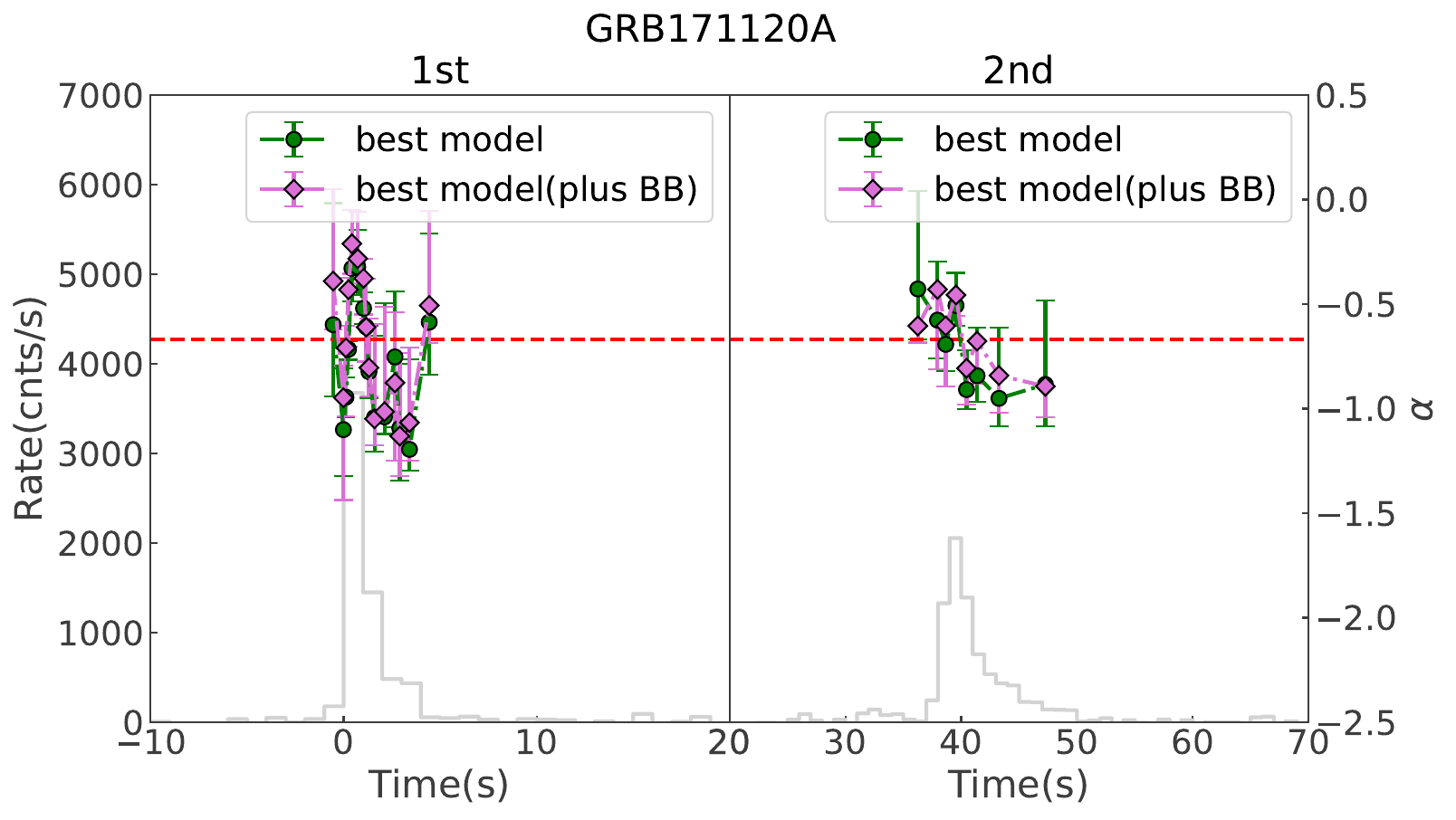}
  \includegraphics [width=8cm,height=4cm]{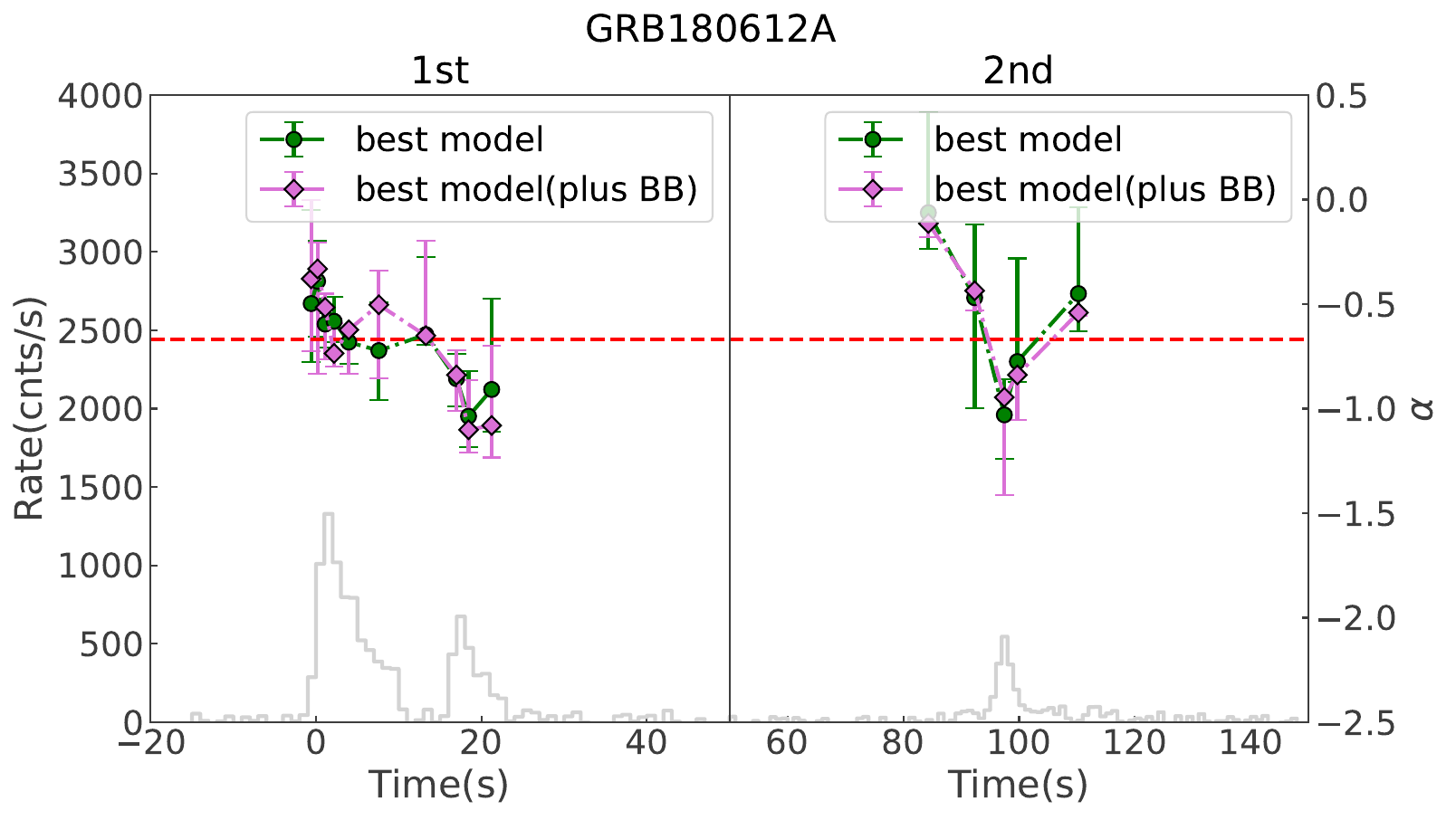}
  \includegraphics [width=8cm,height=4cm]{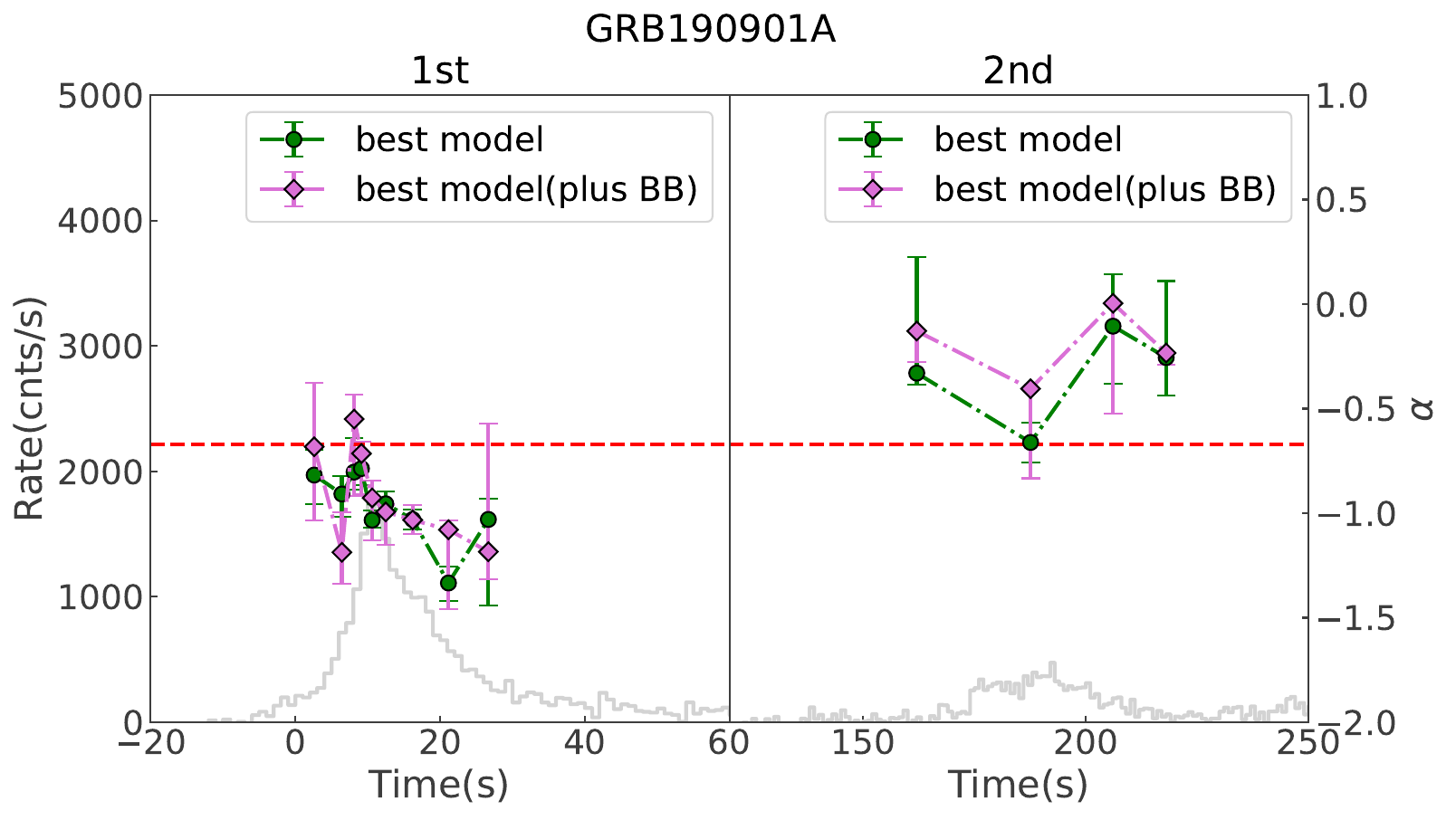}
  \includegraphics [width=8cm,height=4cm]{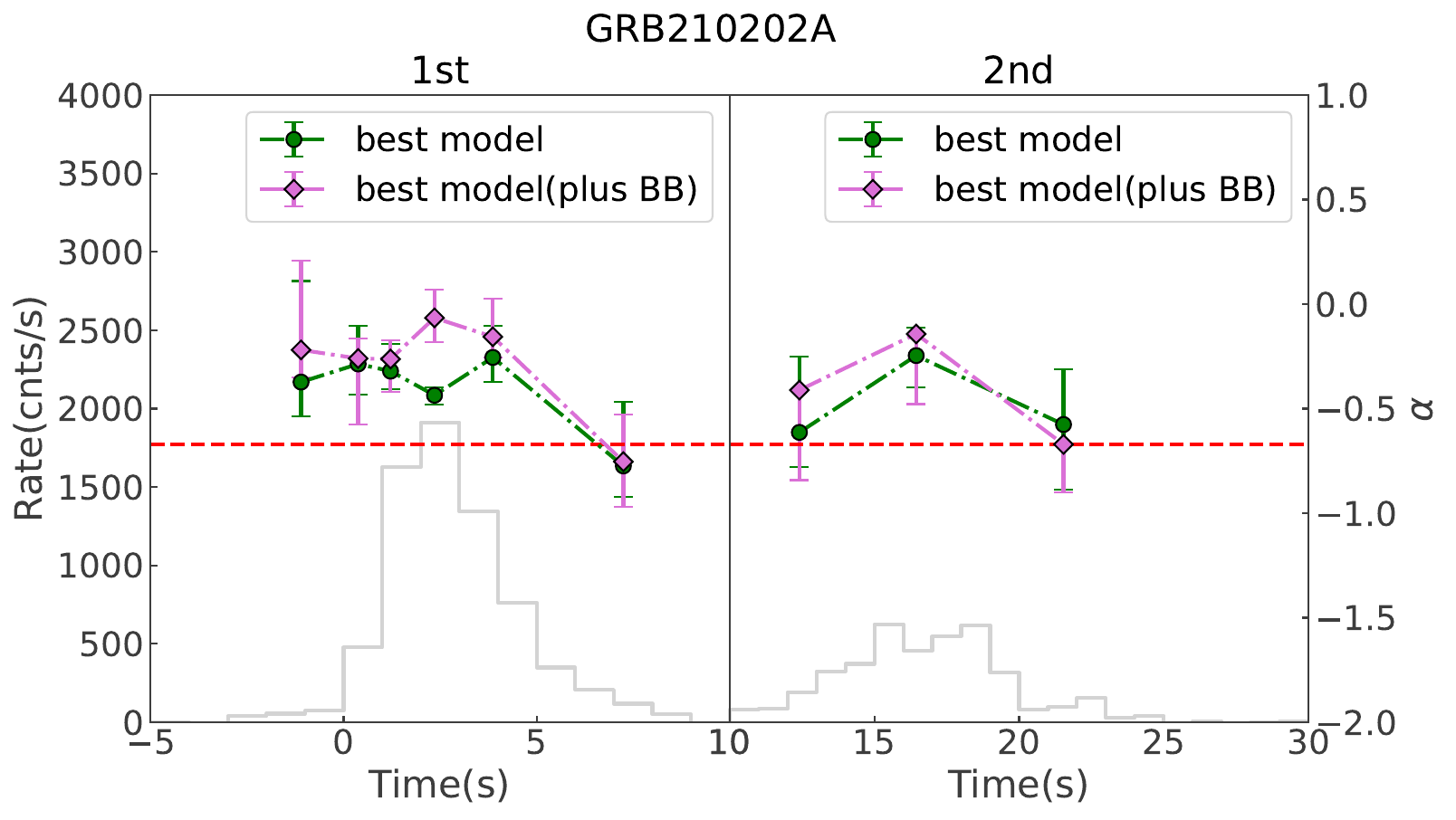}
  \includegraphics [width=8cm,height=4cm]{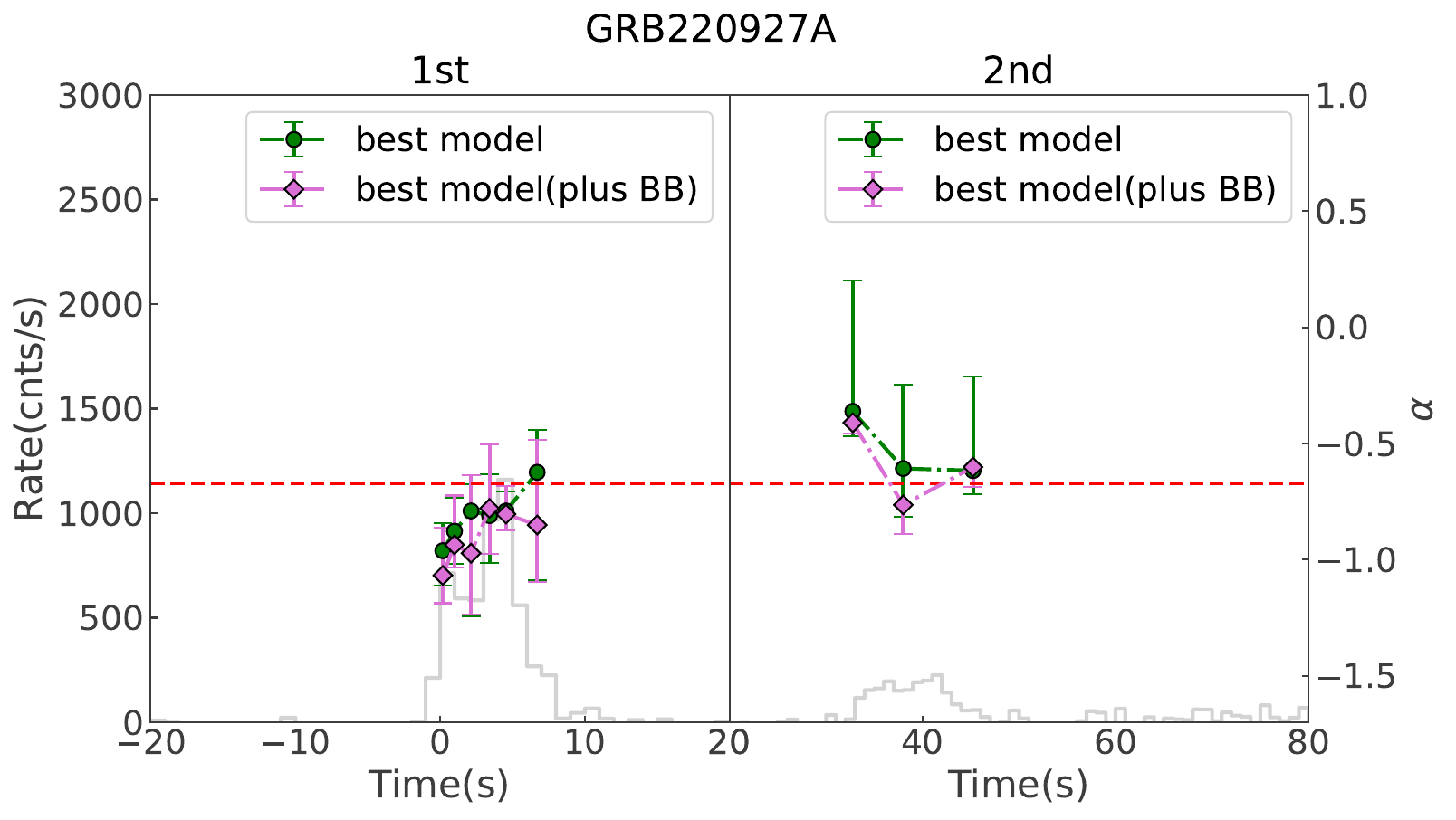}
  \includegraphics [width=8cm,height=4cm]{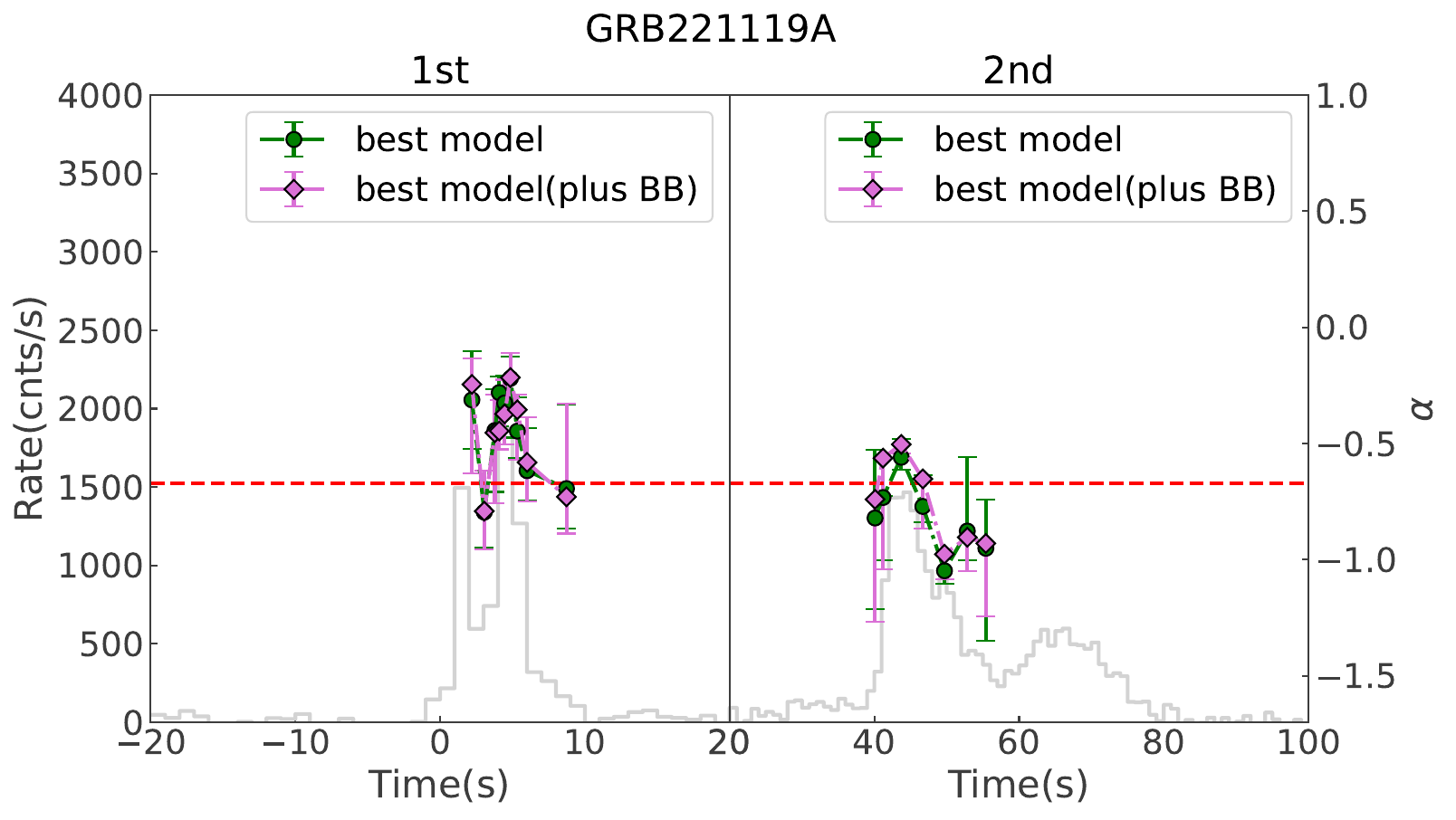}
  \includegraphics [width=8cm,height=4cm]{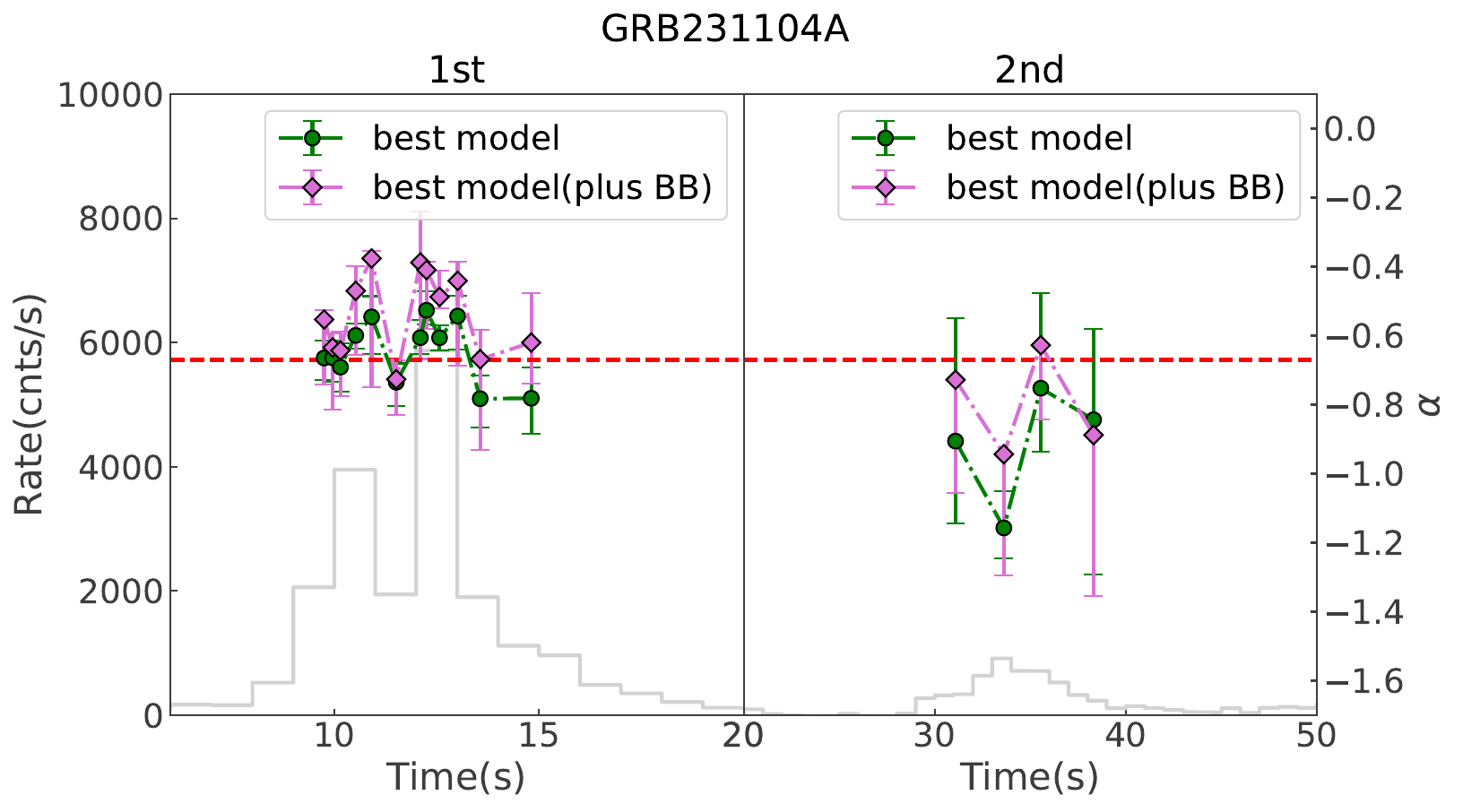}
  \includegraphics [width=8cm,height=4cm]{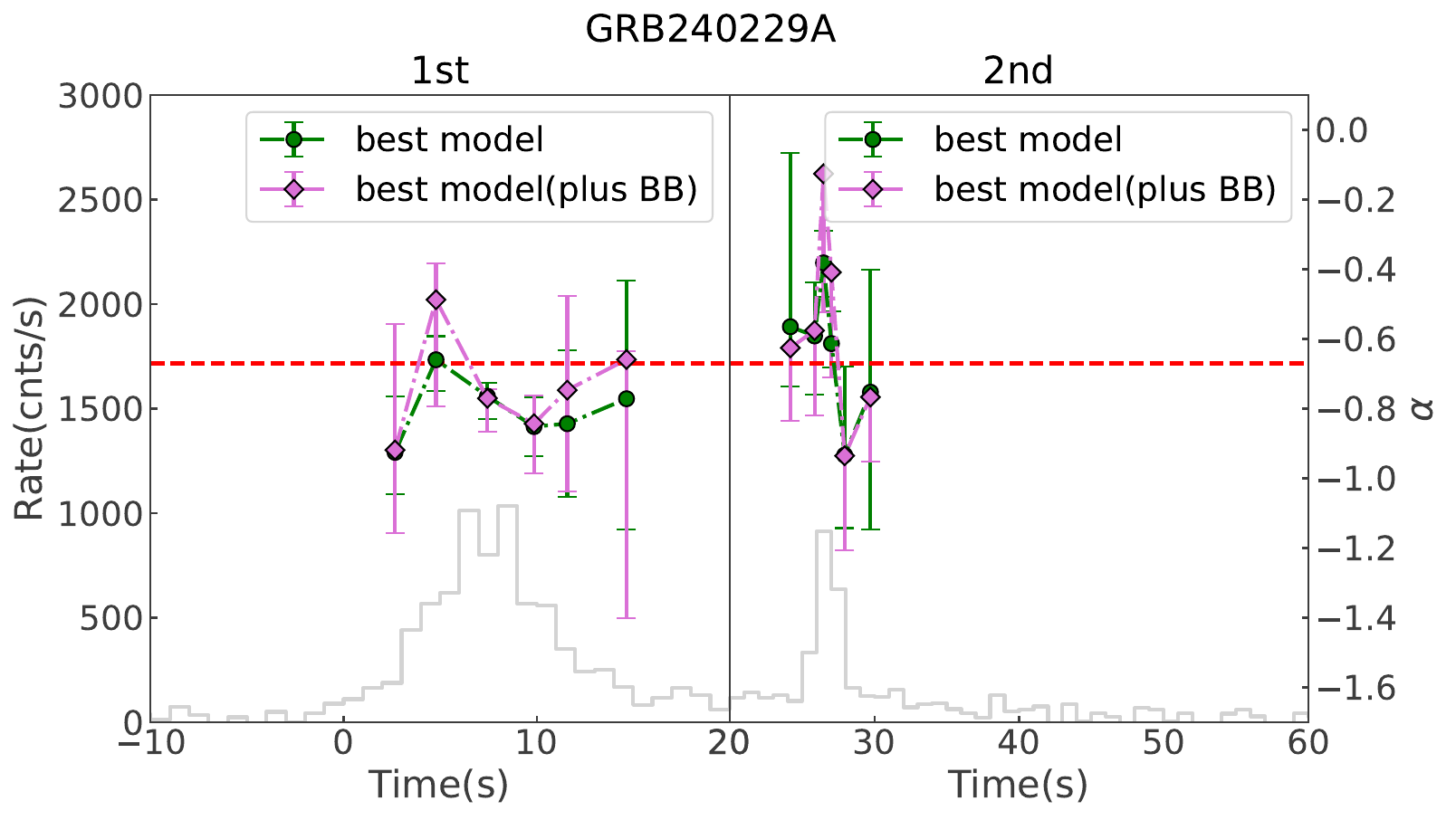}
  \caption{(Continued.) \label{fig 10}}
   
\end{figure}

\setcounter{figure}{10}  
\begin{figure}[H]
\centering
\includegraphics [width=8cm,height=4cm]{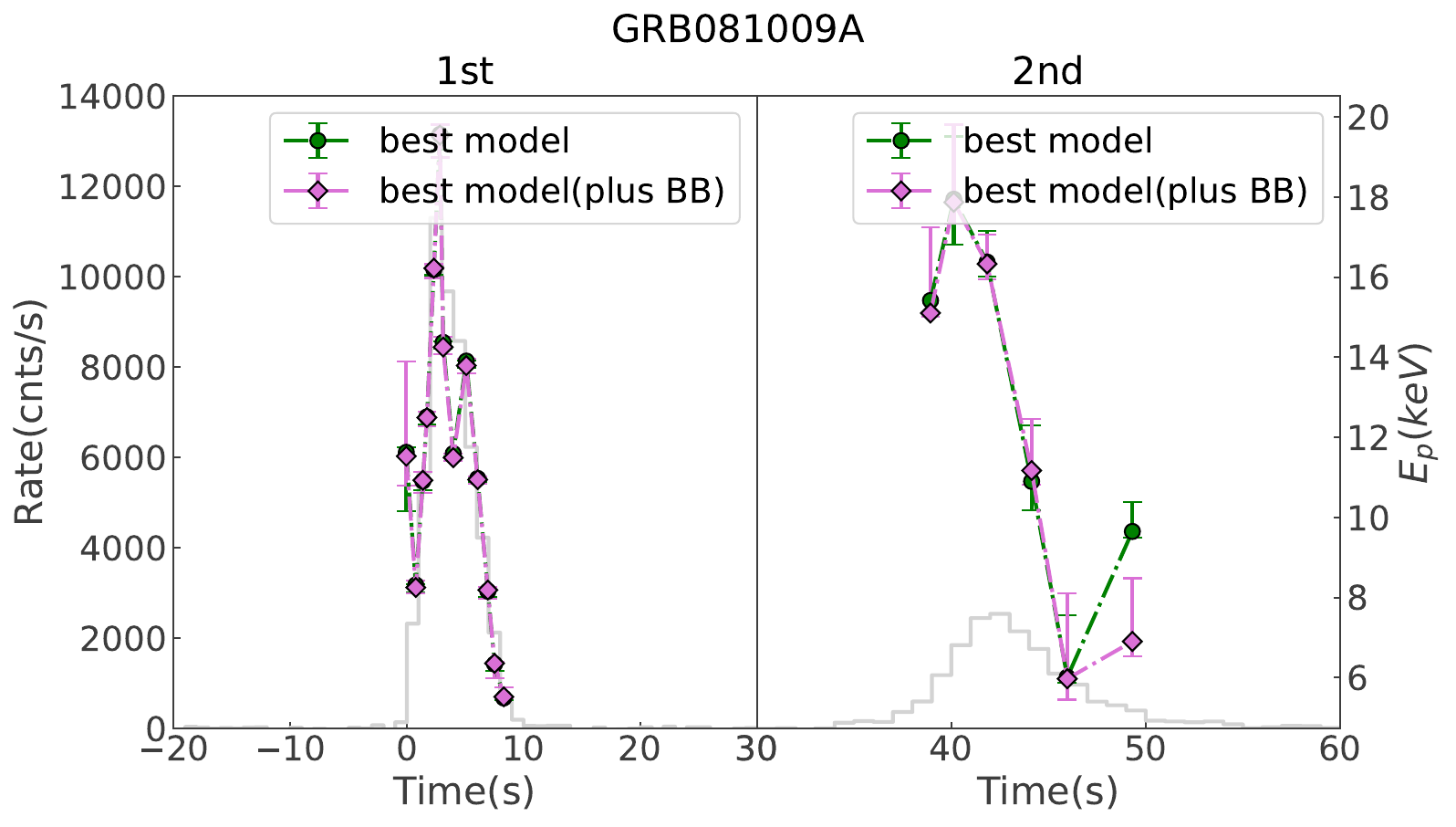}  
\includegraphics [width=8cm,height=4cm]{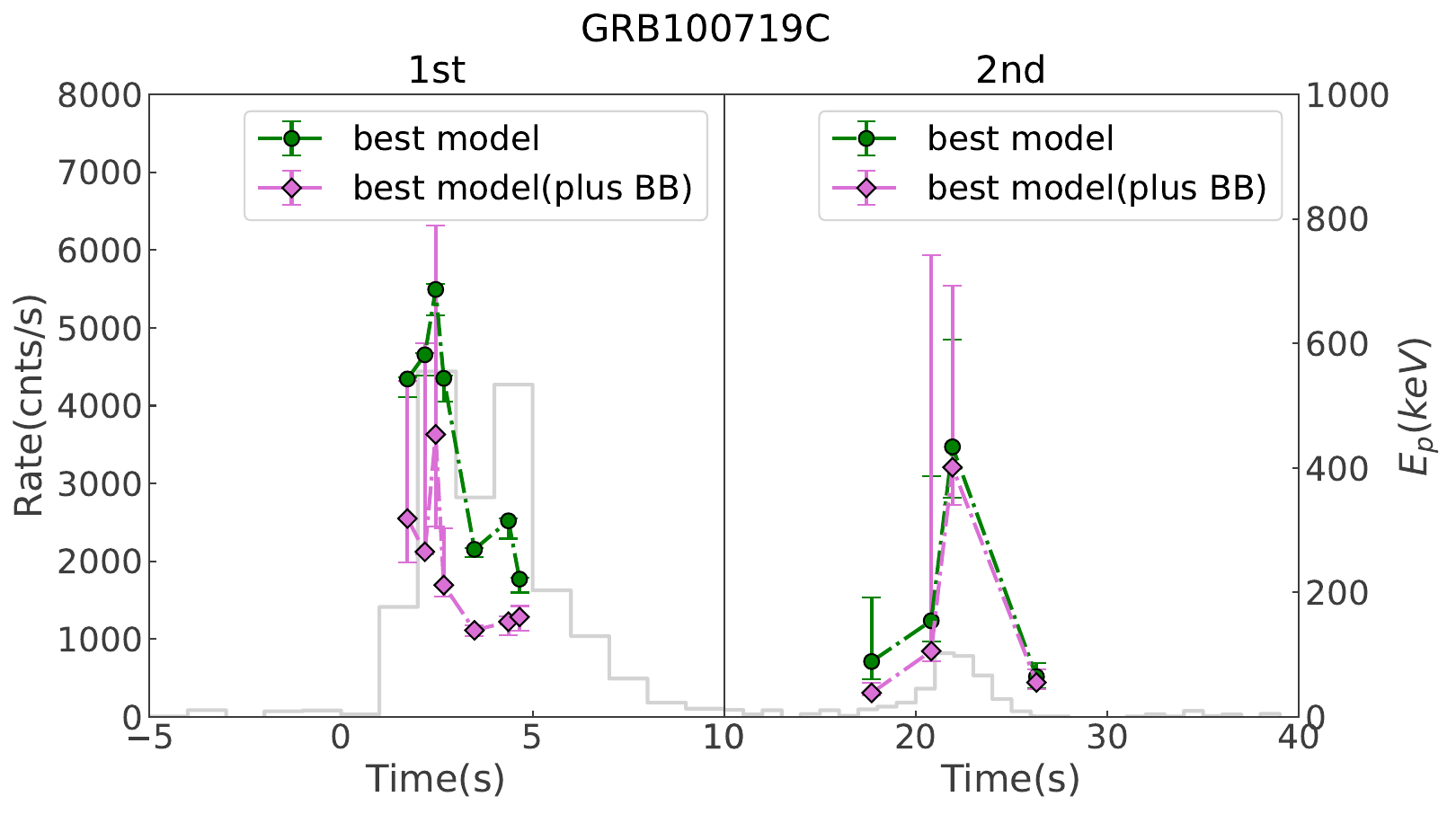}
\includegraphics [width=8cm,height=4cm]{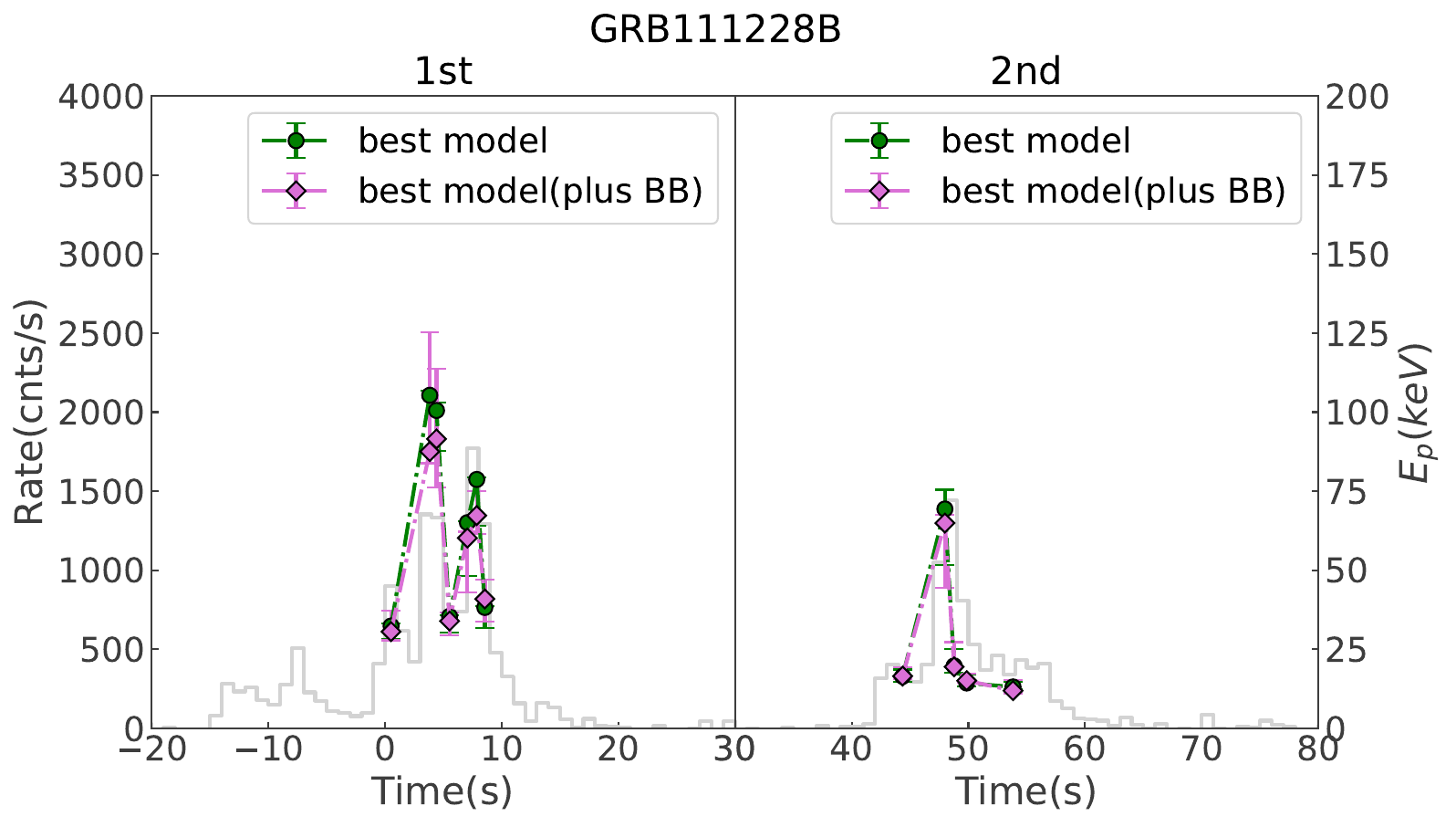}
\includegraphics [width=8cm,height=4cm]{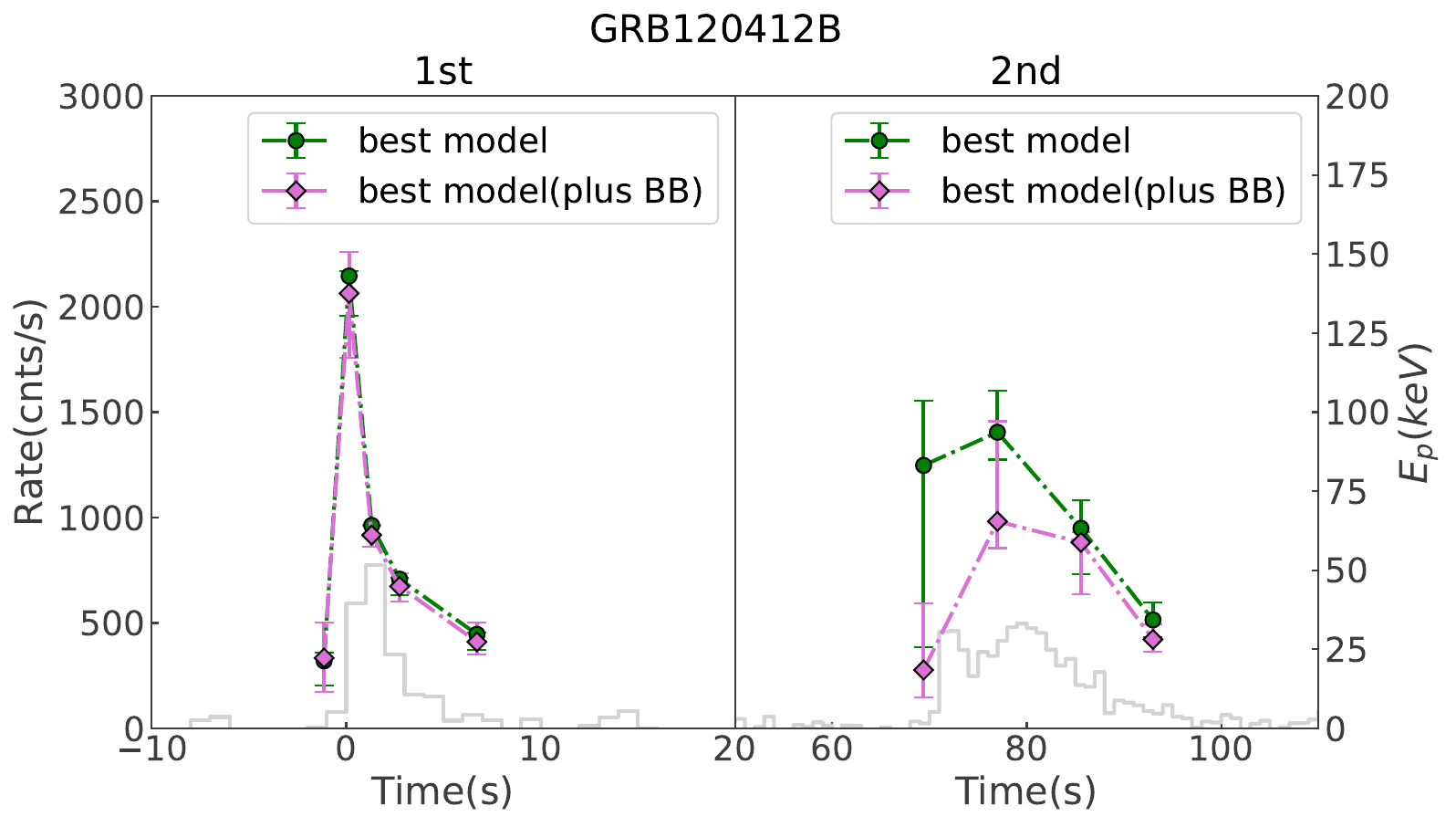}
\includegraphics [width=8cm,height=4cm]{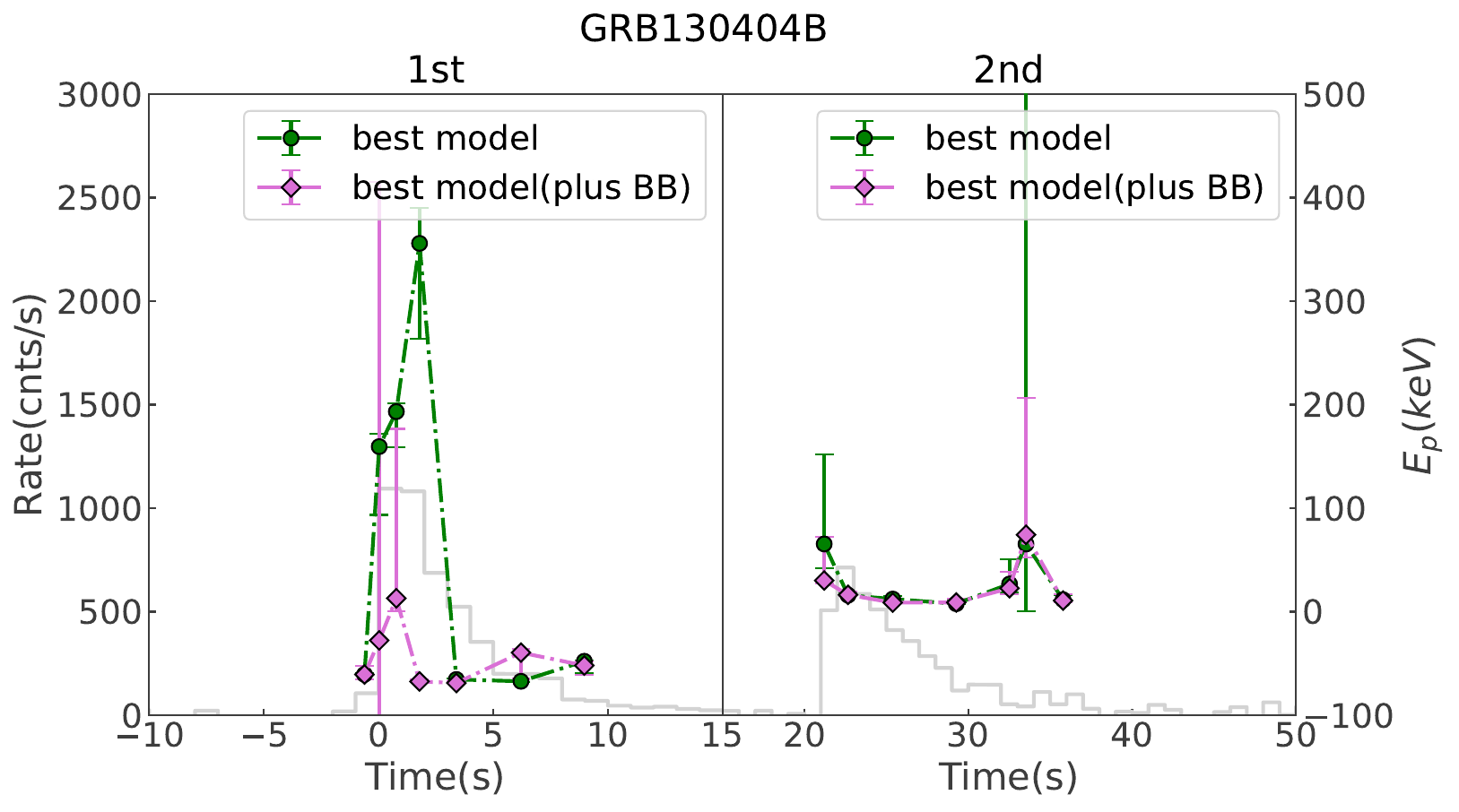}
\includegraphics [width=8cm,height=4cm]{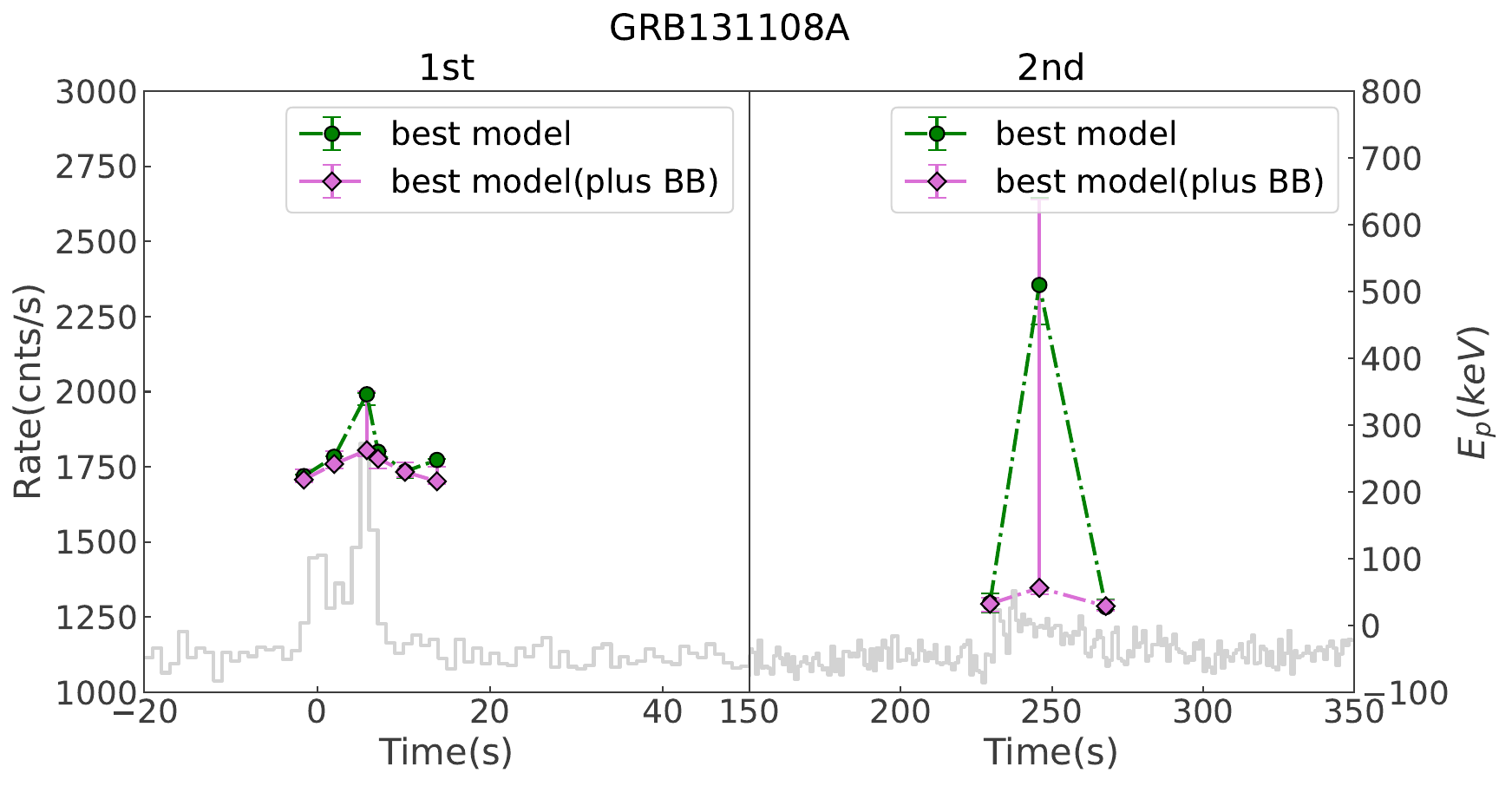}
\includegraphics [width=8cm,height=4cm]{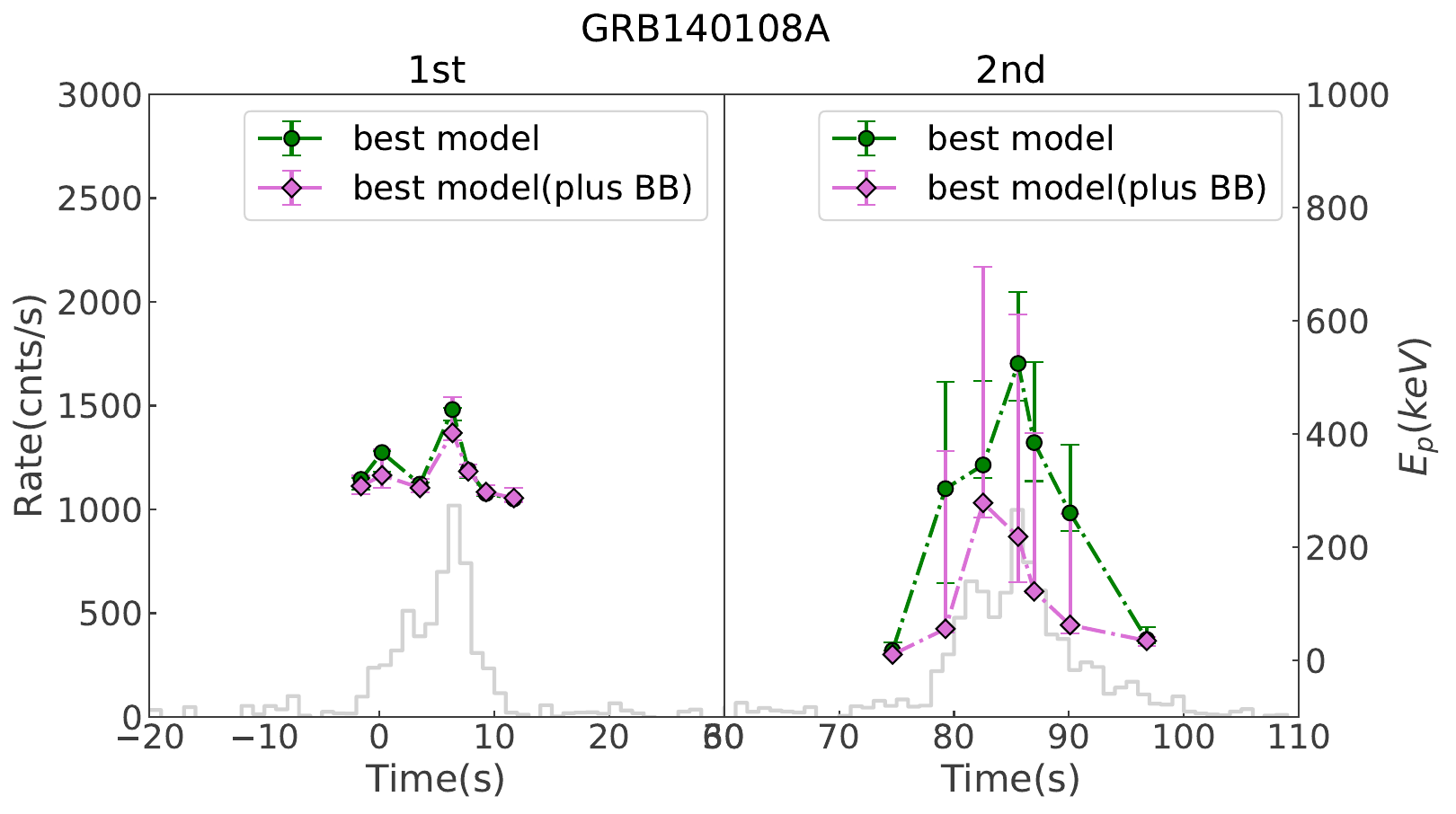}
\includegraphics [width=8cm,height=4cm]{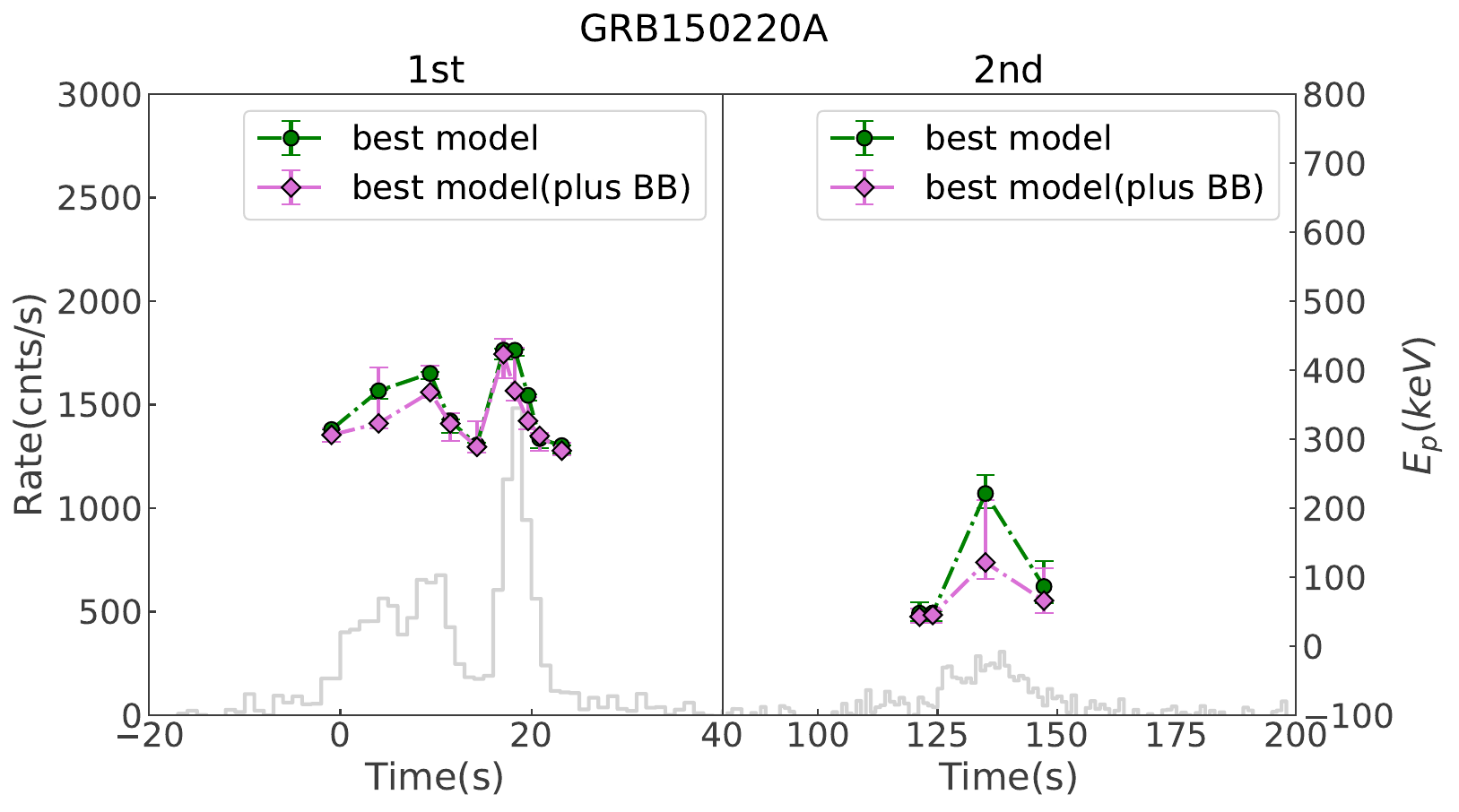}
\includegraphics [width=8cm,height=4cm]{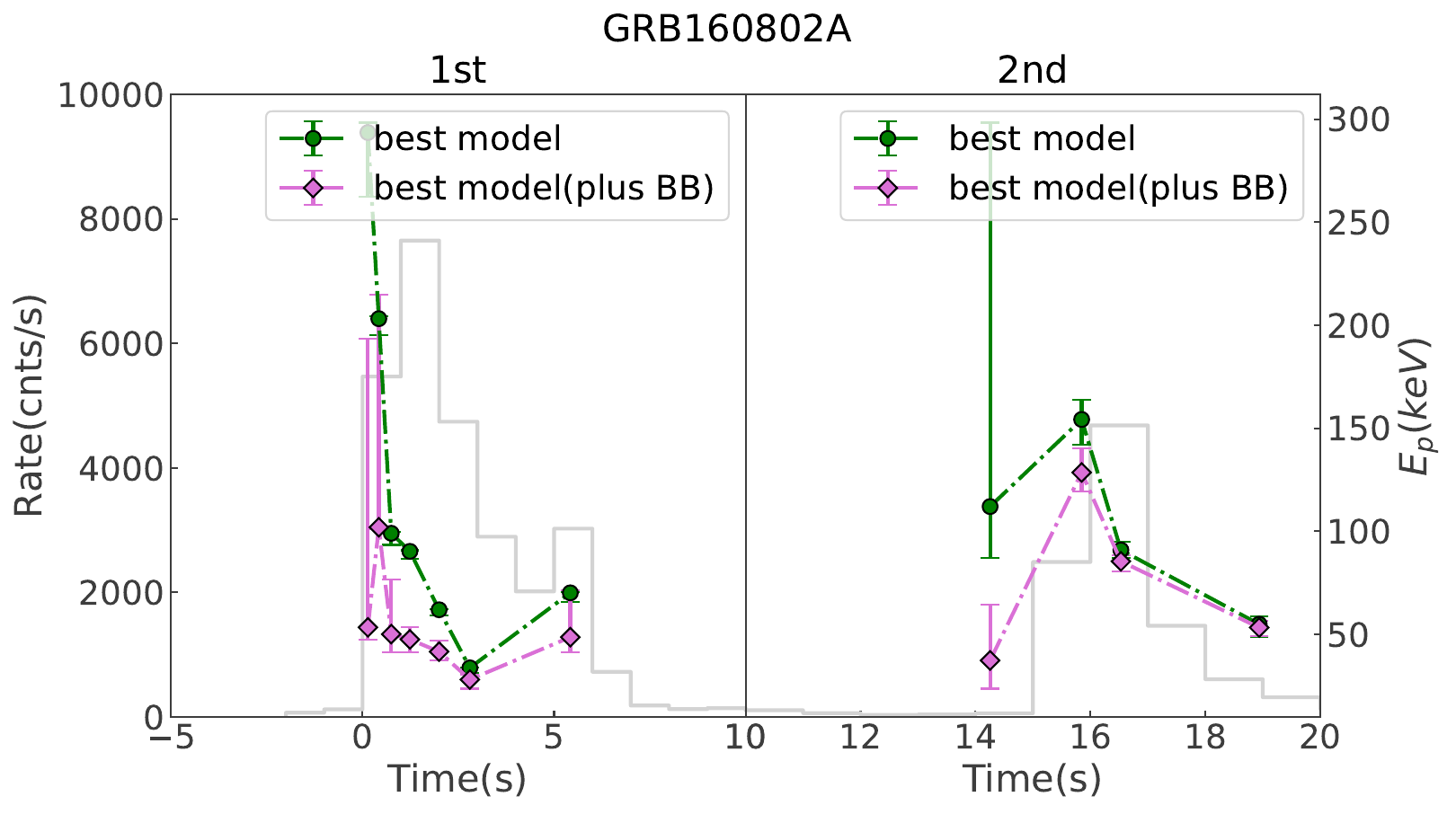}
\includegraphics [width=8cm,height=4cm]{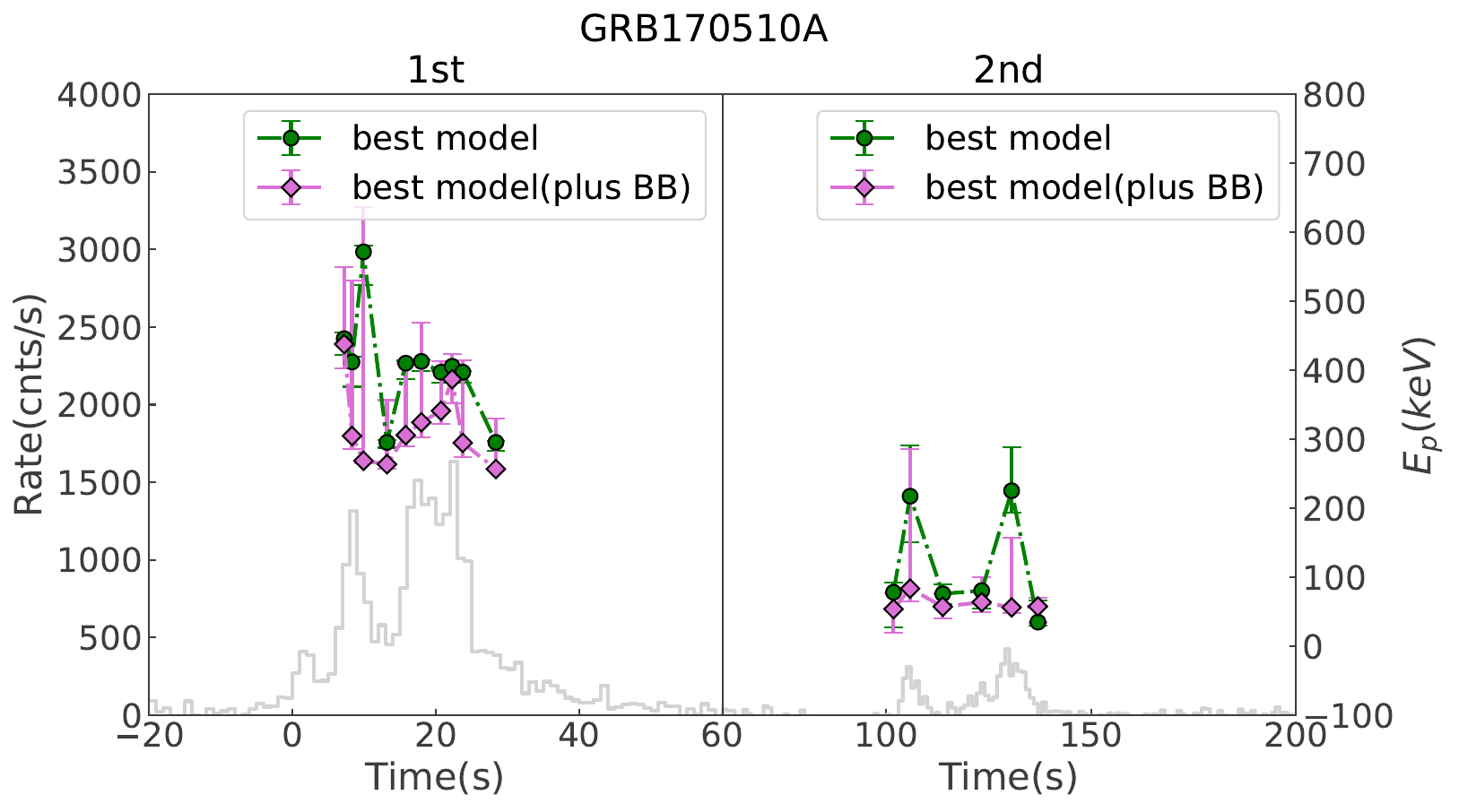}
   \figcaption{The evolution of the spectral parameter $E_{p}$ over time, fitted with the best model. Similar to Figure \ref{fig 10}. \label{fig 11}}   

\end{figure} 

\setcounter{figure}{10}  
\begin{figure}[H]

\centering
\includegraphics [width=8cm,height=4cm]{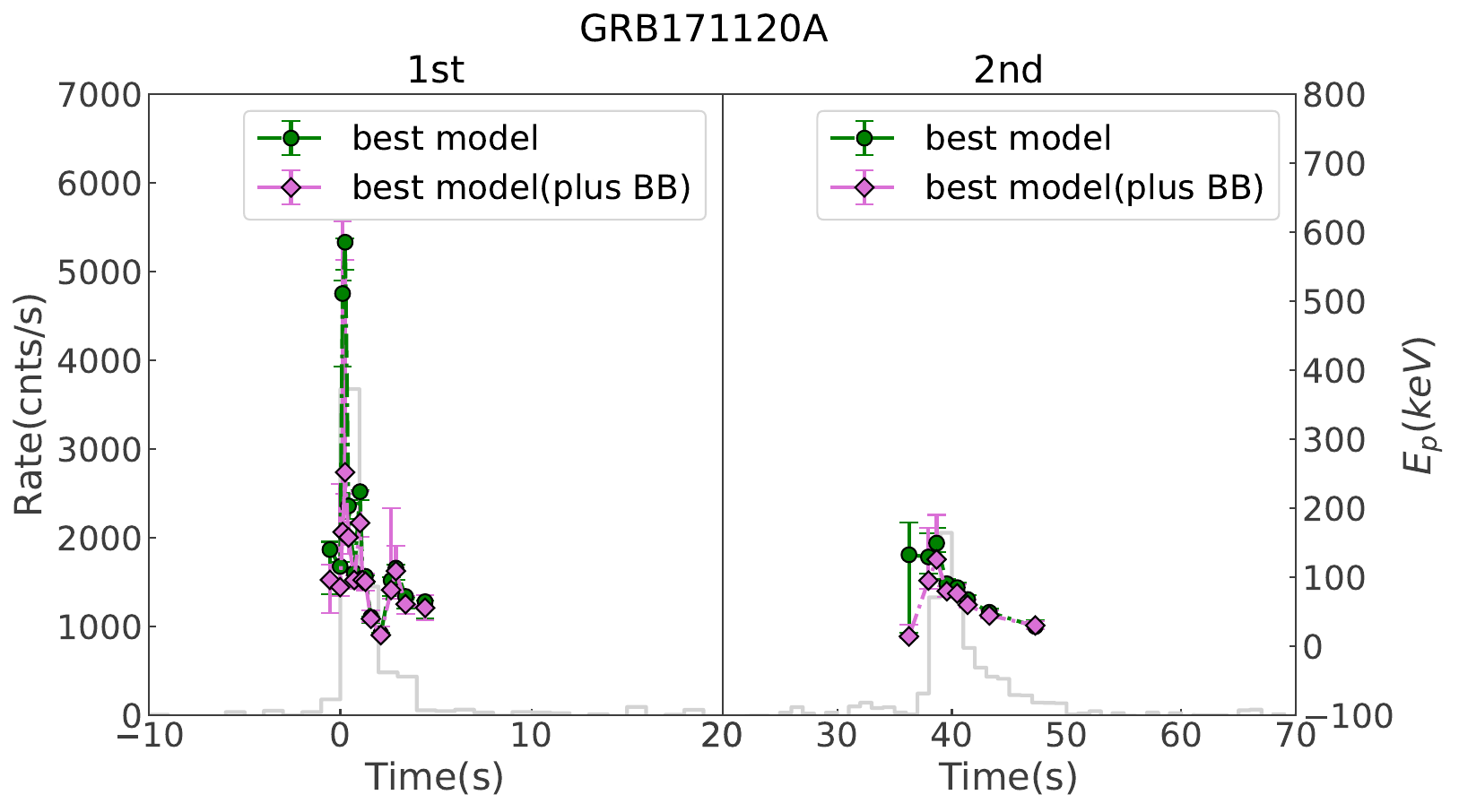}
\includegraphics [width=8cm,height=4cm]{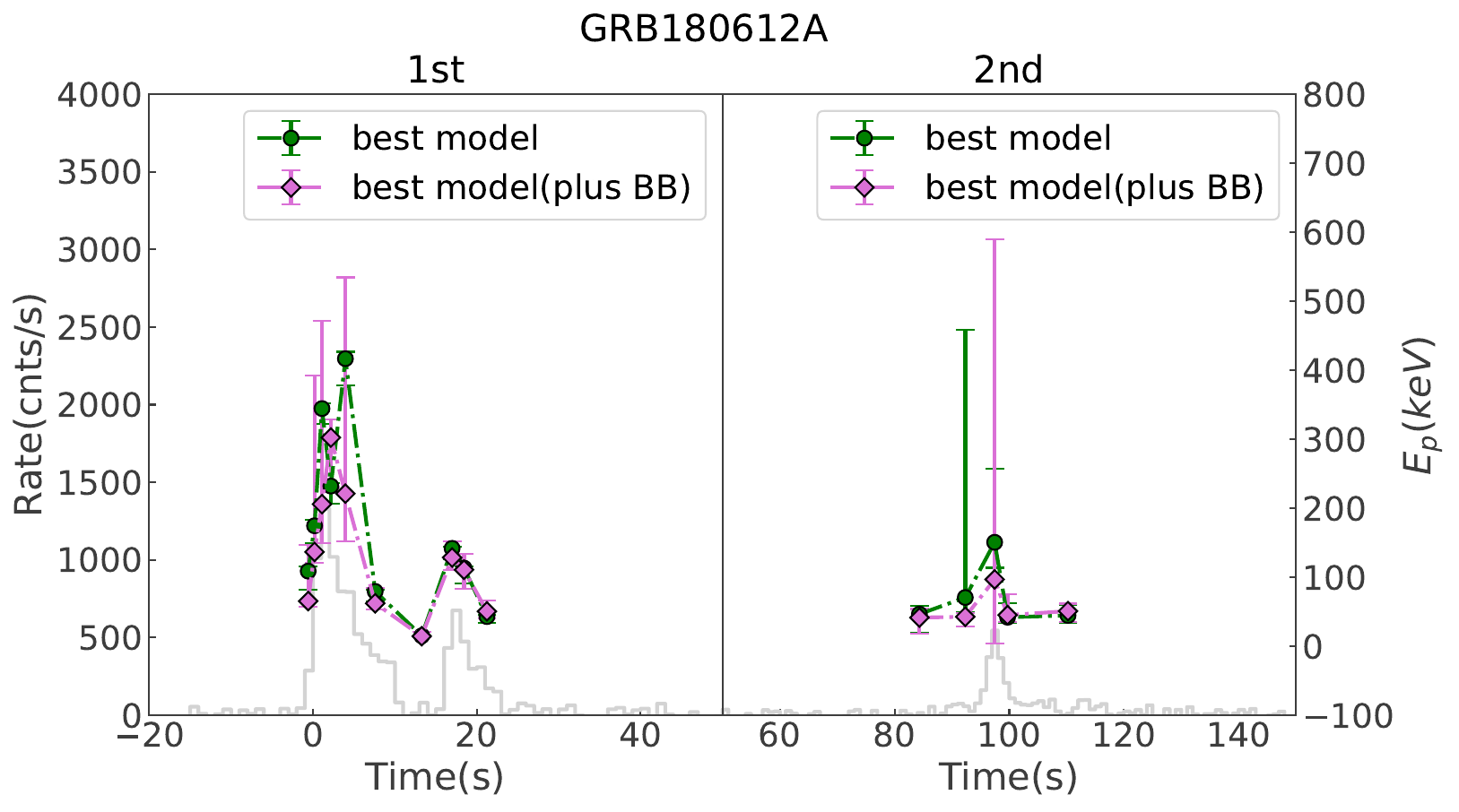}
\includegraphics [width=8cm,height=4cm]{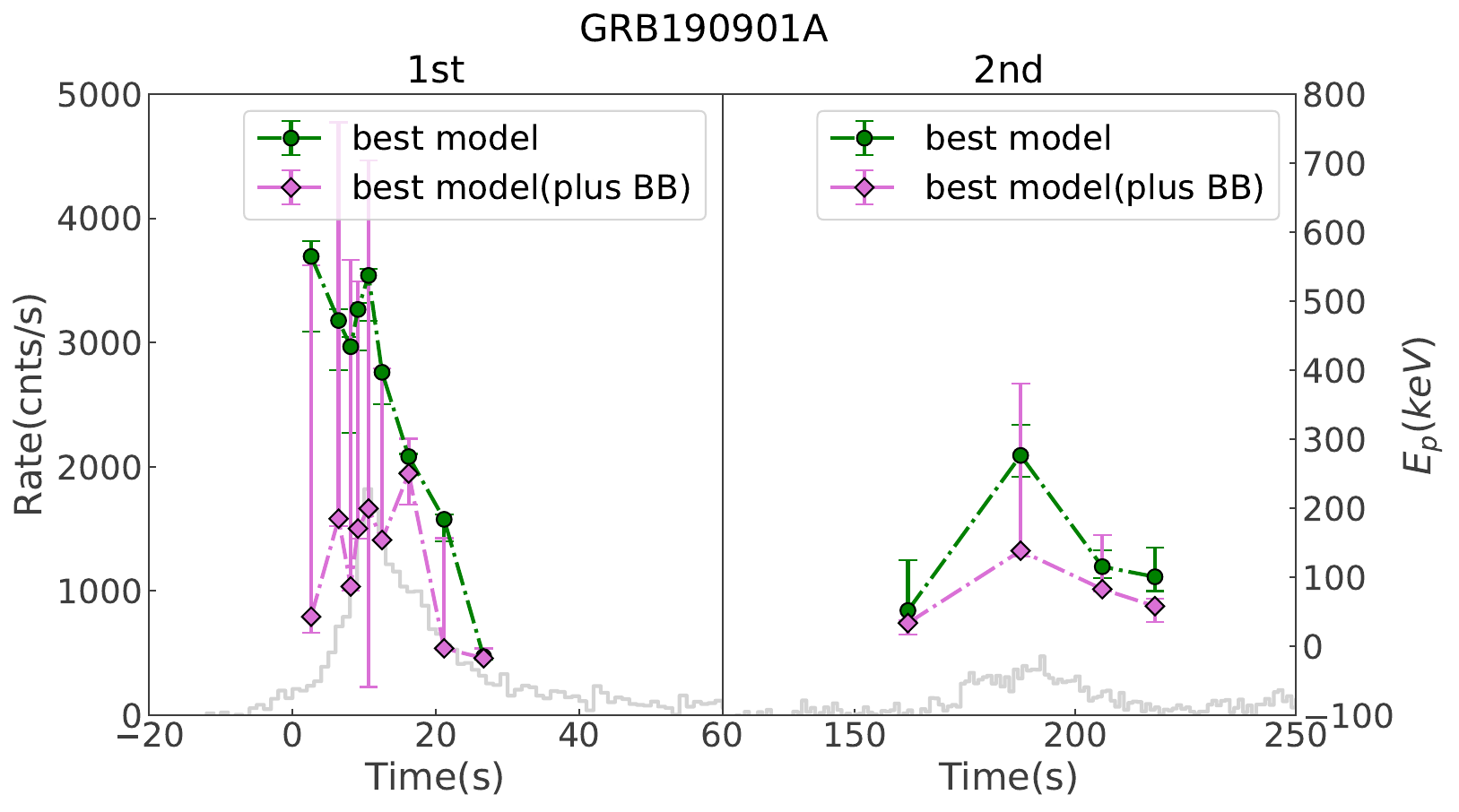}
\includegraphics [width=8cm,height=4cm]{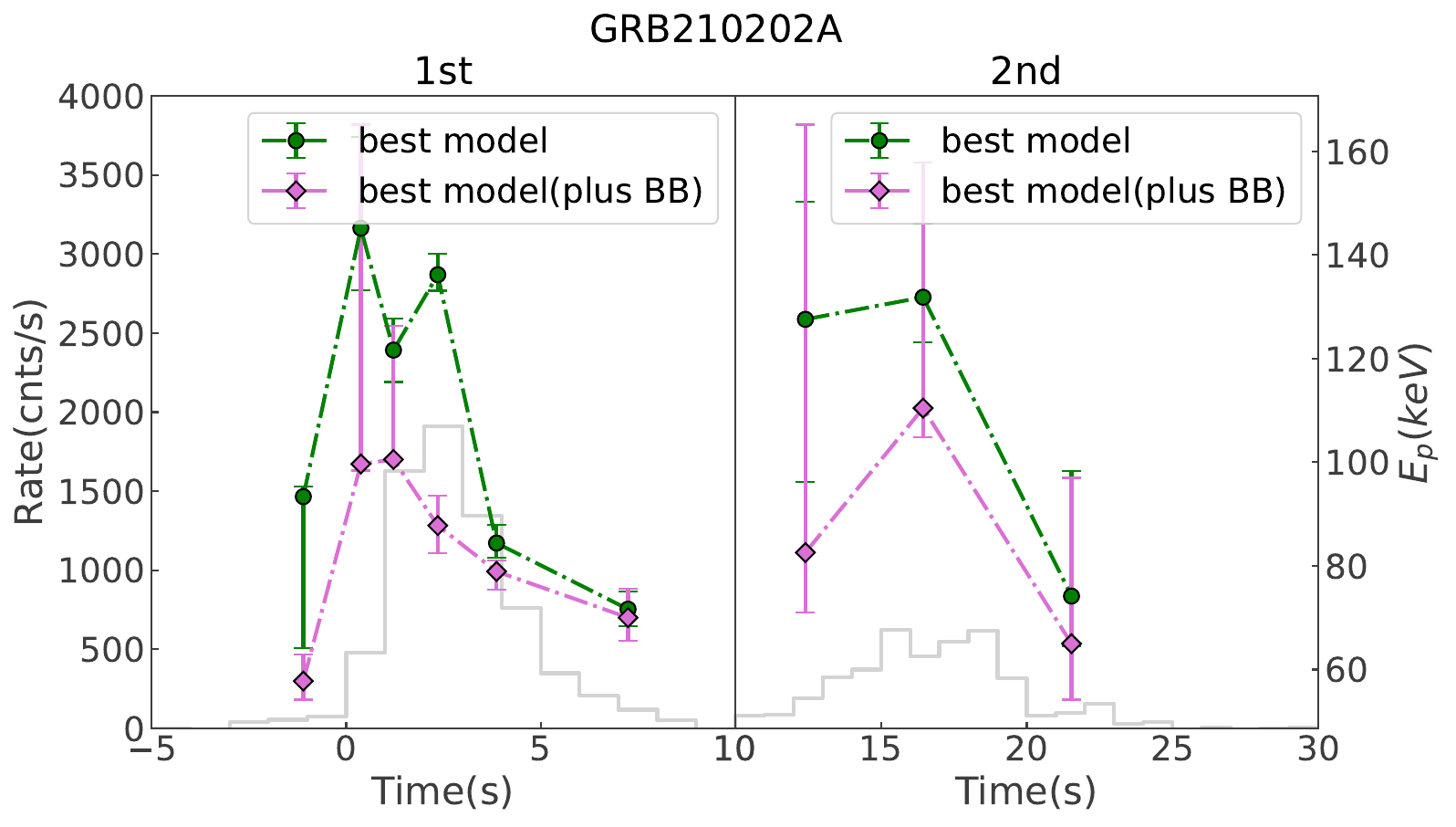}
\includegraphics [width=8cm,height=4cm]{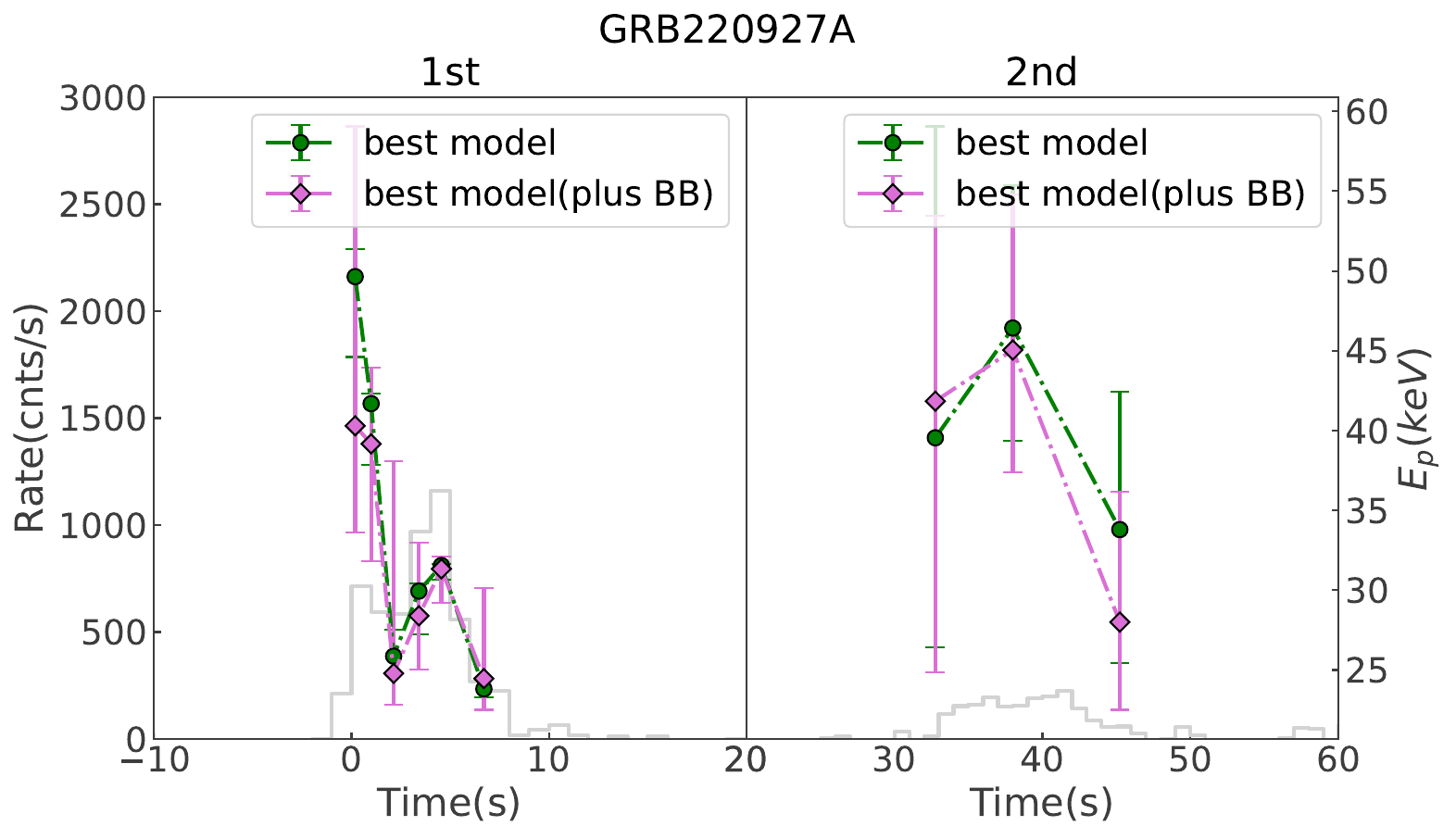}
\includegraphics [width=8cm,height=4cm]{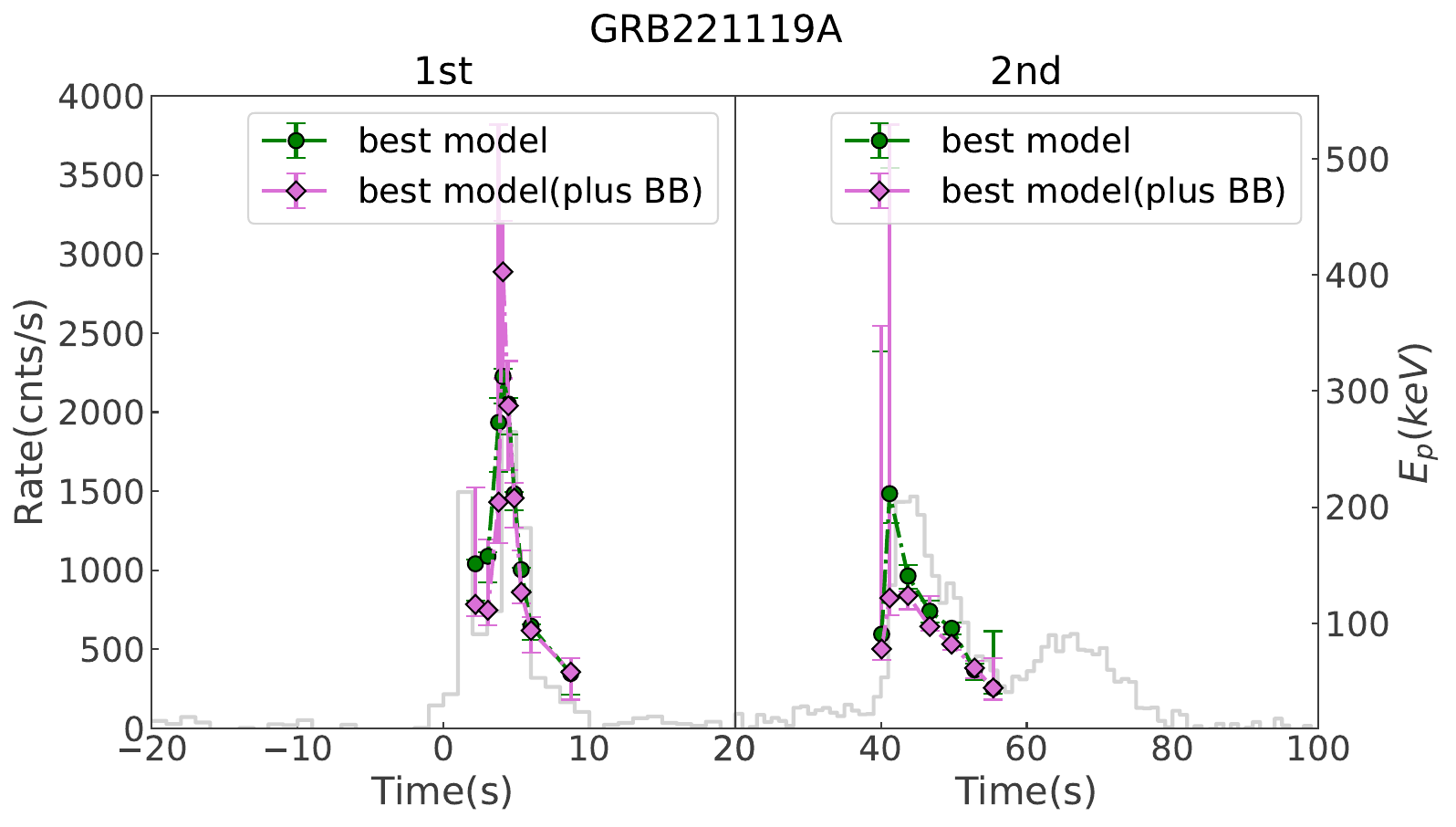}
\includegraphics [width=8cm,height=4cm]{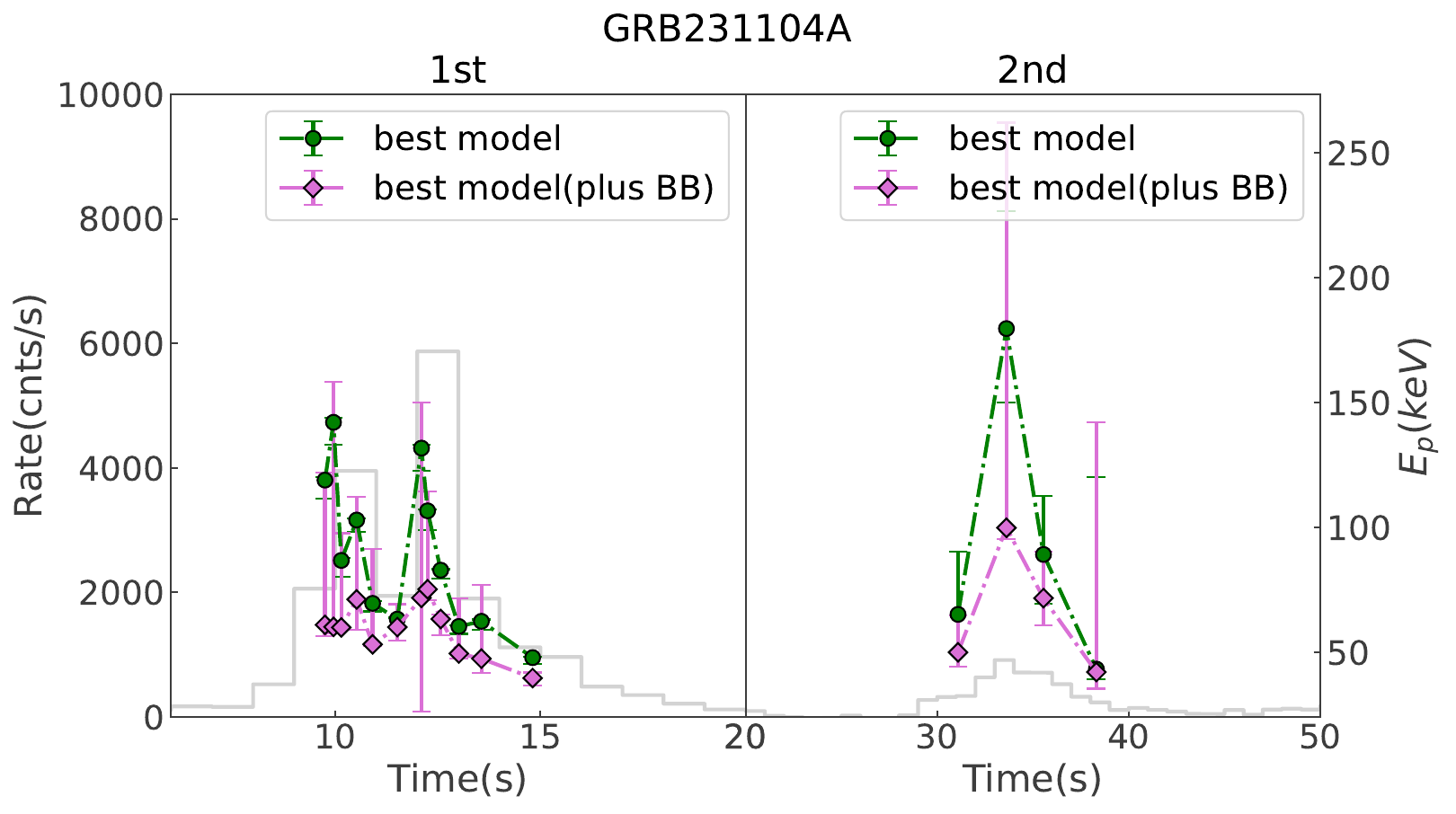}
\includegraphics [width=8cm,height=4cm]{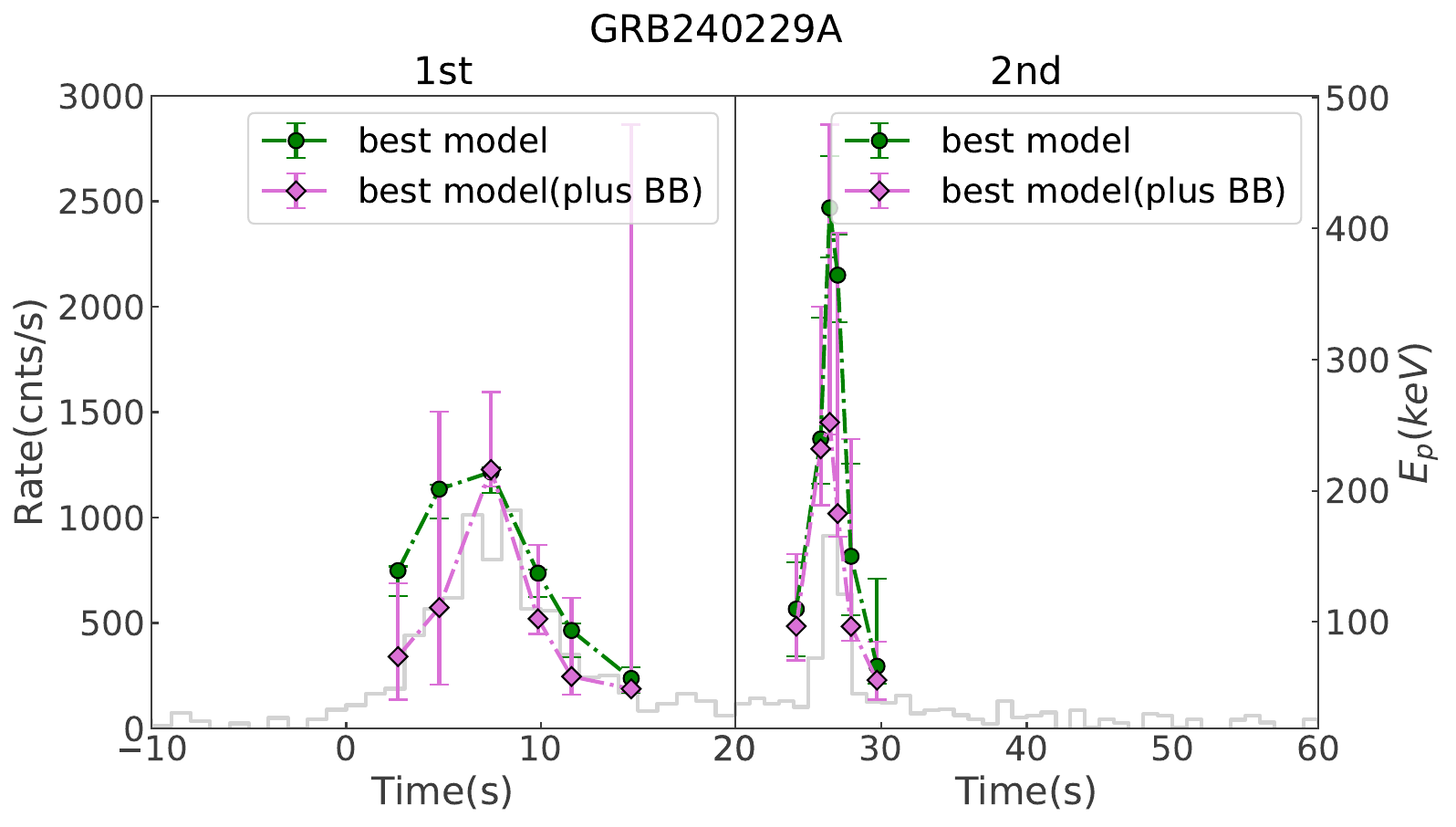}
   \figcaption{(Continued.) \label{fig 11}}   

\end{figure} 

\setcounter{figure}{11}  
\begin{figure}[H]
\centering
\includegraphics [width=8cm,height=4cm]{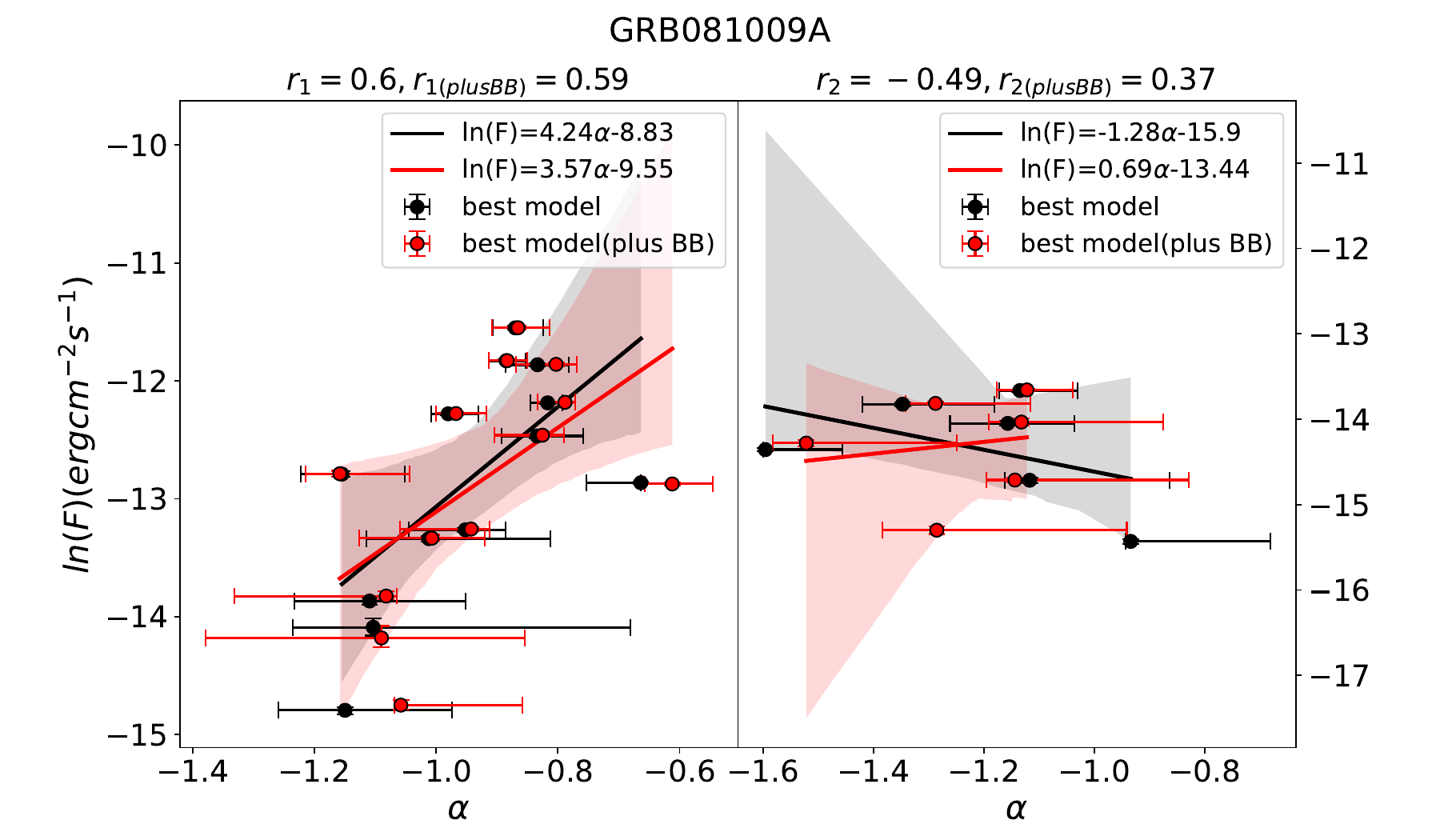}
\includegraphics [width=8cm,height=4cm]{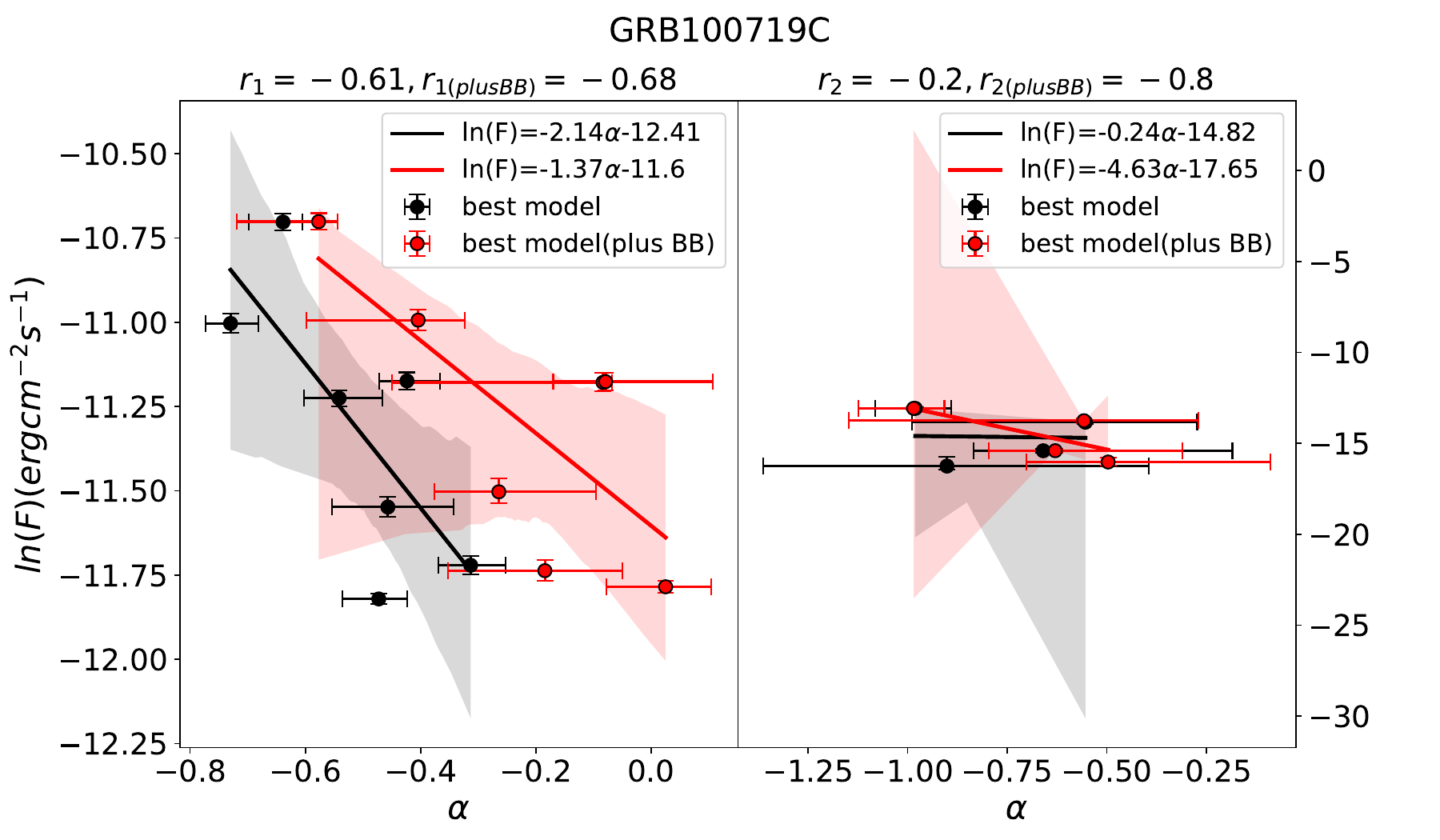}
\includegraphics [width=8cm,height=4cm]{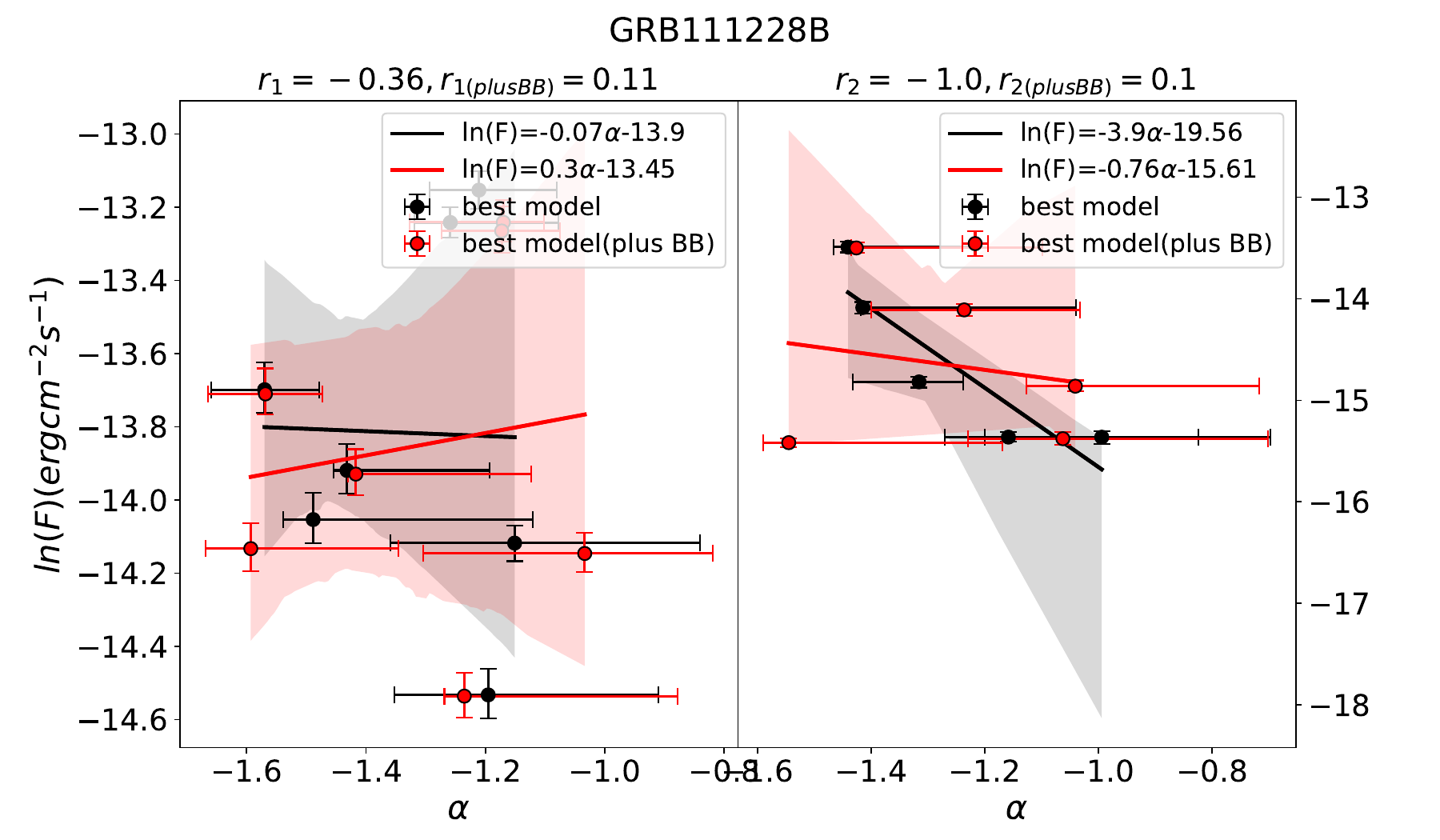}
\includegraphics [width=8cm,height=4cm]{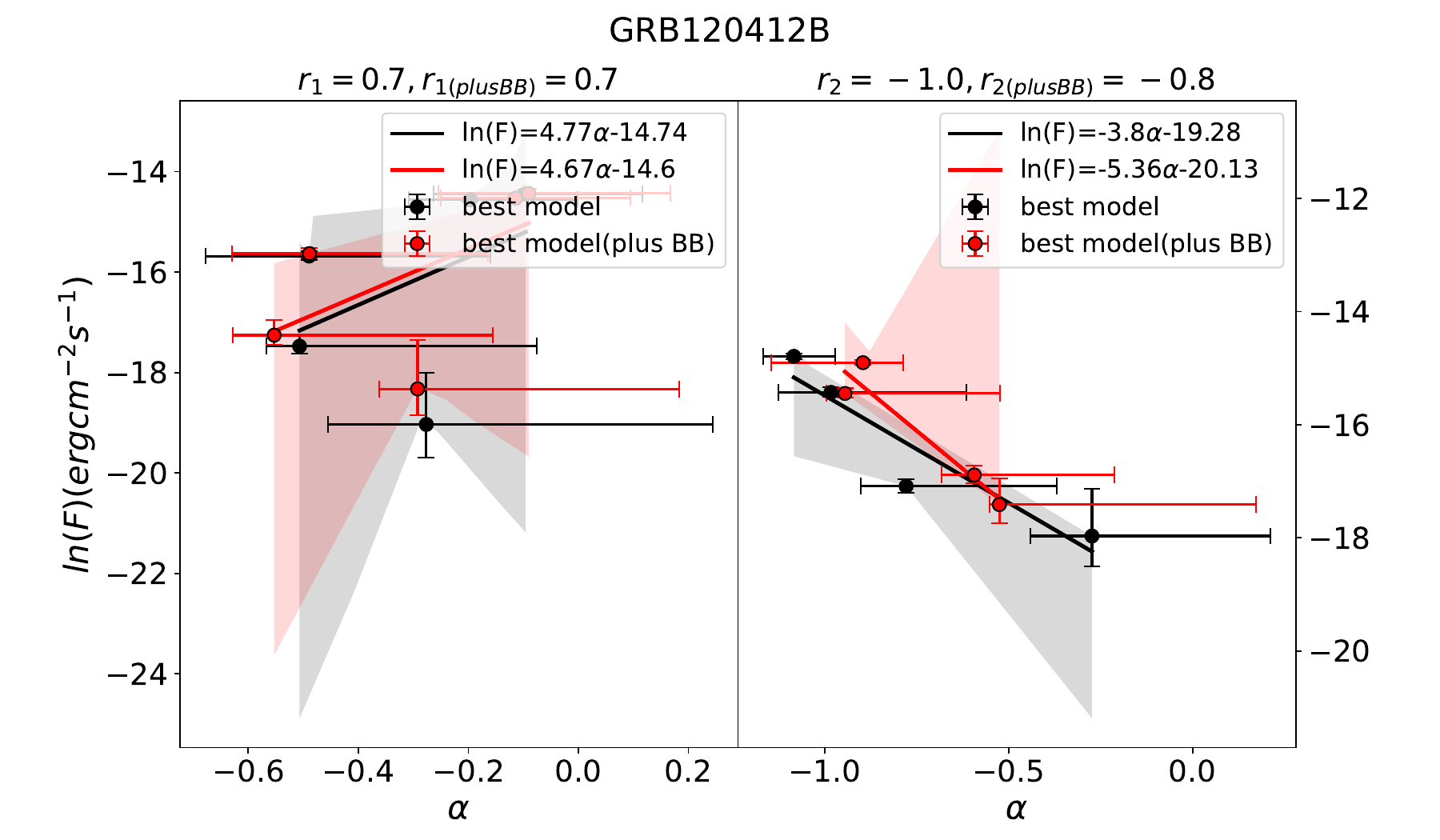}
\includegraphics [width=8cm,height=4cm]{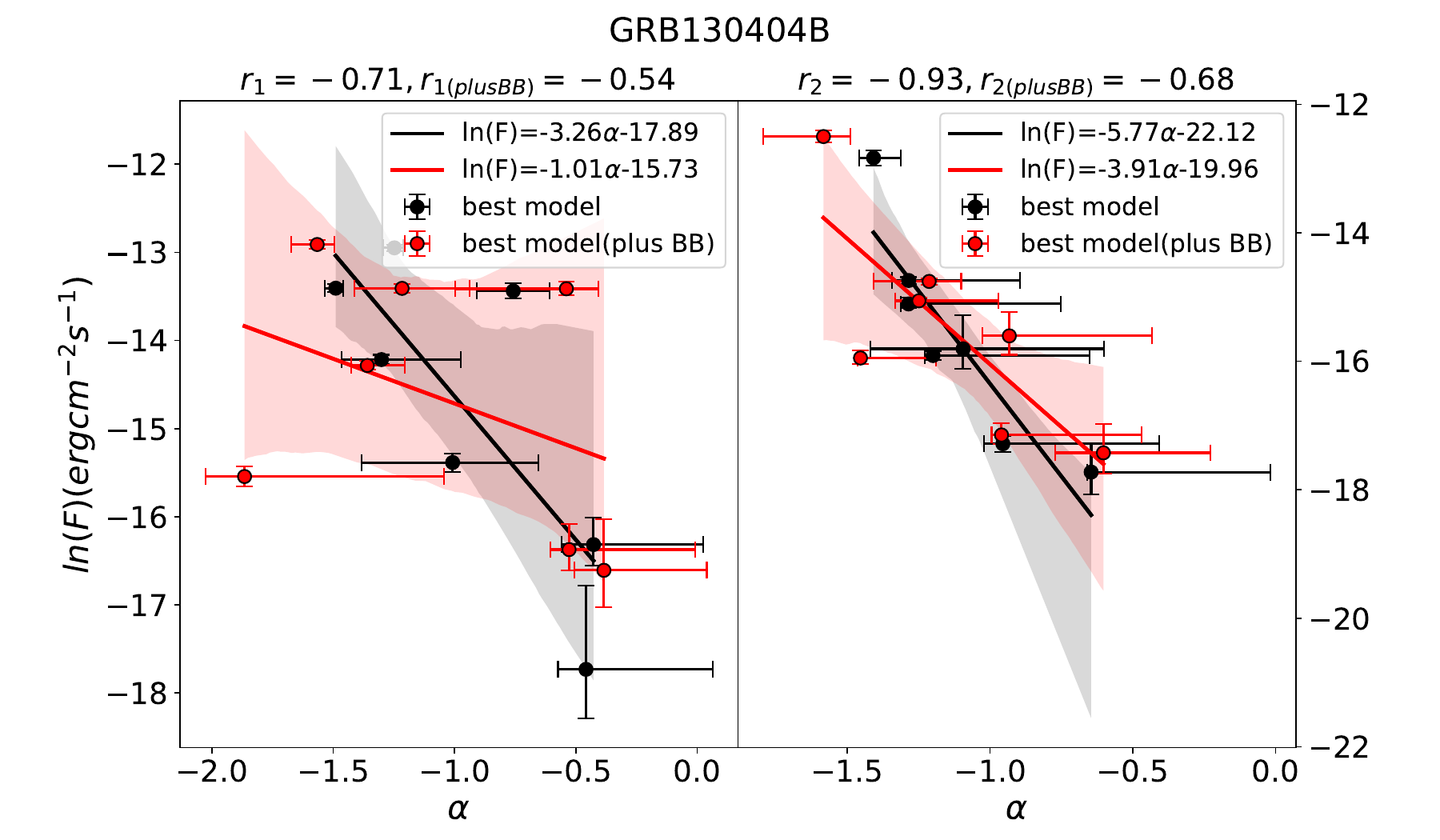}
\includegraphics [width=8cm,height=4cm]{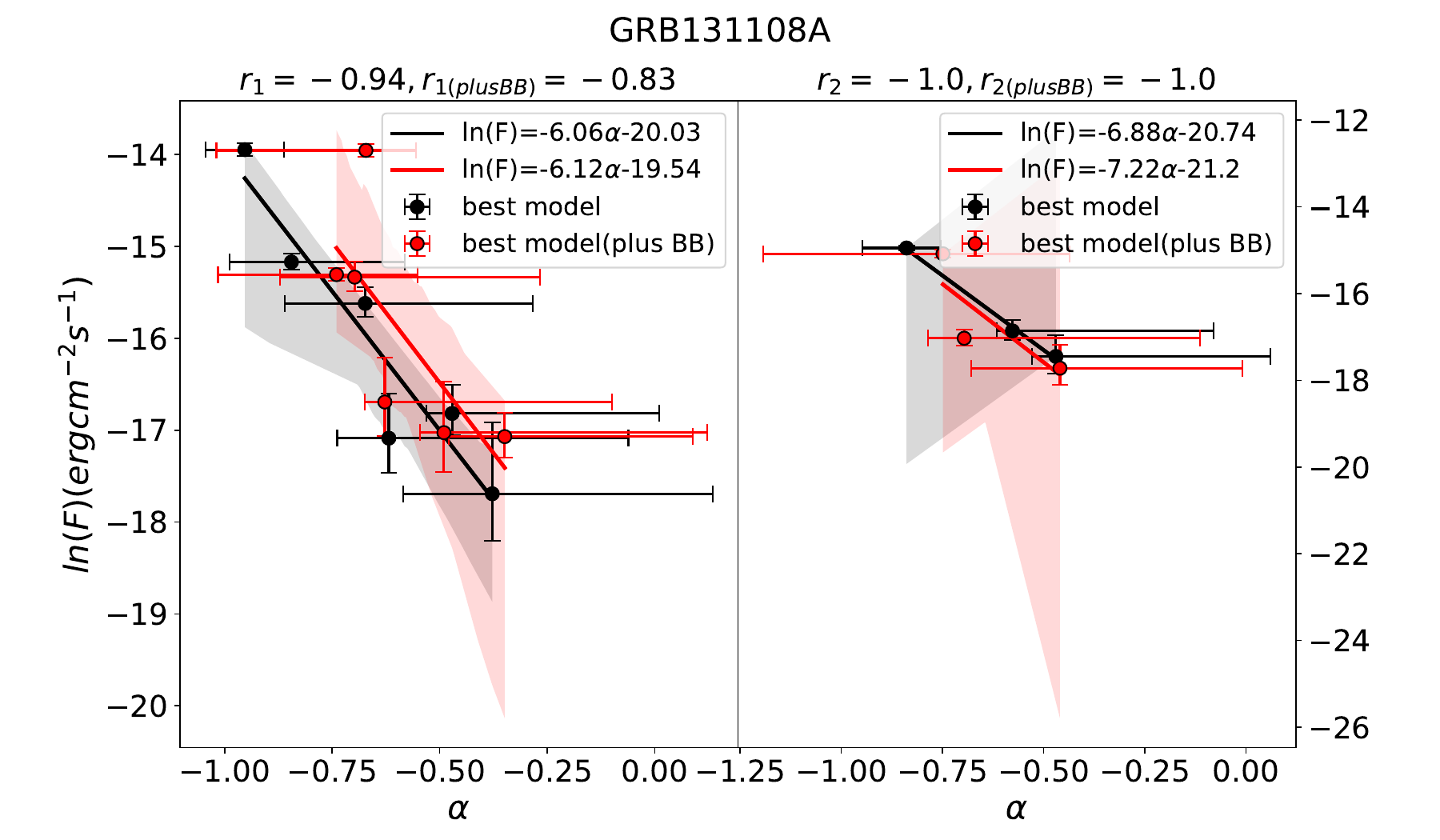}
\includegraphics [width=8cm,height=4cm]{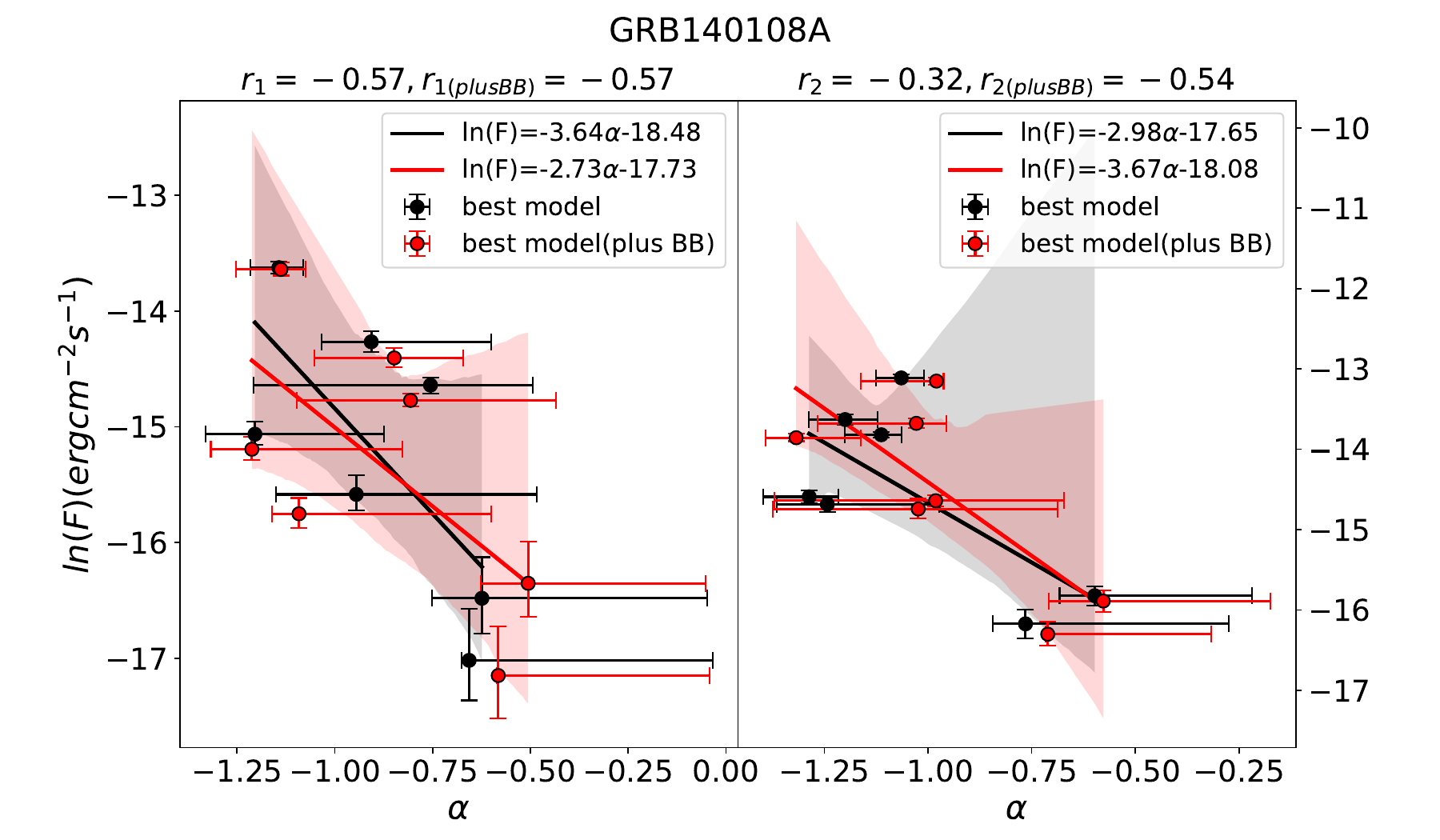}
\includegraphics [width=8cm,height=4cm]{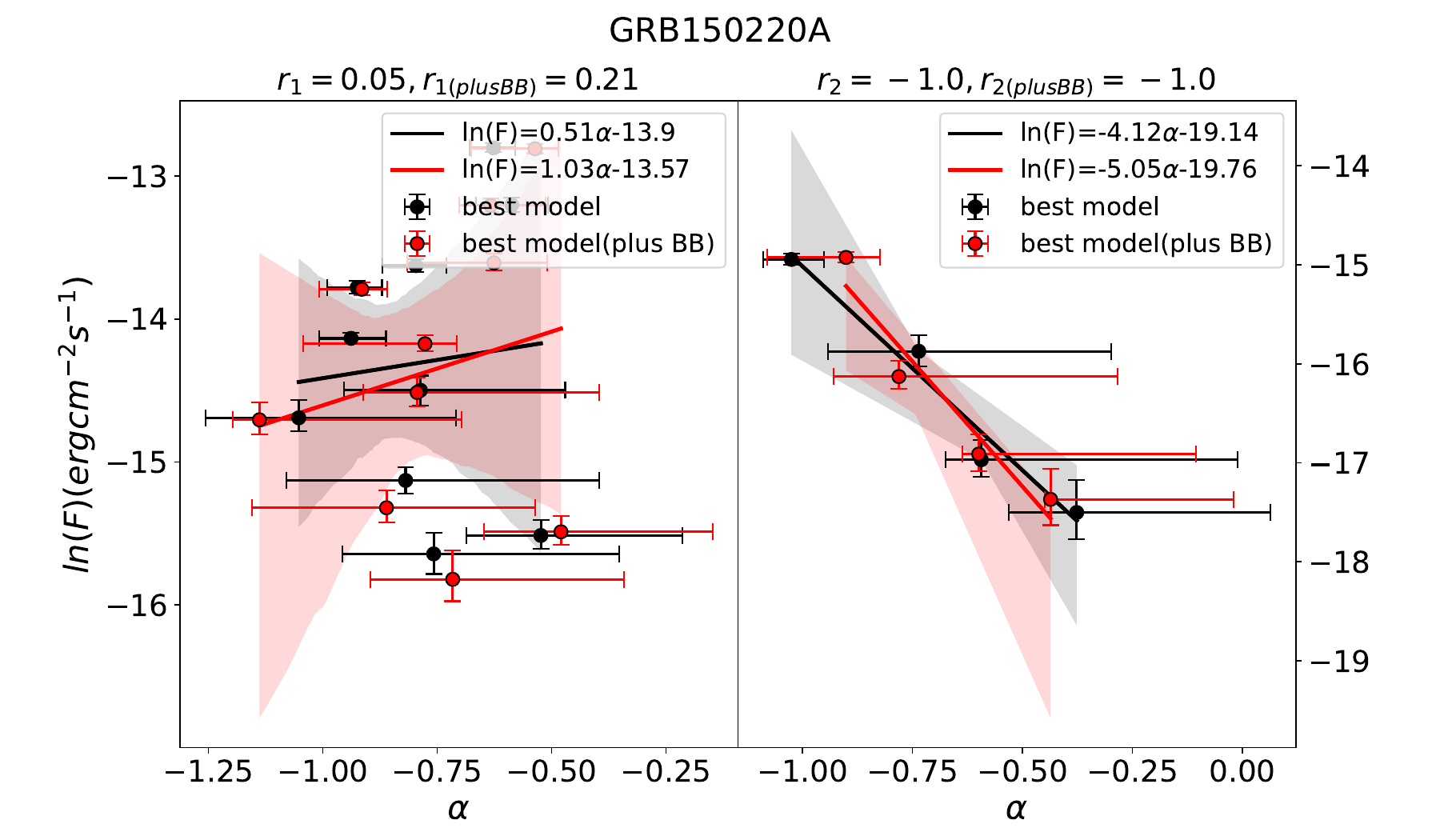}
\includegraphics [width=8cm,height=4cm]{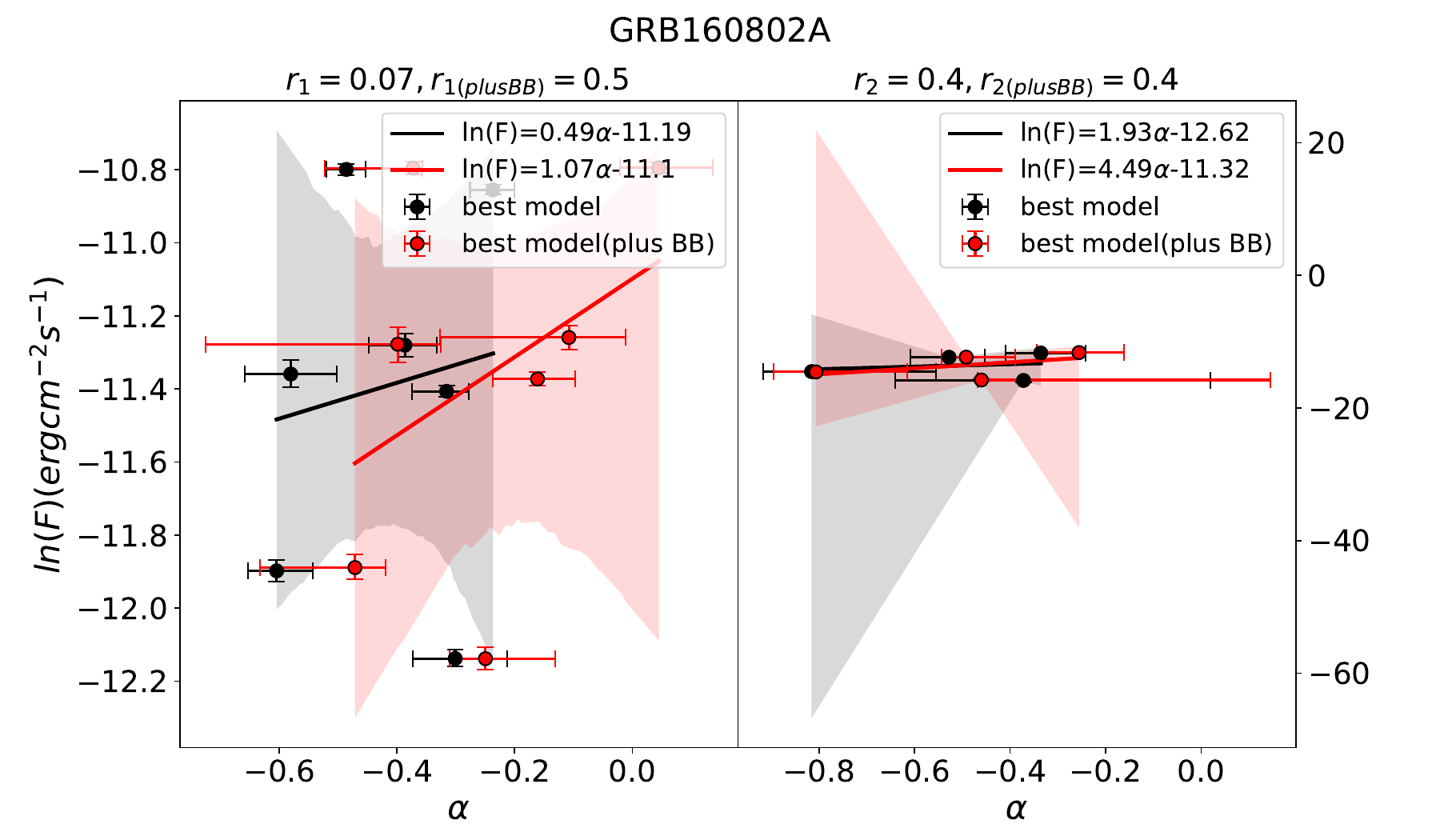}
\includegraphics [width=8cm,height=4cm]{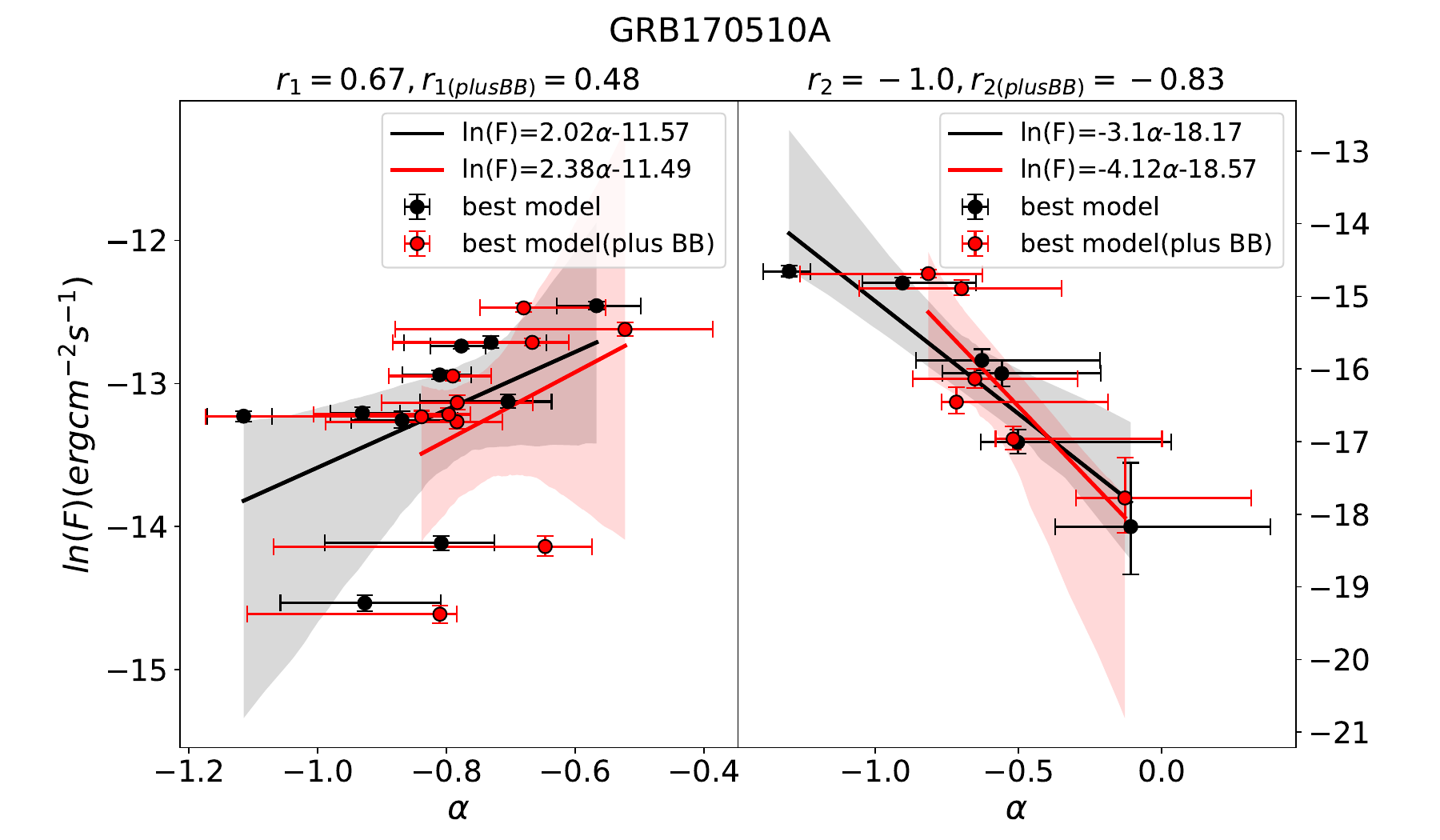}
   \figcaption{The correlation between $F$ and $\alpha$. The black and red dots represent the data points fitted using the best model and the best model $+$ BB, respectively. $r_{1}$ and $r_{2}$ represent the correlation coefficients between $F$ and $\alpha$ in the  main and second bursts, respectively. \label{fig 12} }
      
\end{figure}

\setcounter{figure}{11}  
\begin{figure}[H]

\centering
\includegraphics [width=8cm,height=4cm]{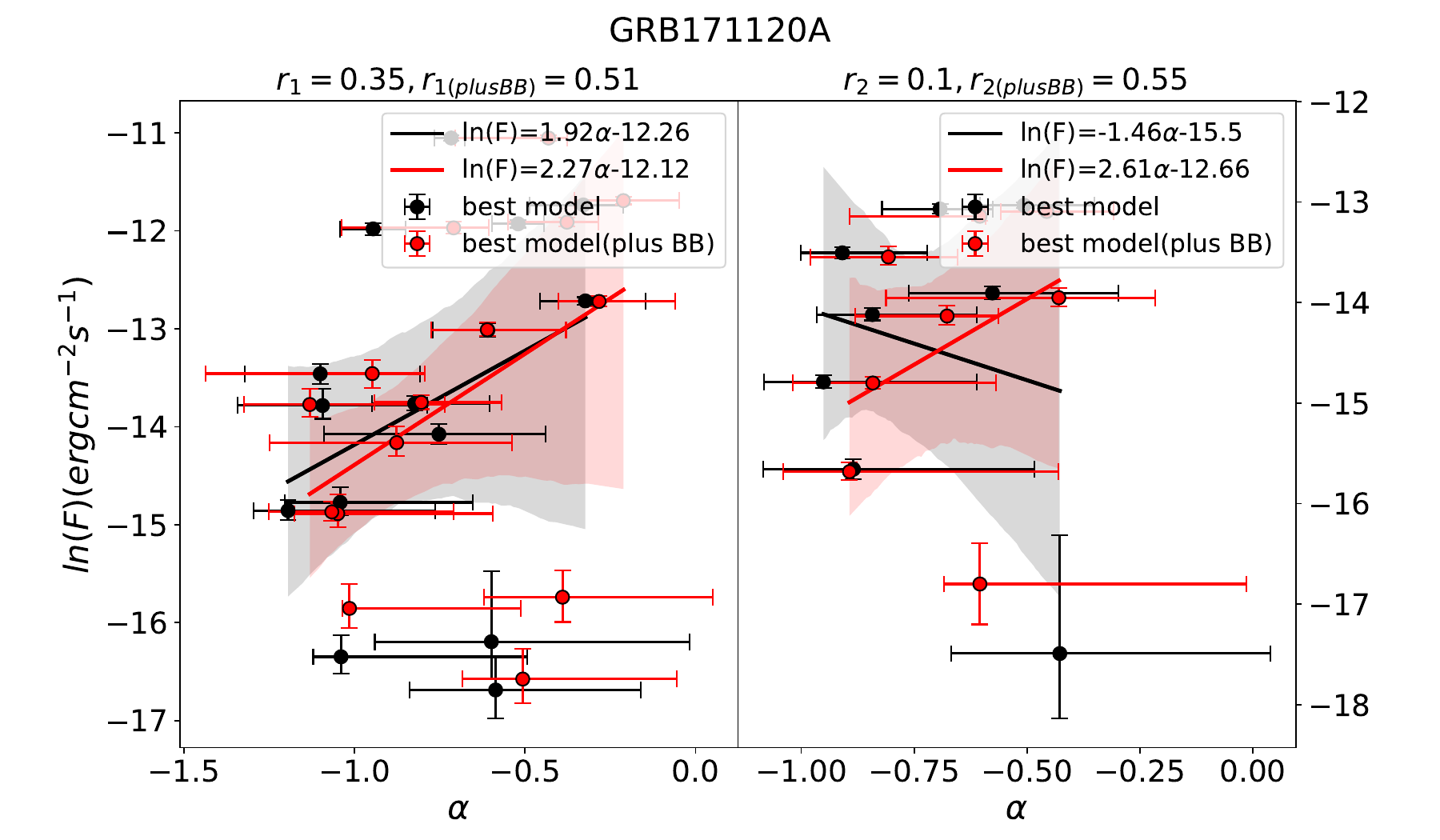}
\includegraphics [width=8cm,height=4cm]{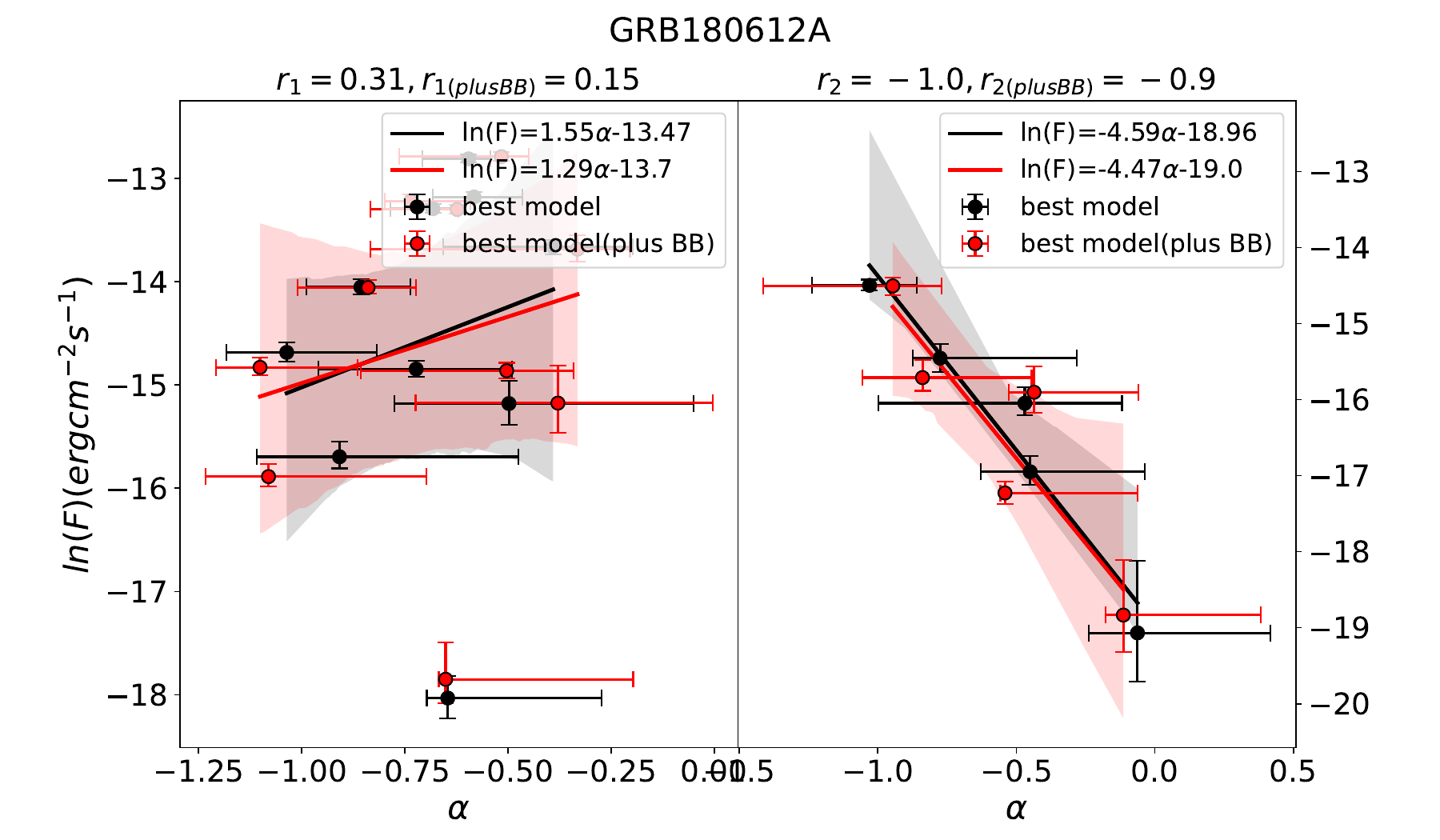}
\includegraphics [width=8cm,height=4cm]{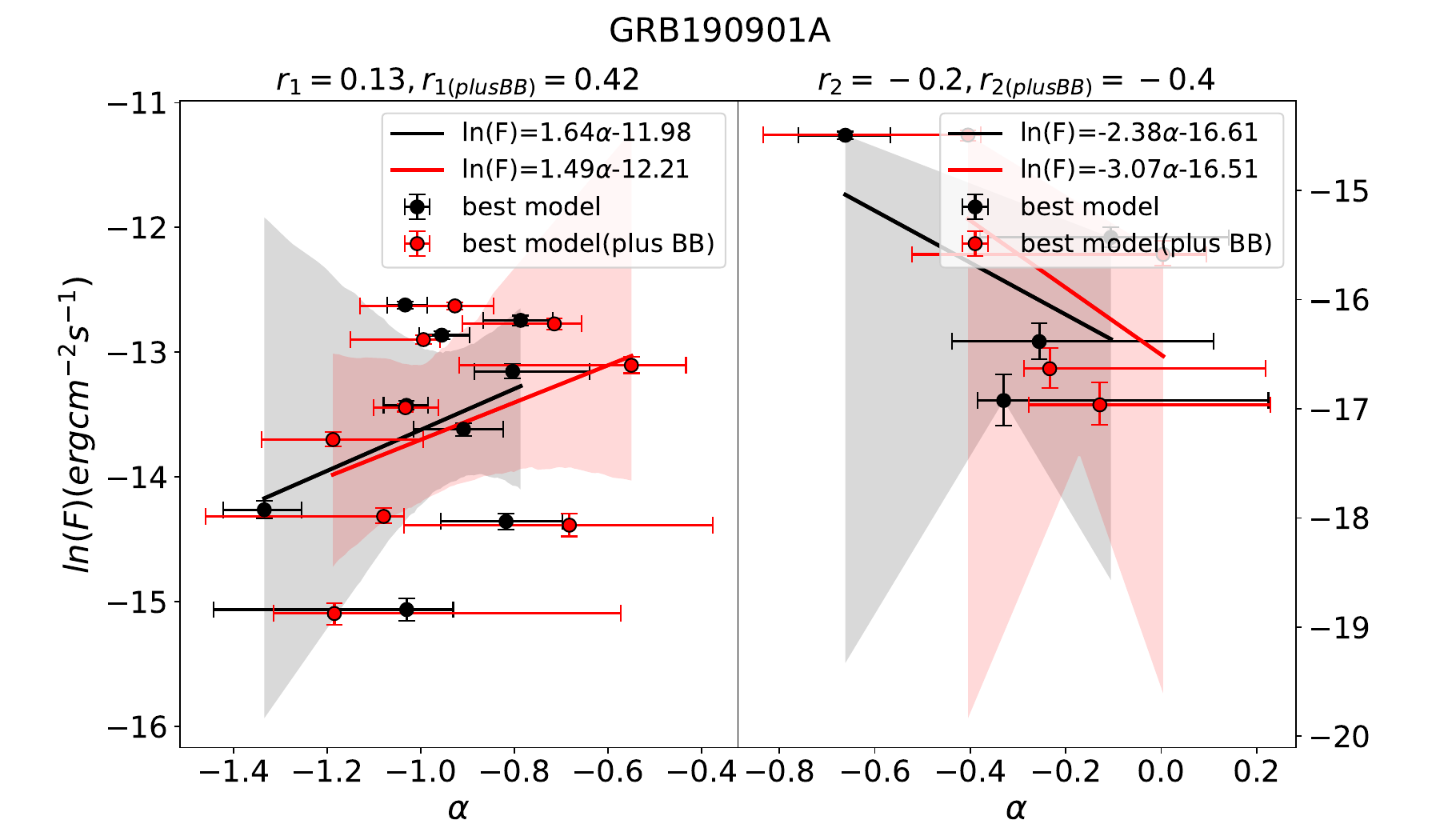}
\includegraphics [width=8cm,height=4cm]{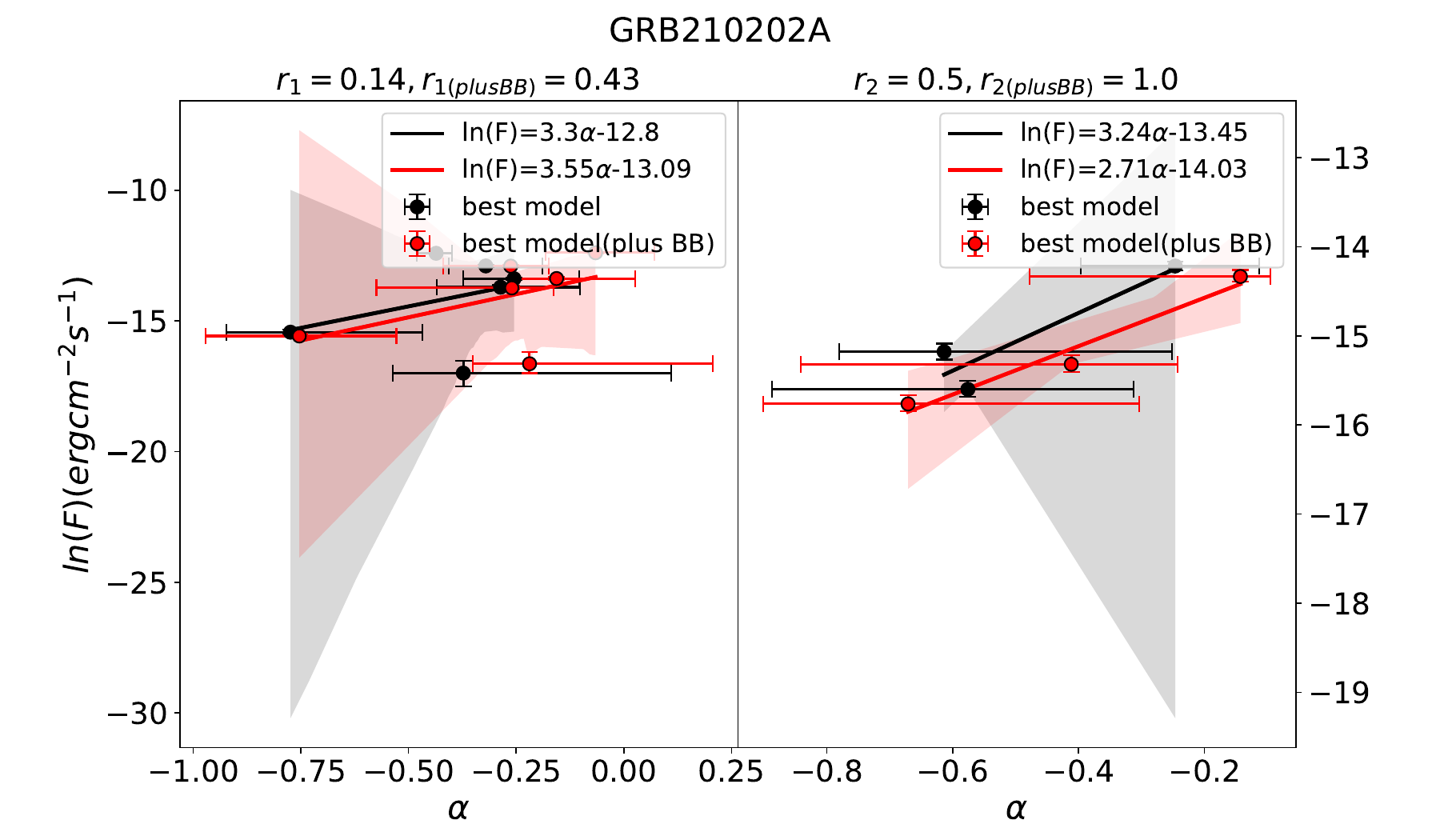}
\includegraphics [width=8cm,height=4cm]{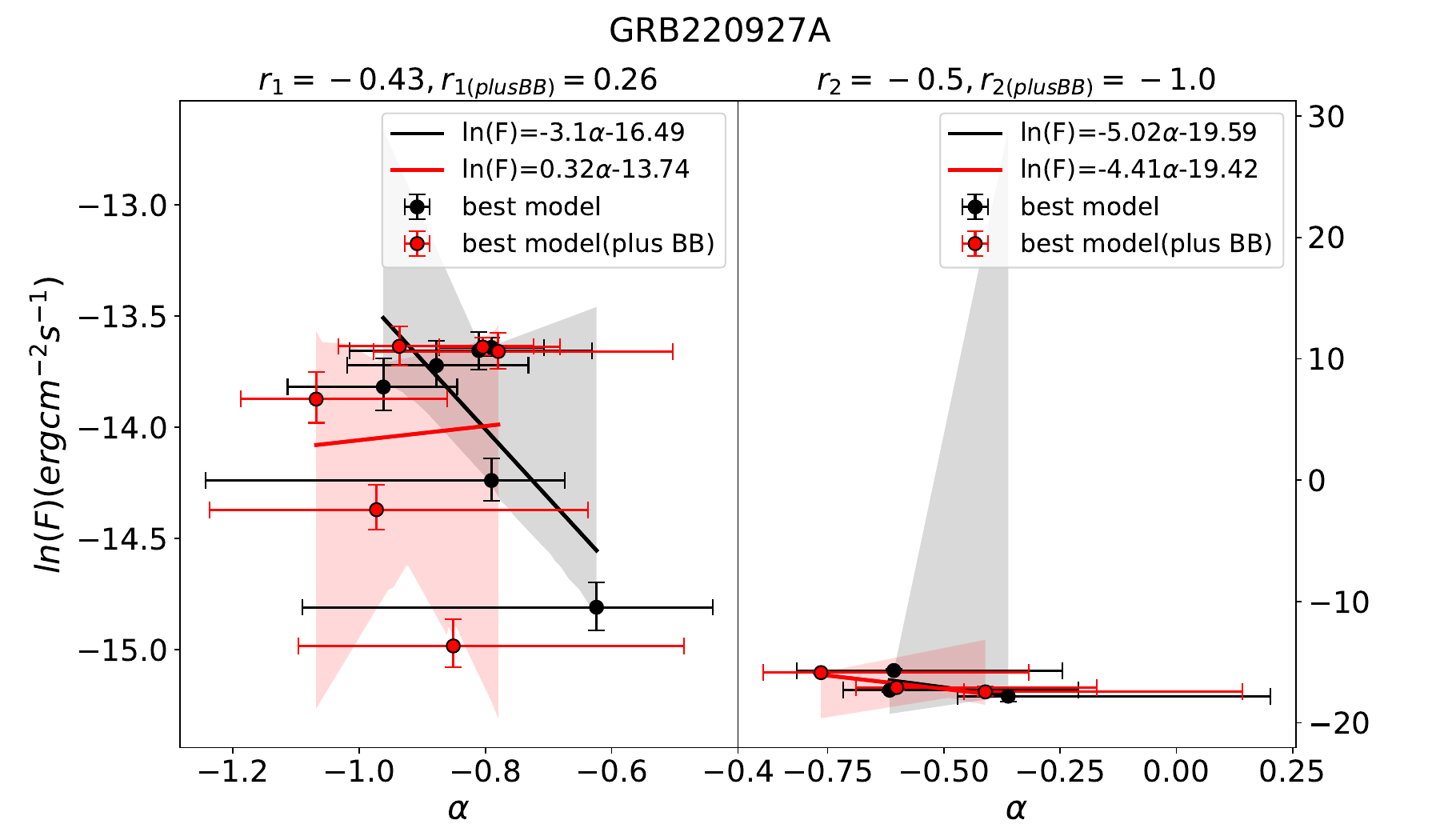}
\includegraphics [width=8cm,height=4cm]{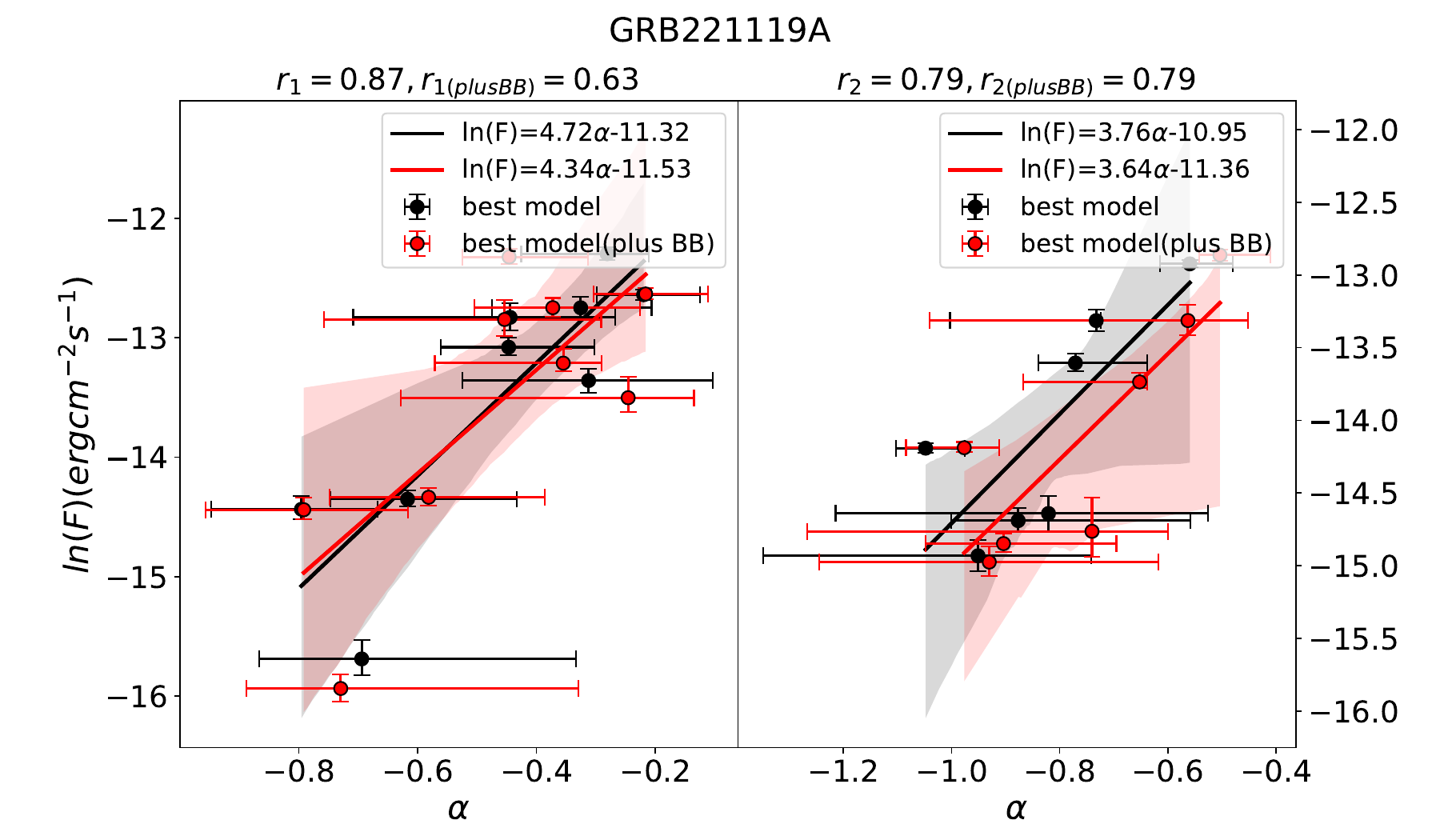}
\includegraphics [width=8cm,height=4cm]{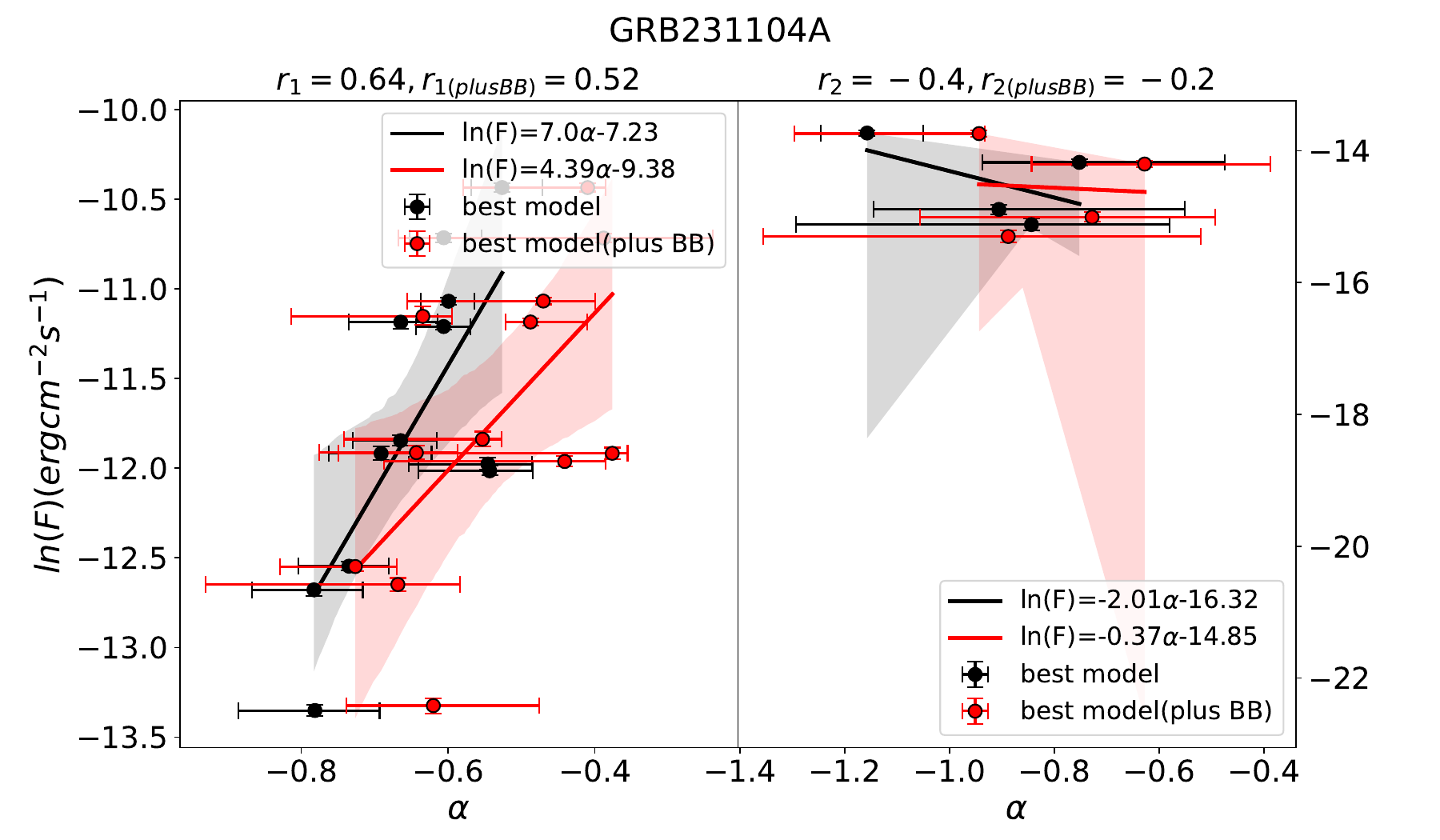}
\includegraphics [width=8cm,height=4cm]{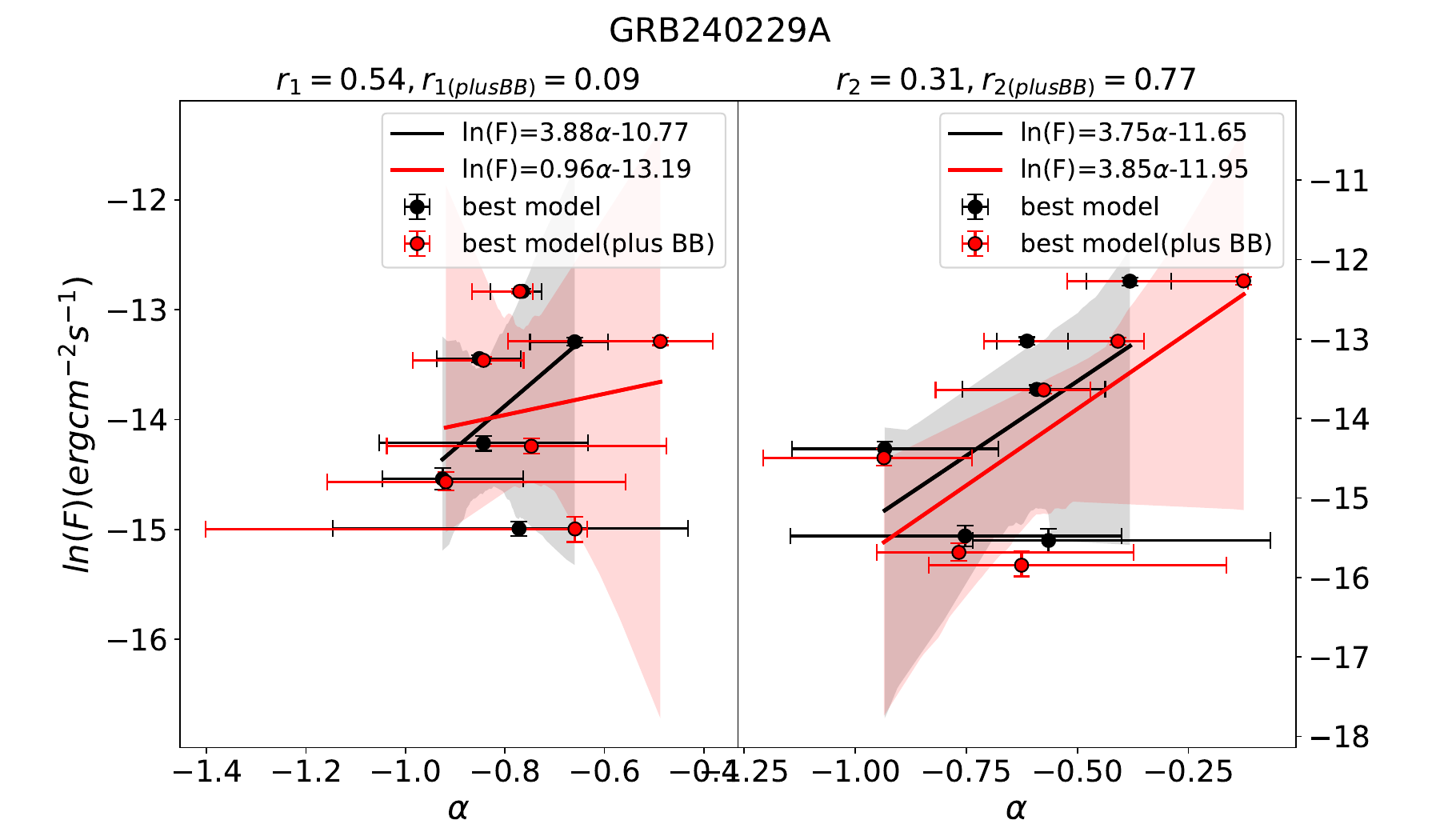}
   \figcaption{(Continued.) \label{fig 12} }
      
\end{figure}

\setcounter{figure}{12}  
\begin{figure}[H]
\centering
\includegraphics[width=8cm,height=4cm]{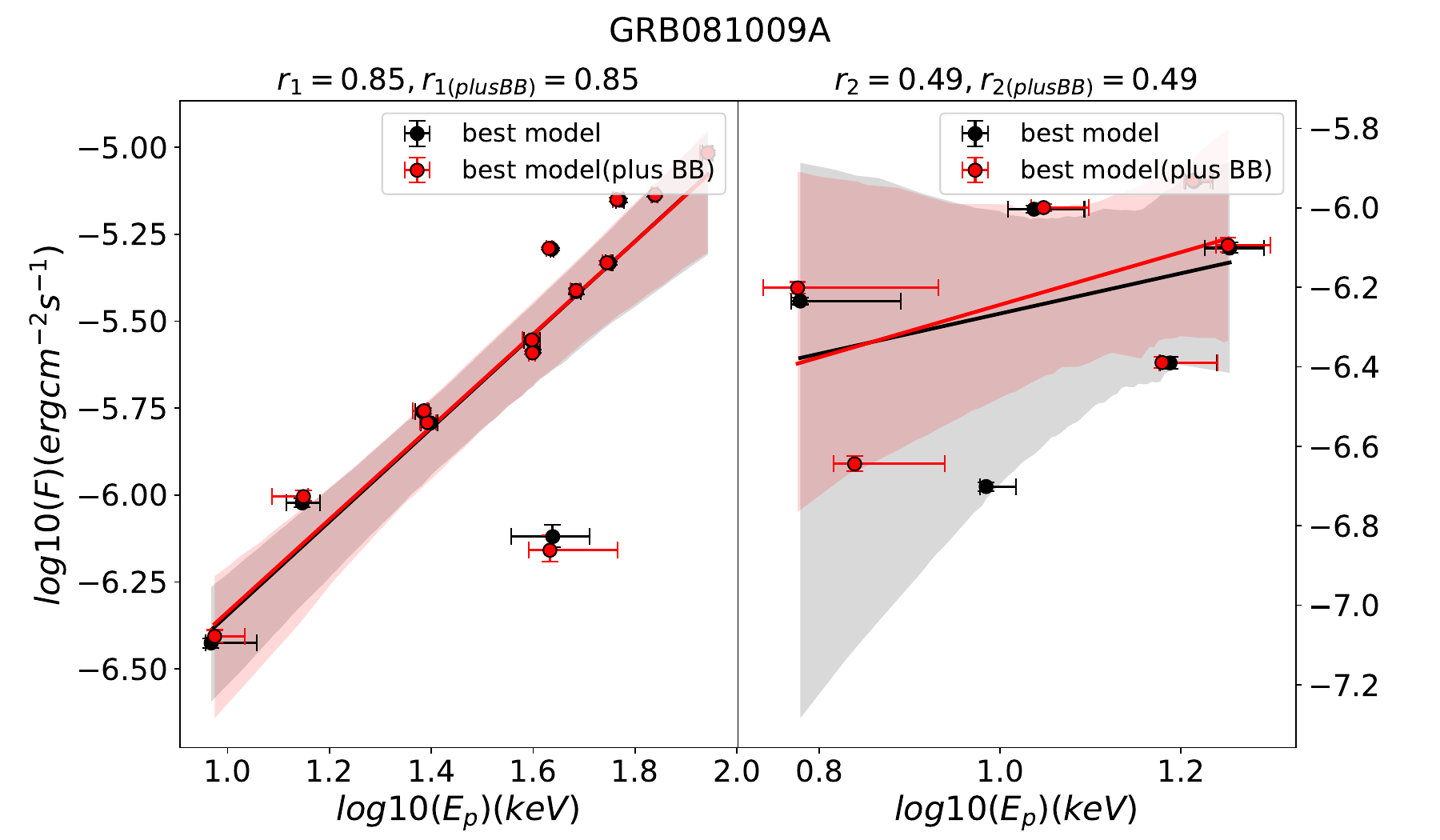}
\includegraphics[width=8cm,height=4cm]{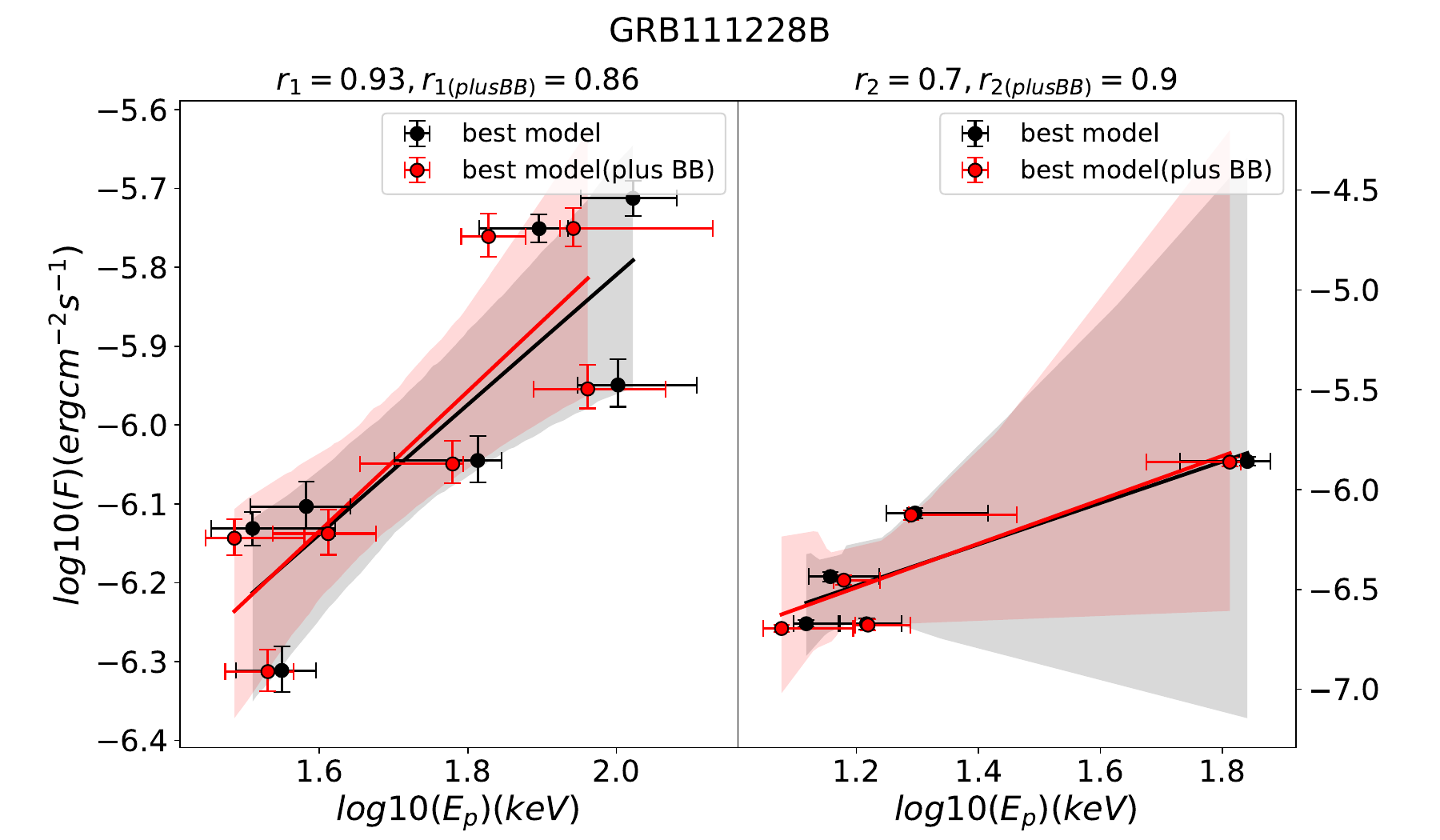}
\includegraphics[width=8cm,height=4cm]{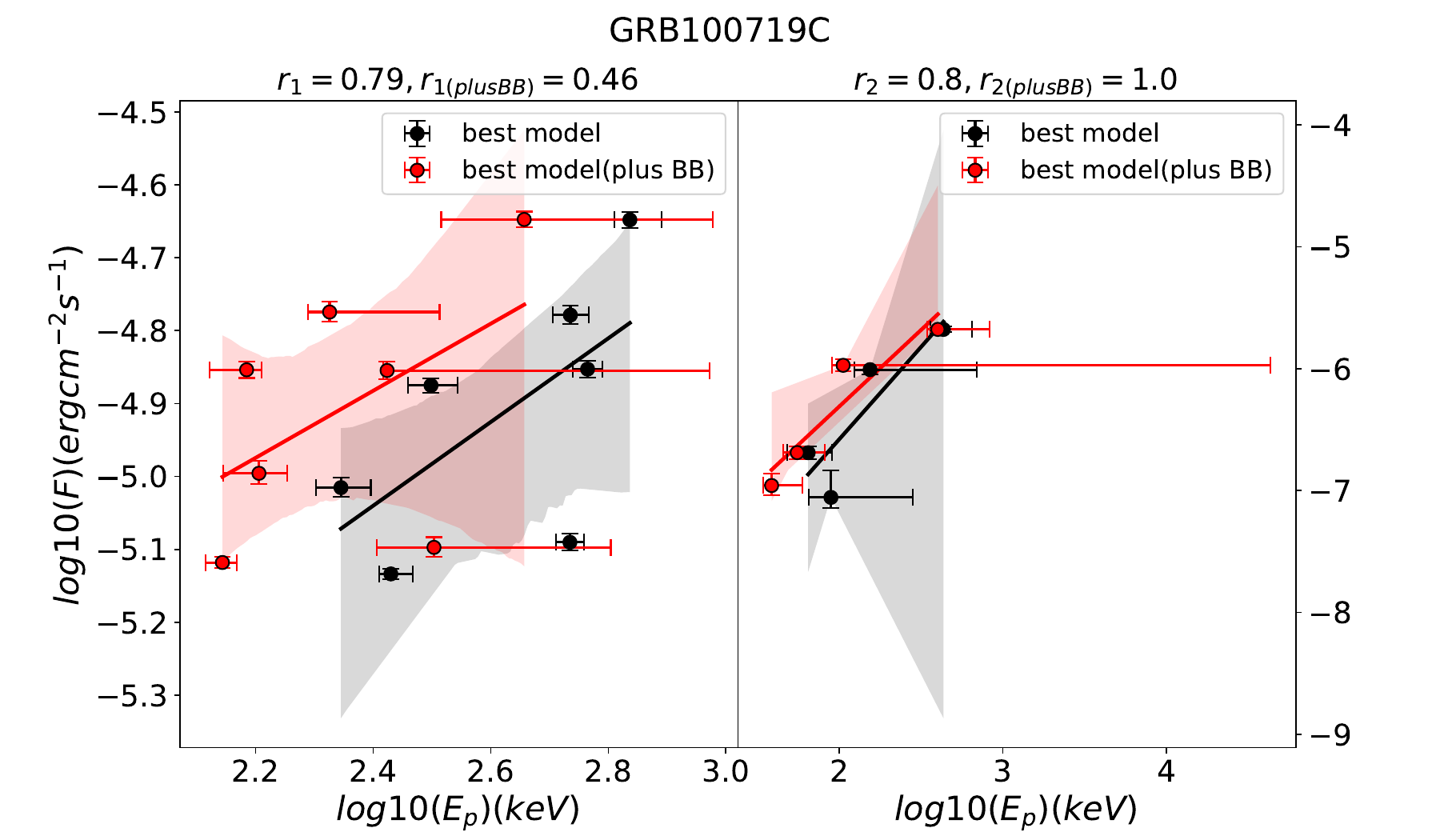}
\includegraphics[width=8cm,height=4cm]{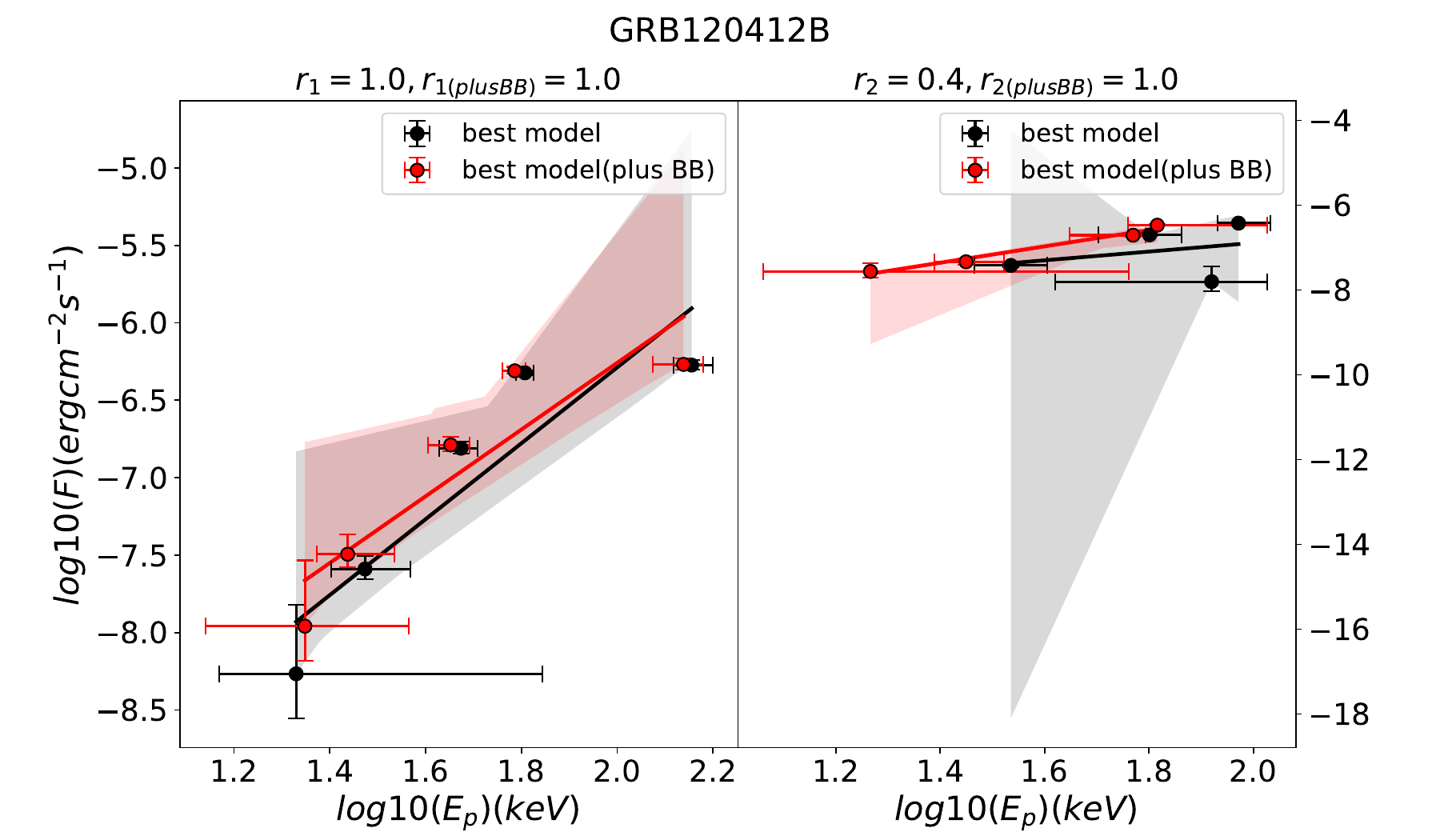}
\includegraphics[width=8cm,height=4cm]{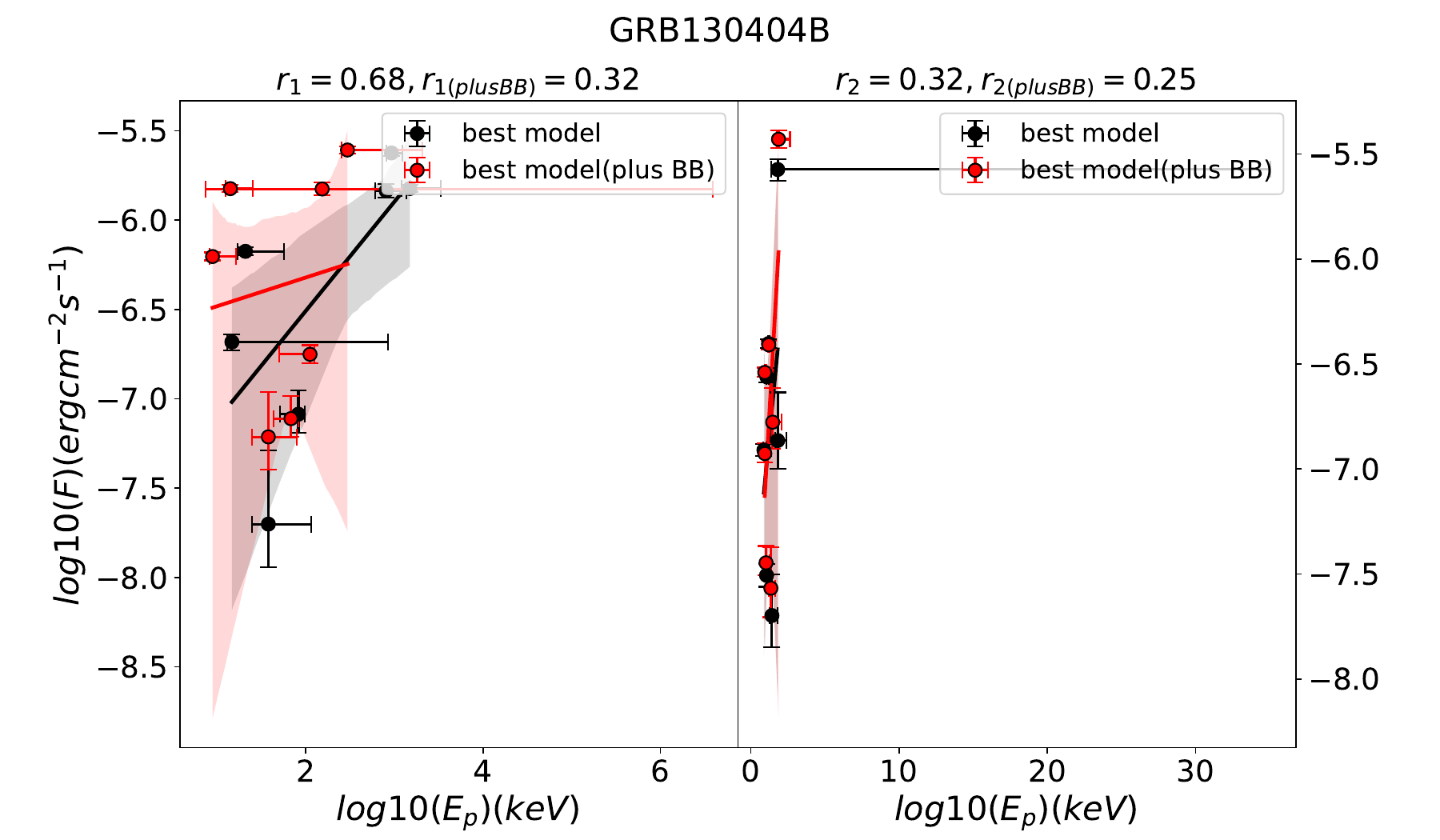}
\includegraphics[width=8cm,height=4cm]{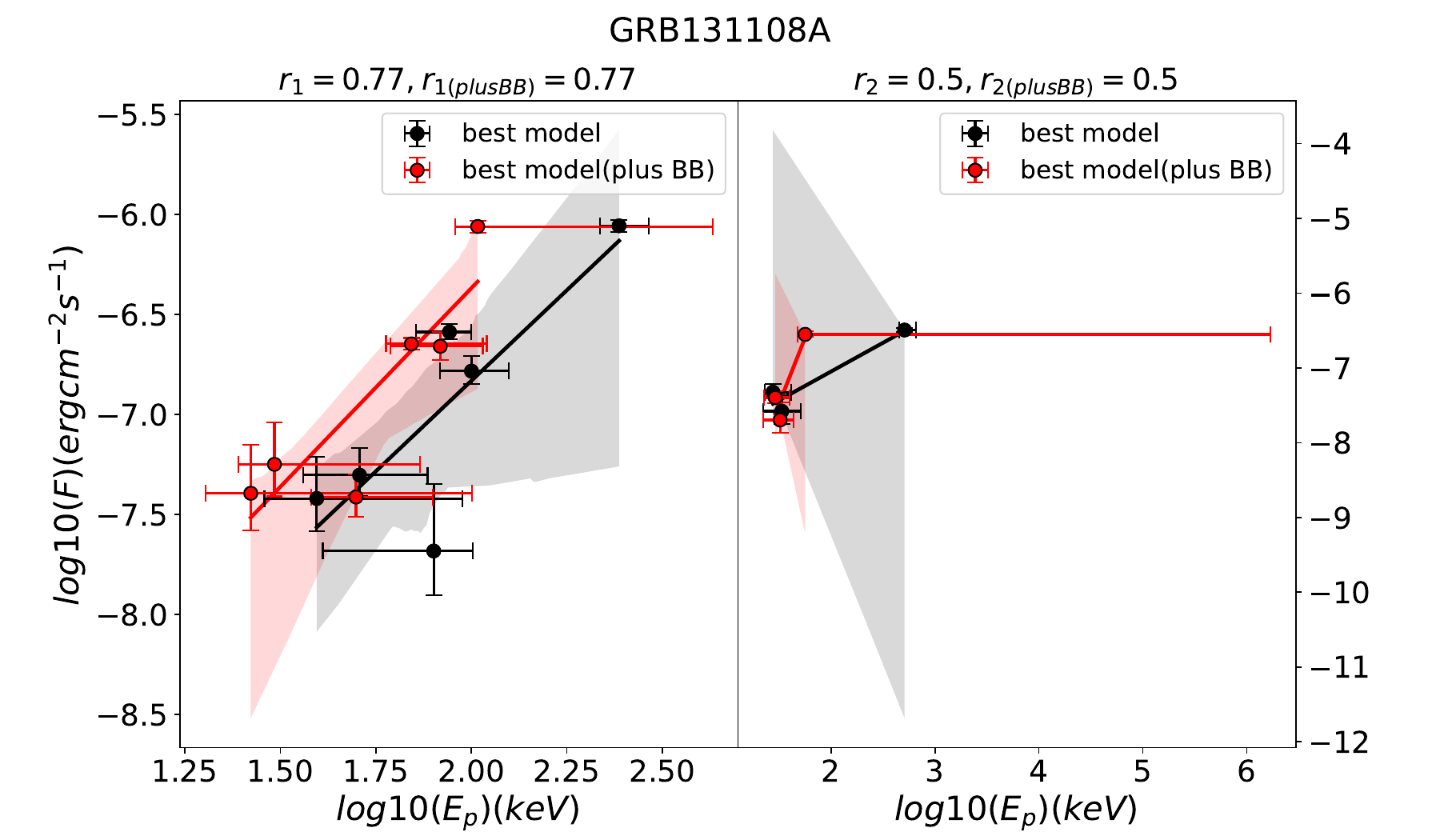}
\includegraphics[width=8cm,height=4cm]{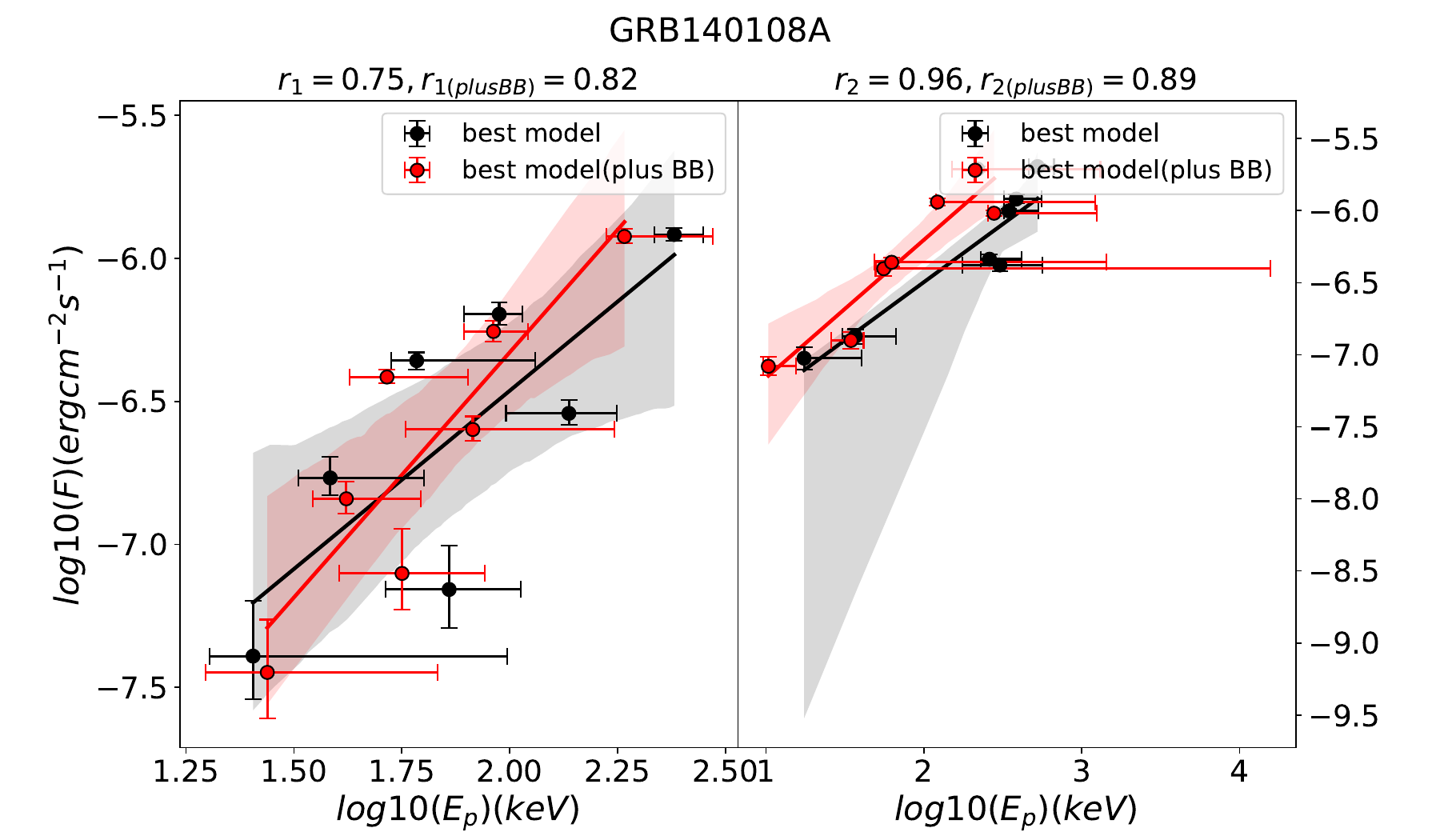}
\includegraphics[width=8cm,height=4cm]{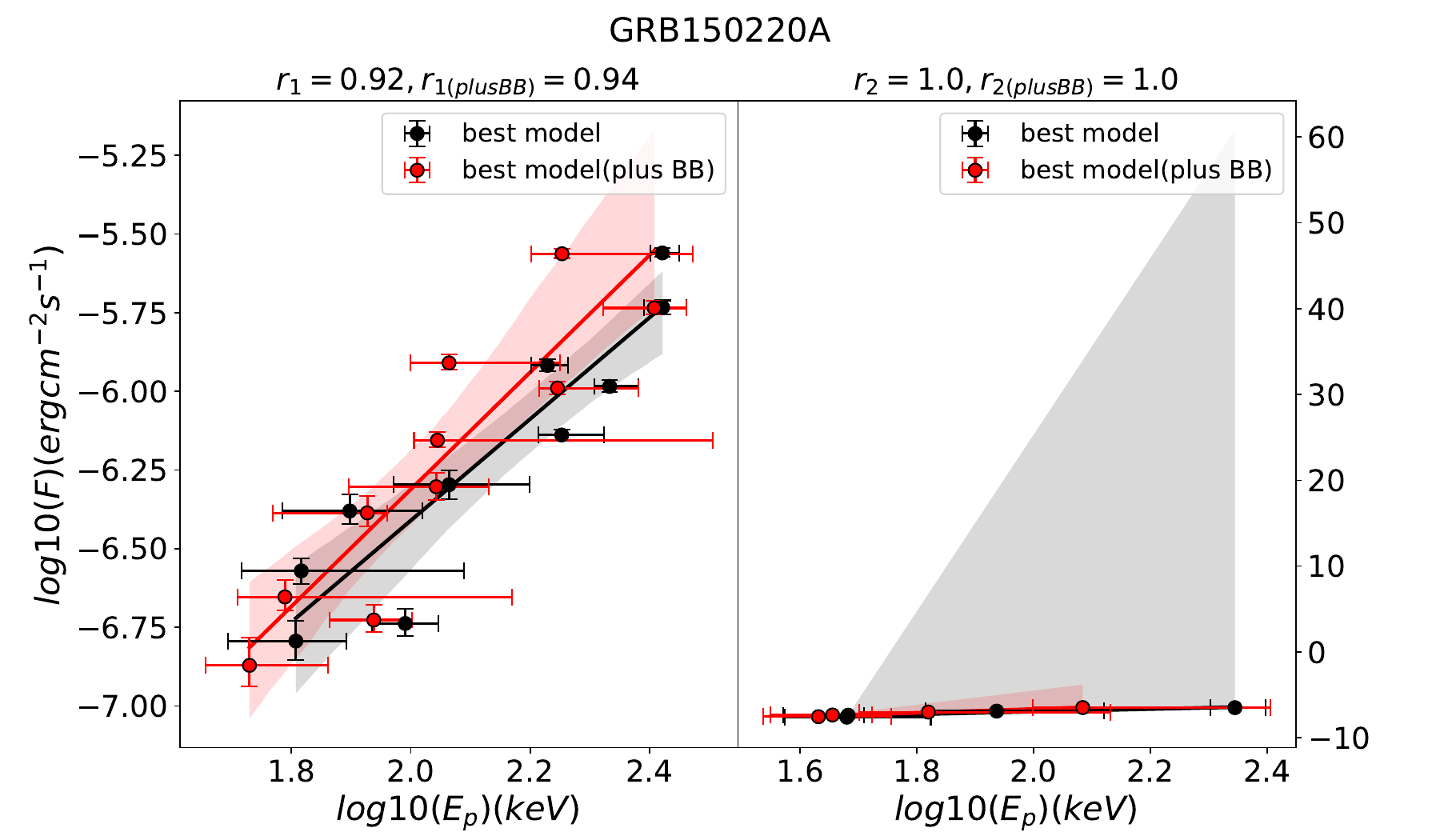}
\includegraphics [width=8cm,height=4cm]{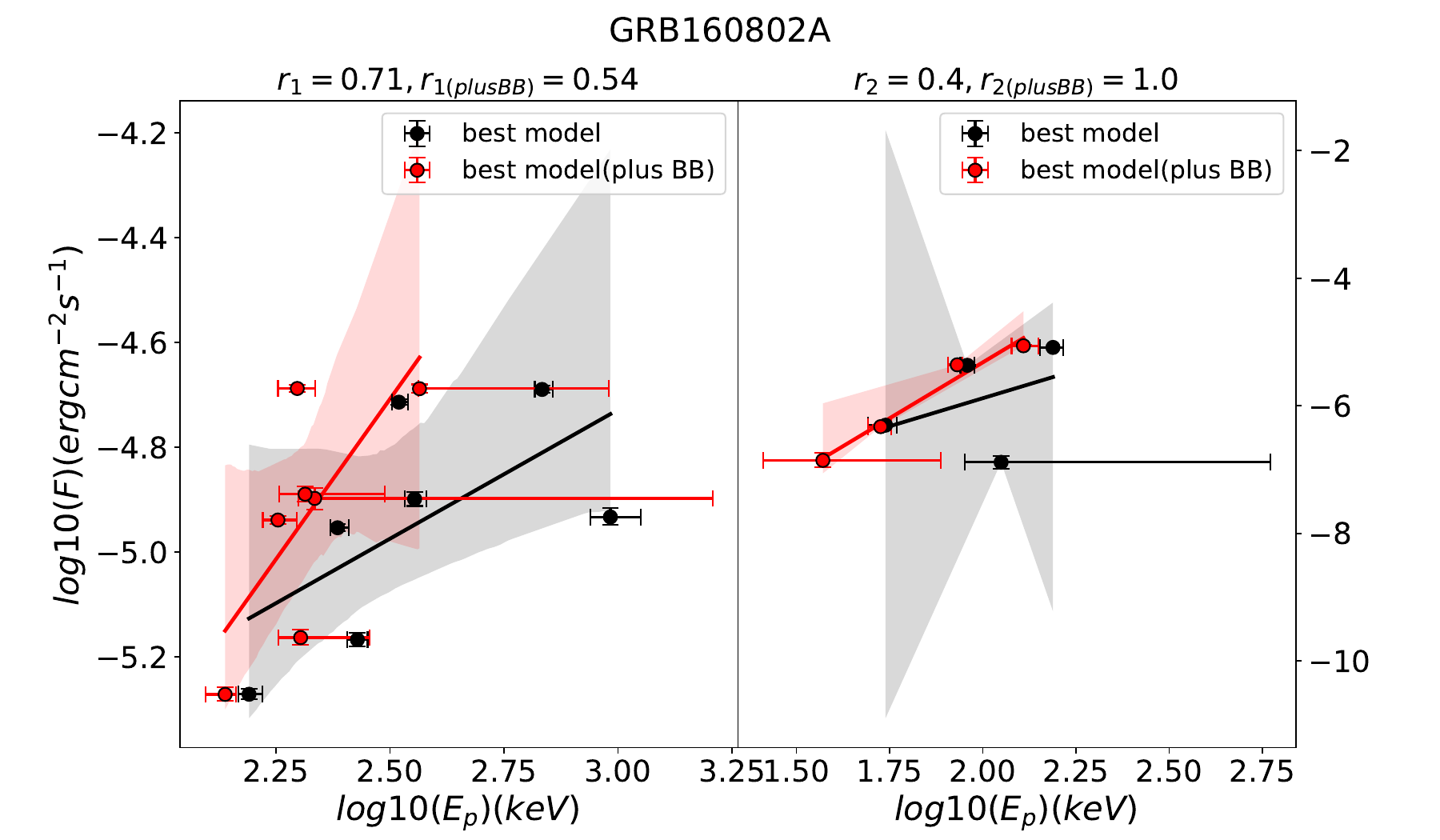}
\includegraphics [width=8cm,height=4cm]{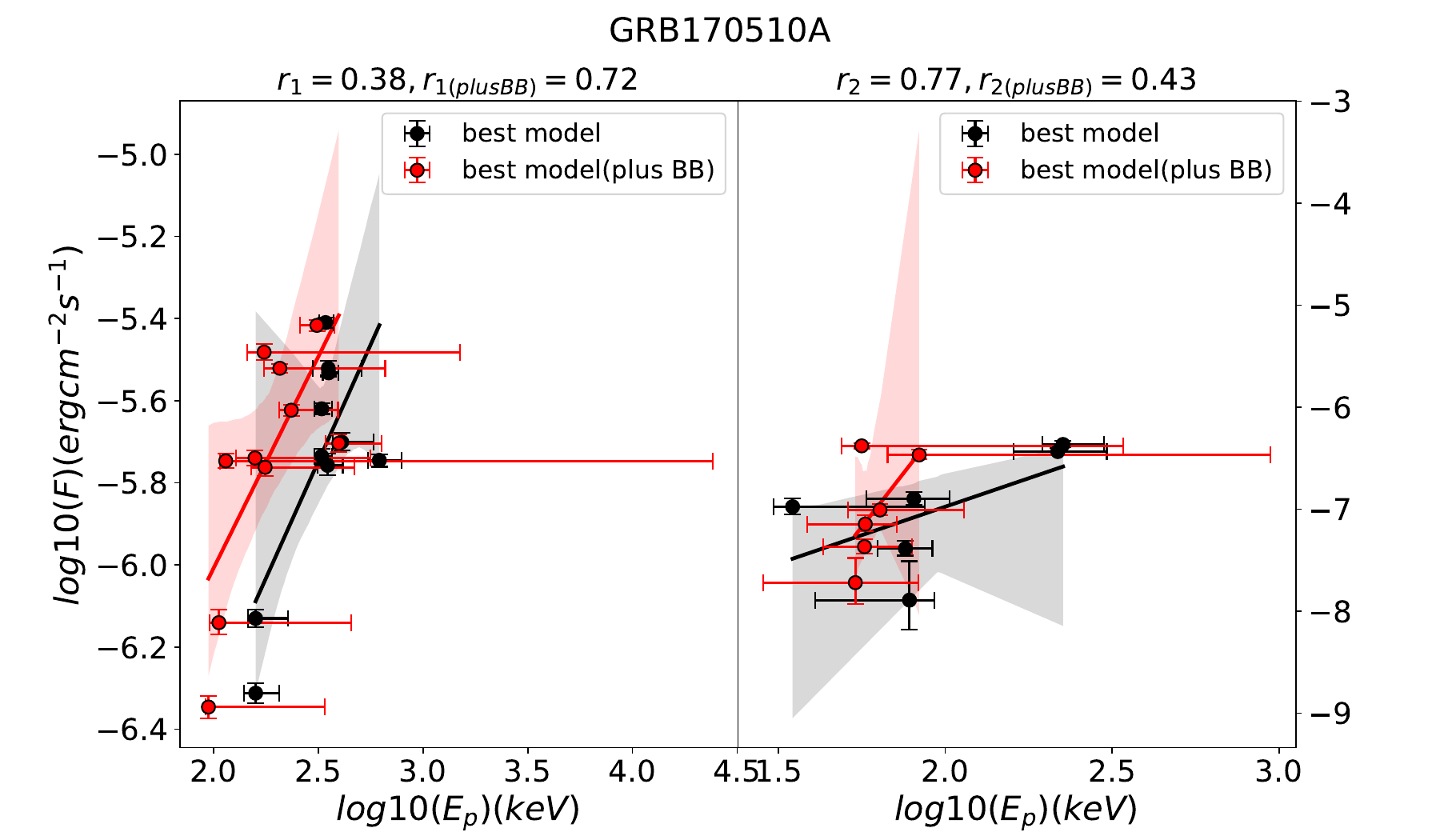}
   \caption{The correlation between $F$ and $E_{p}$. The labels are similar to those in Figure \ref{fig 12}. \label{fig 13}}

\end{figure}

\setcounter{figure}{12}  
\begin{figure}[H]

\centering
\includegraphics[width=8cm,height=4cm]{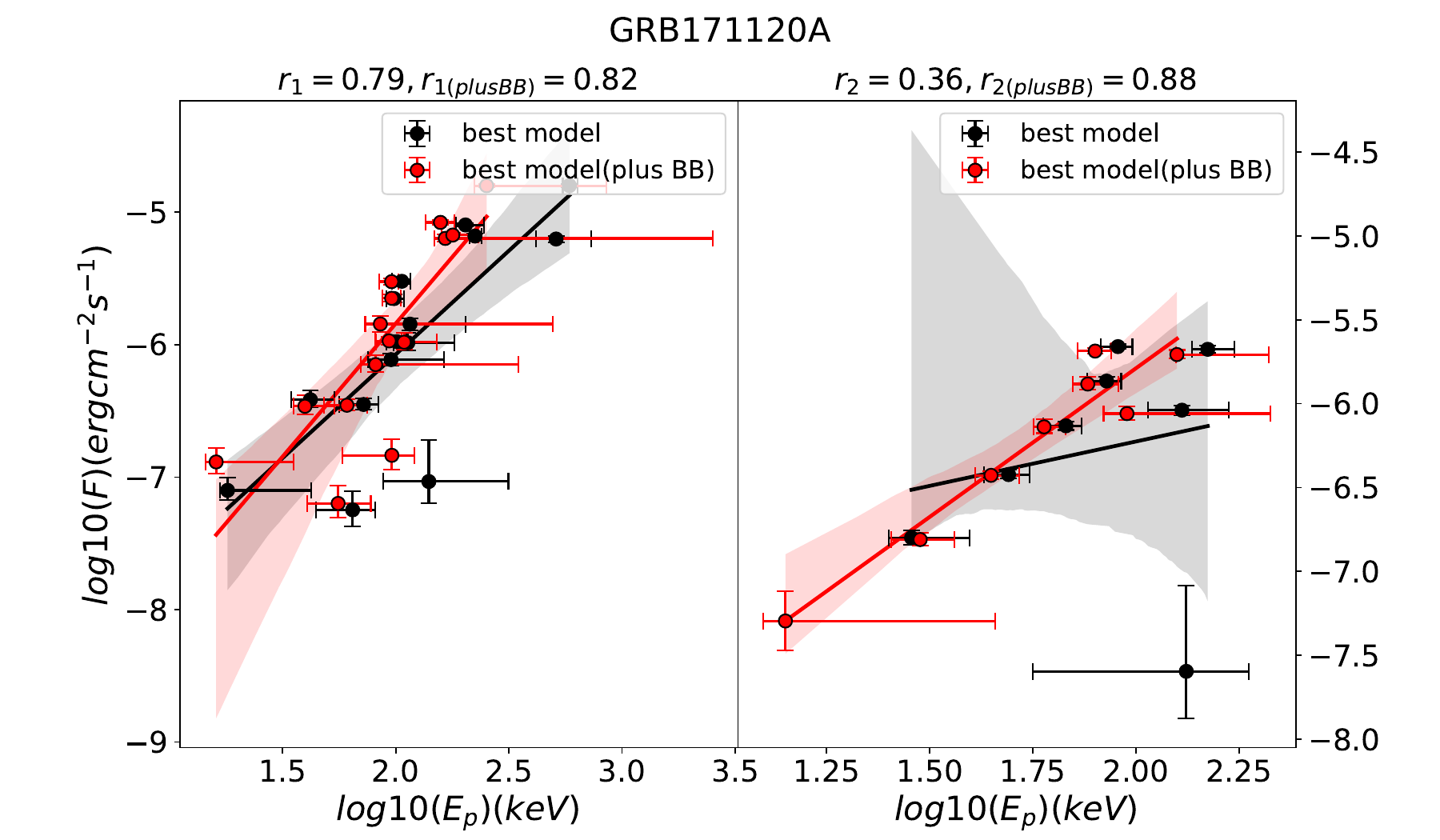}
\includegraphics[width=8cm,height=4cm]{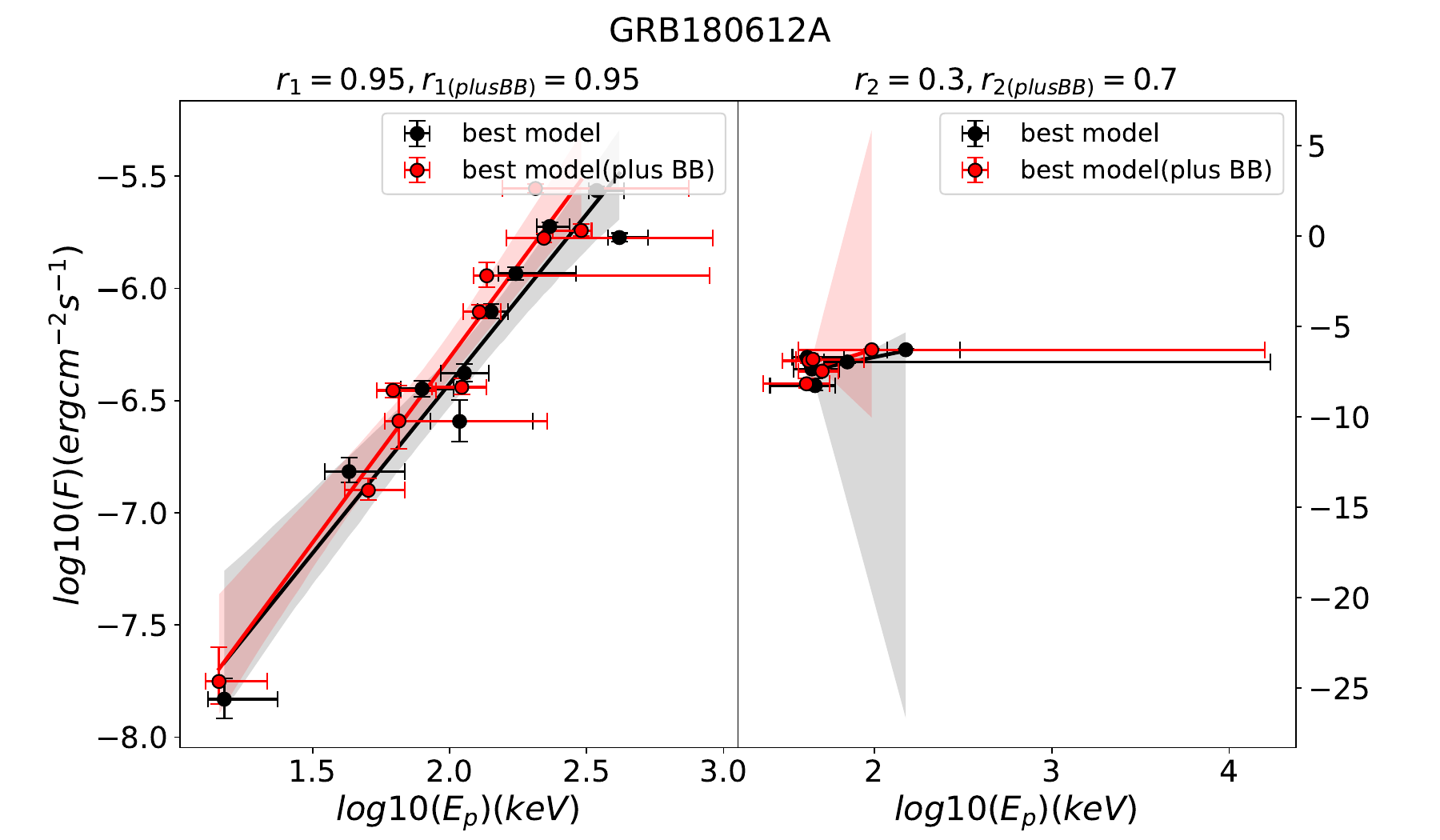}
\includegraphics[width=8cm,height=4cm]{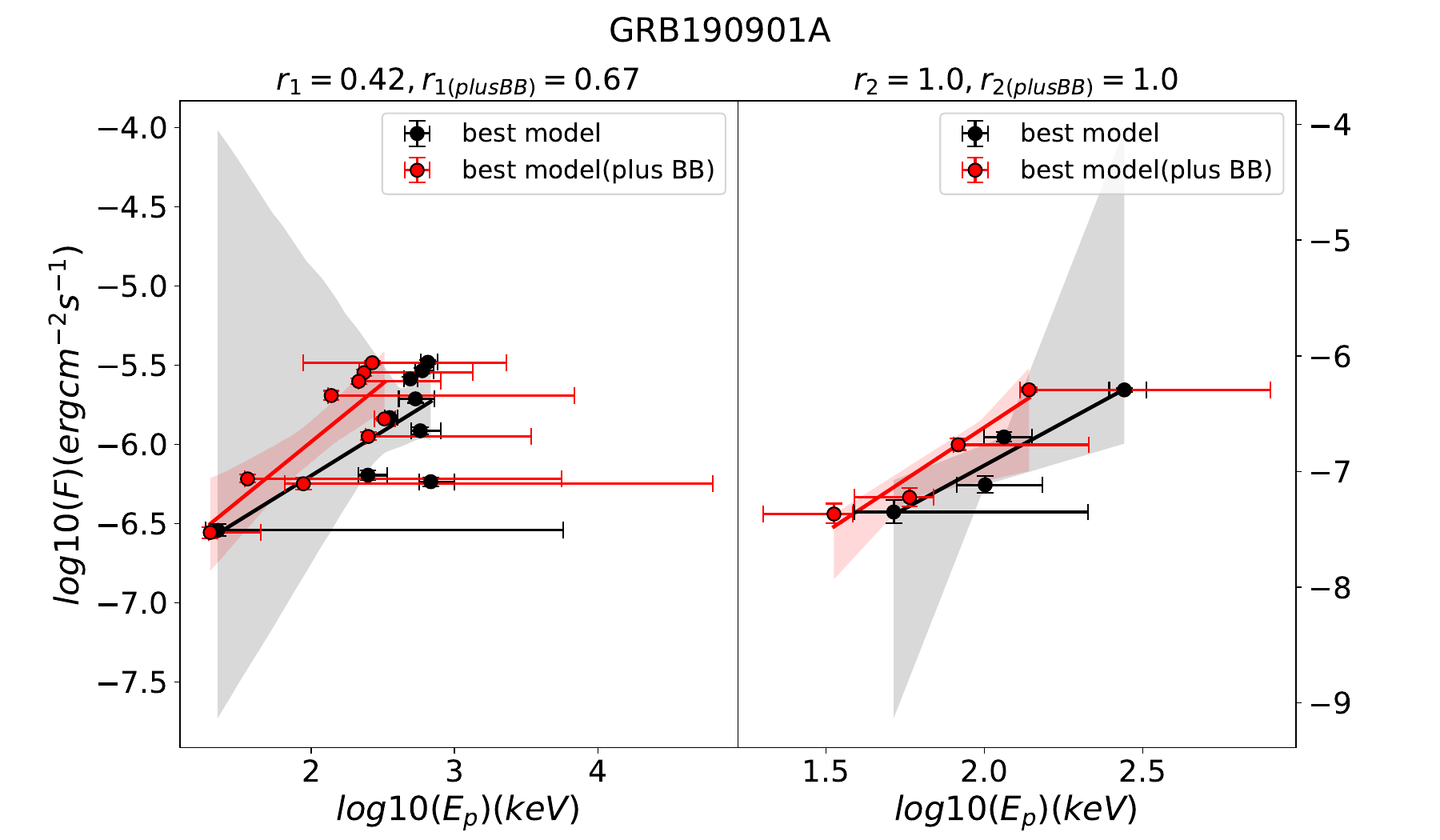}
\includegraphics[width=8cm,height=4cm]{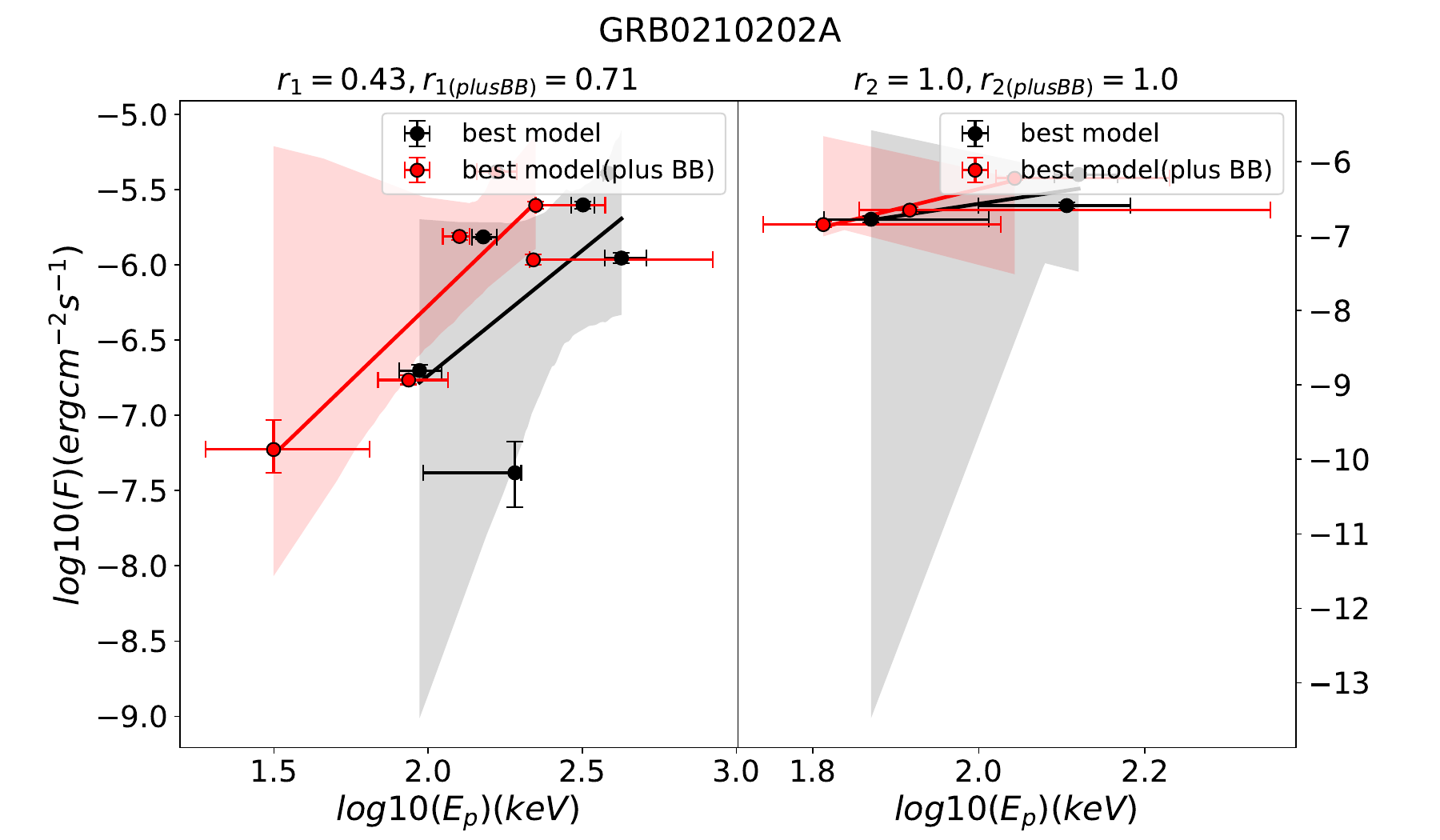}
\includegraphics[width=8cm,height=4cm]{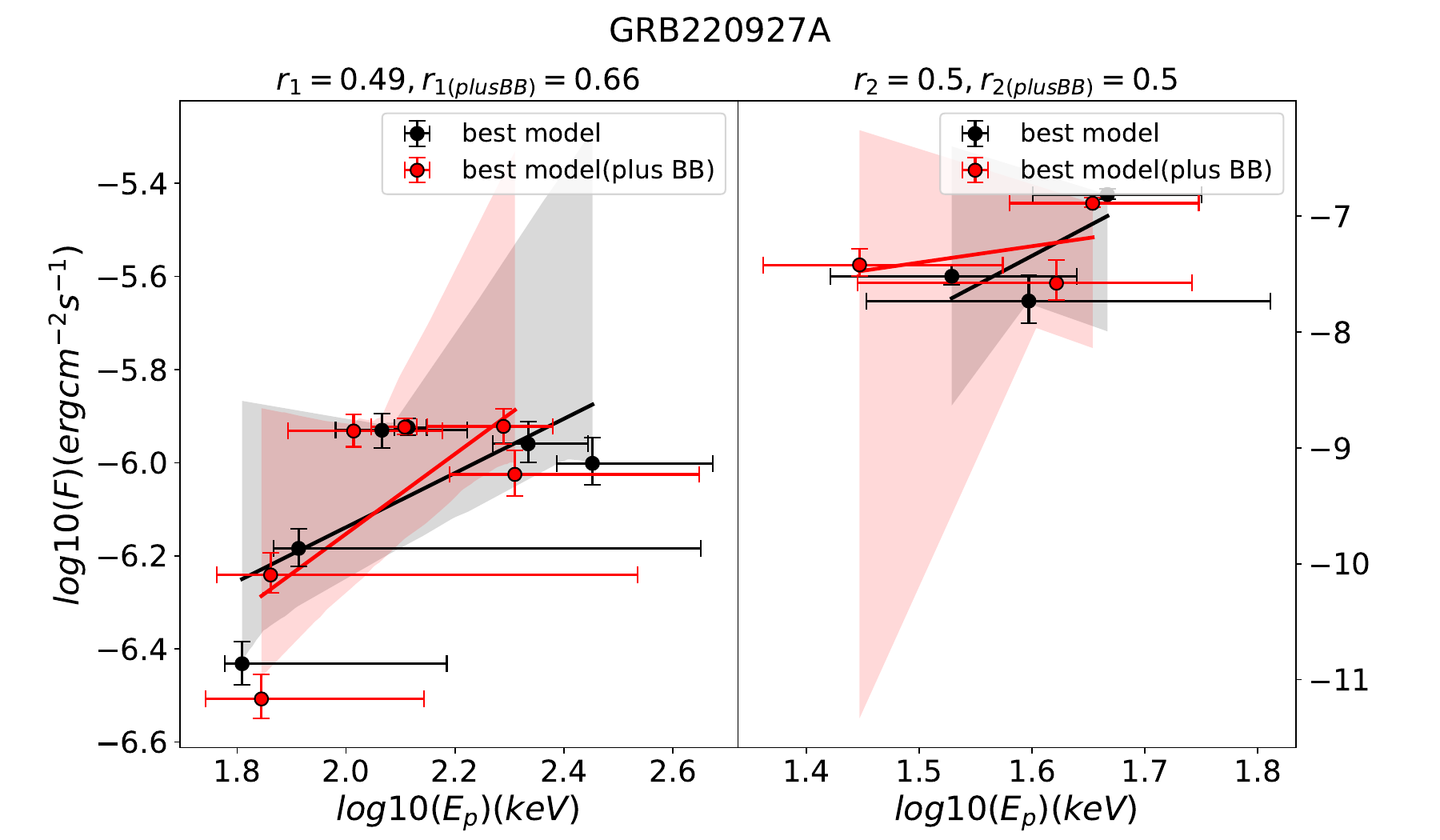}
\includegraphics[width=8cm,height=4cm]{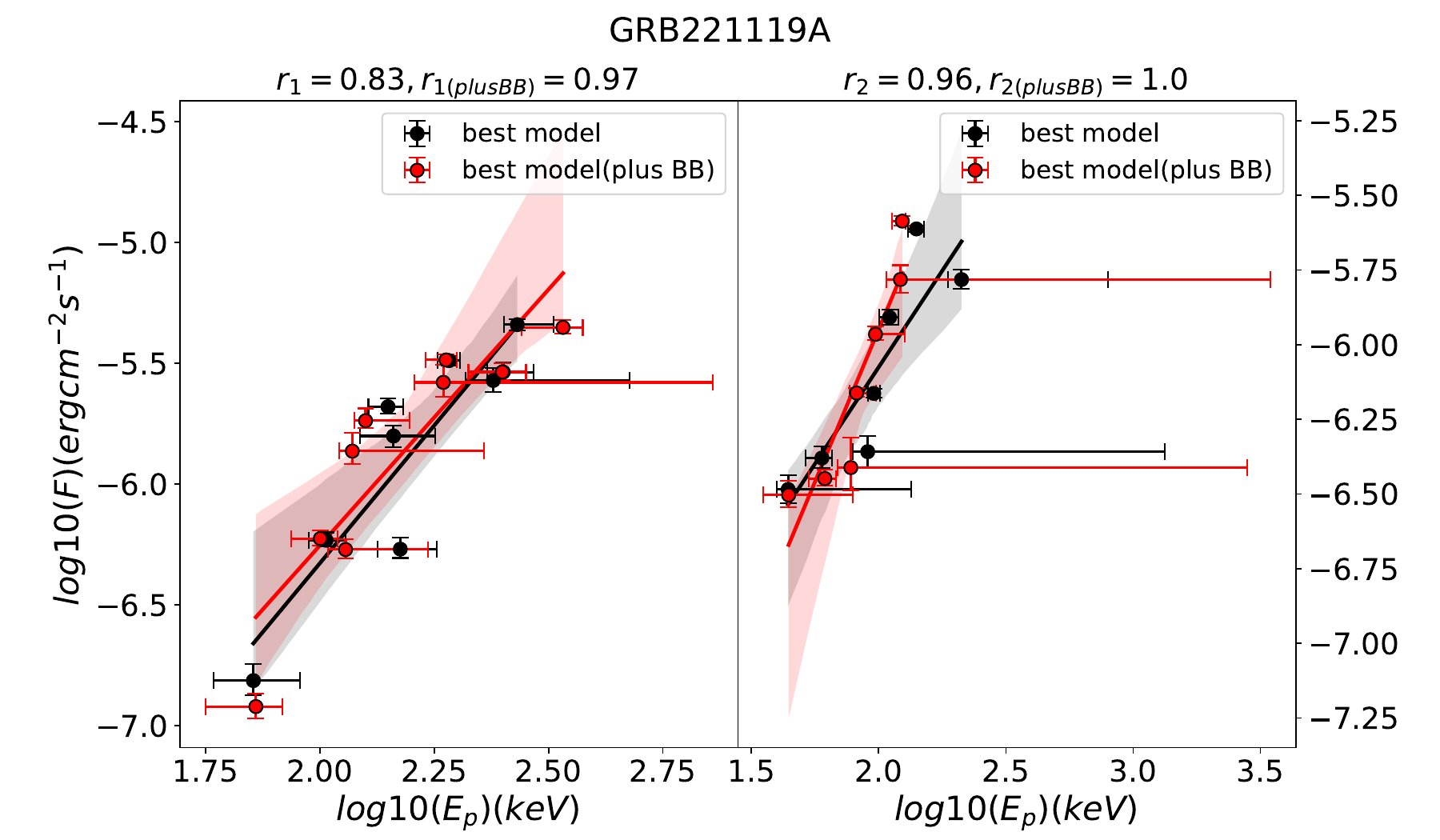}
\includegraphics [width=8cm,height=4cm]{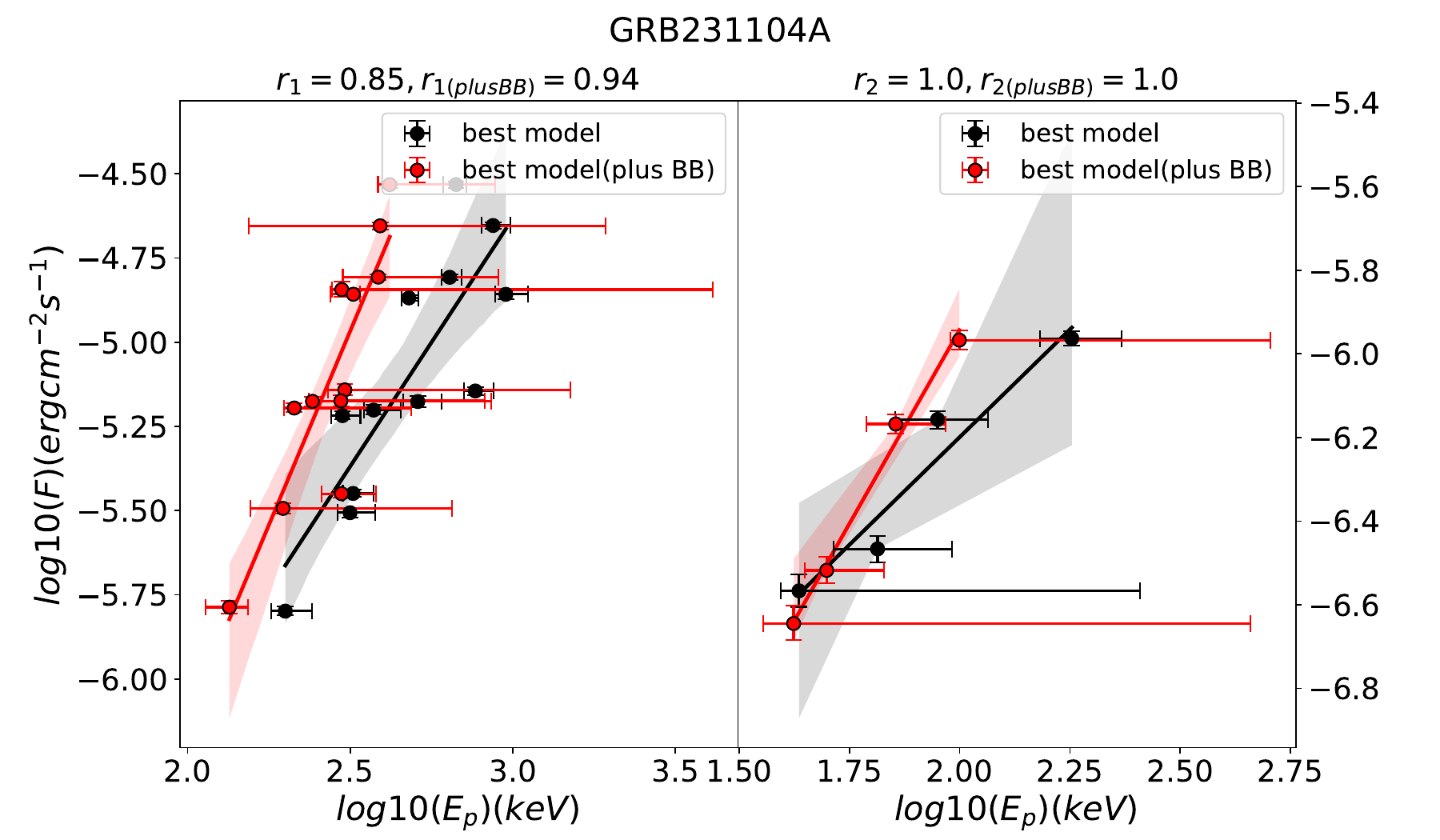}
\includegraphics [width=8cm,height=4cm]{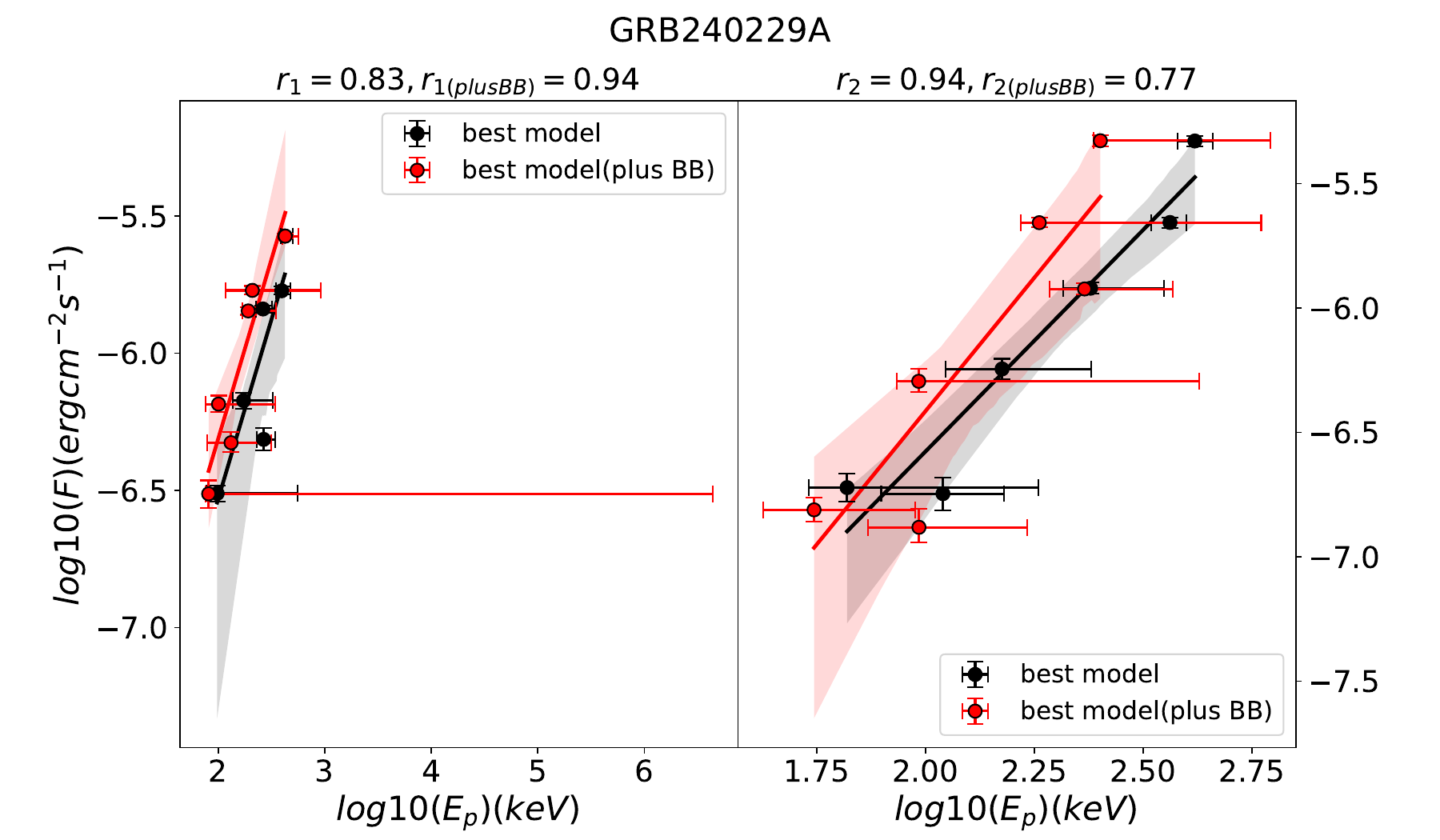}
   \caption{(Continued.) \label{fig 13}}

\end{figure}

\setcounter{figure}{13}  
\begin{figure}[H]
\centering
\includegraphics [width=8cm,height=4cm]{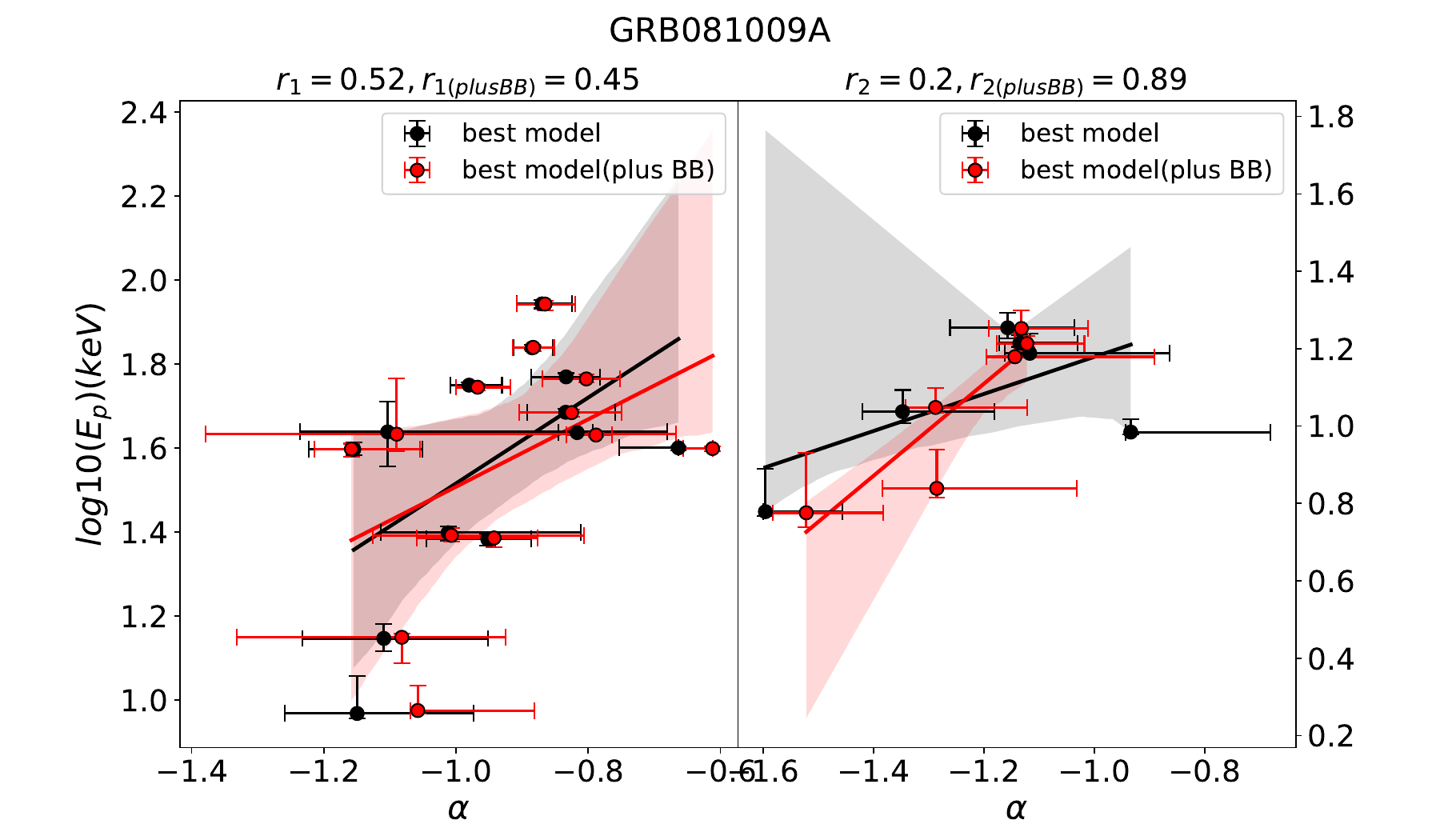}
\includegraphics [width=8cm,height=4cm]{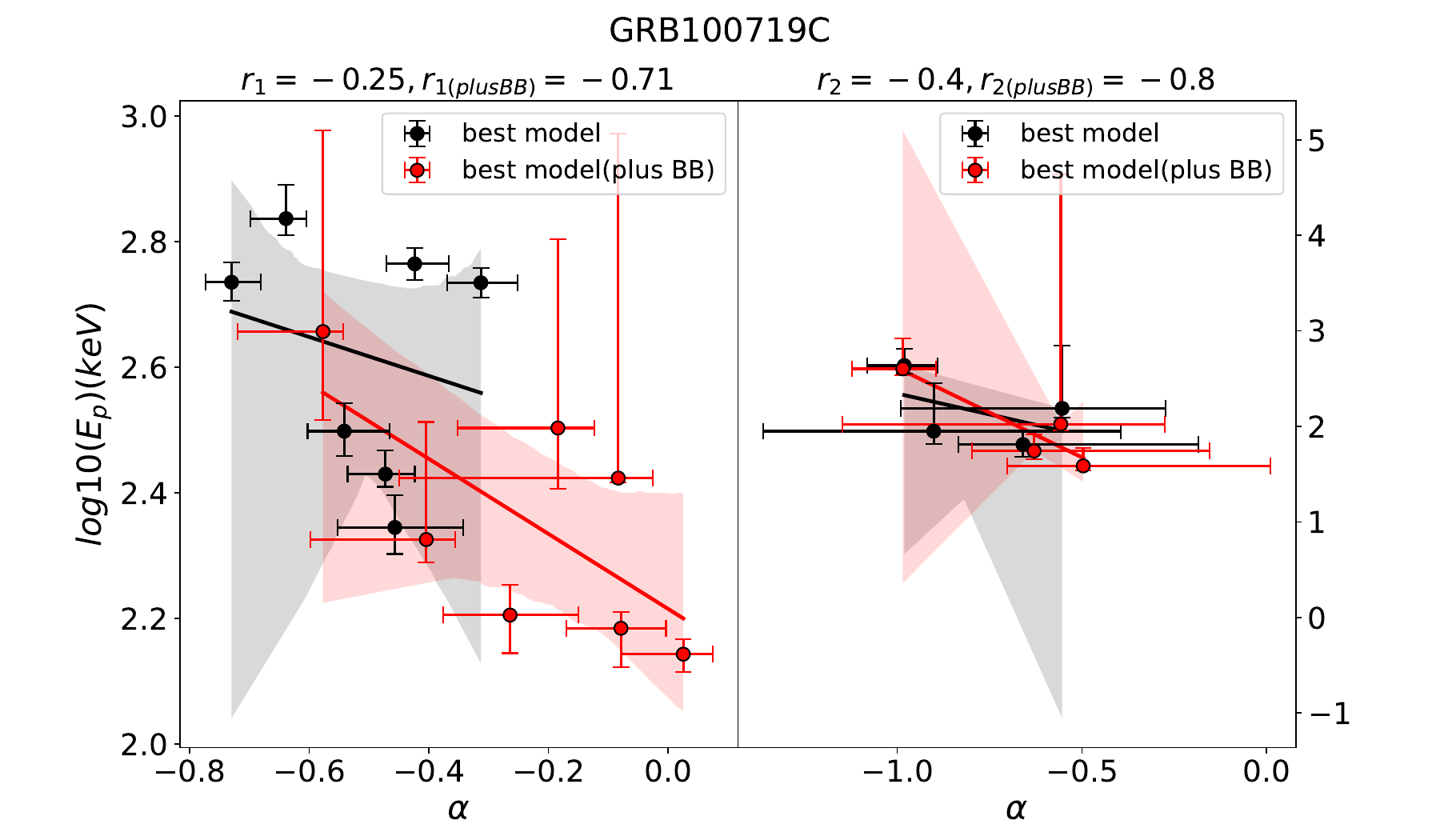}
\includegraphics [width=8cm,height=4cm]{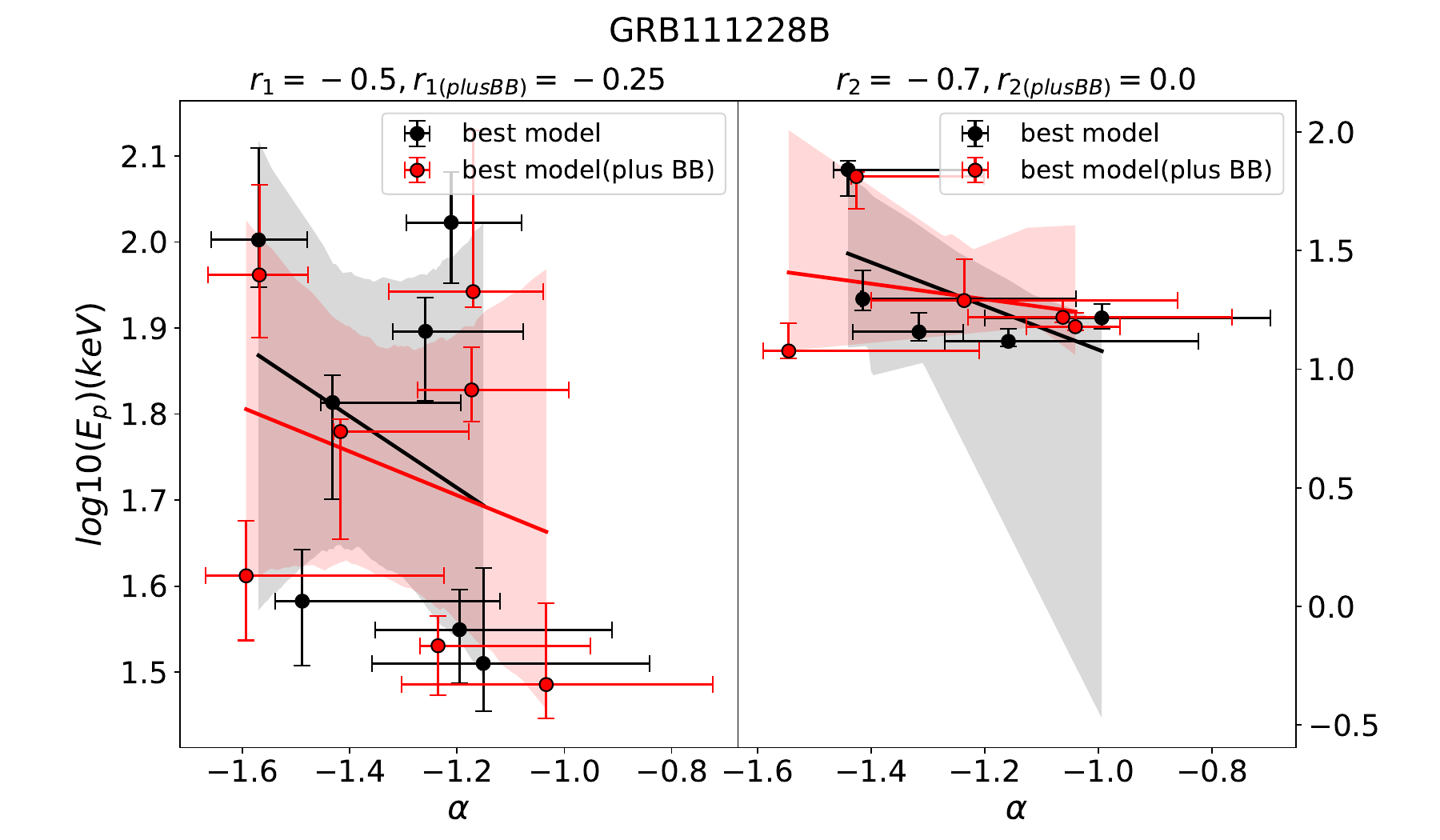}
\includegraphics [width=8cm,height=4cm]{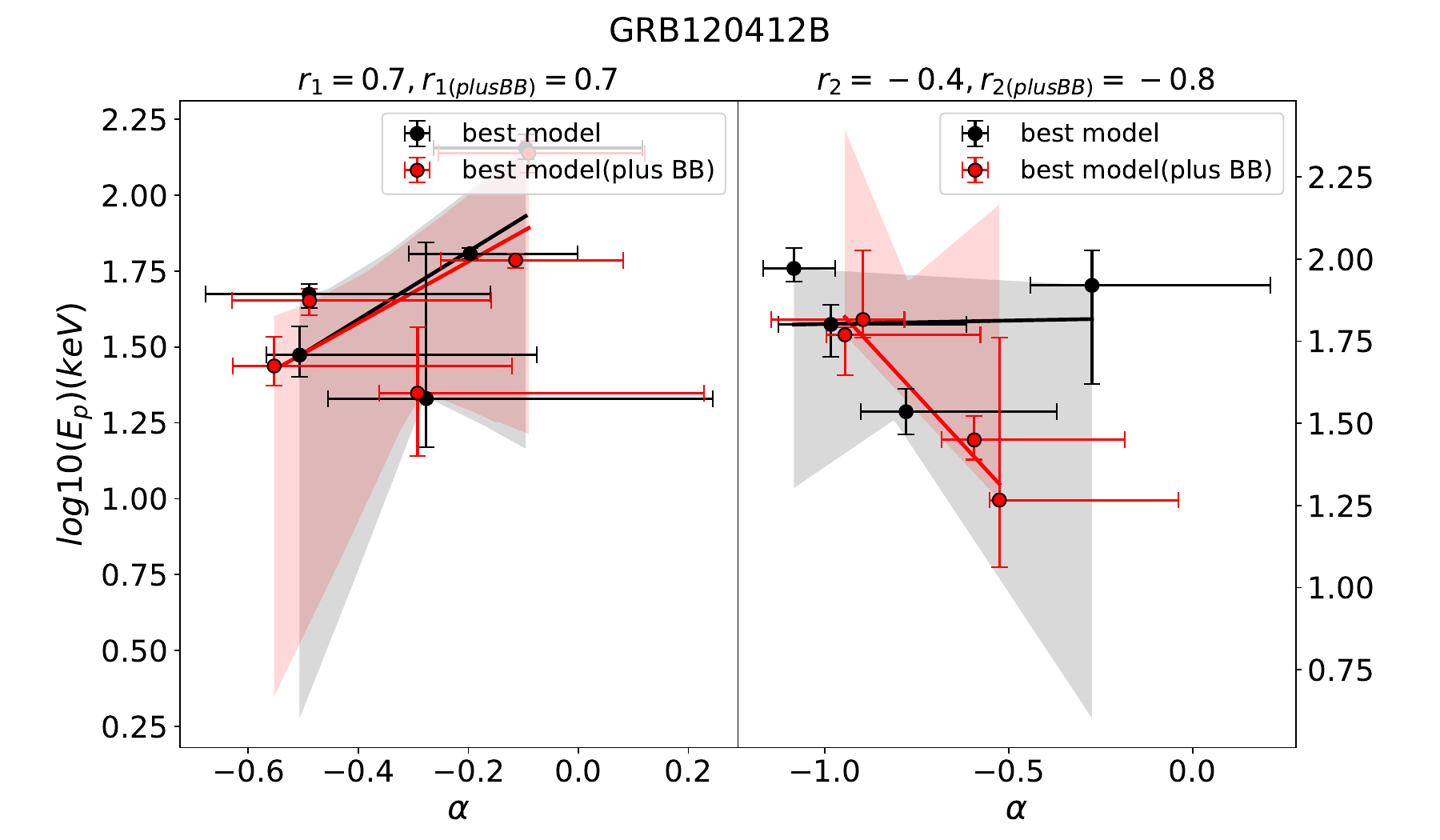}
\includegraphics [width=8cm,height=4cm]{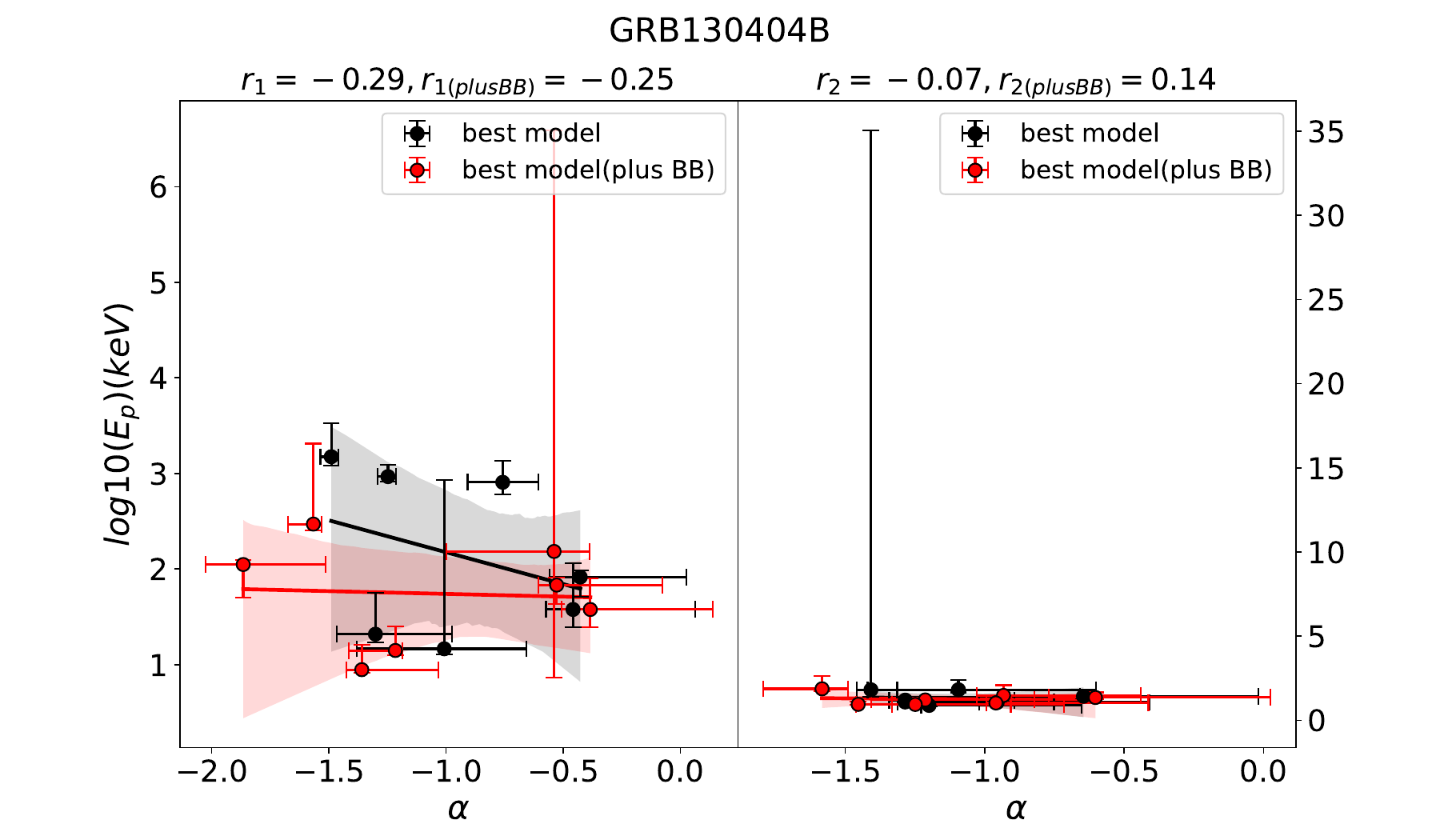}
\includegraphics [width=8cm,height=4cm]{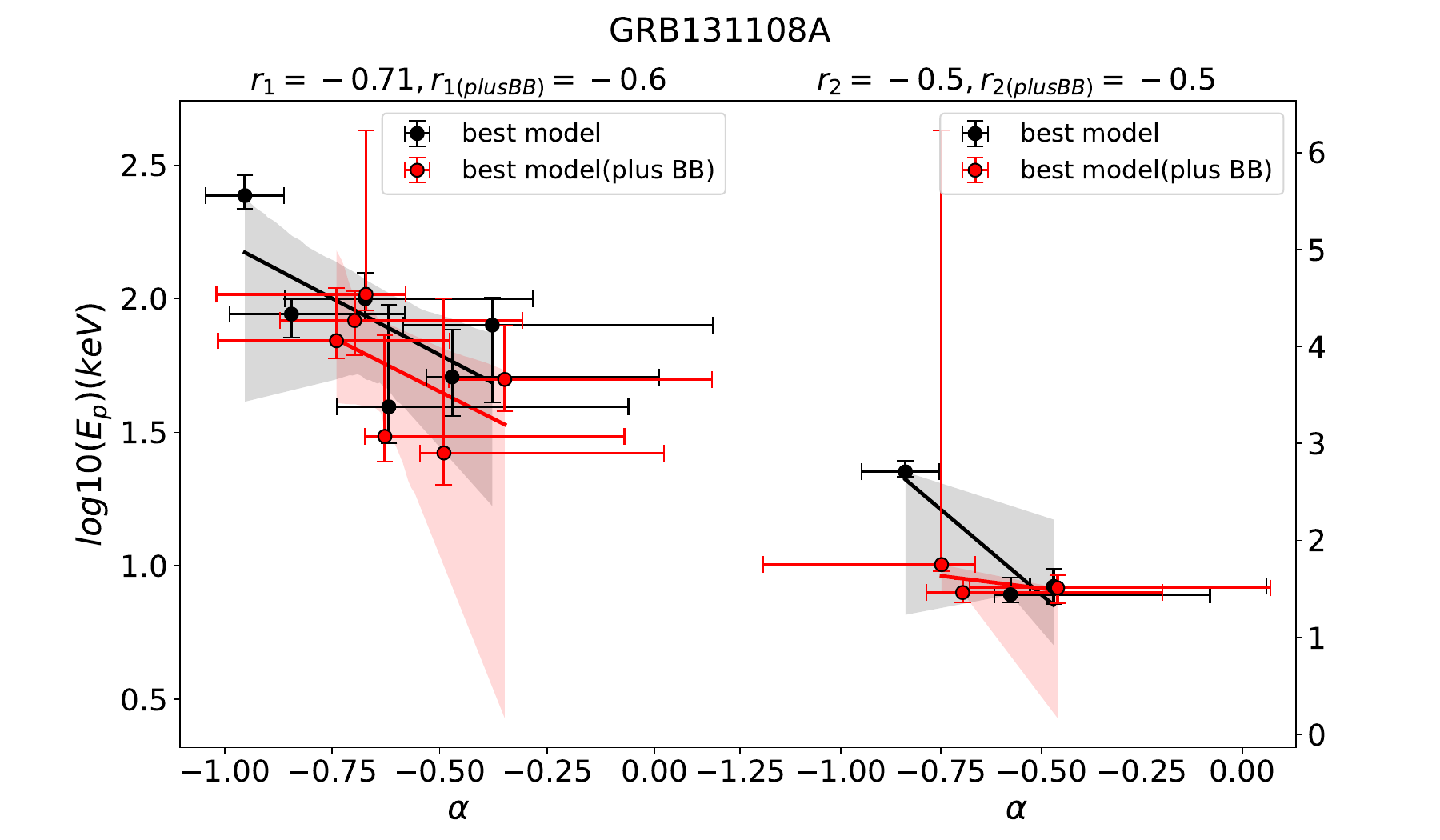}
\includegraphics [width=8cm,height=4cm]{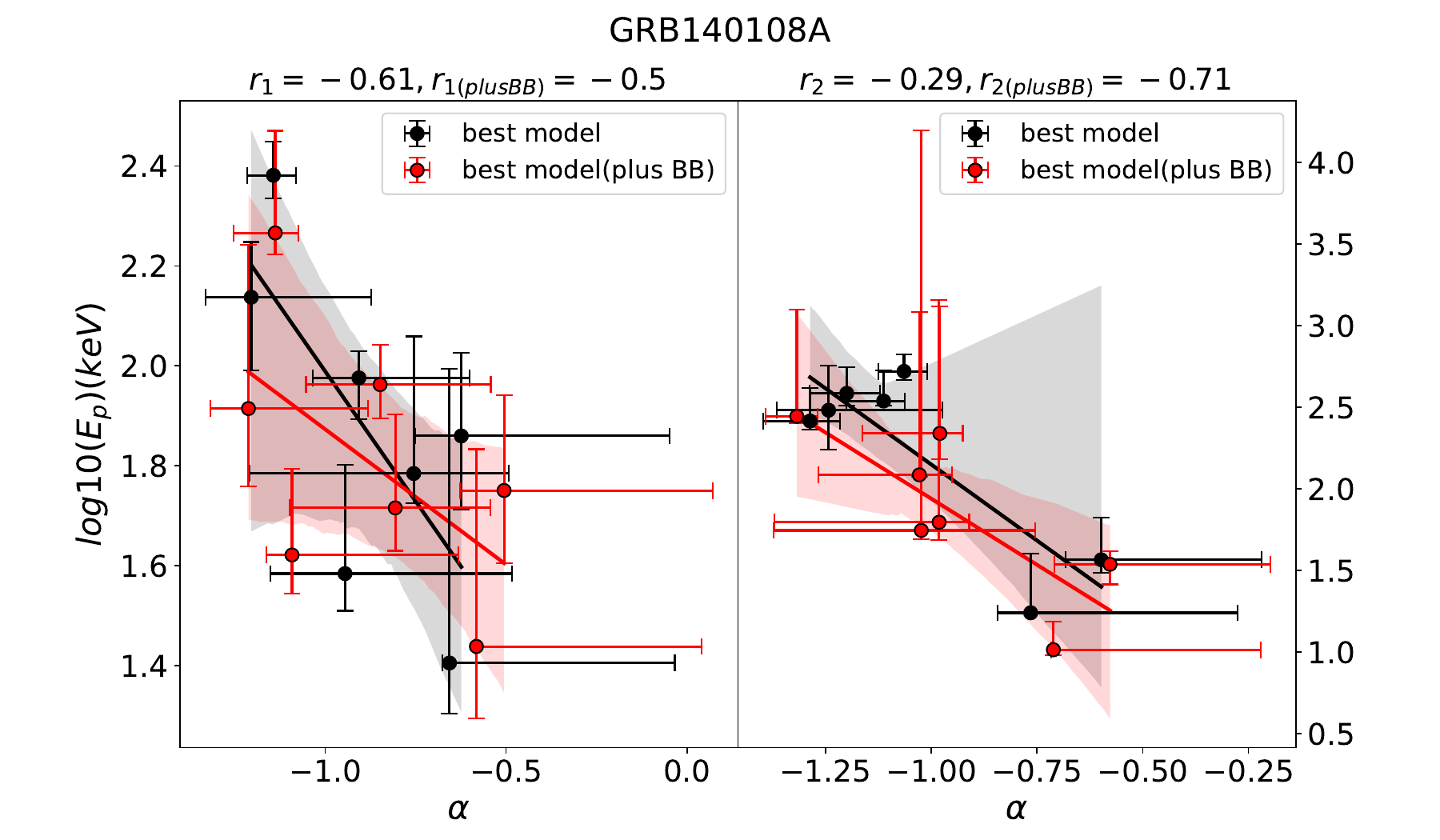}
\includegraphics [width=8cm,height=4cm]{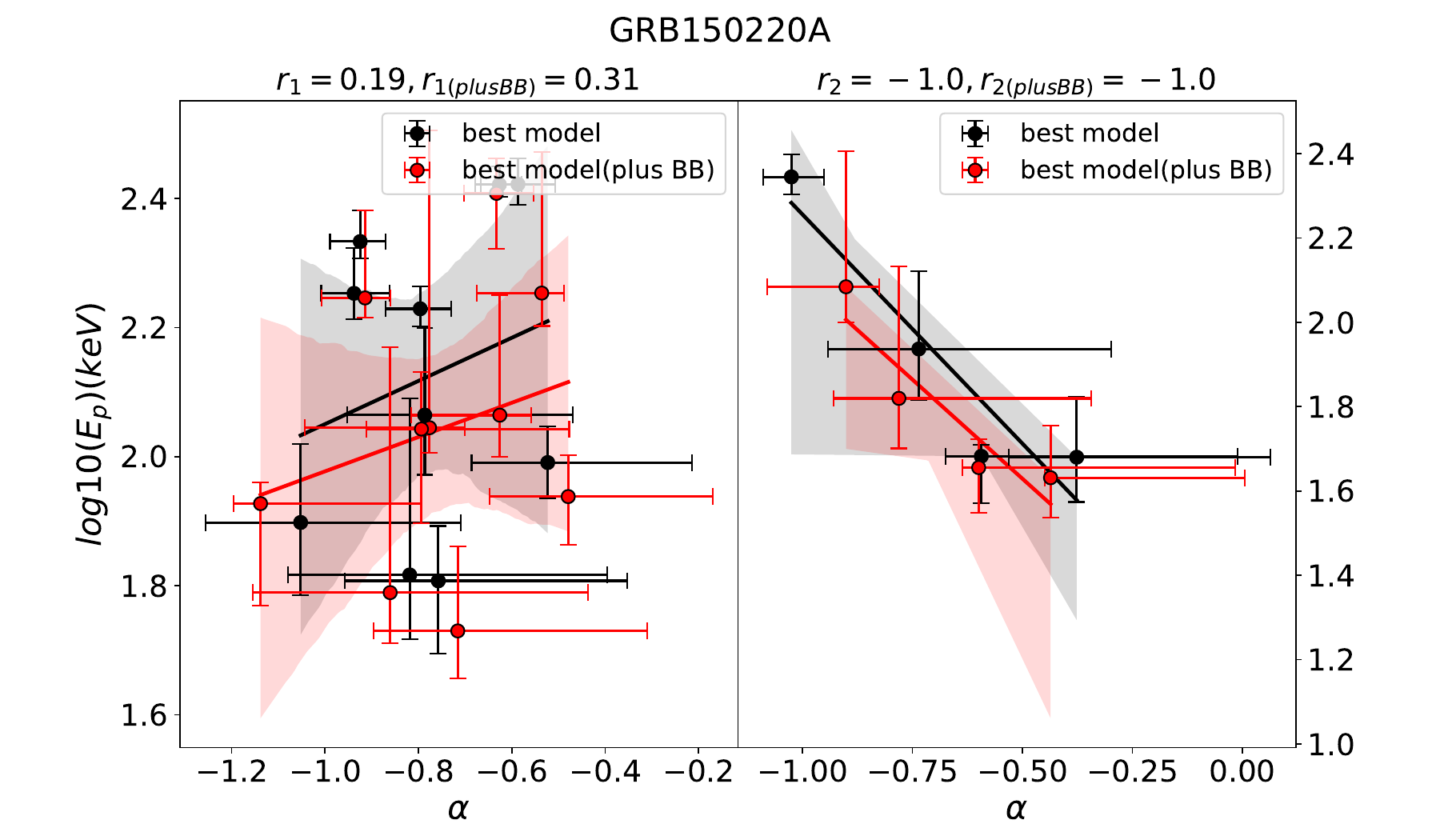}
\includegraphics [width=8cm,height=4cm]{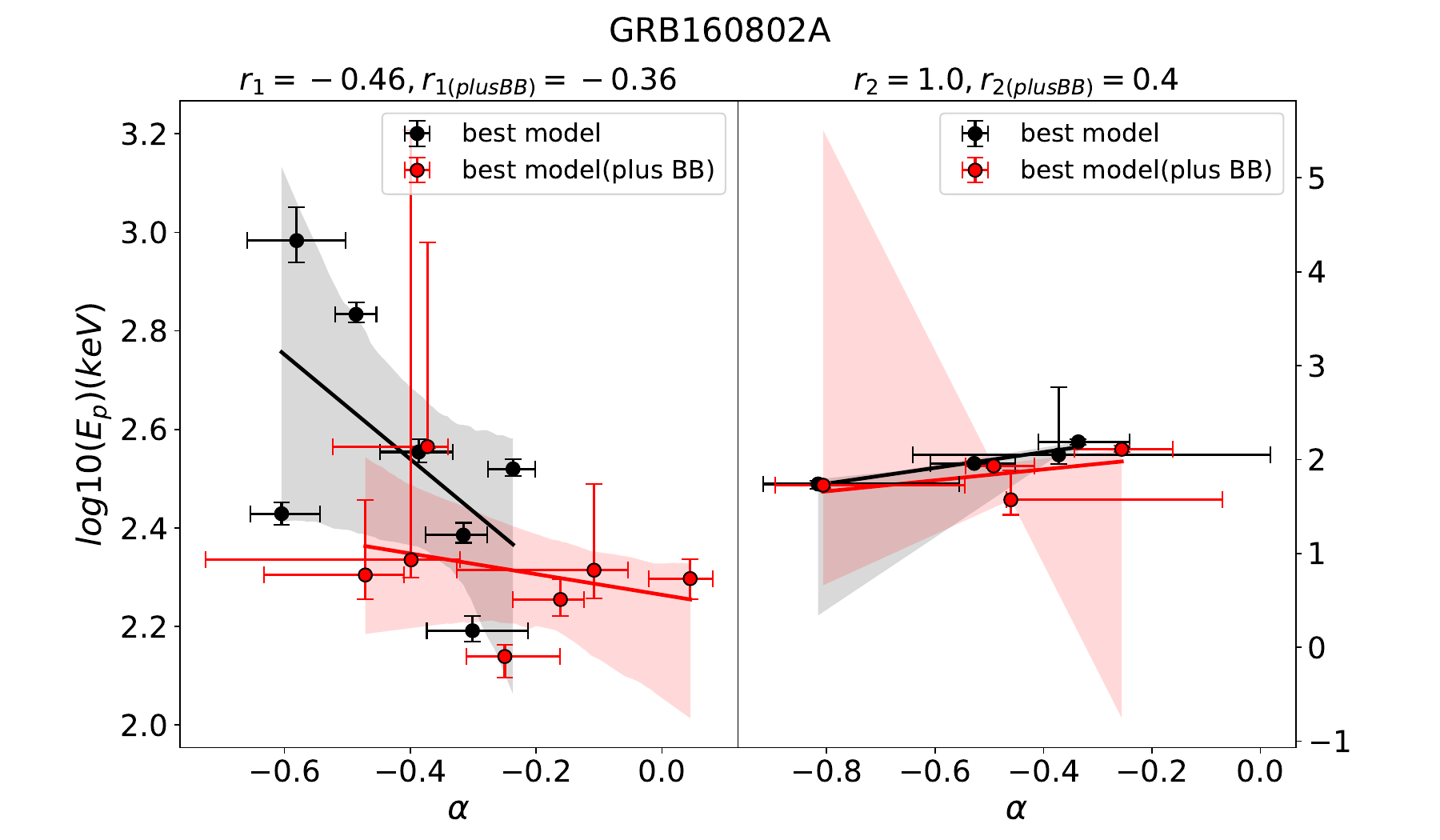}
\includegraphics [width=8cm,height=4cm]{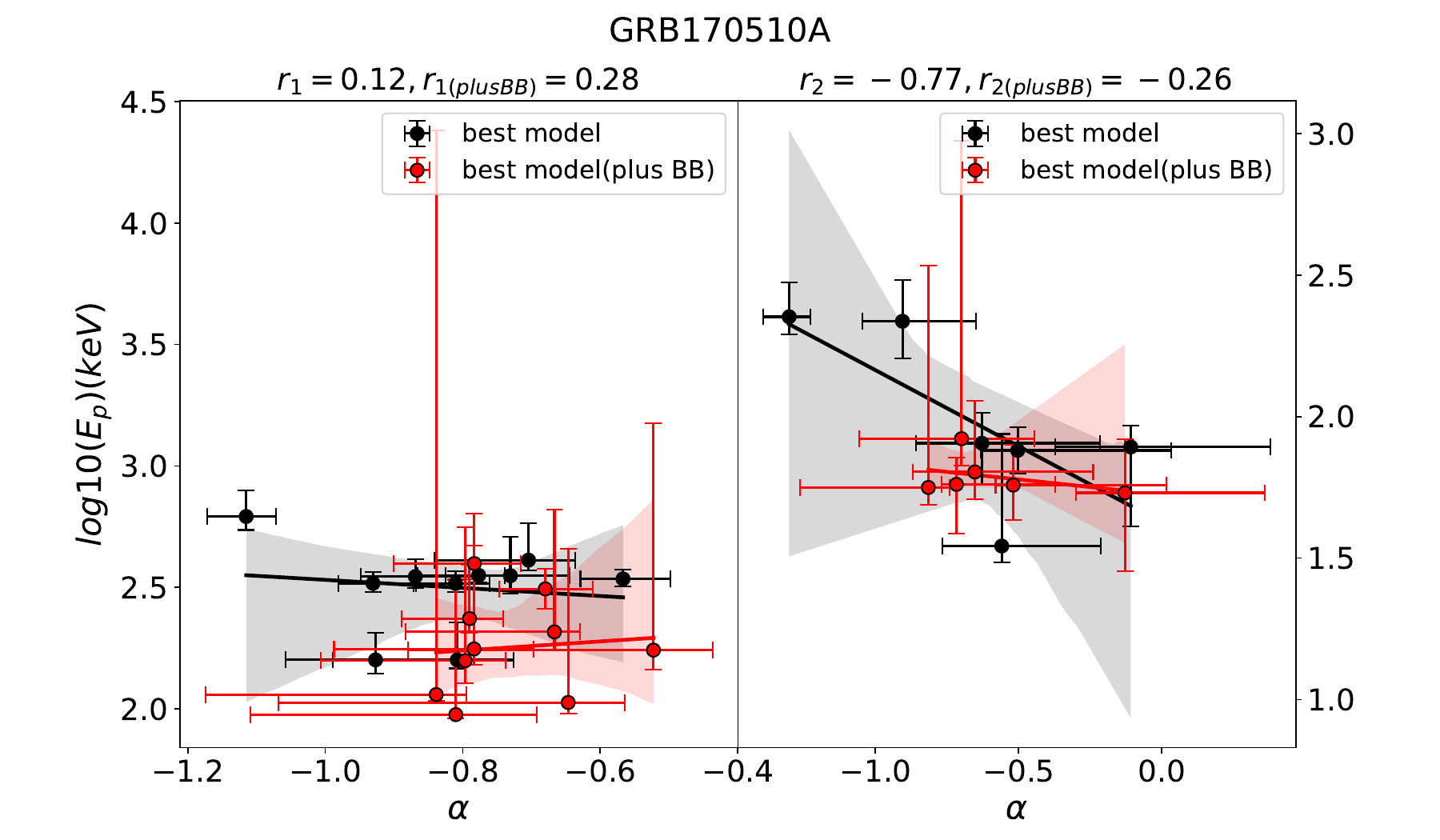}
   \figcaption{The correlation between $E_{p}$ and $\alpha$. The label symbols are similar to those in Figure \ref{fig 12}. \label{fig 14}}

\end{figure}

\setcounter{figure}{13}  
\begin{figure}[H]

\centering
\includegraphics [width=8cm,height=4cm]{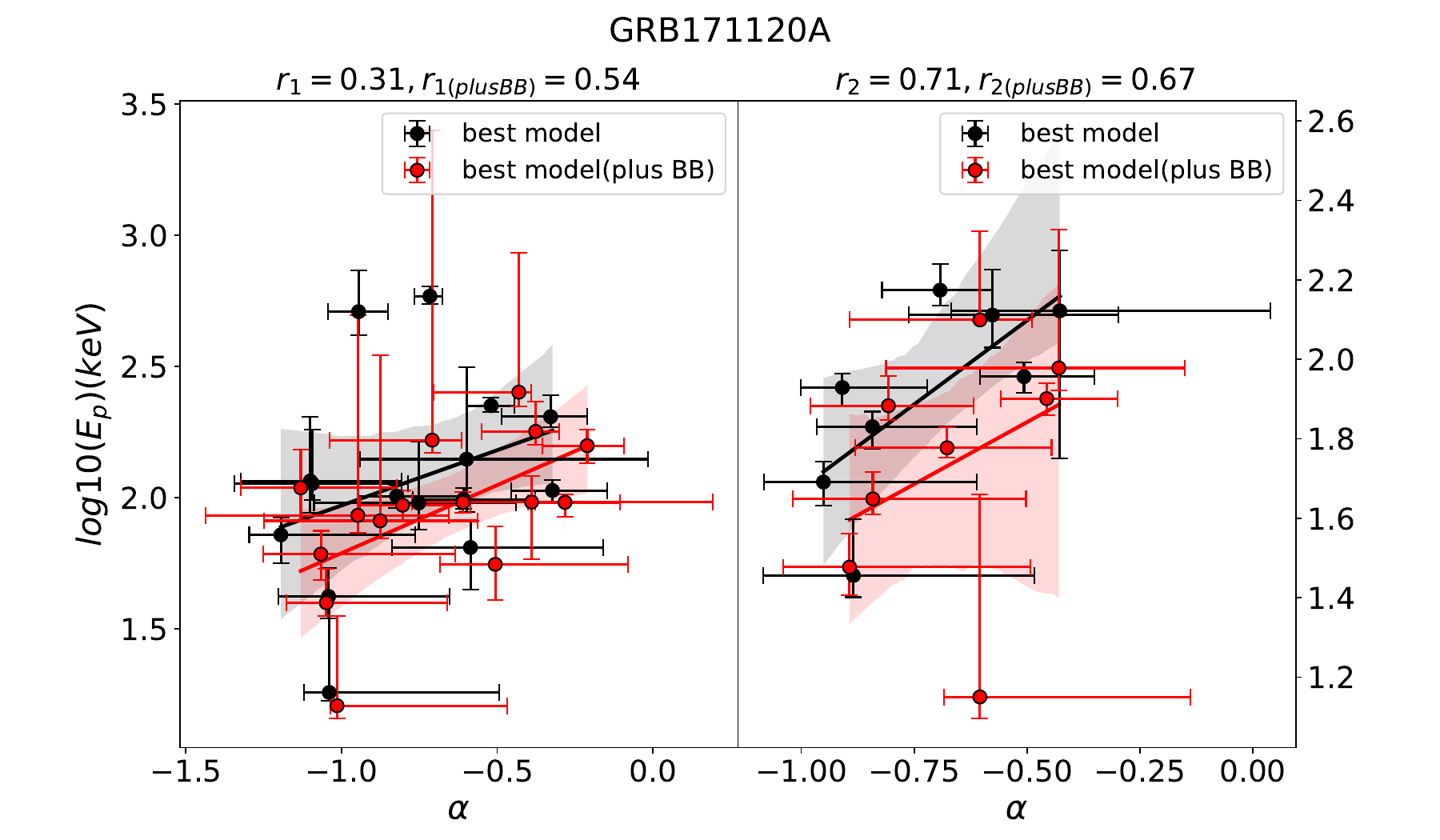}
\includegraphics [width=8cm,height=4cm]{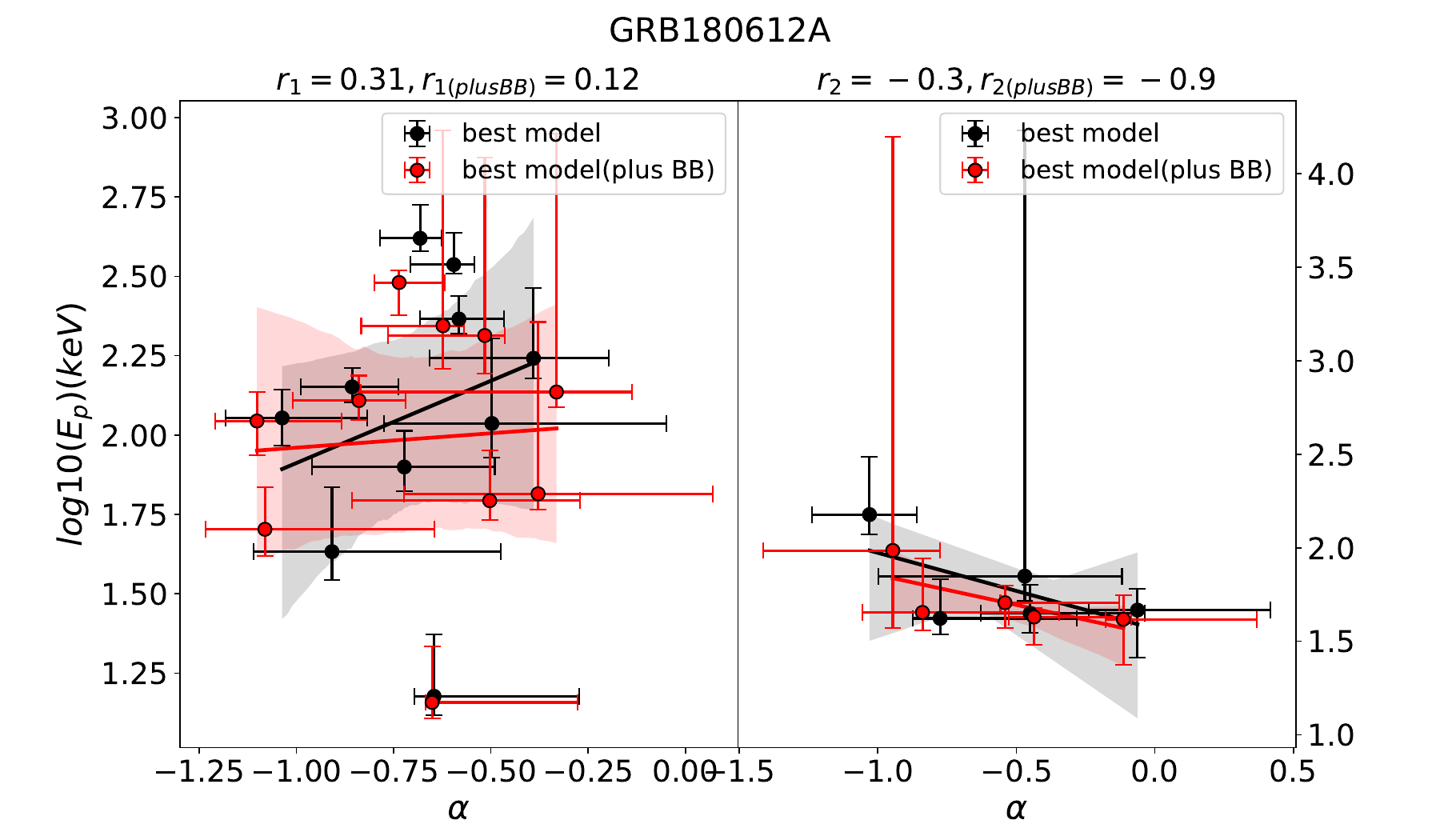}
\includegraphics [width=8cm,height=4cm]{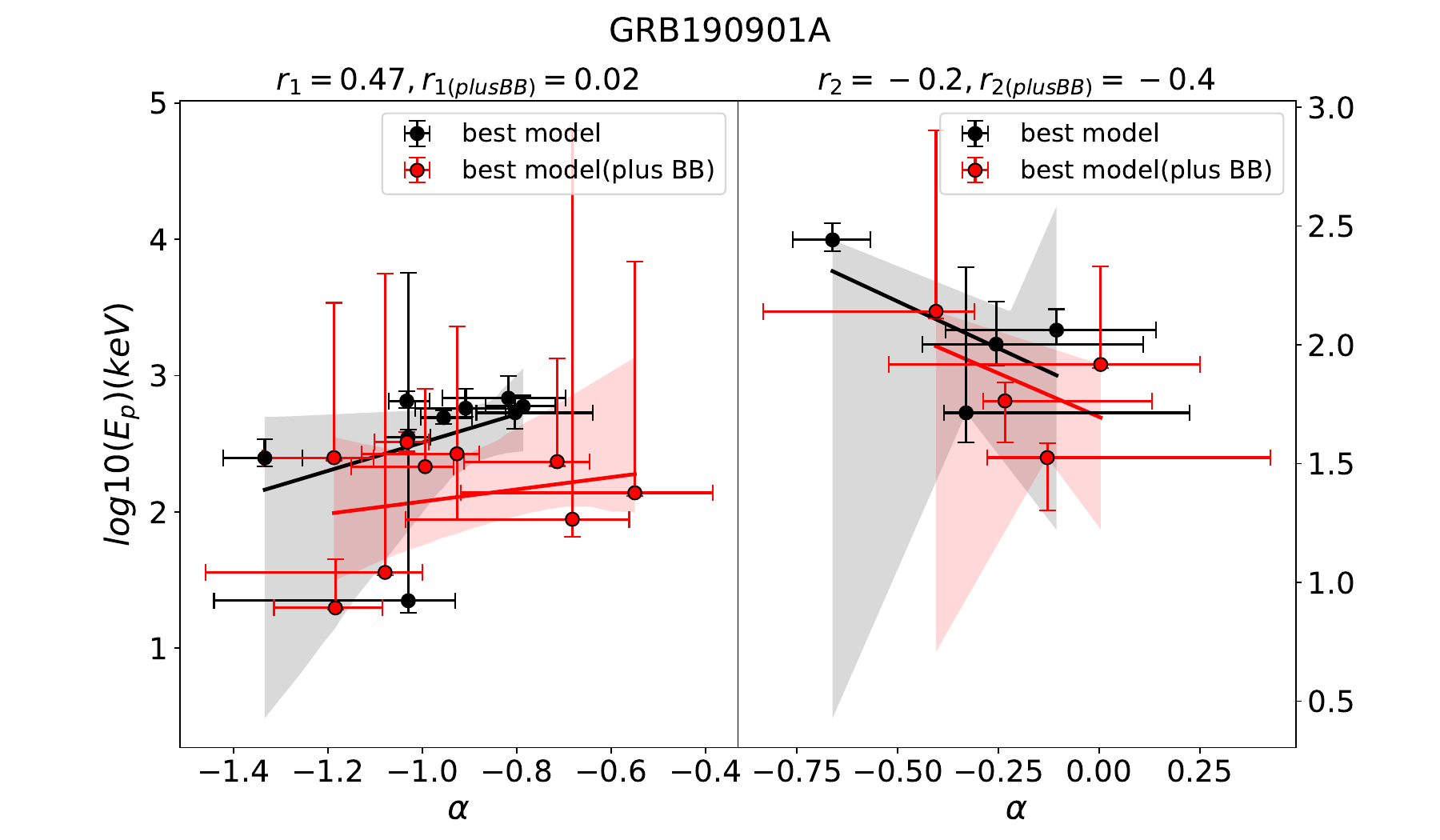}
\includegraphics [width=8cm,height=4cm]{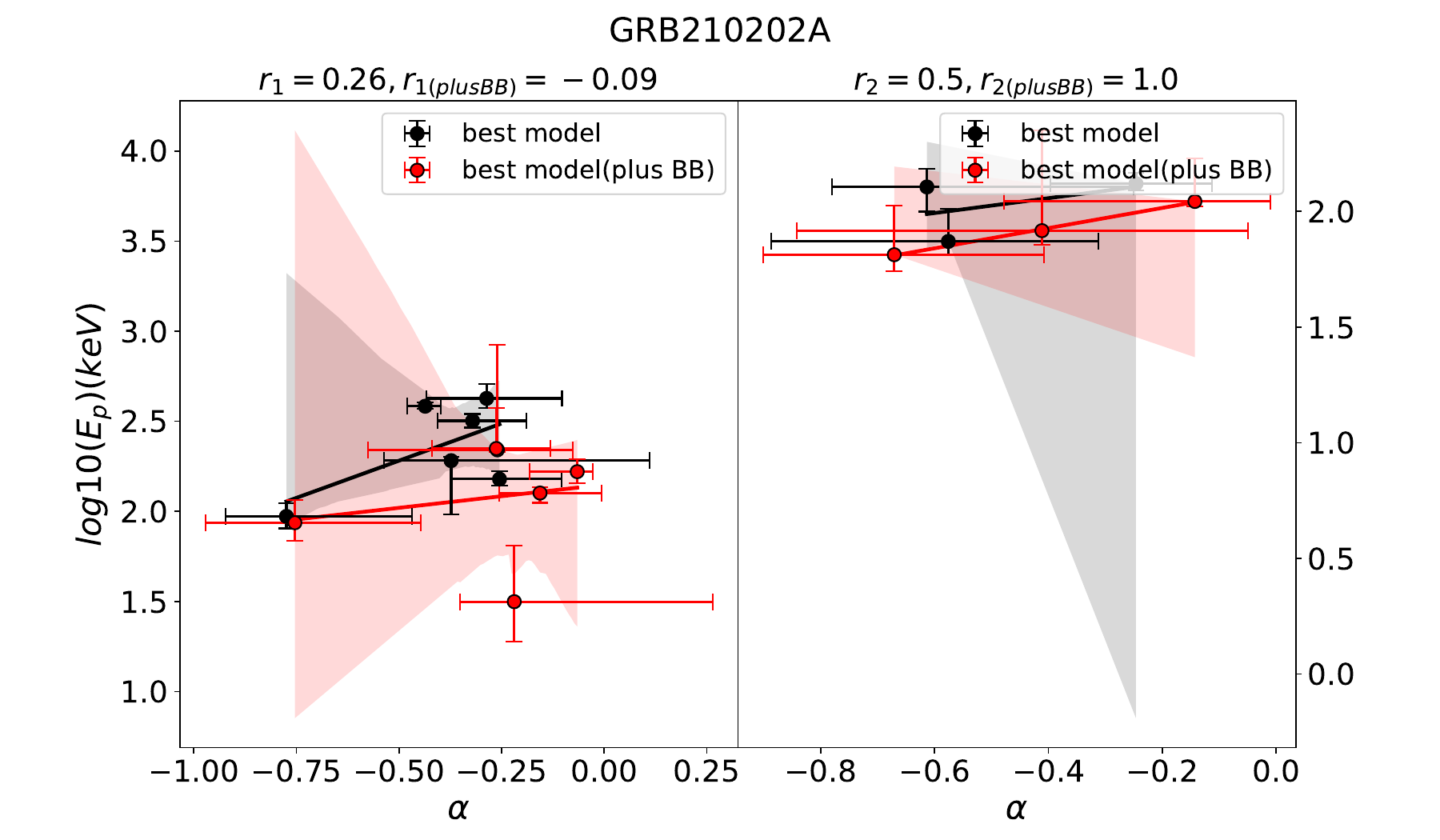}
\includegraphics [width=8cm,height=4cm]{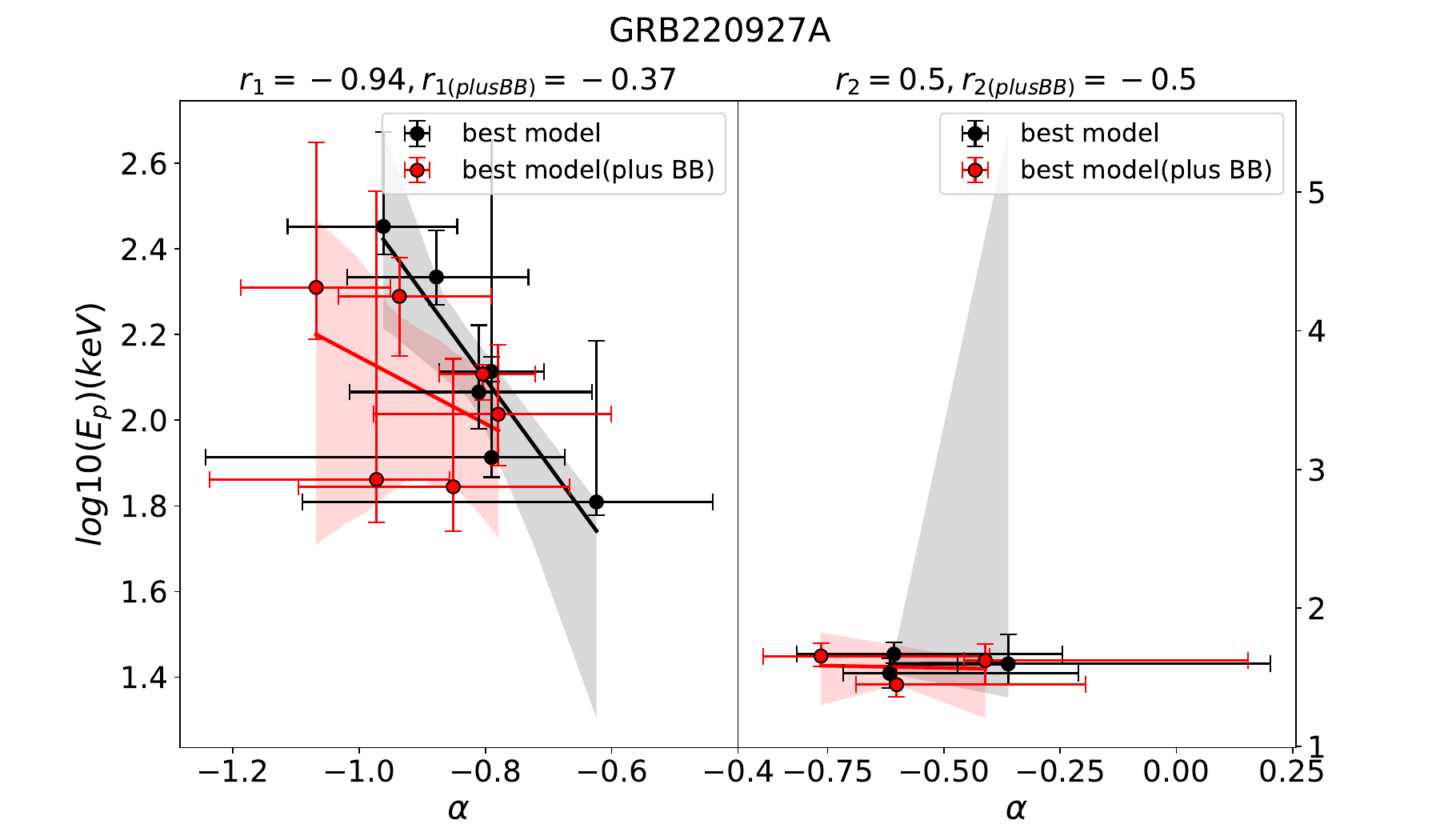}
\includegraphics [width=8cm,height=4cm]{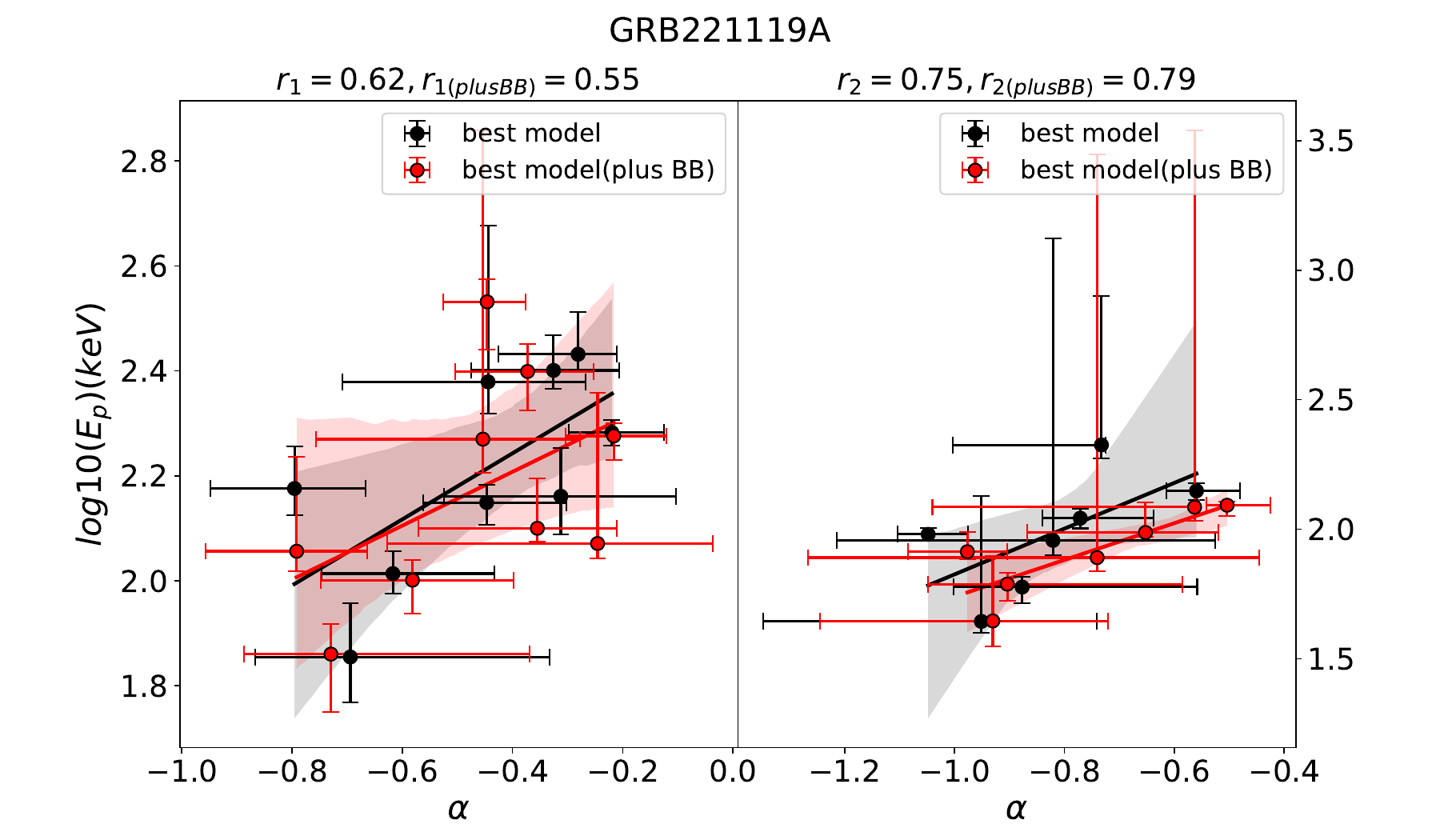}
\includegraphics [width=8cm,height=4cm]{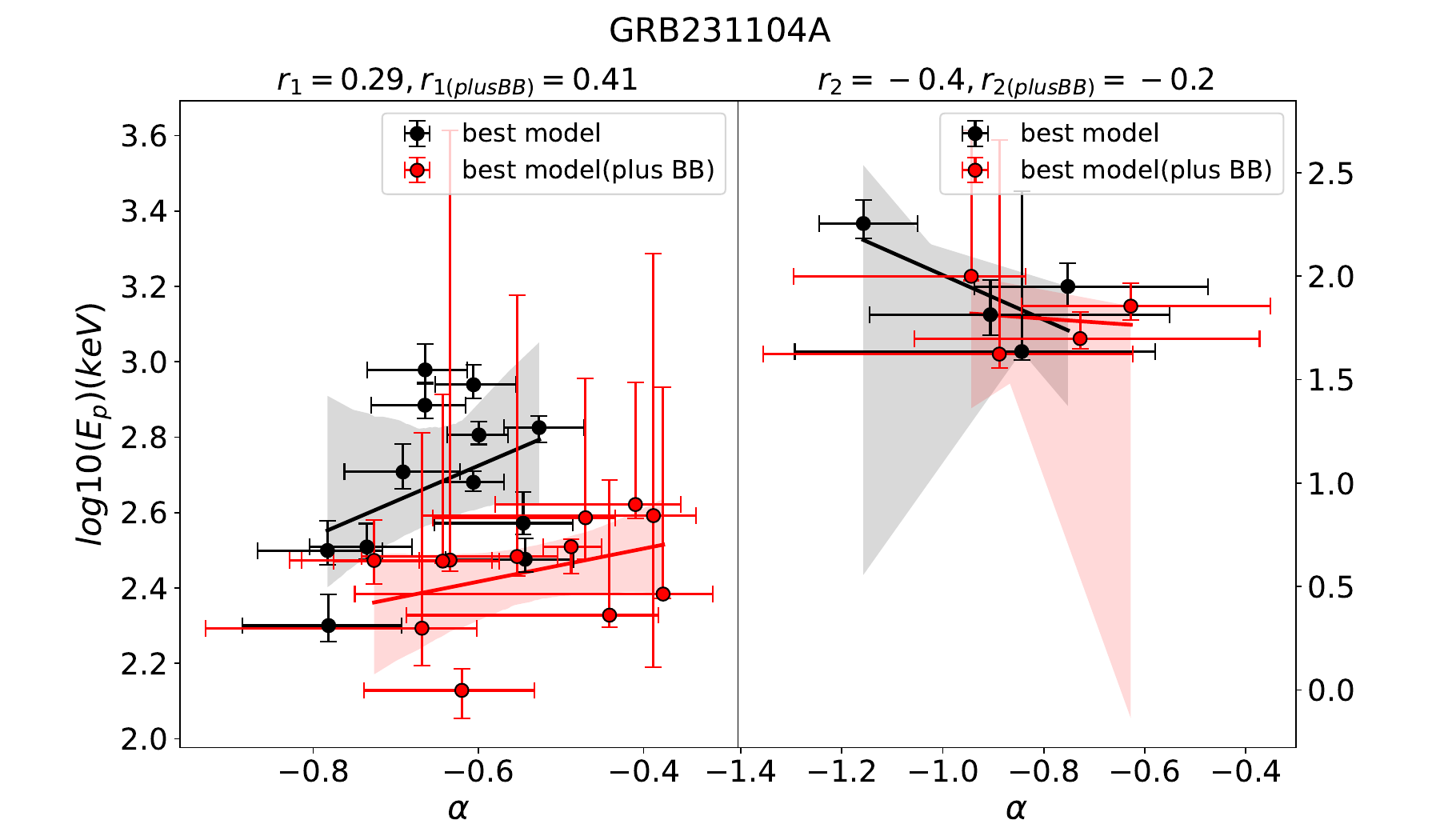}
\includegraphics [width=8cm,height=4cm]{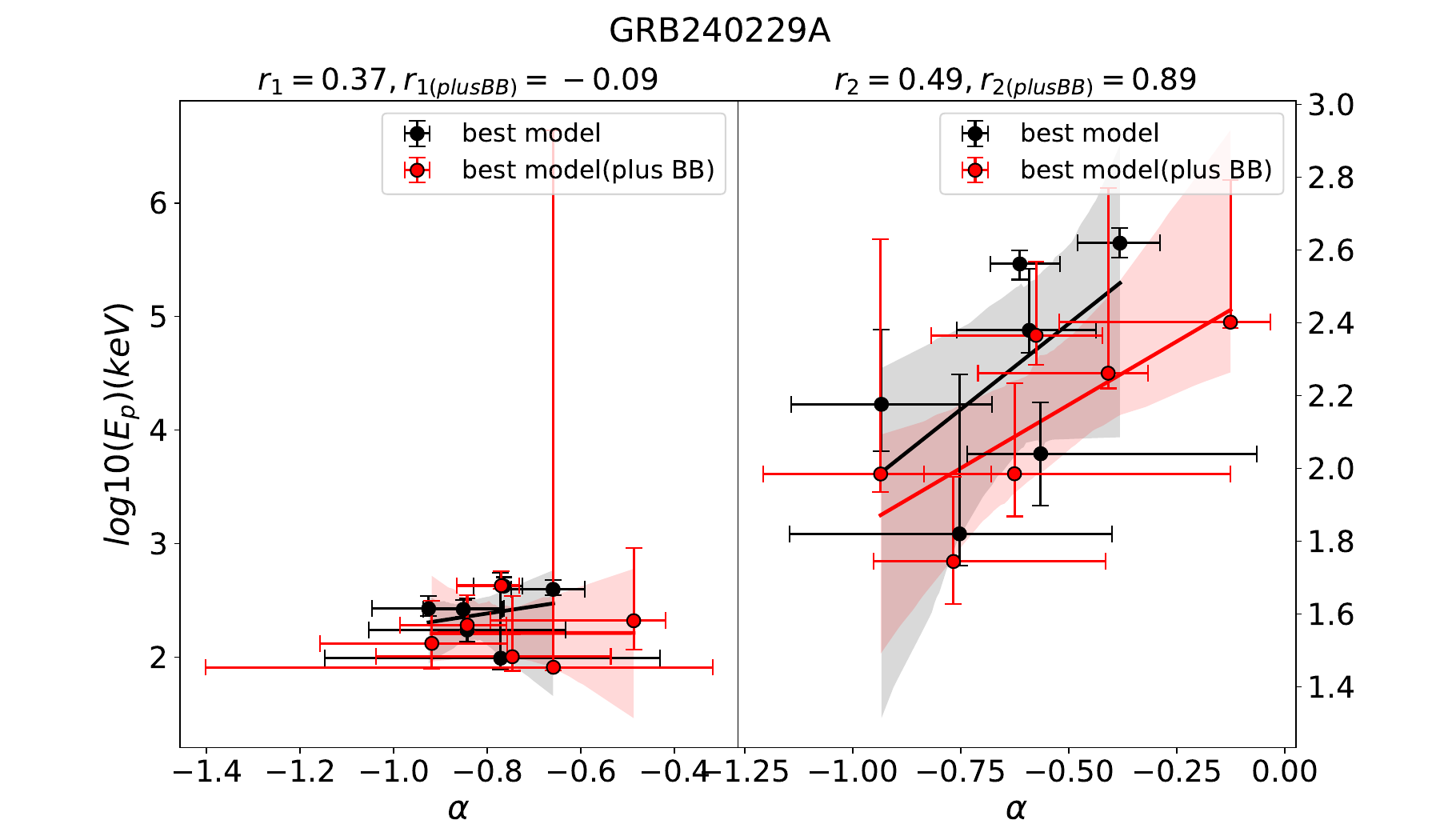}
   \figcaption{(Continued.) \label{fig 14}}
      
\end{figure}

\setcounter{figure}{14}  
\begin{figure}[H]
\centering
\includegraphics [width=8cm,height=4cm]{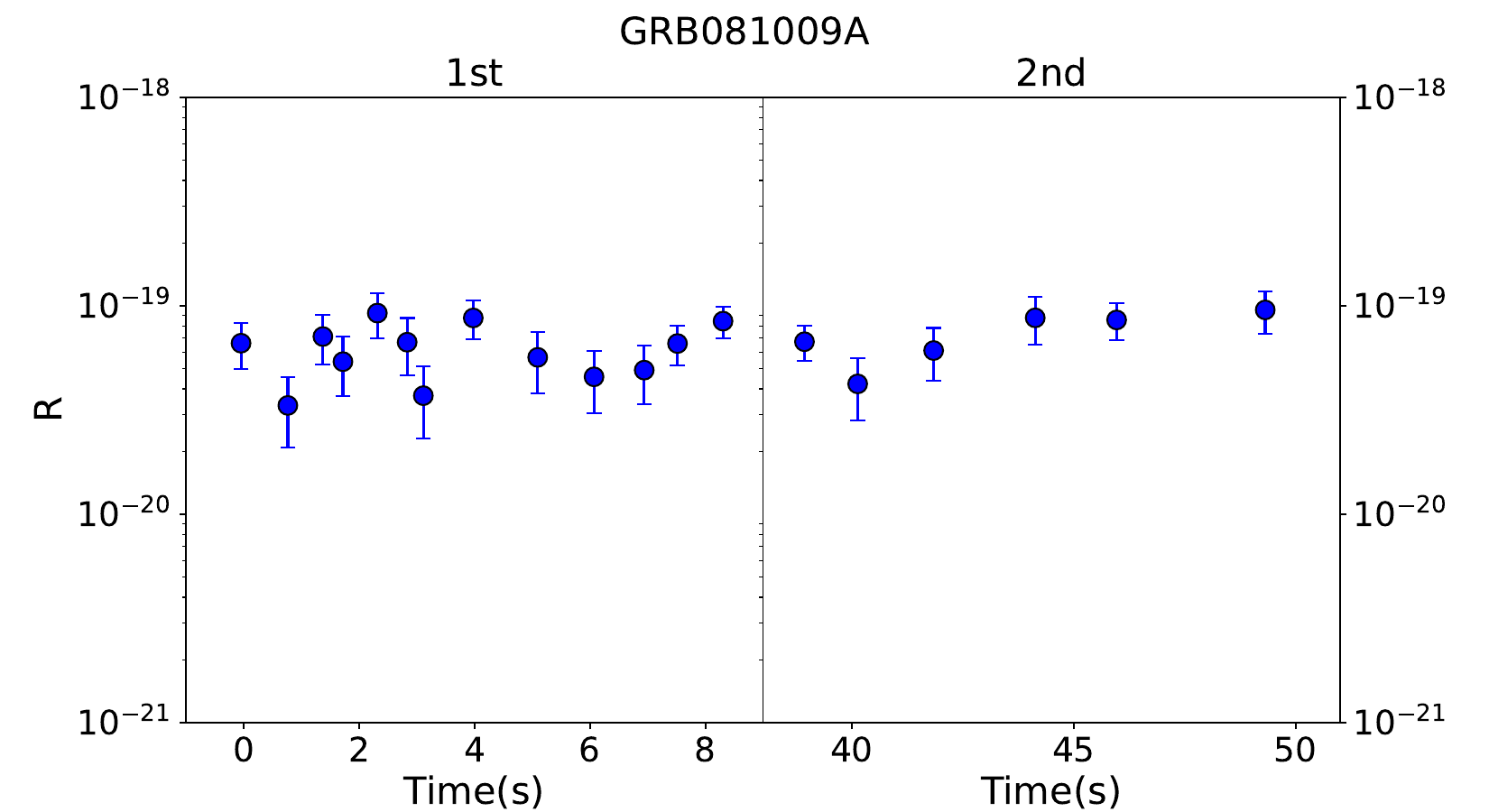}
\includegraphics [width=8cm,height=4cm]{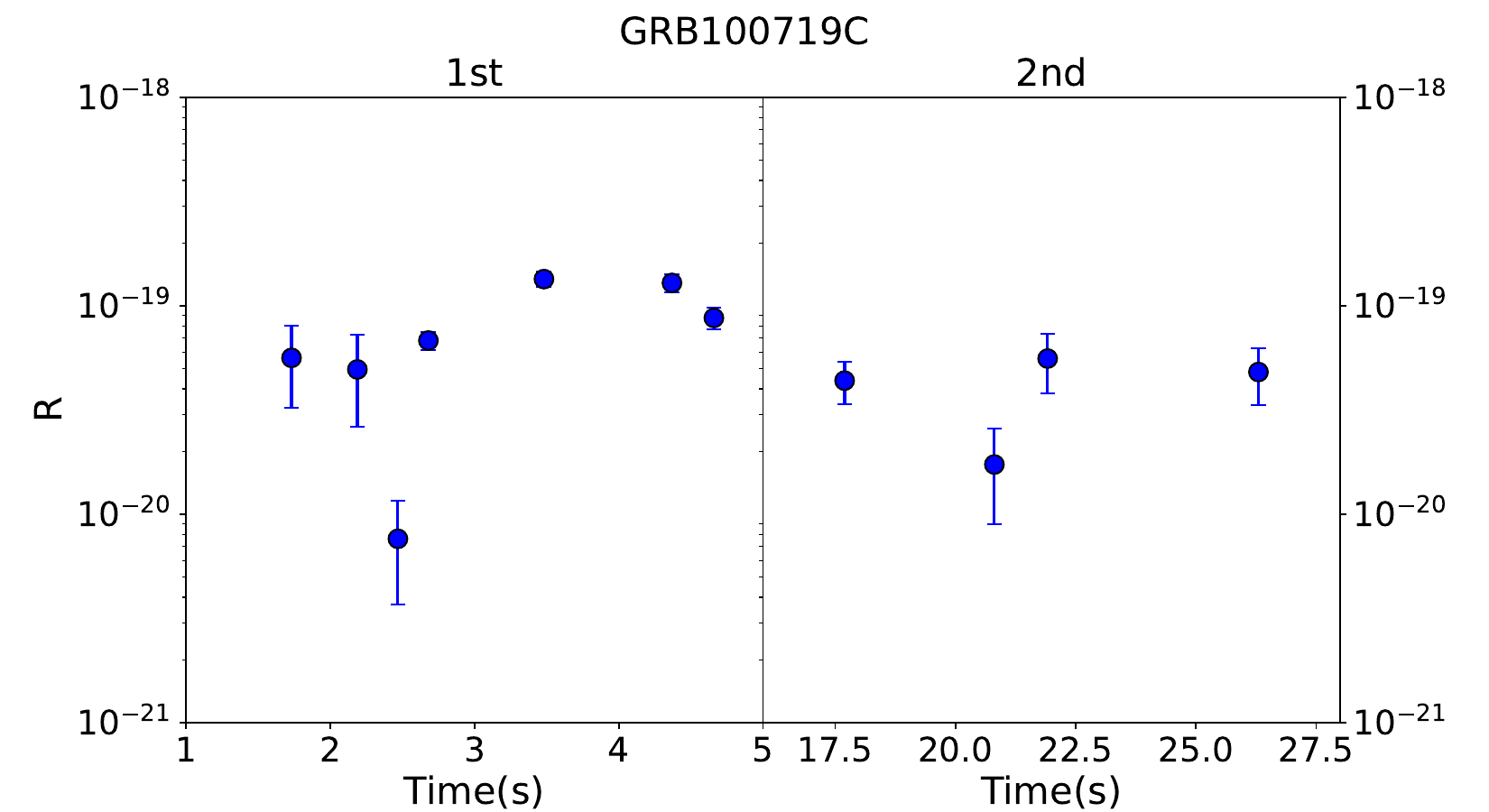}
\includegraphics [width=8cm,height=4cm]{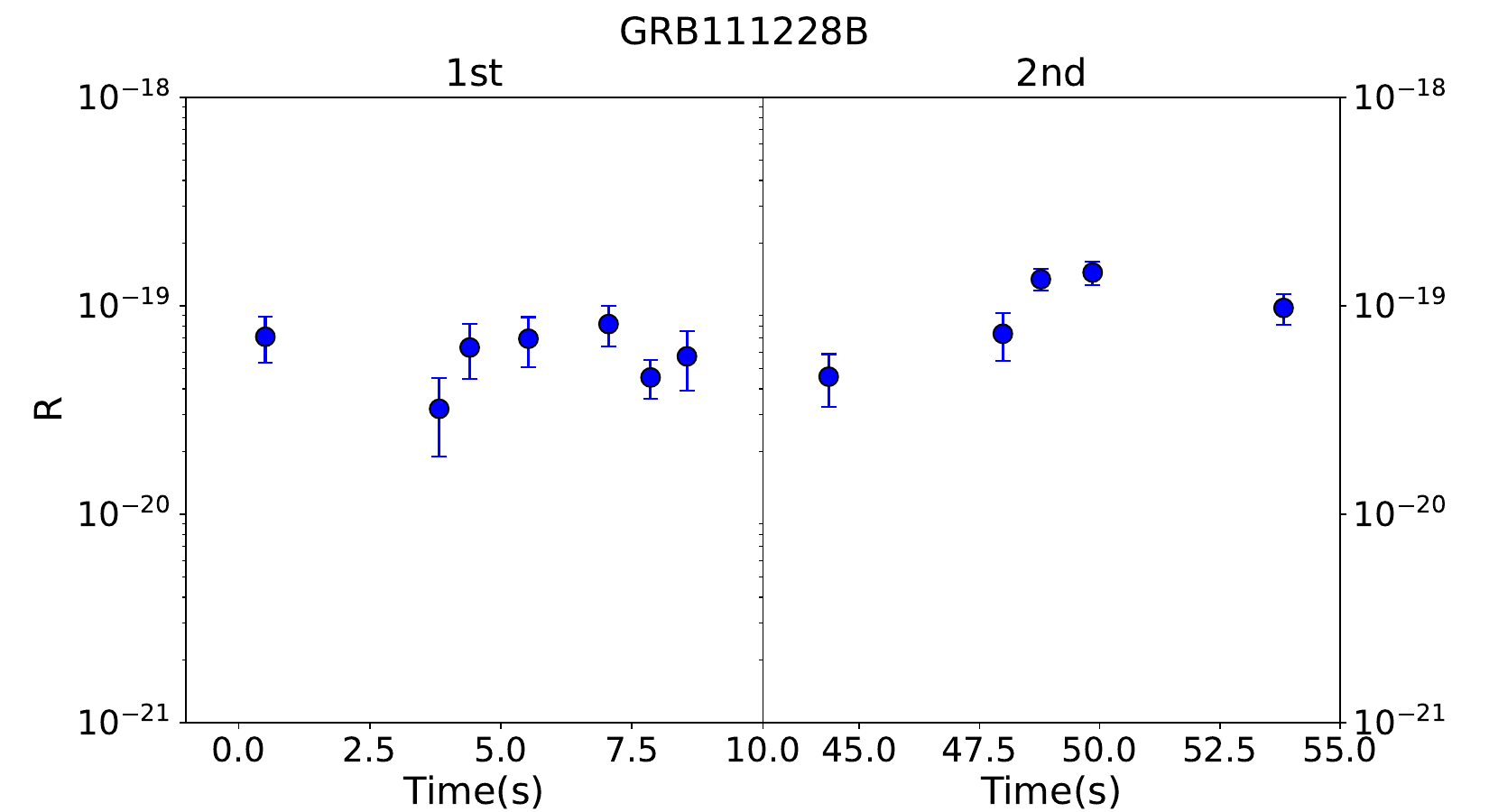}
\includegraphics [width=8cm,height=4cm]{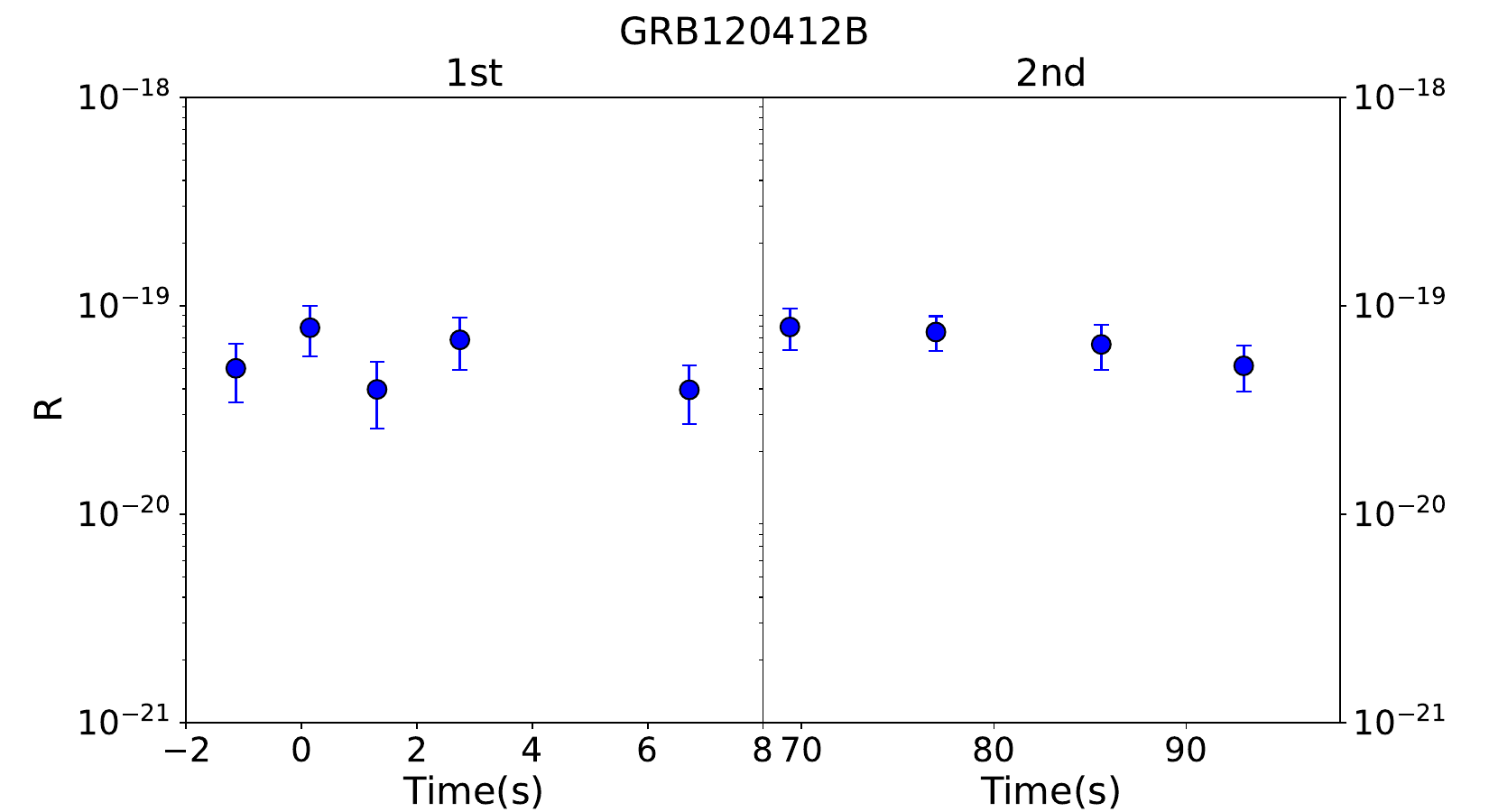}
\includegraphics [width=8cm,height=4cm]{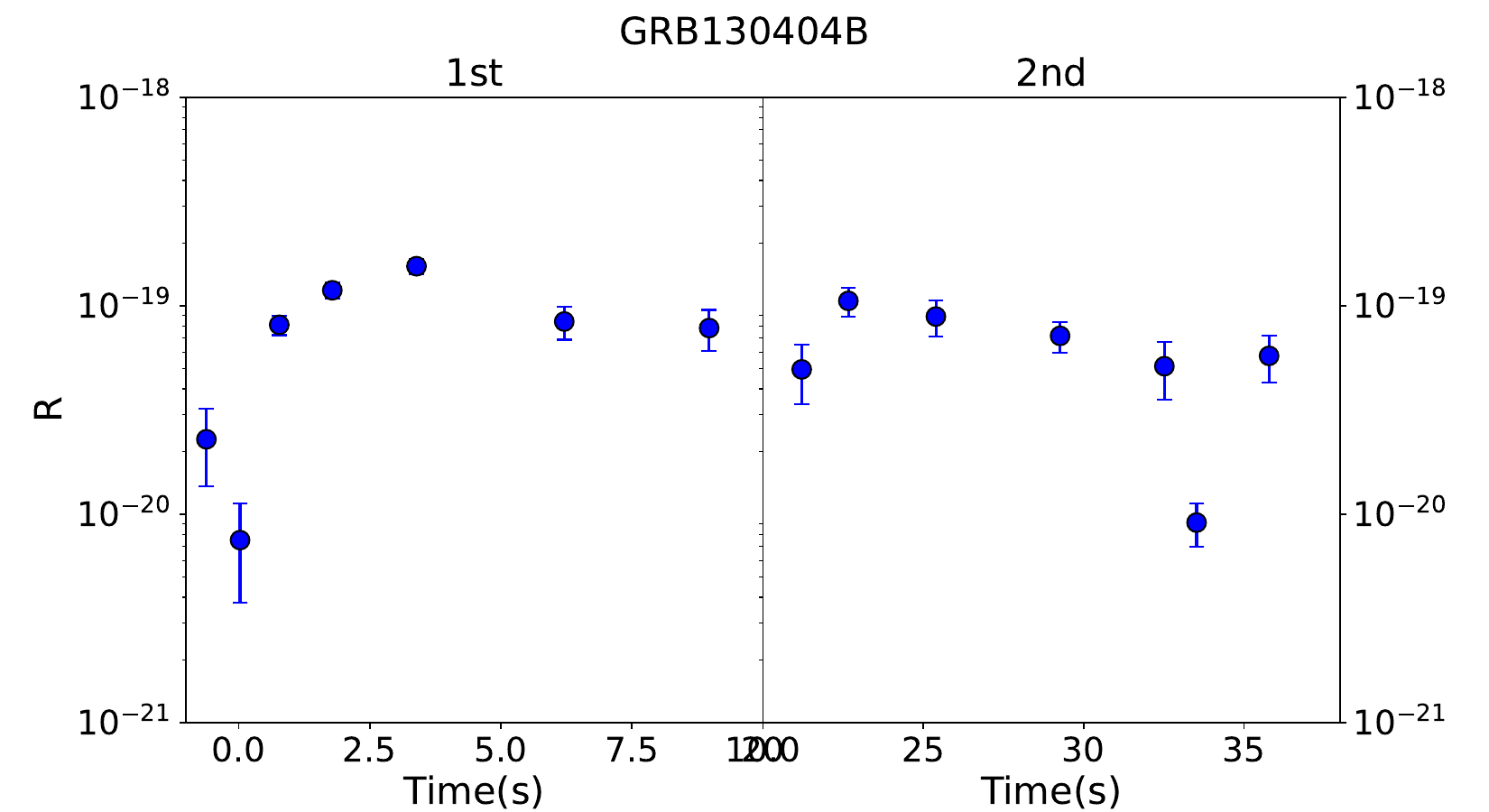}
\includegraphics [width=8cm,height=4cm]{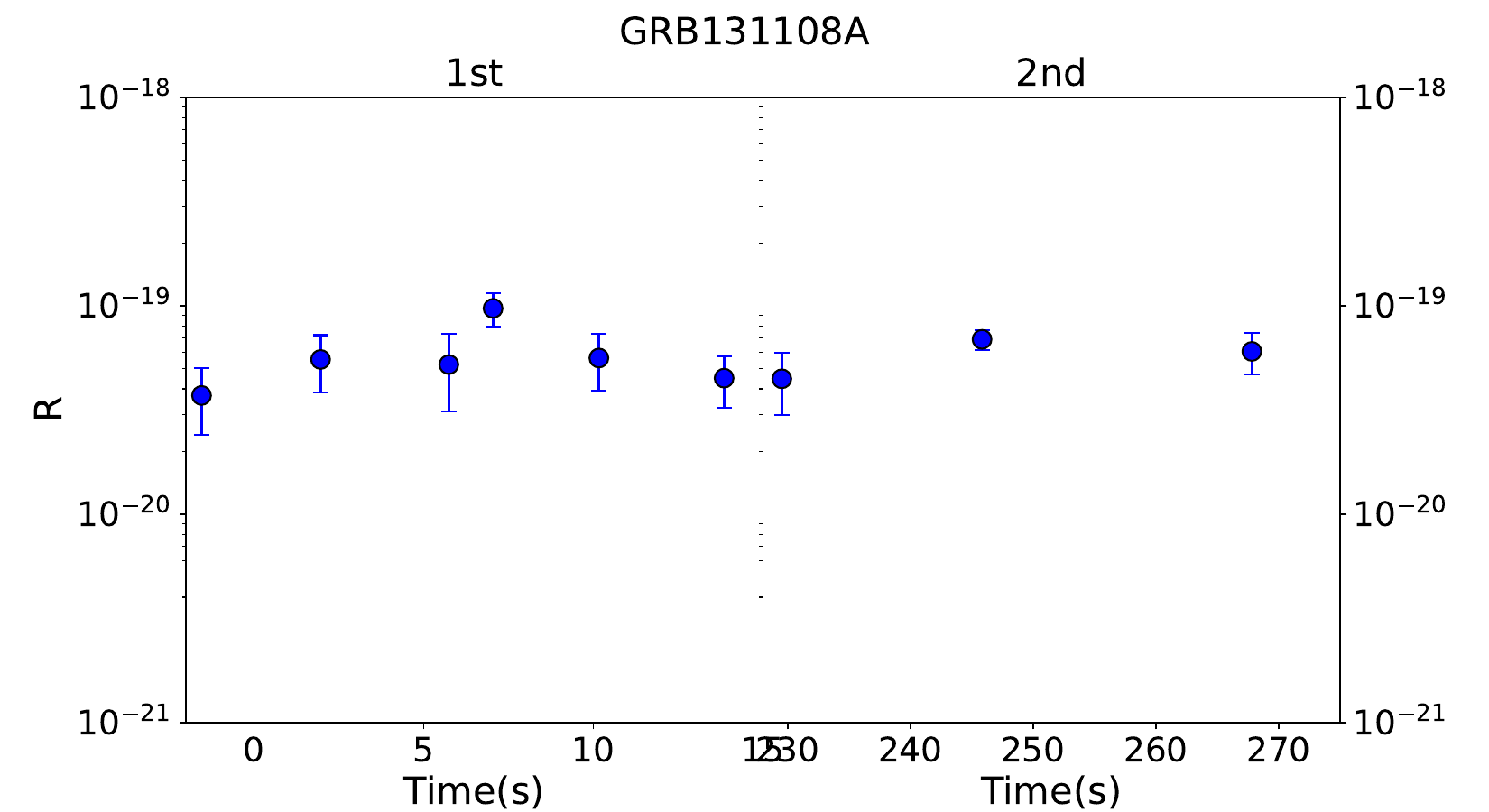}
\includegraphics [width=8cm,height=4cm]{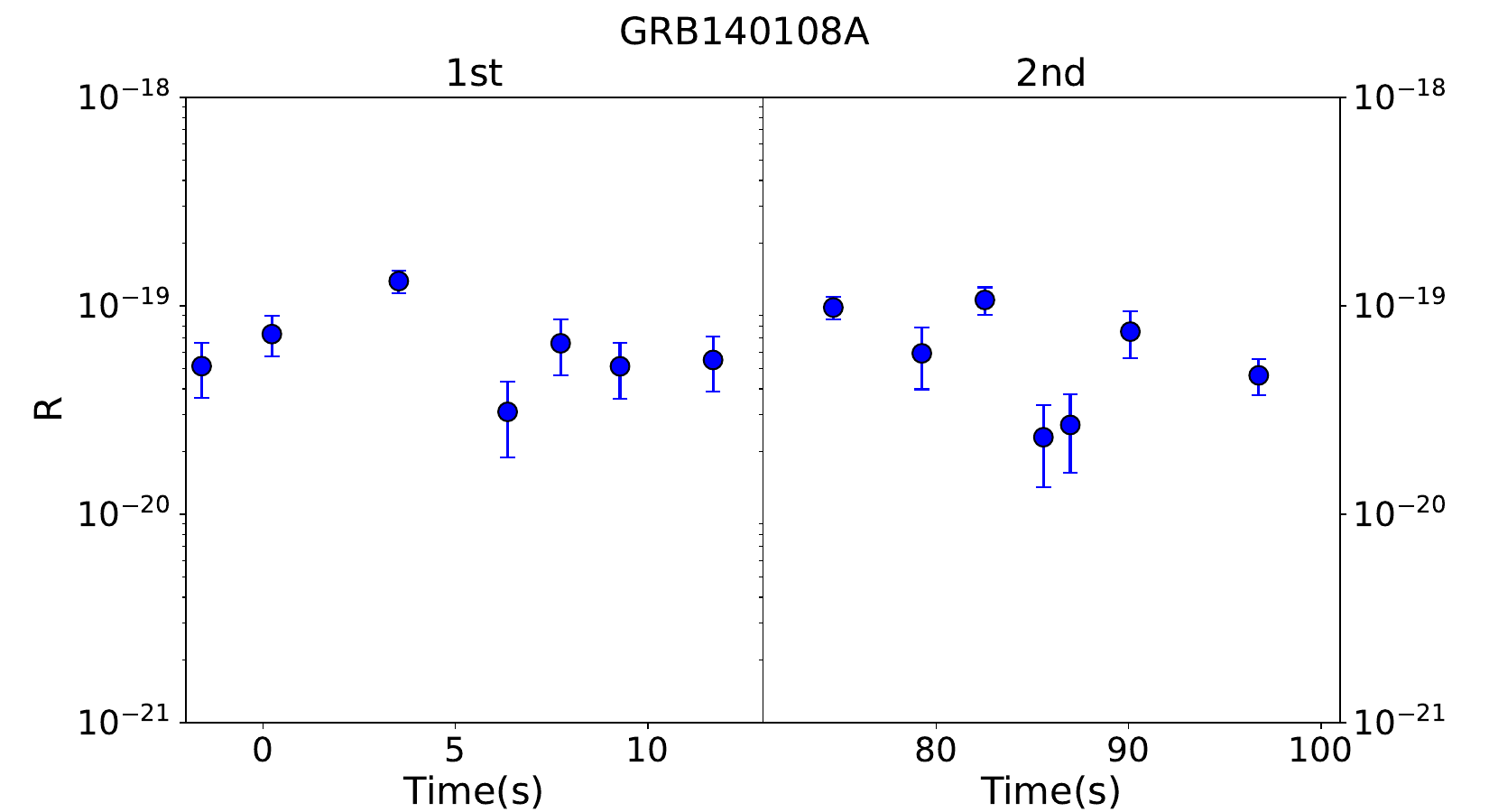}
\includegraphics [width=8cm,height=4cm]{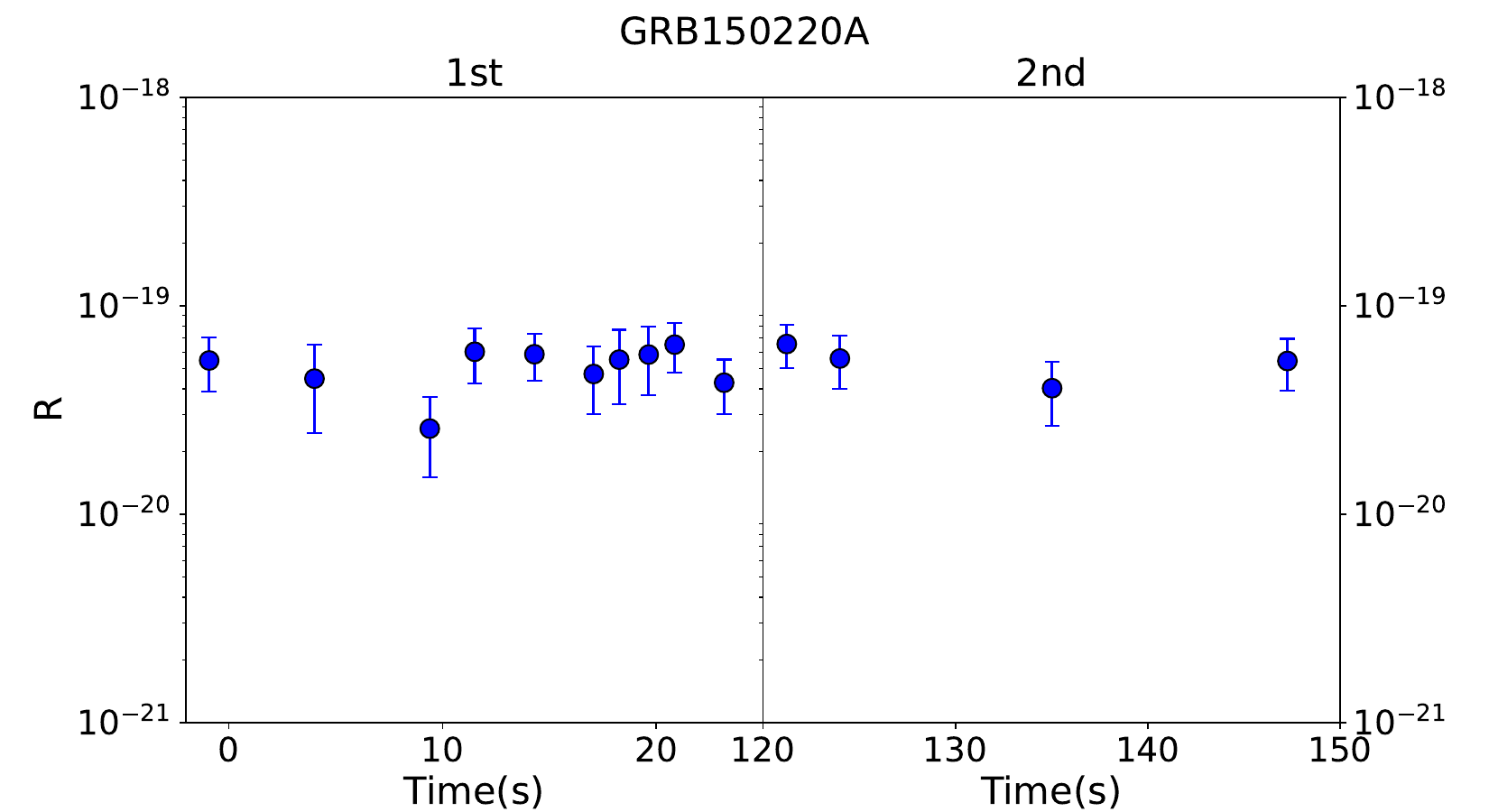}
\includegraphics [width=8cm,height=4cm]{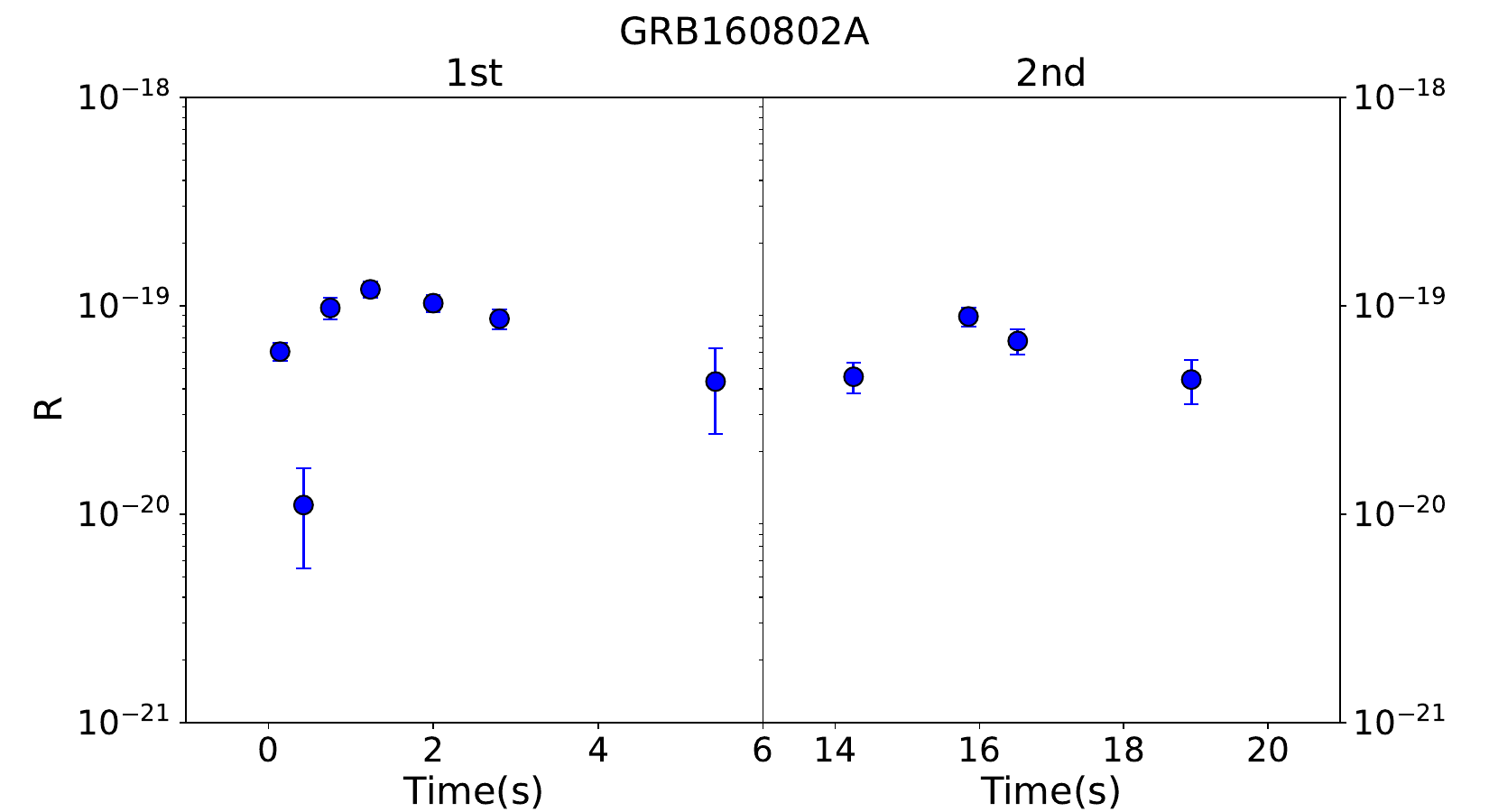}
\includegraphics [width=8cm,height=4cm]{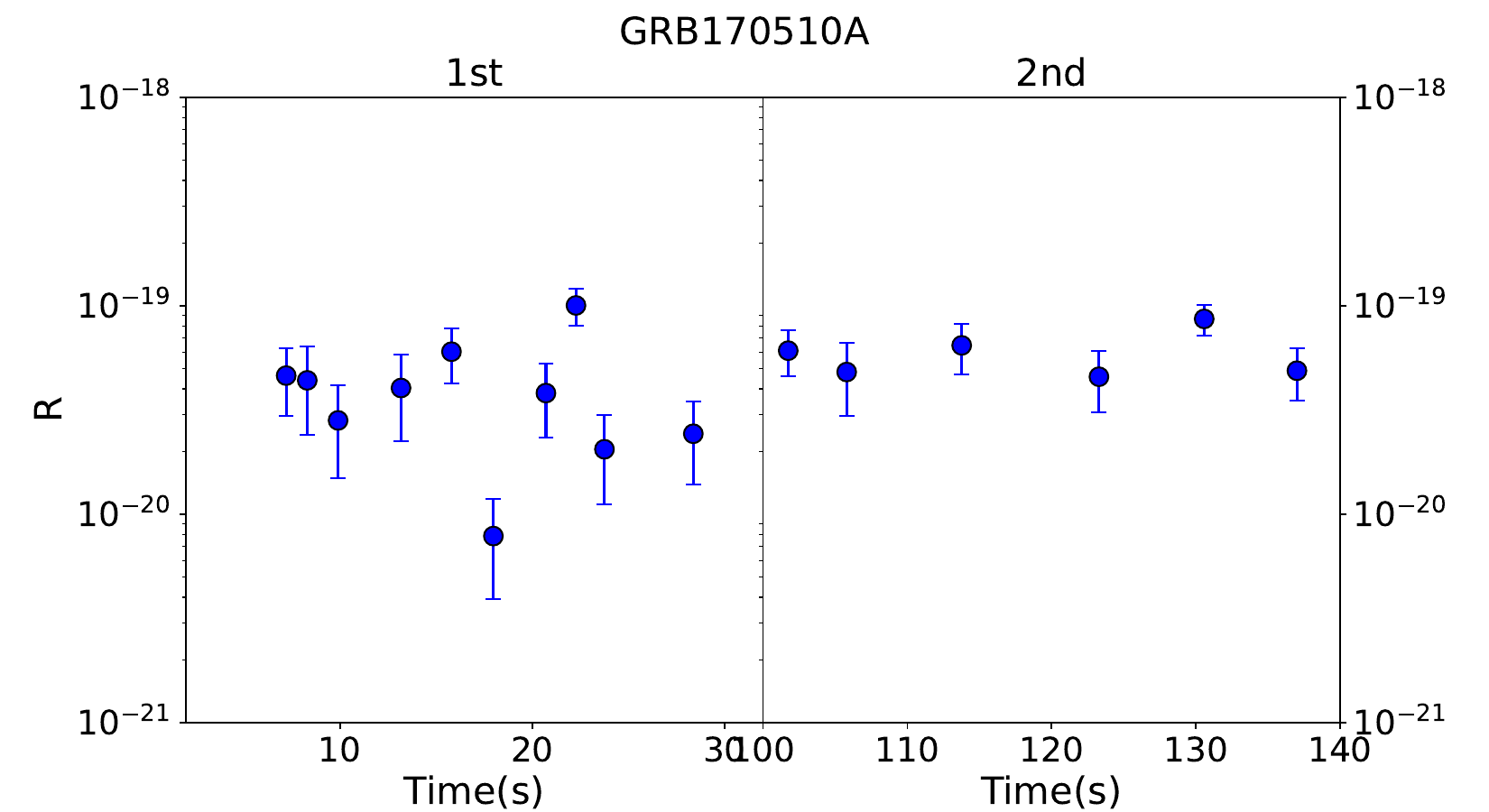}
   \figcaption{Evolution of $\Re $. ``1st'' refers to the main burst, and ``2nd'' refers to the second burst. \label{fig 15} }

\end{figure}

\setcounter{figure}{14}  
\begin{figure}[H]

\centering
\includegraphics [width=8cm,height=4cm]{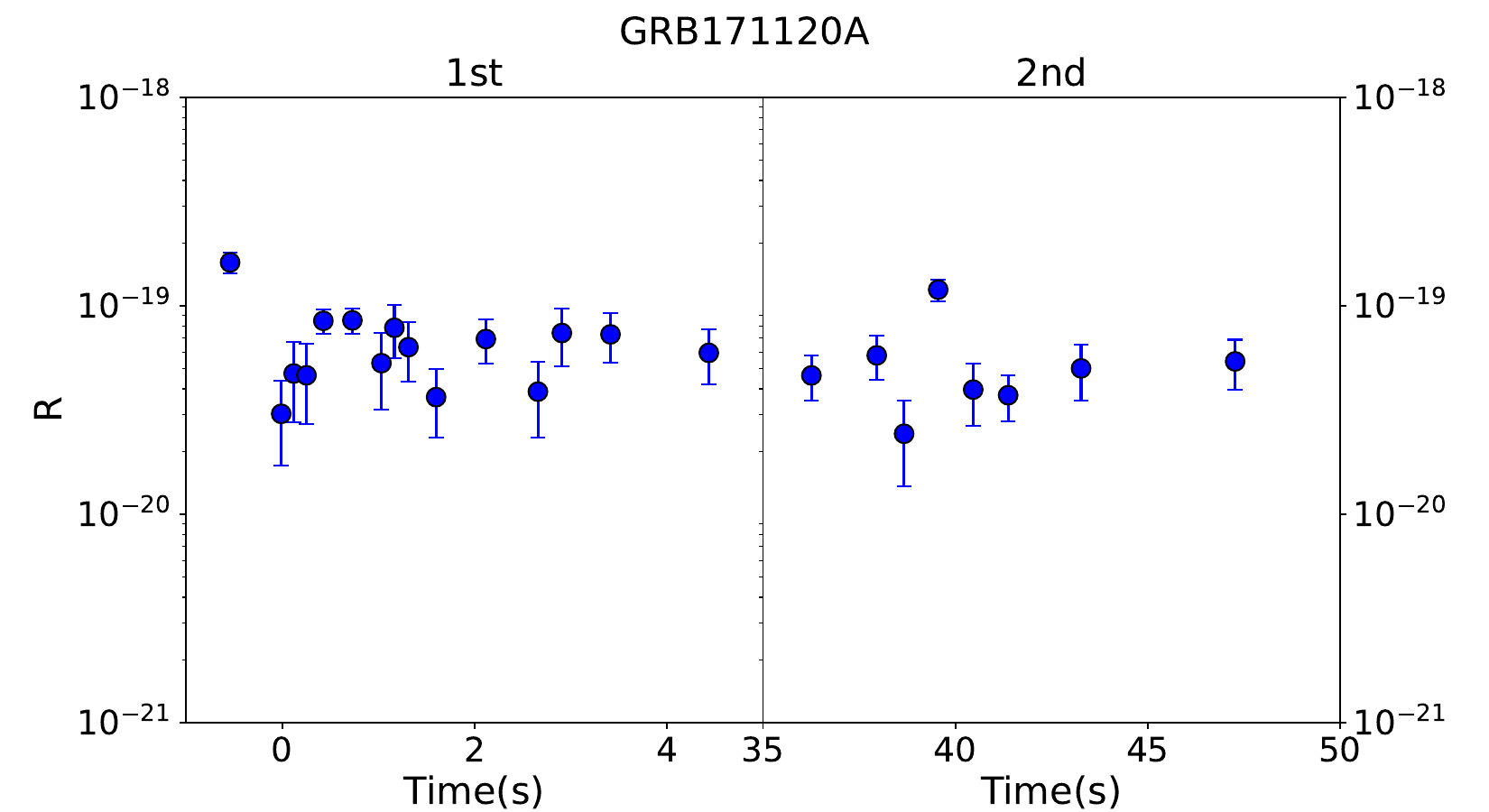}
\includegraphics [width=8cm,height=4cm]{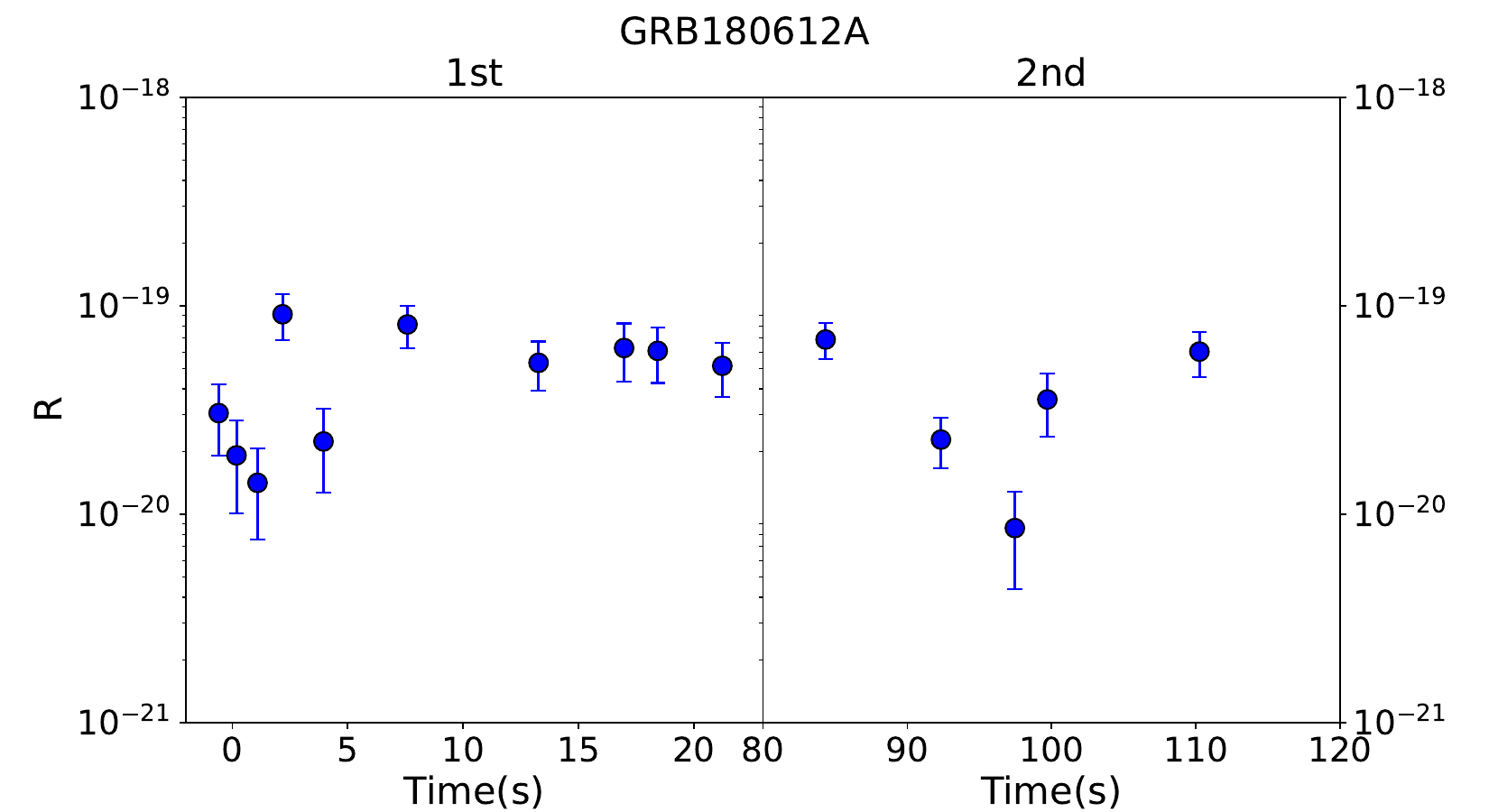}
\includegraphics [width=8cm,height=4cm]{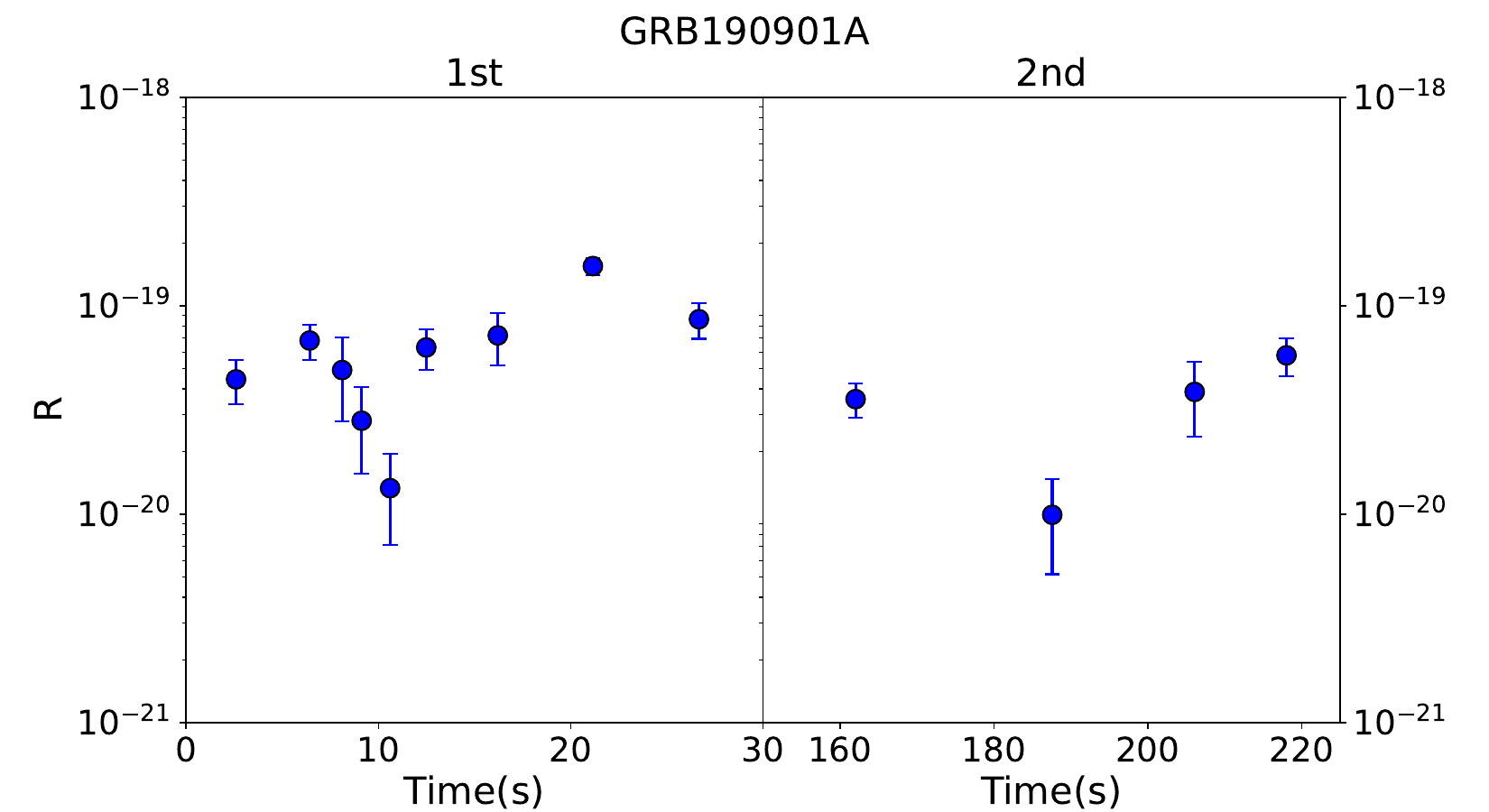}
\includegraphics [width=8cm,height=4cm]{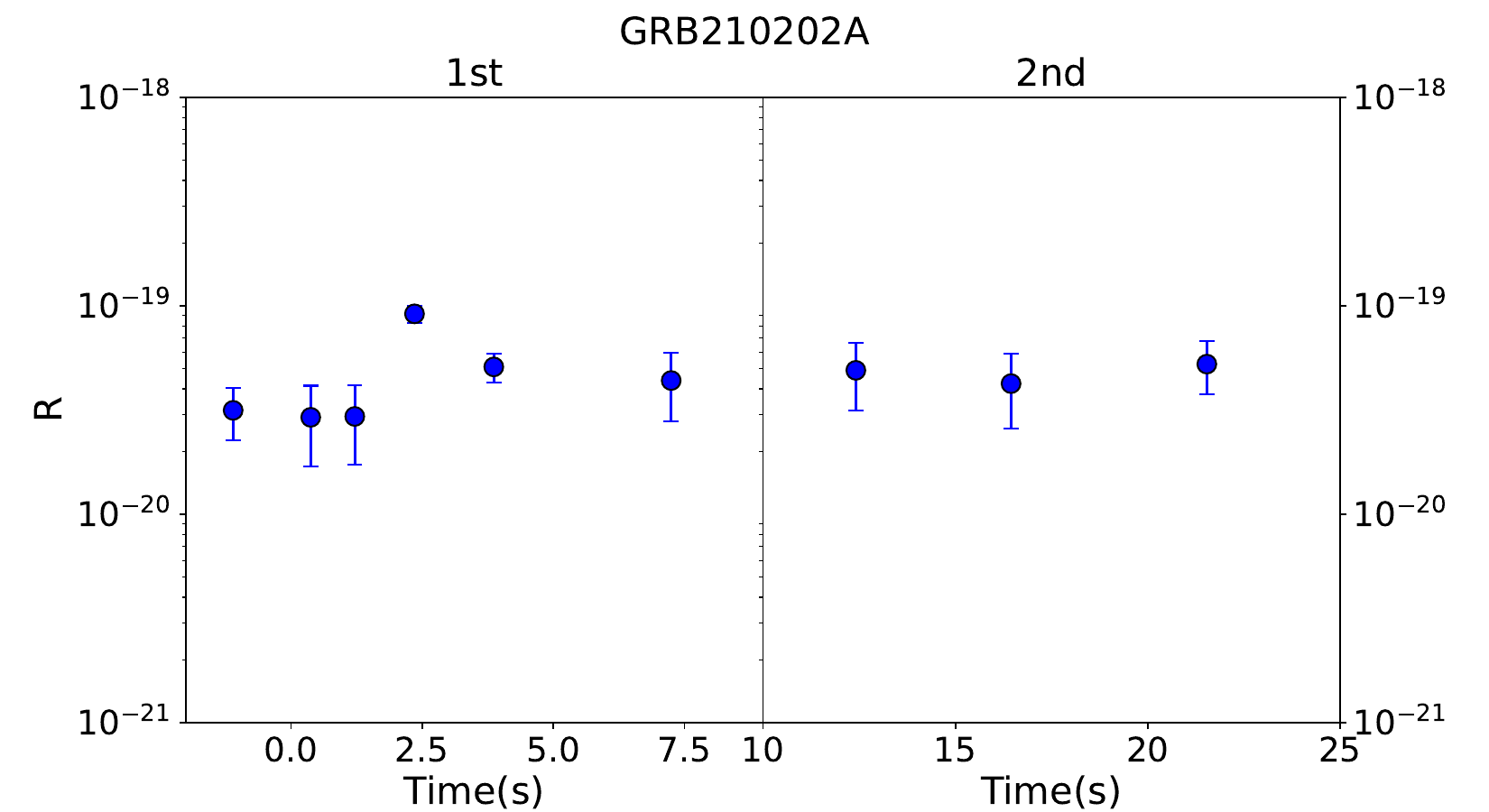}
\includegraphics [width=8cm,height=4cm]{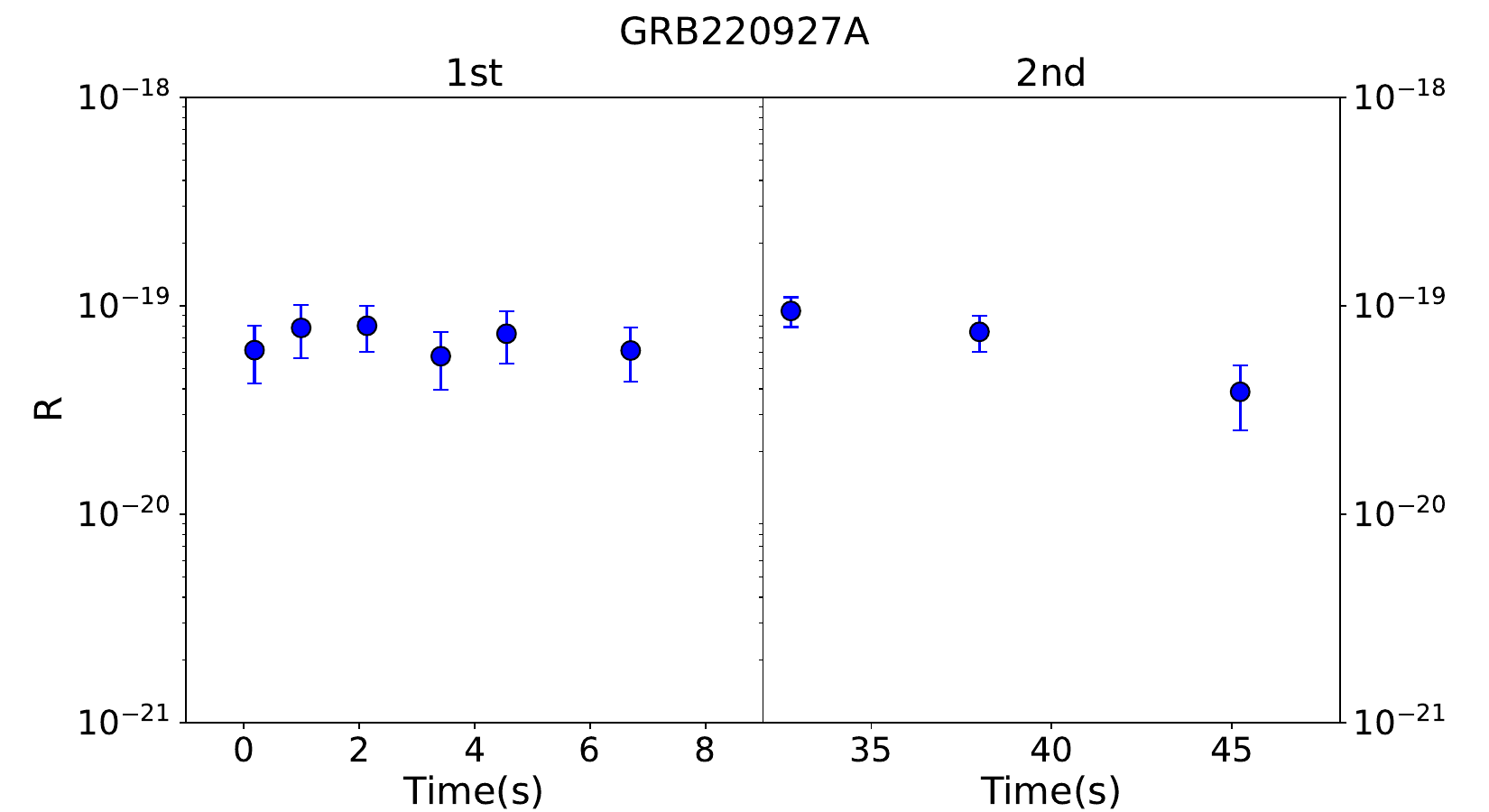}
\includegraphics [width=8cm,height=4cm]{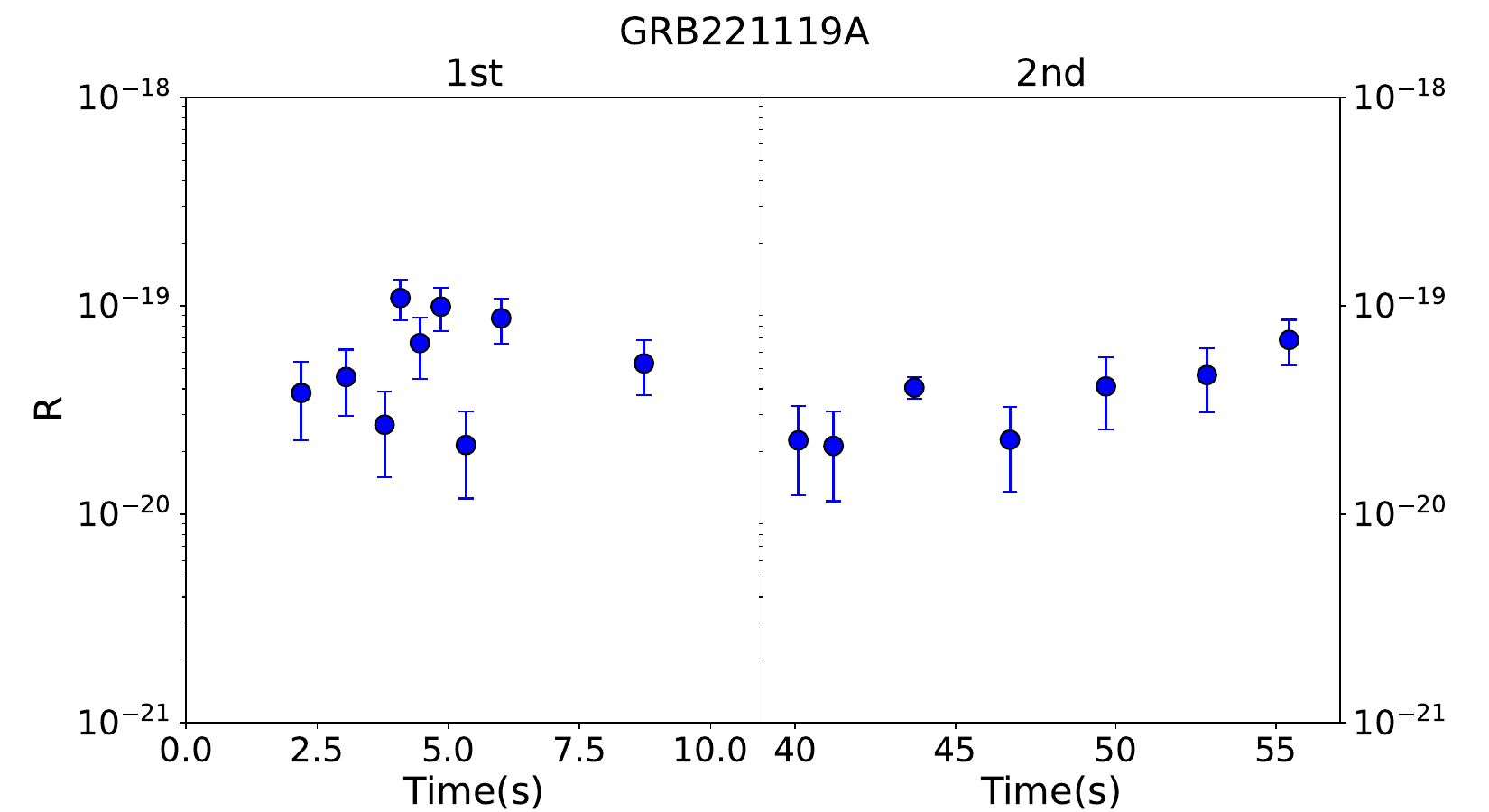}
\includegraphics [width=8cm,height=4cm]{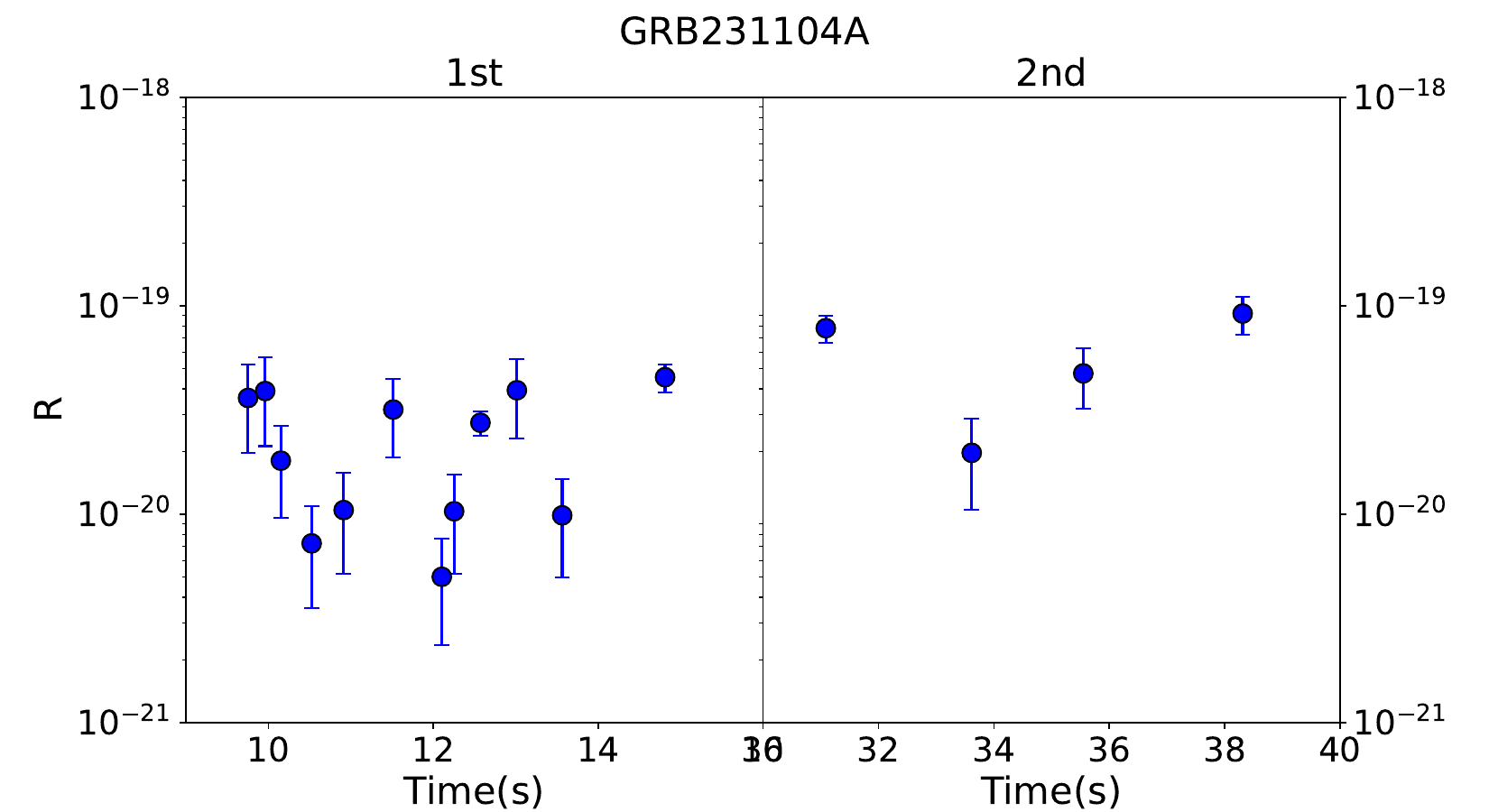}
\includegraphics [width=8cm,height=4cm]{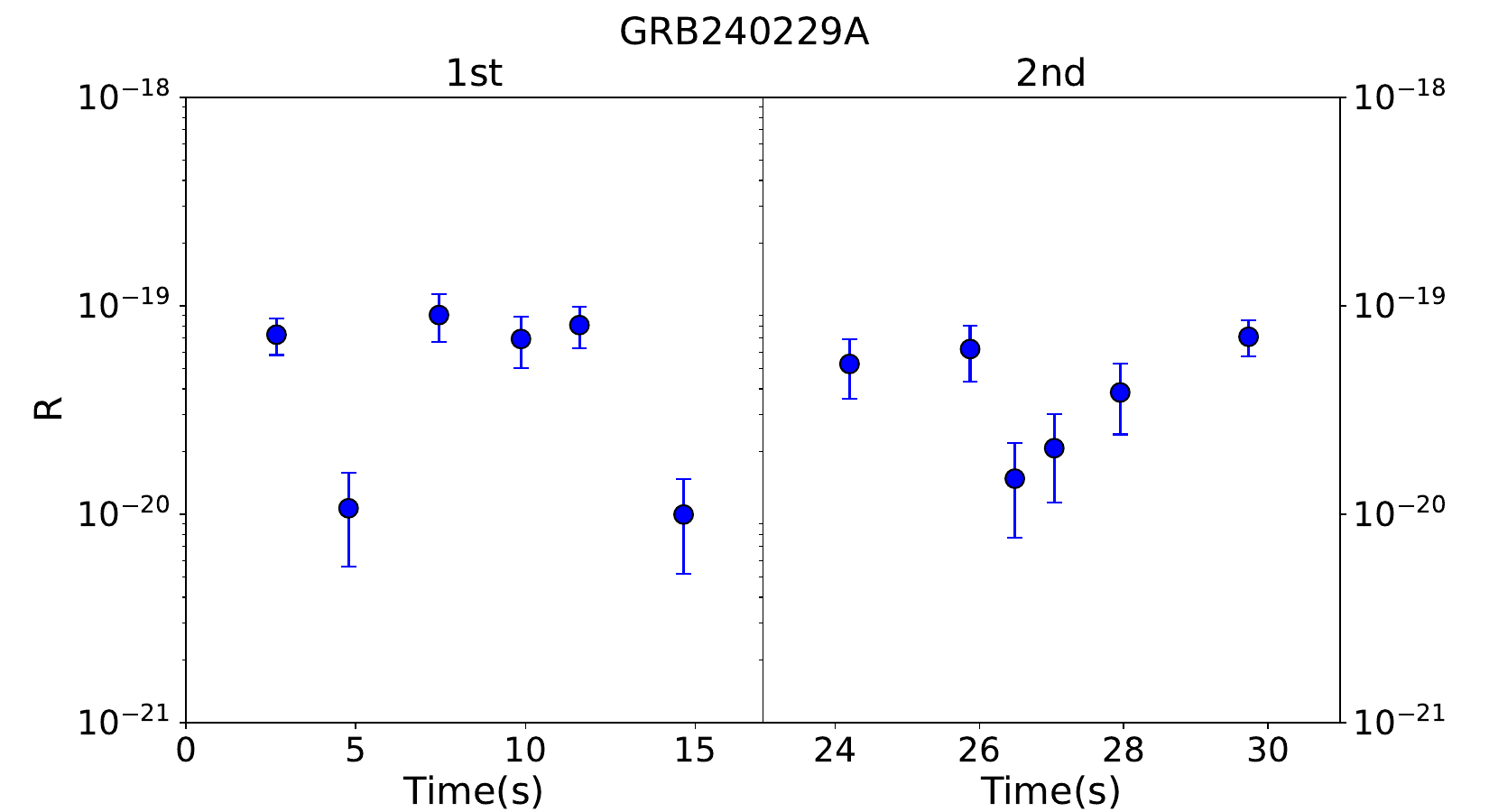}
   \figcaption{(Continued.) \label{fig 15} }
      
\end{figure}

\setcounter{figure}{15}  
\begin{figure}[H]
\centering
\includegraphics [width=8cm,height=4cm]{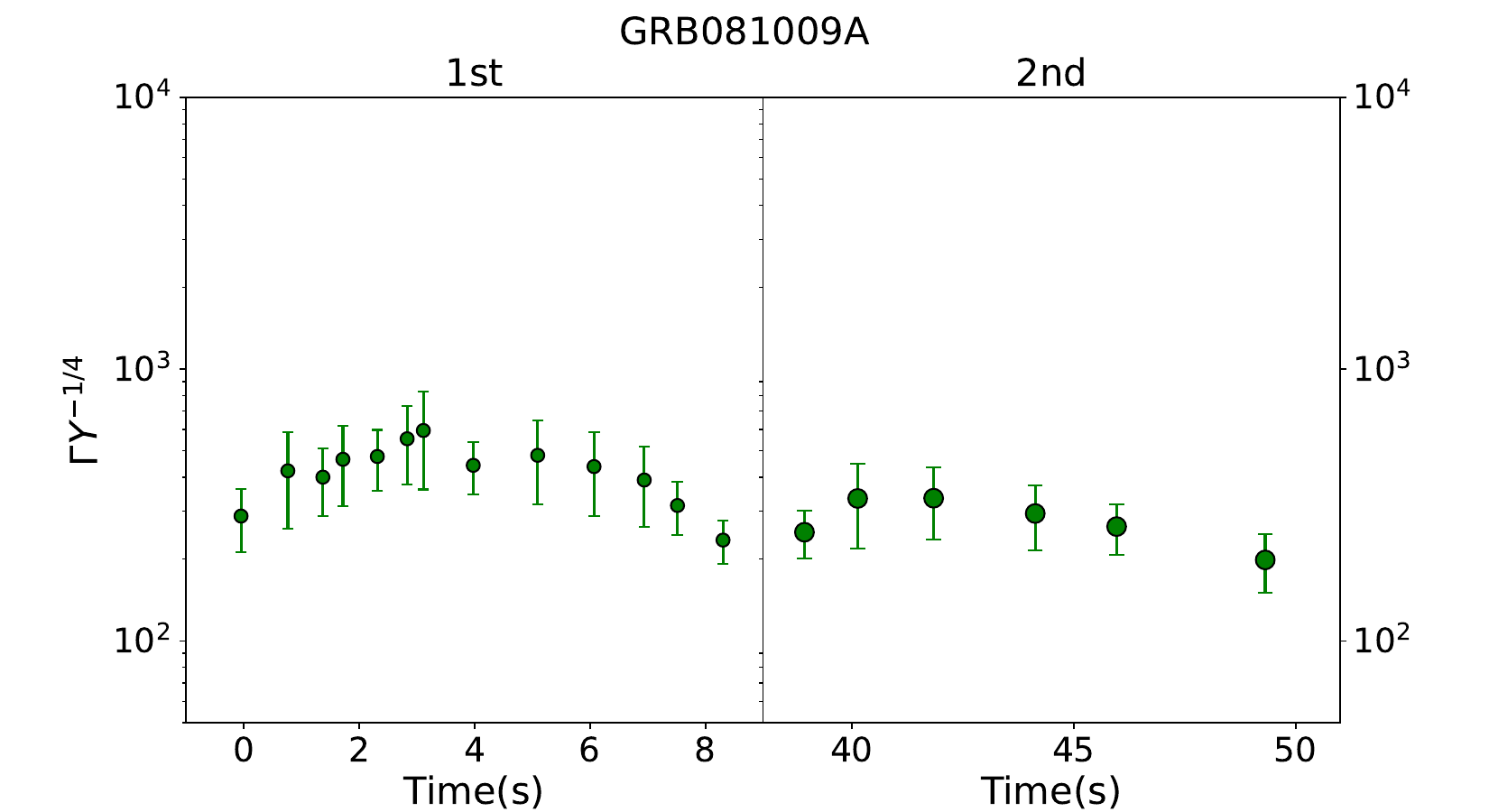}
\includegraphics [width=8cm,height=4cm]{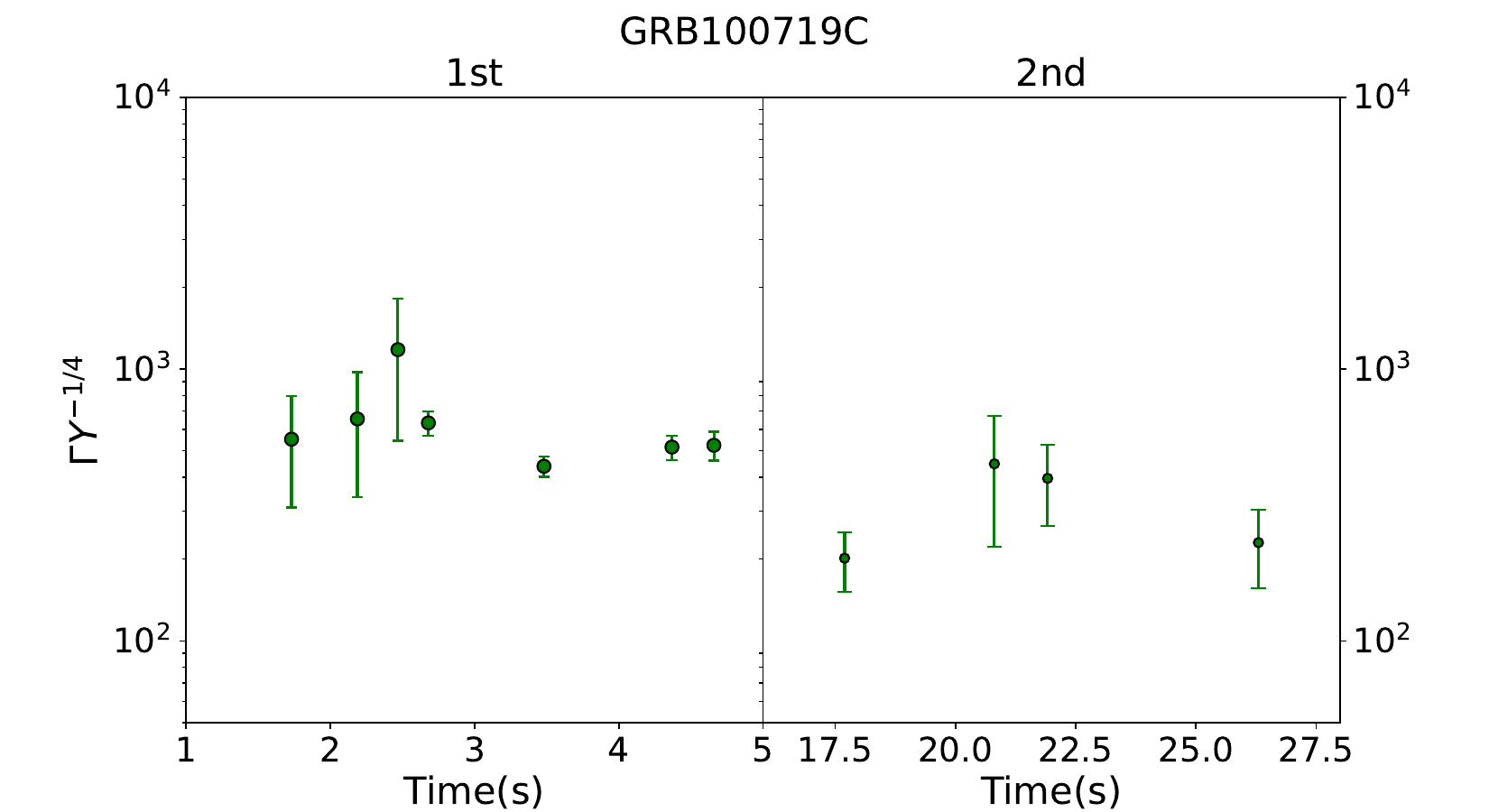}
\includegraphics [width=8cm,height=4cm]{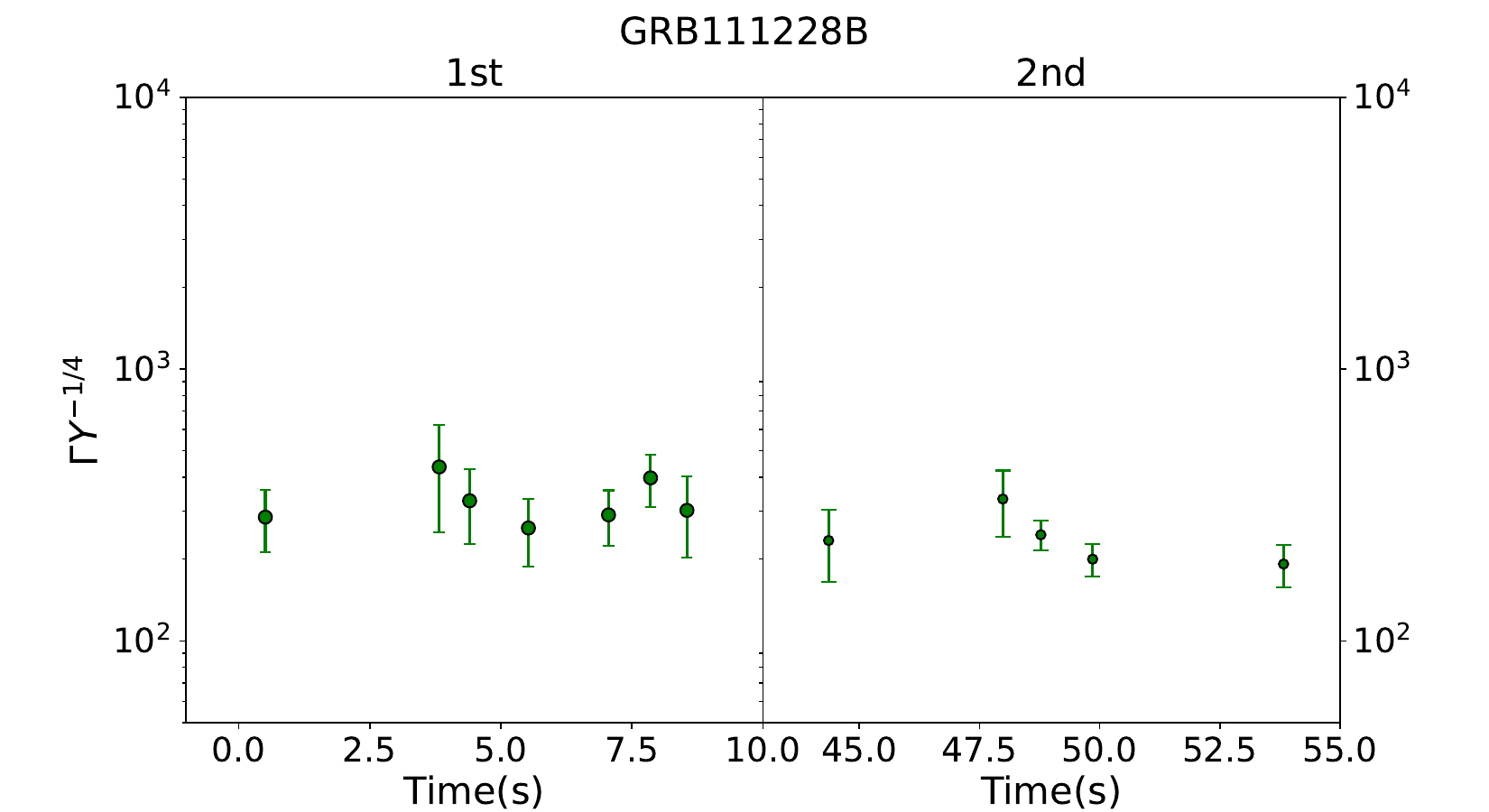}
\includegraphics [width=8cm,height=4cm]{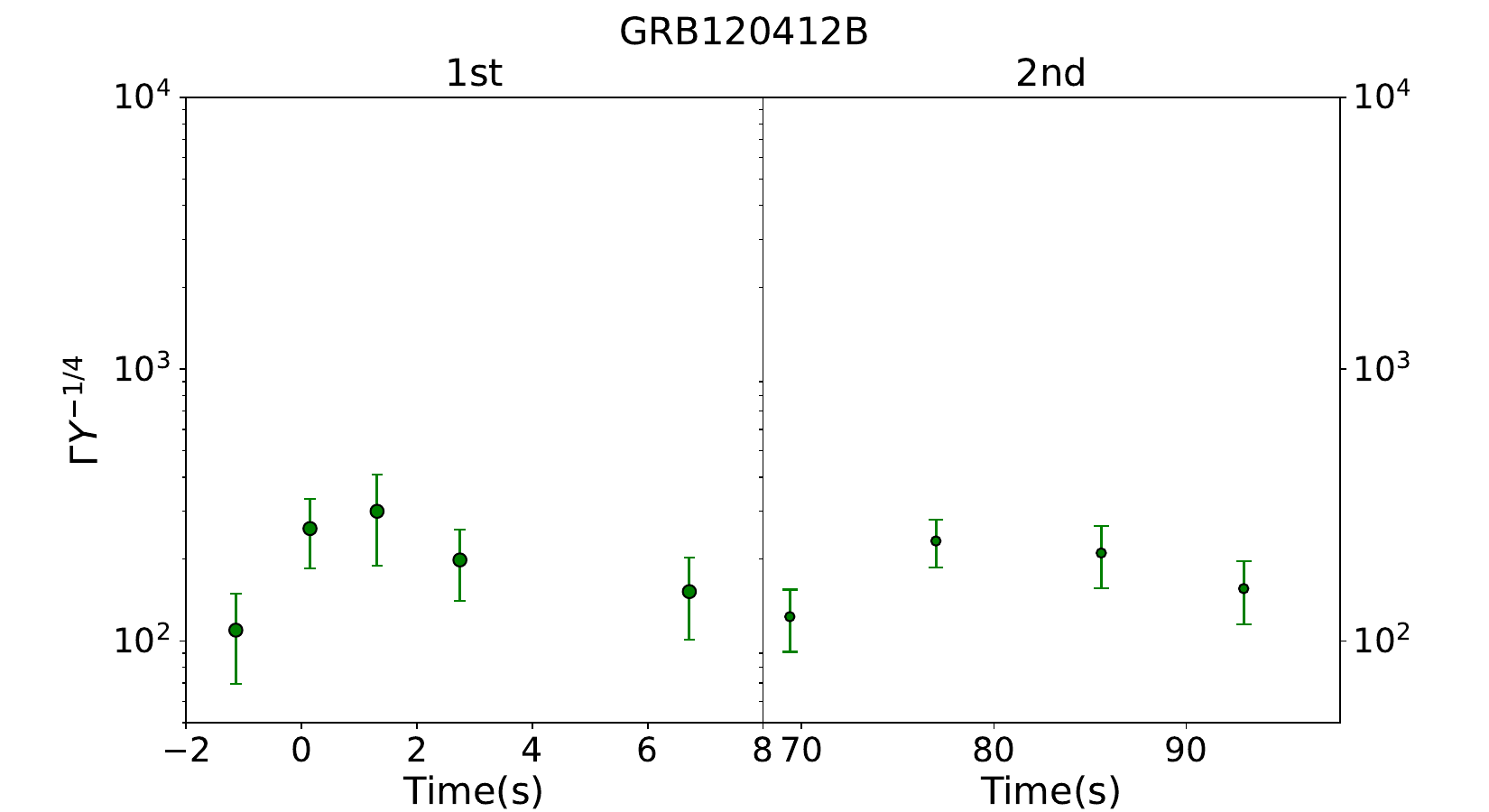}
\includegraphics [width=8cm,height=4cm]{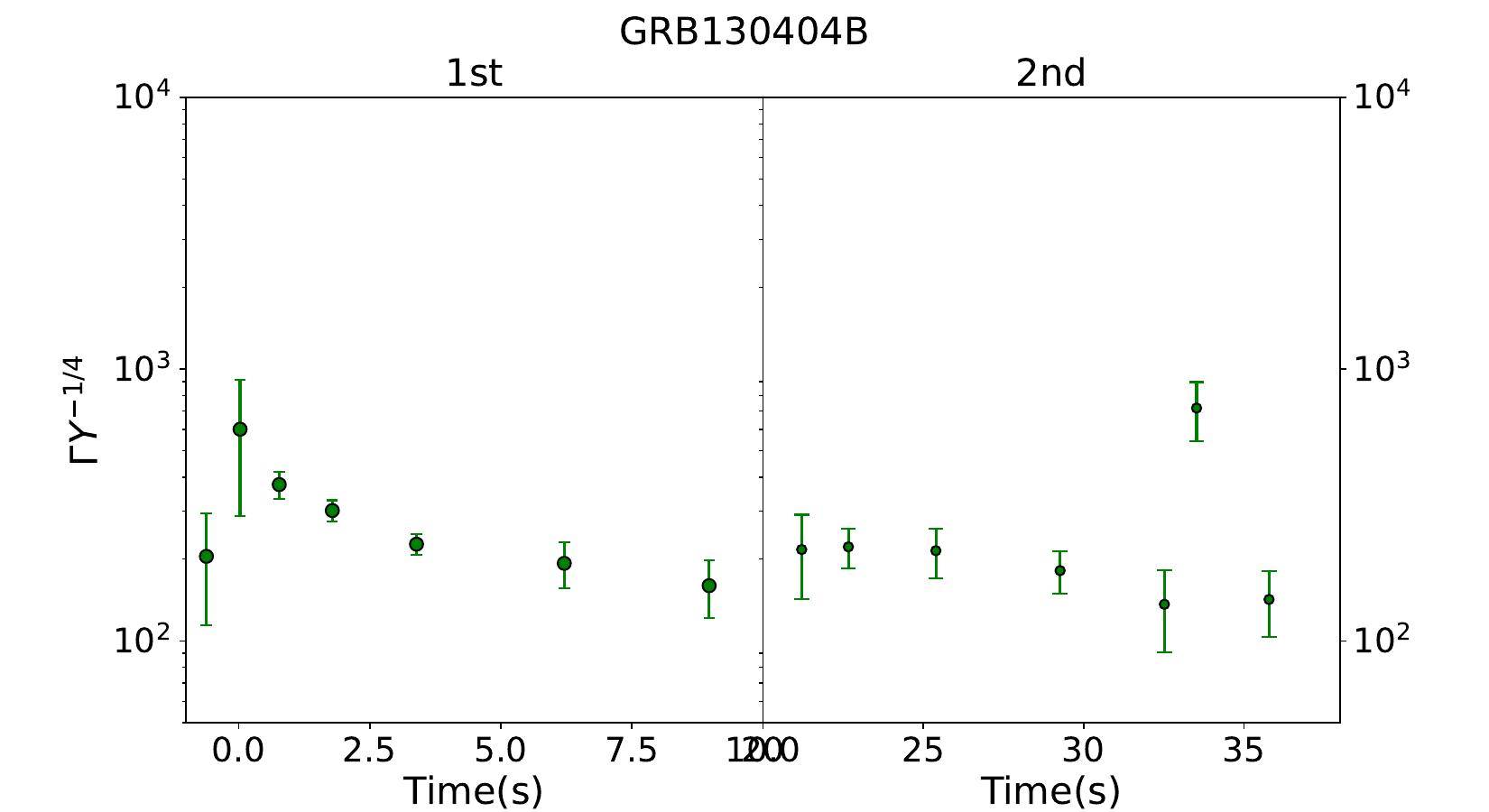}
\includegraphics [width=8cm,height=4cm]{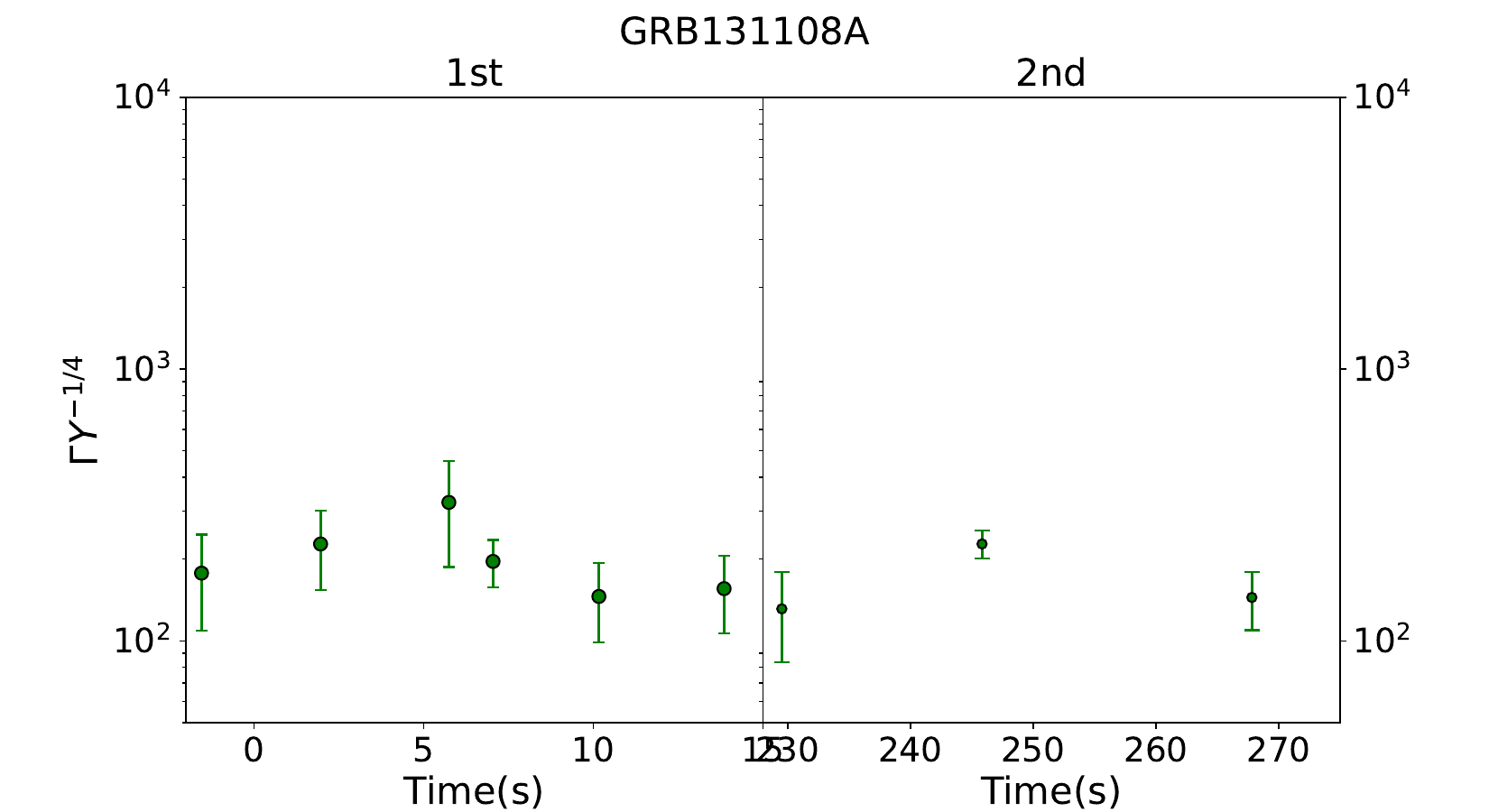}
\includegraphics [width=8cm,height=4cm]{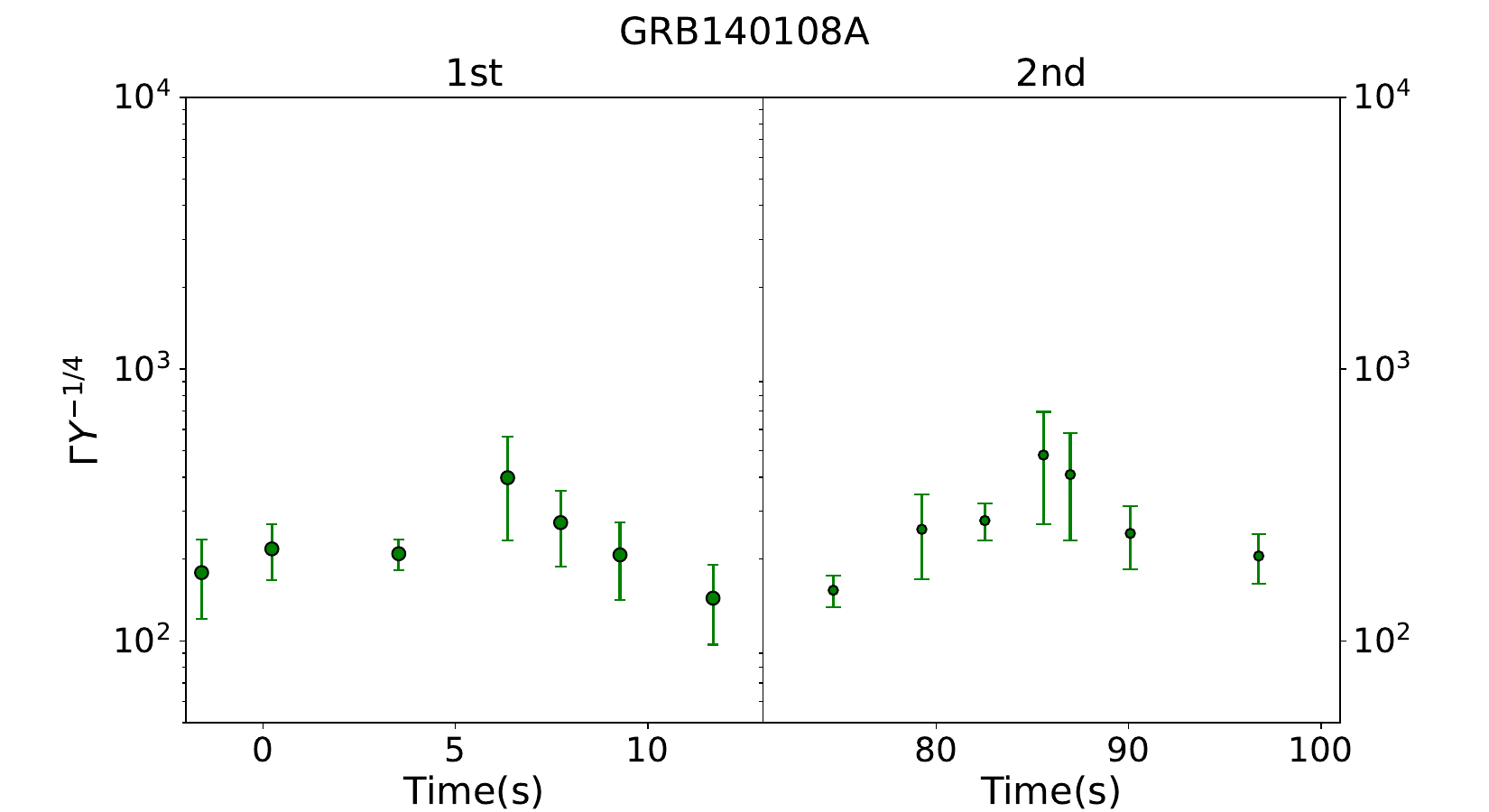}
\includegraphics [width=8cm,height=4cm]{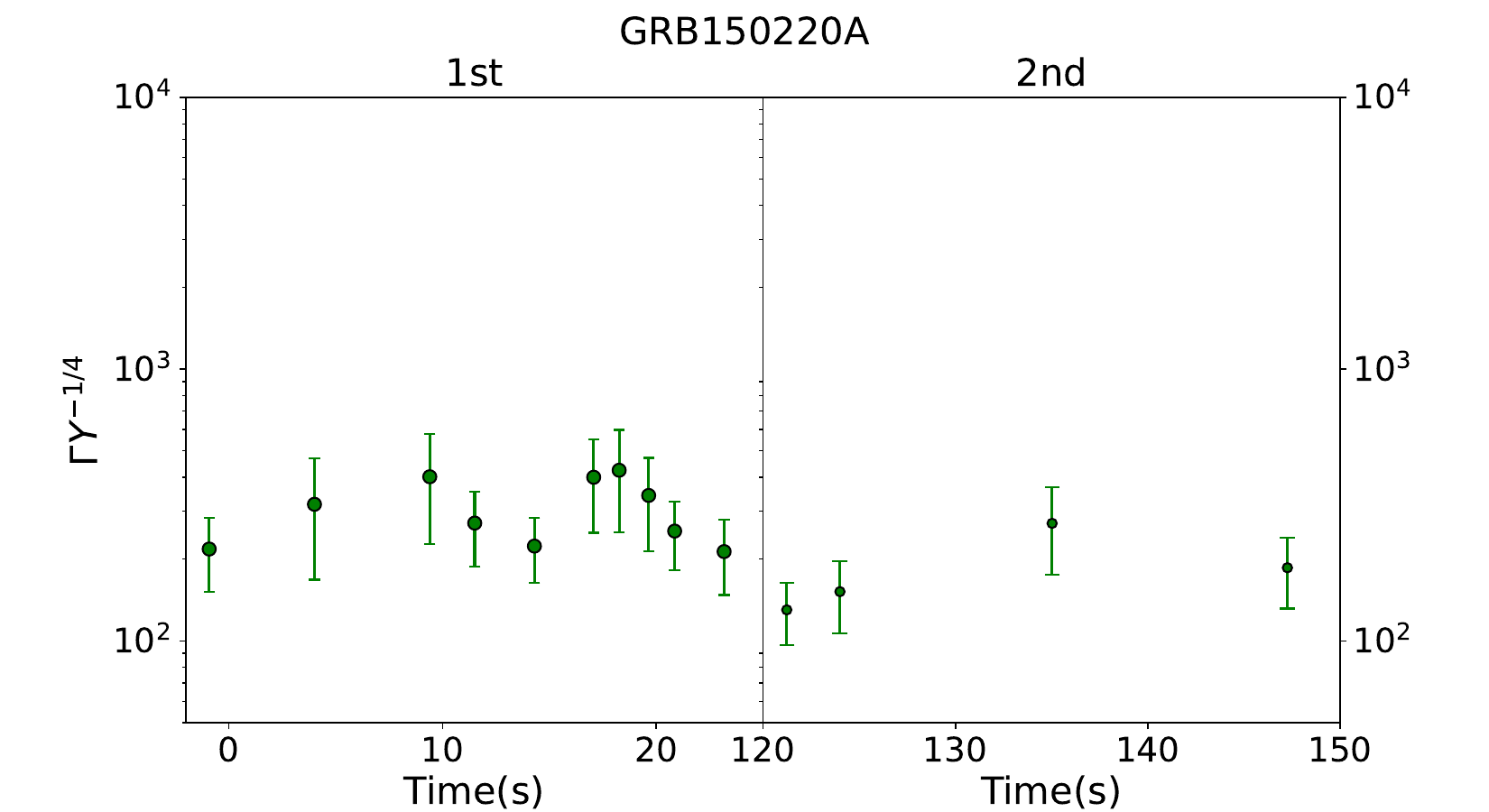}
\includegraphics [width=8cm,height=4cm]{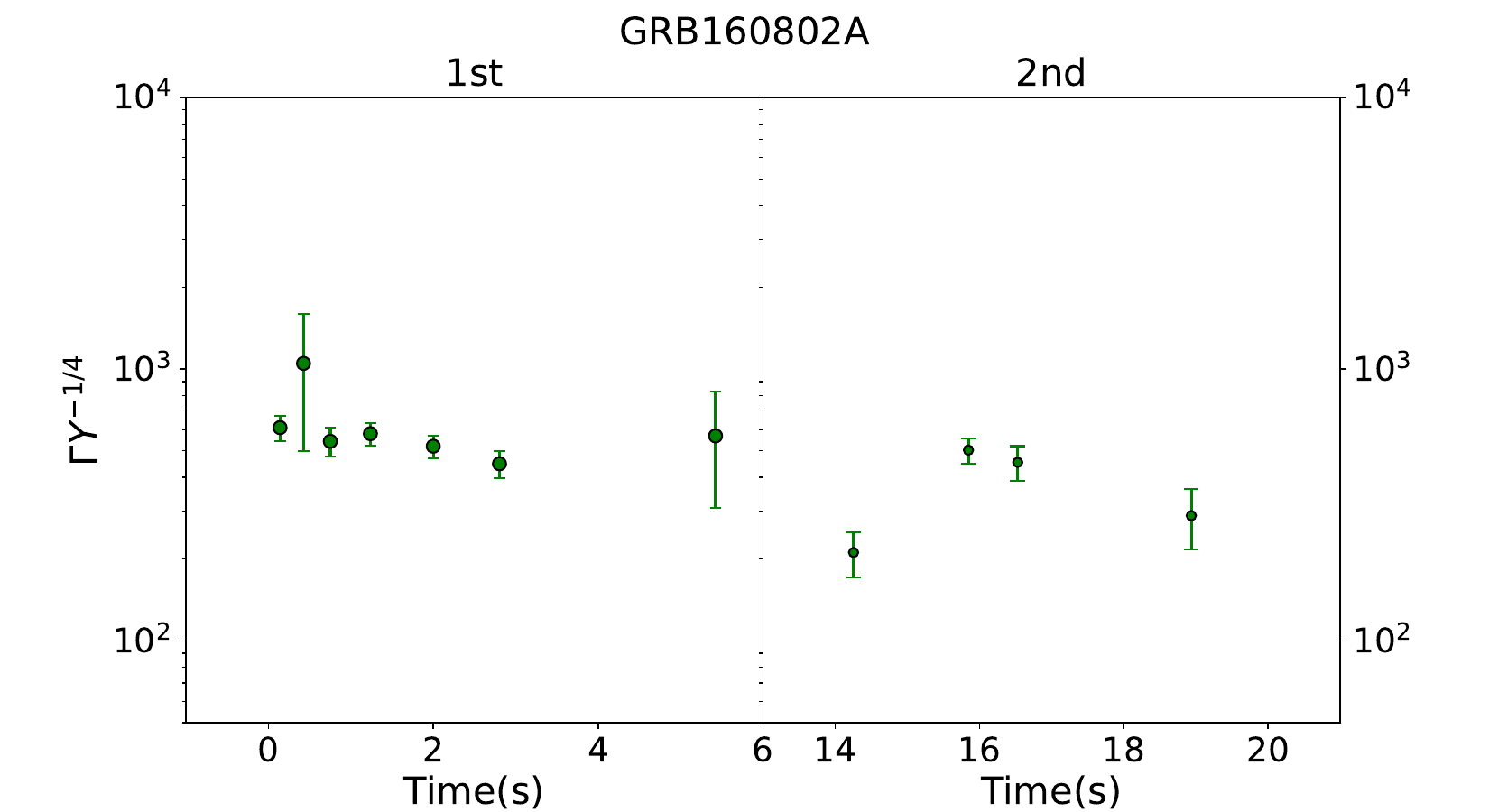}
\includegraphics [width=8cm,height=4cm]{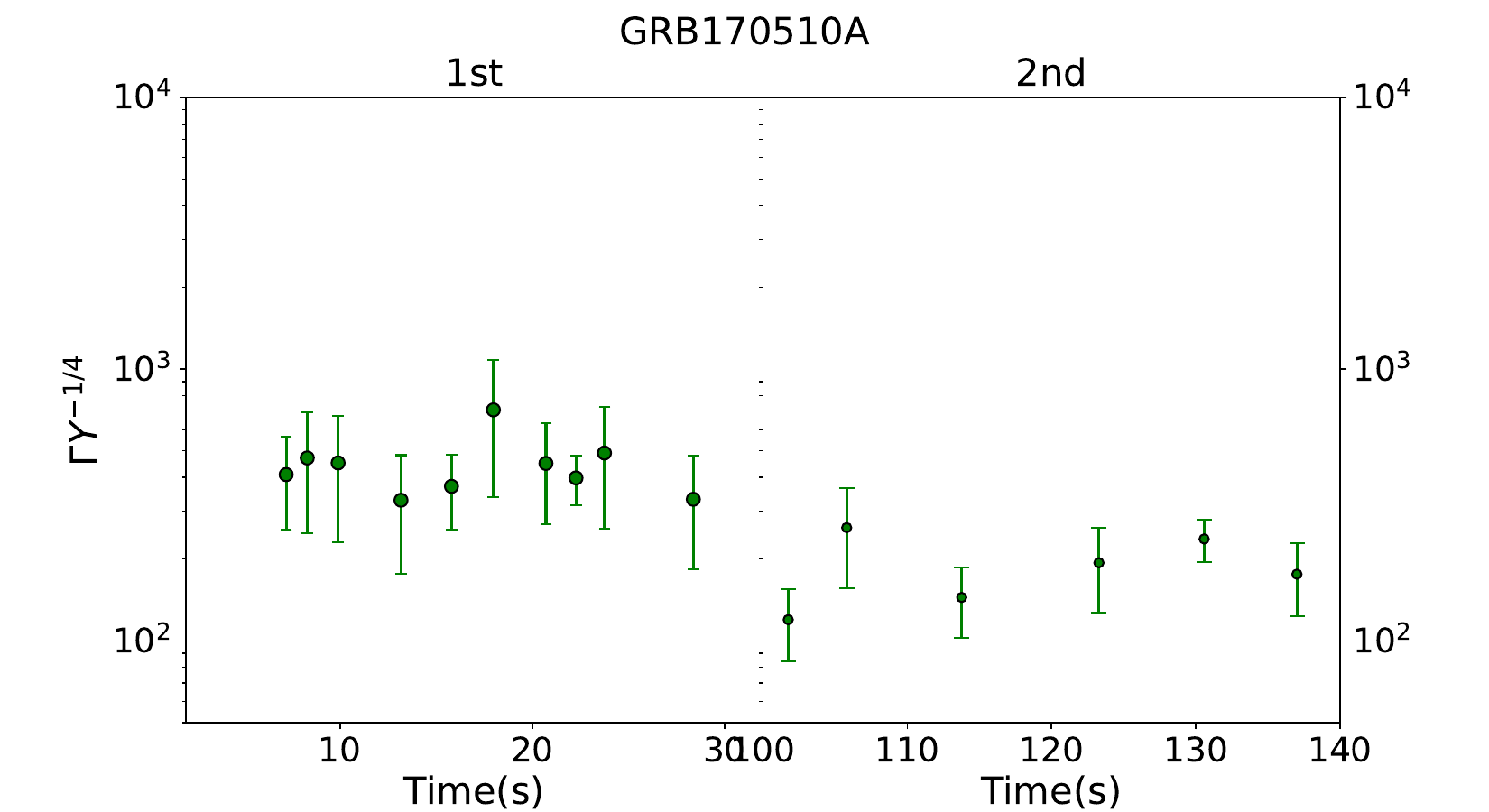}
   \figcaption{Evolution of $\Gamma $. ``1st'' denotes main burst, and ``2nd'' denotes second burst. \label{fig 16}}
      
\end{figure}

\setcounter{figure}{15}  
\begin{figure}[H]

\centering
\includegraphics [width=8cm,height=4cm]{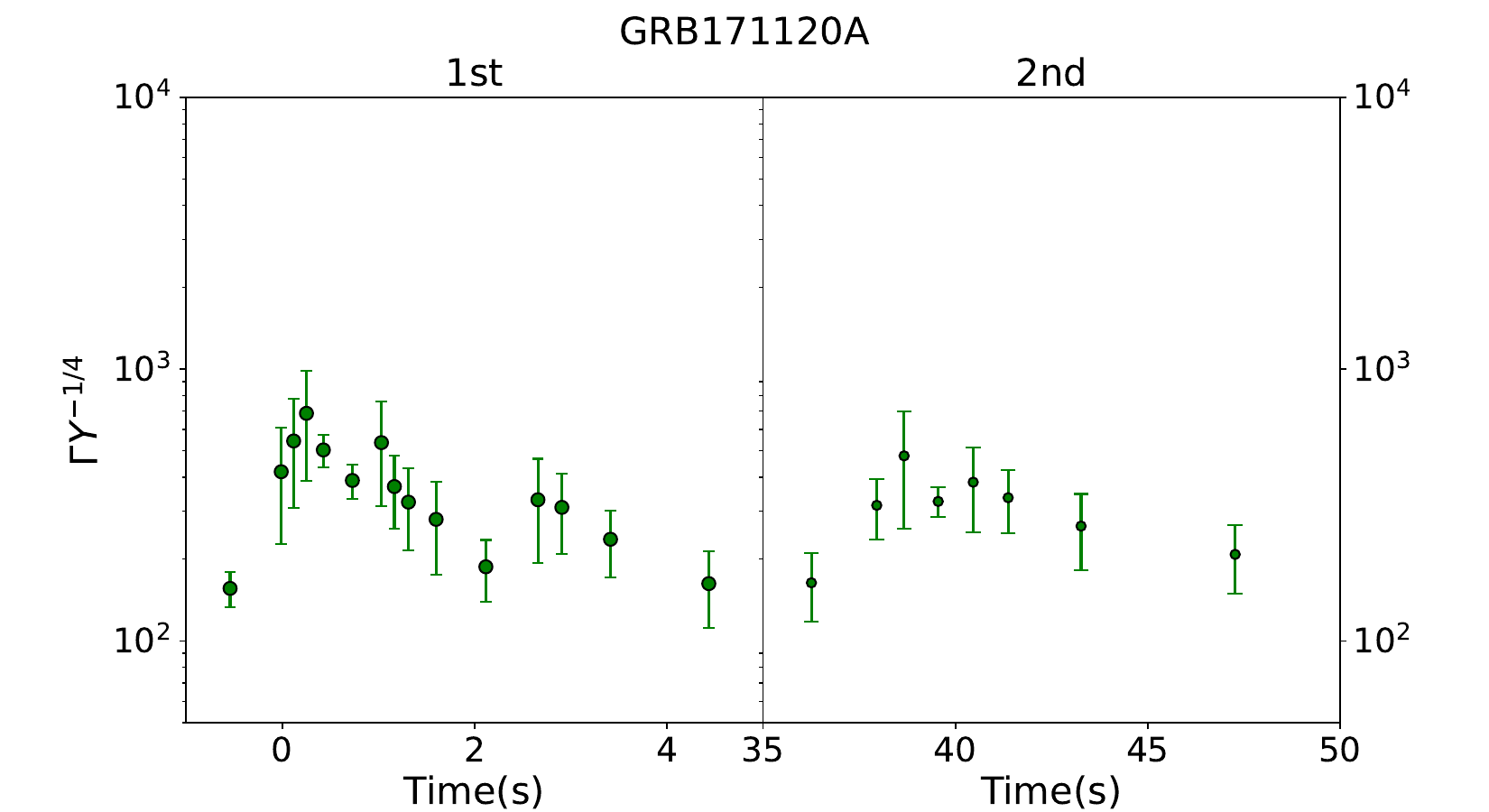}
\includegraphics [width=8cm,height=4cm]{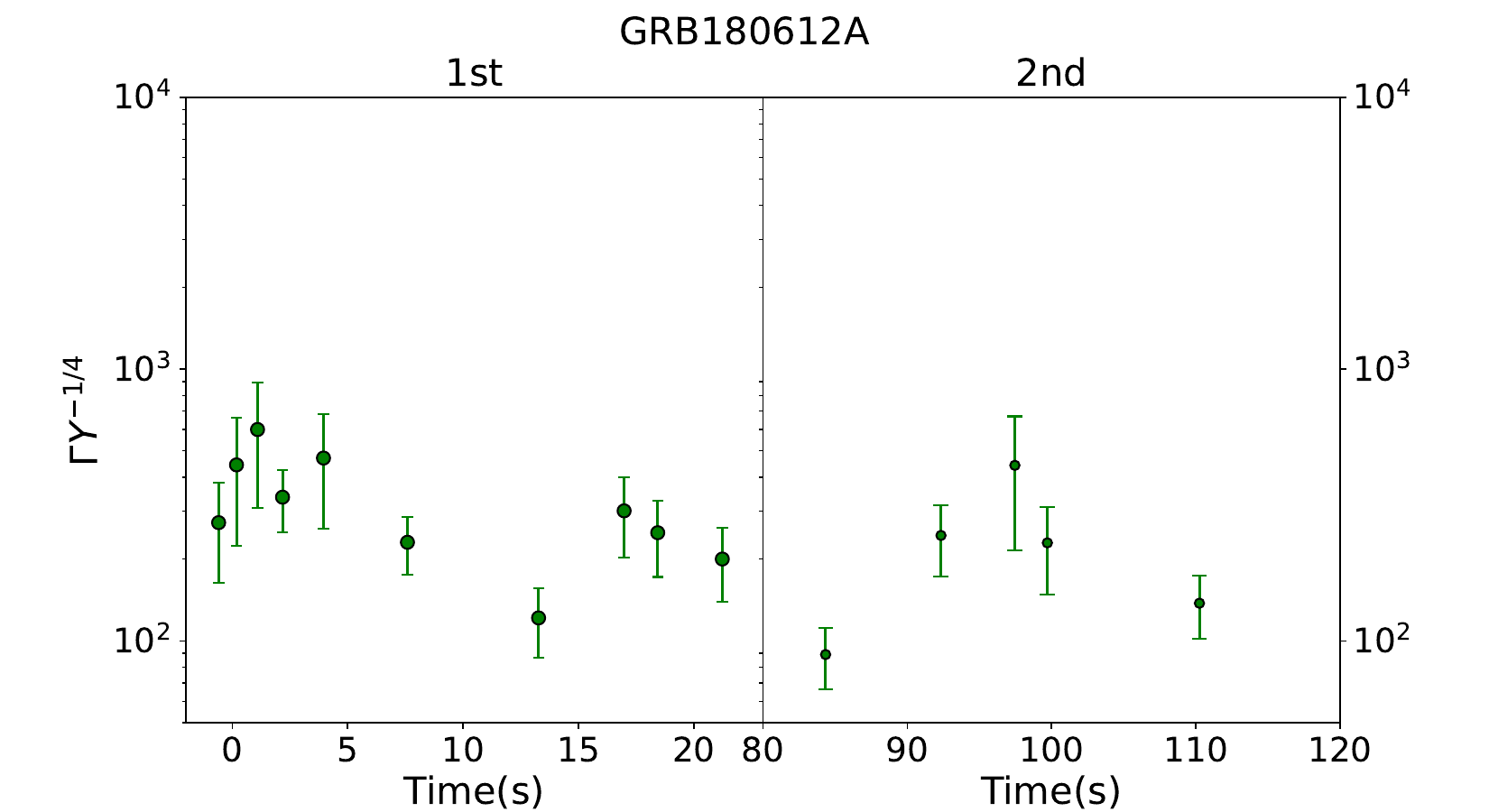}
\includegraphics [width=8cm,height=4cm]{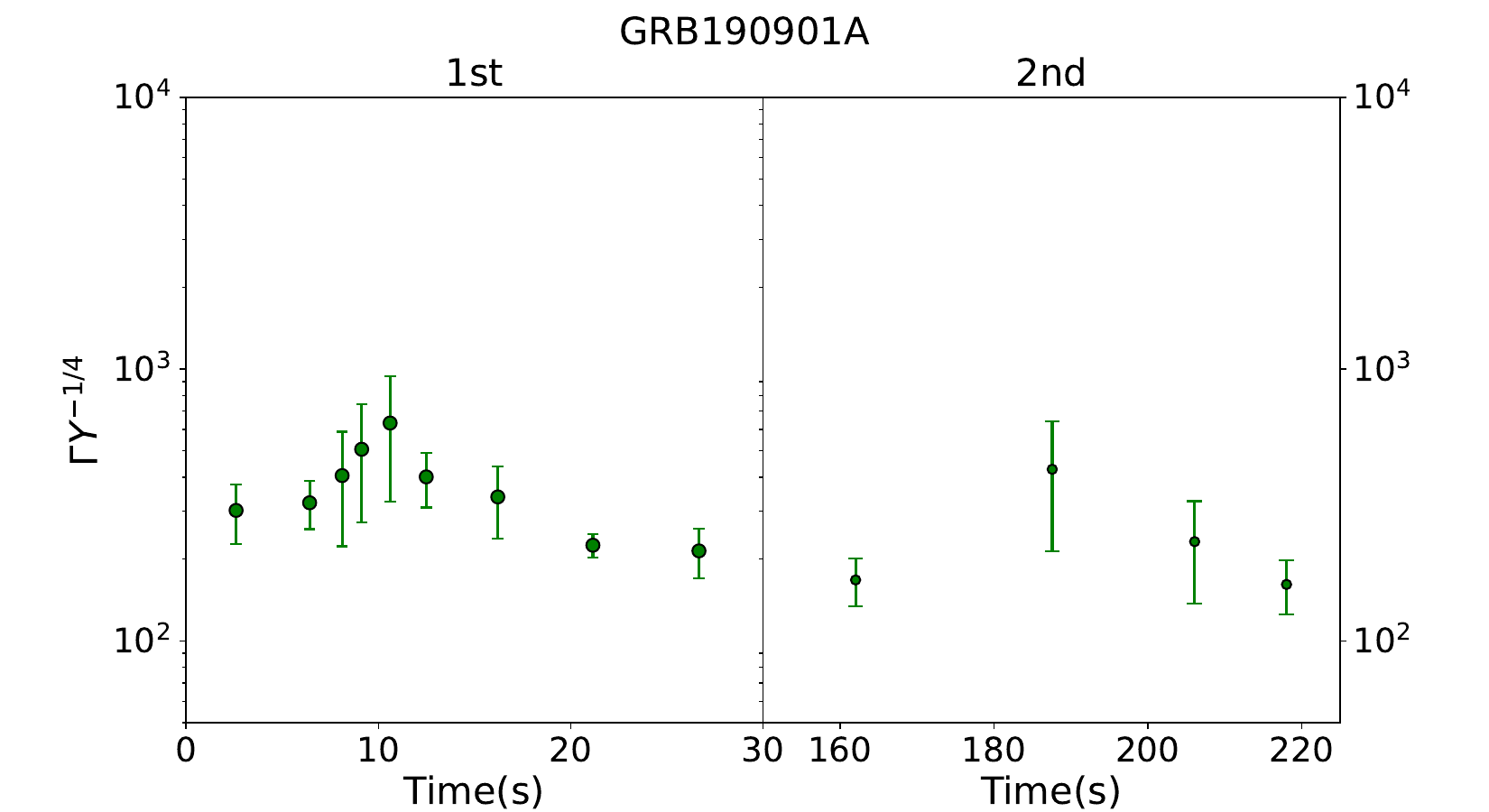}
\includegraphics [width=8cm,height=4cm]{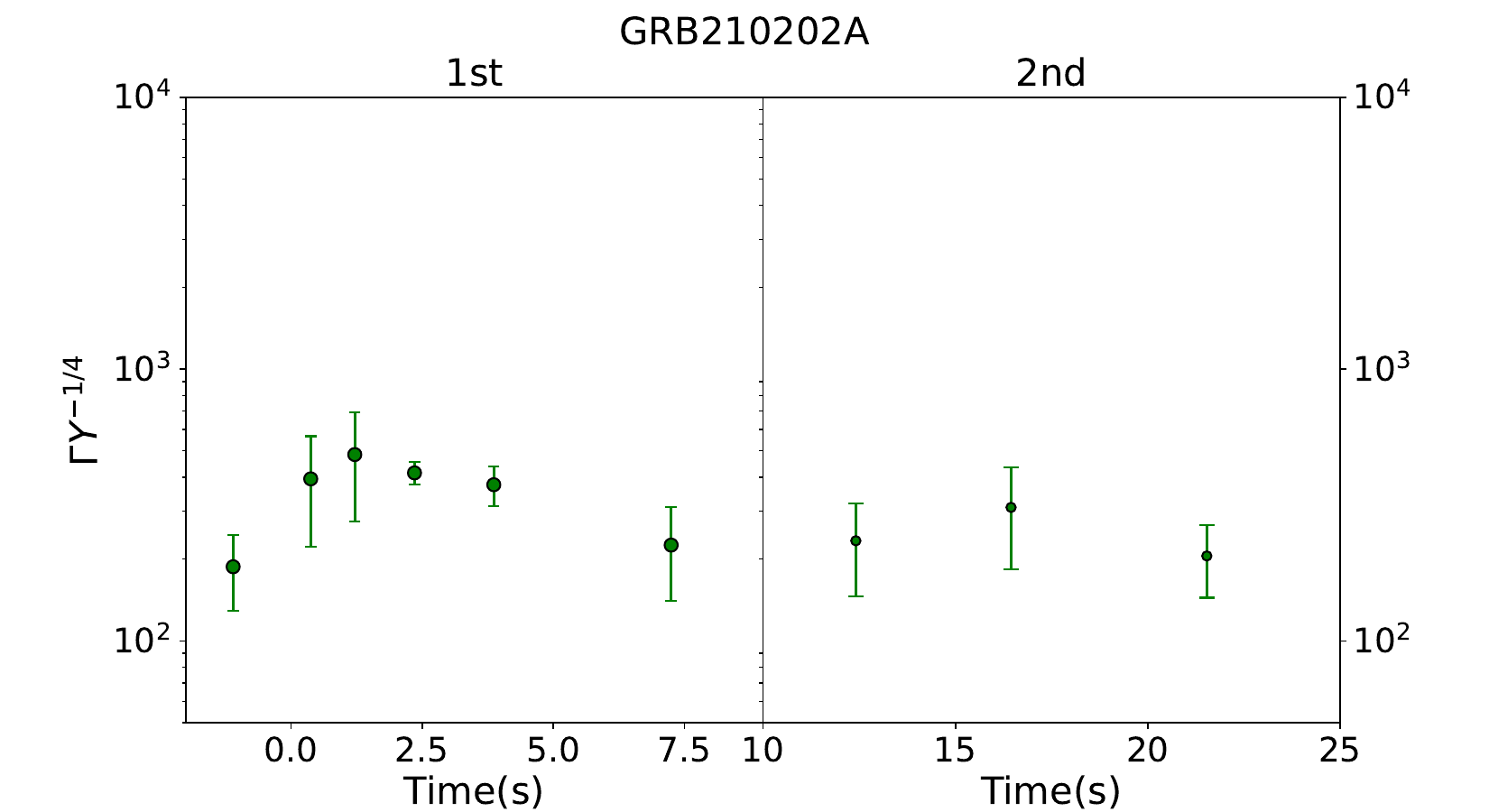}
\includegraphics [width=8cm,height=4cm]{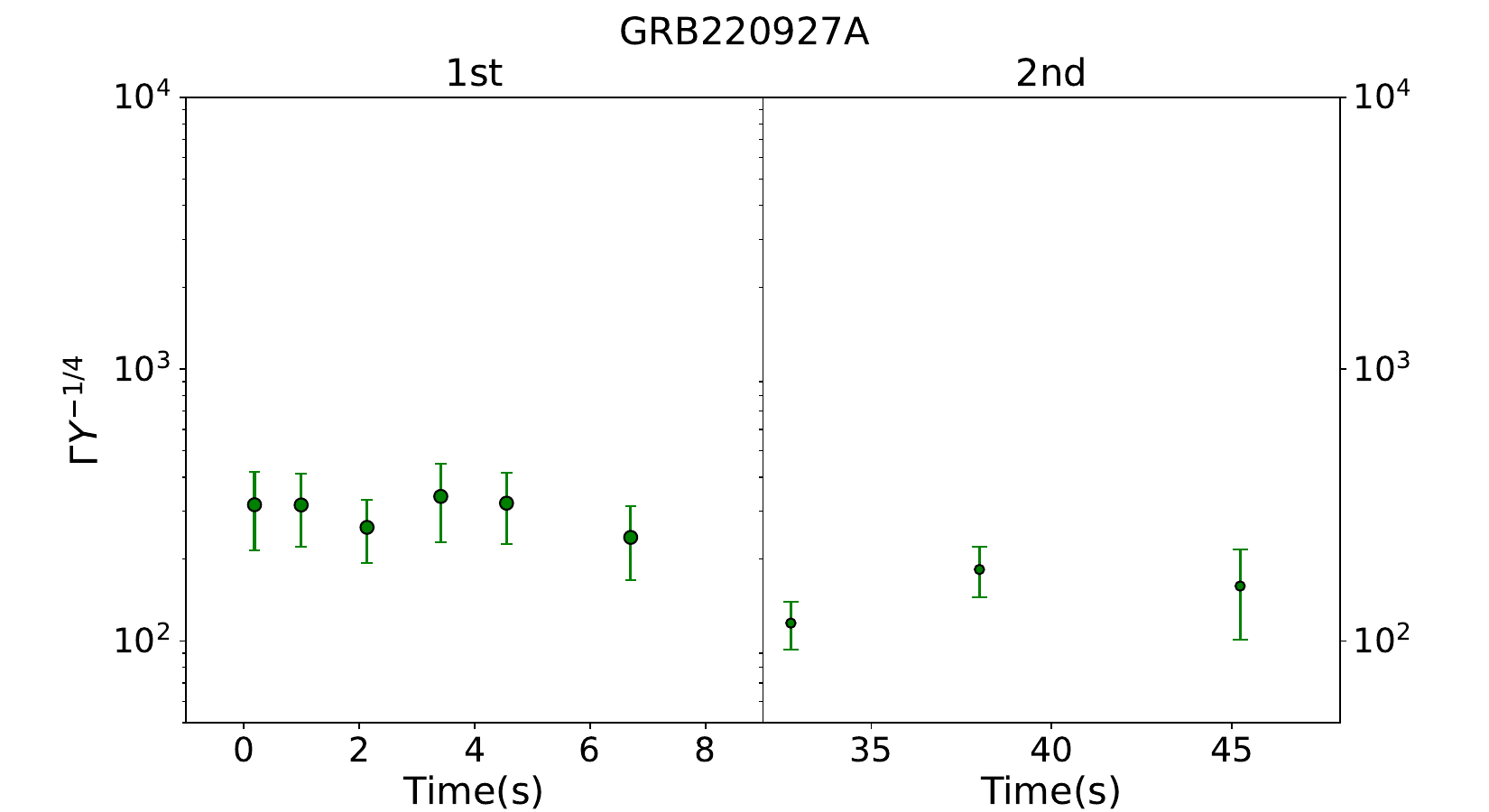}
\includegraphics [width=8cm,height=4cm]{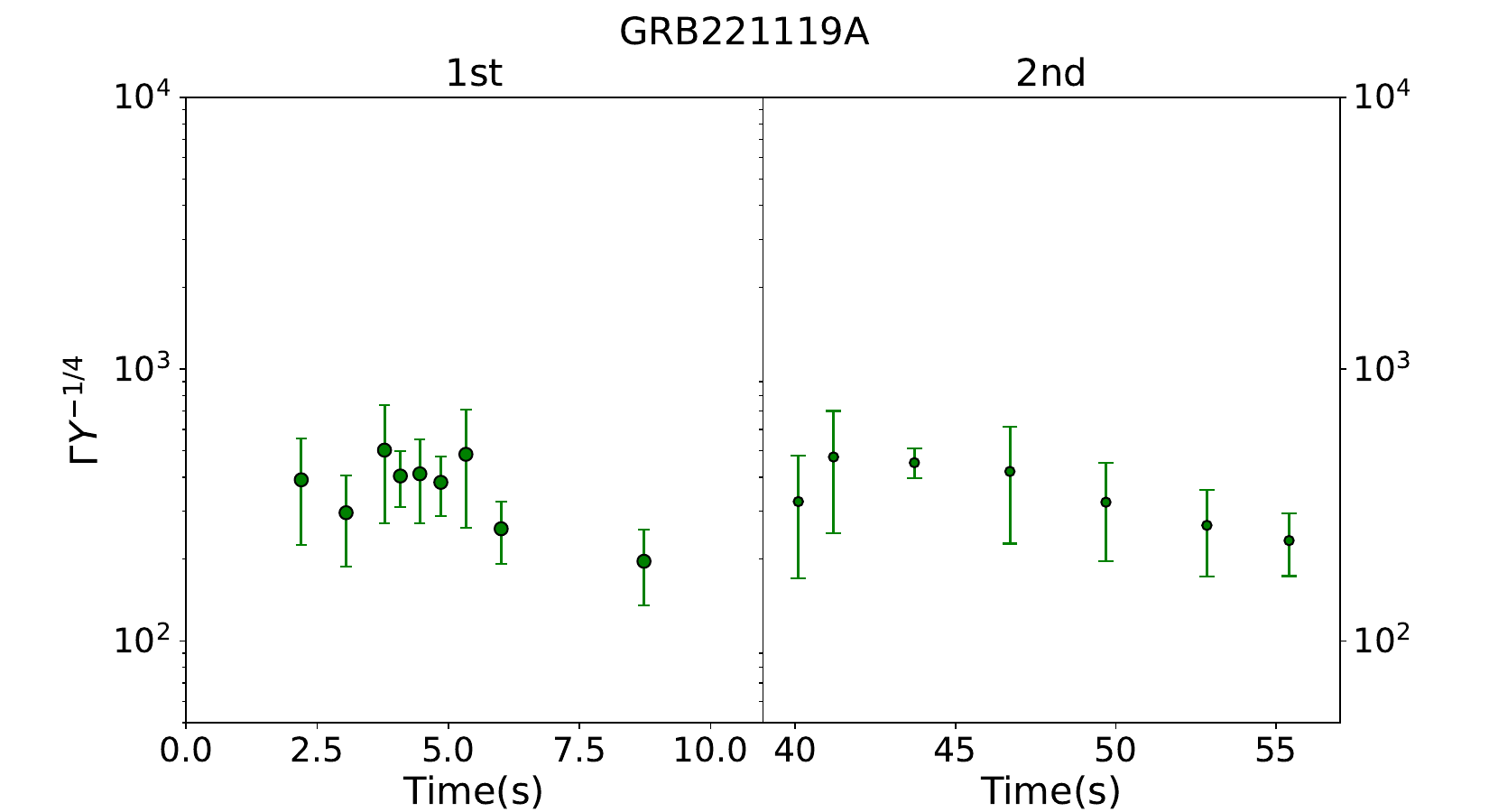}
\includegraphics [width=8cm,height=4cm]{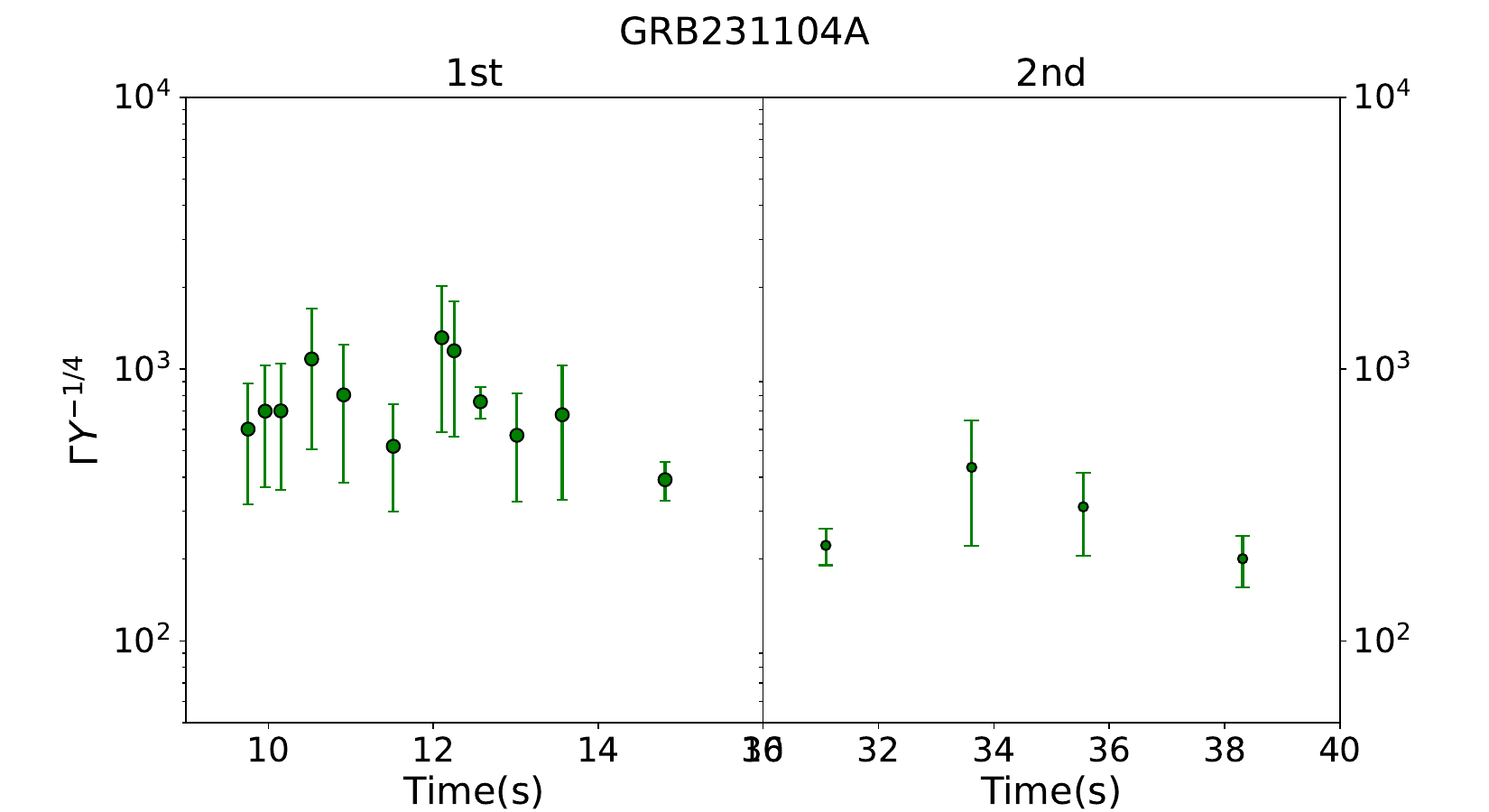}
\includegraphics [width=8cm,height=4cm]{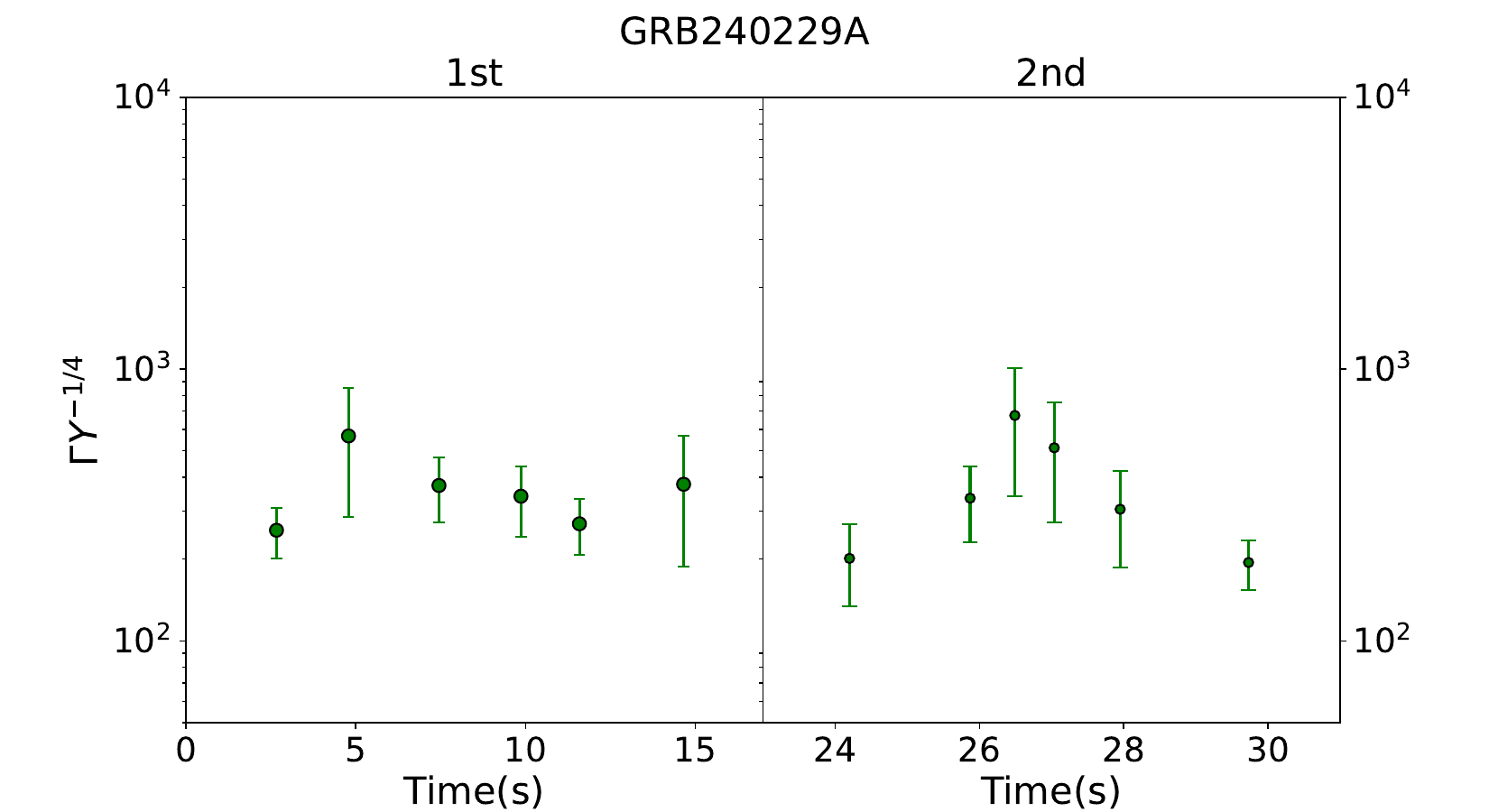}
   \figcaption{(Continued.) \label{fig 16}}
      
\end{figure}

\setcounter{figure}{16}  
\begin{figure}[H]
\centering
\includegraphics [width=8cm,height=4cm]{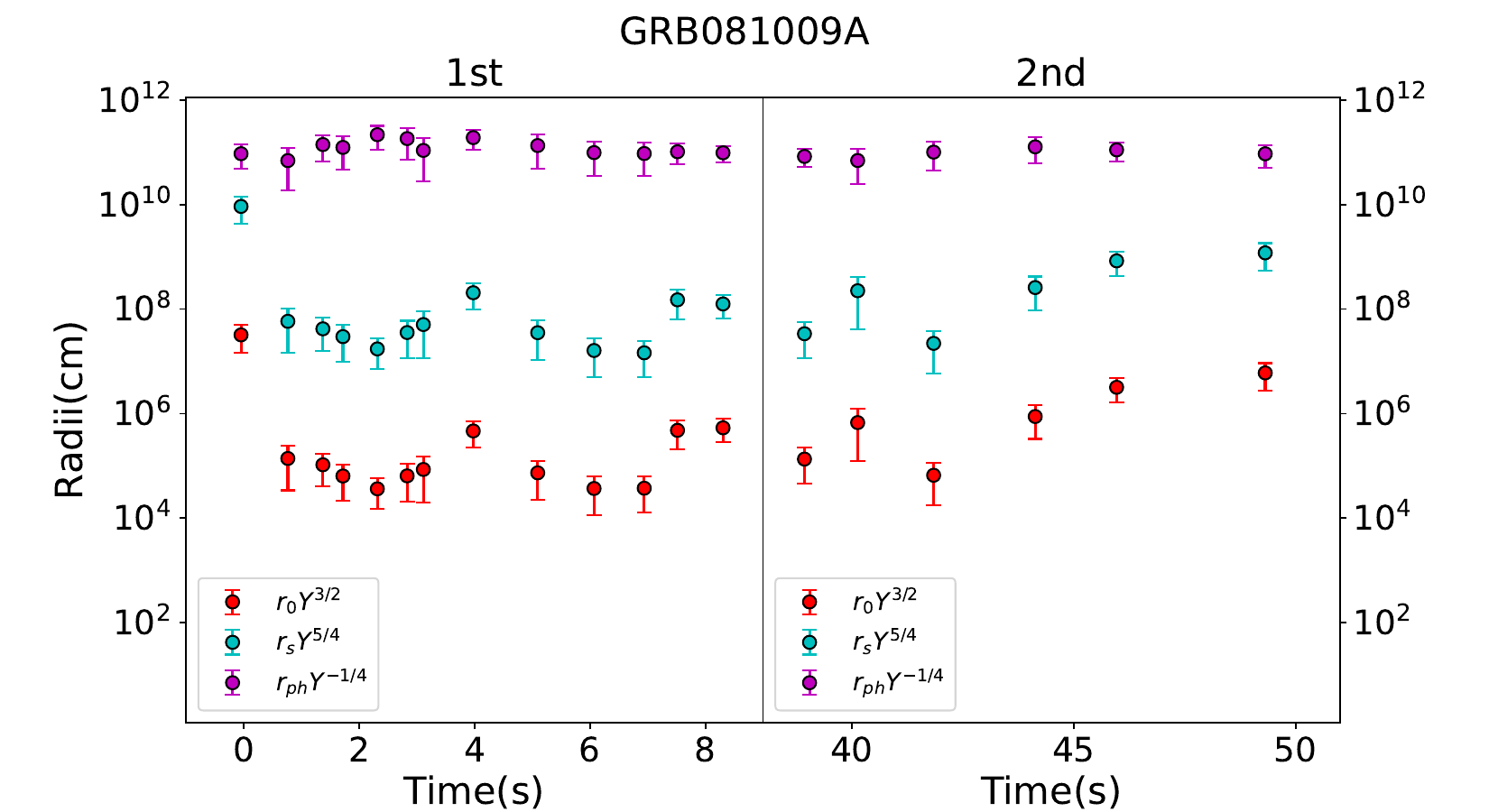}
\includegraphics [width=8cm,height=4cm]{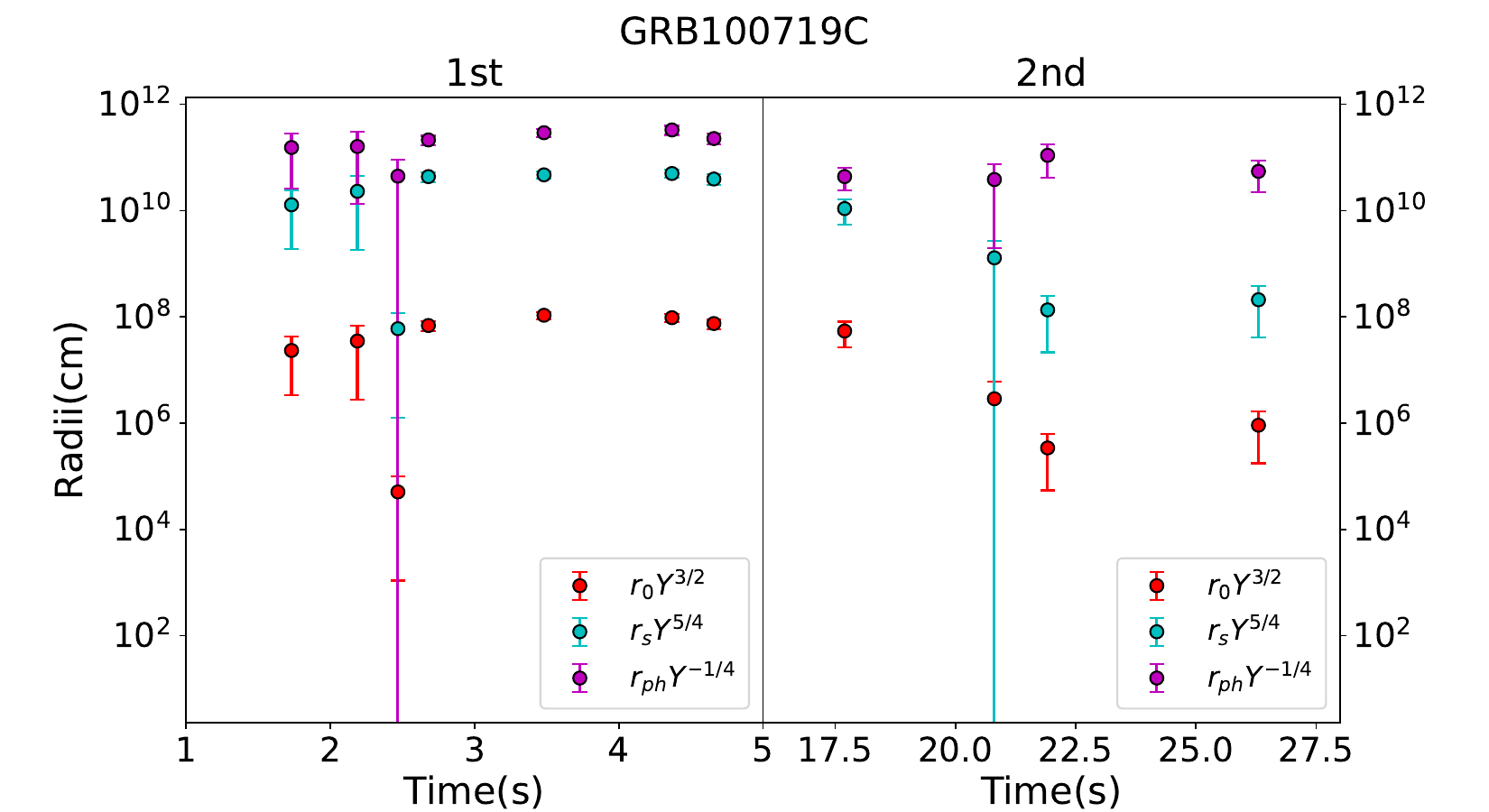}
\includegraphics [width=8cm,height=4cm]{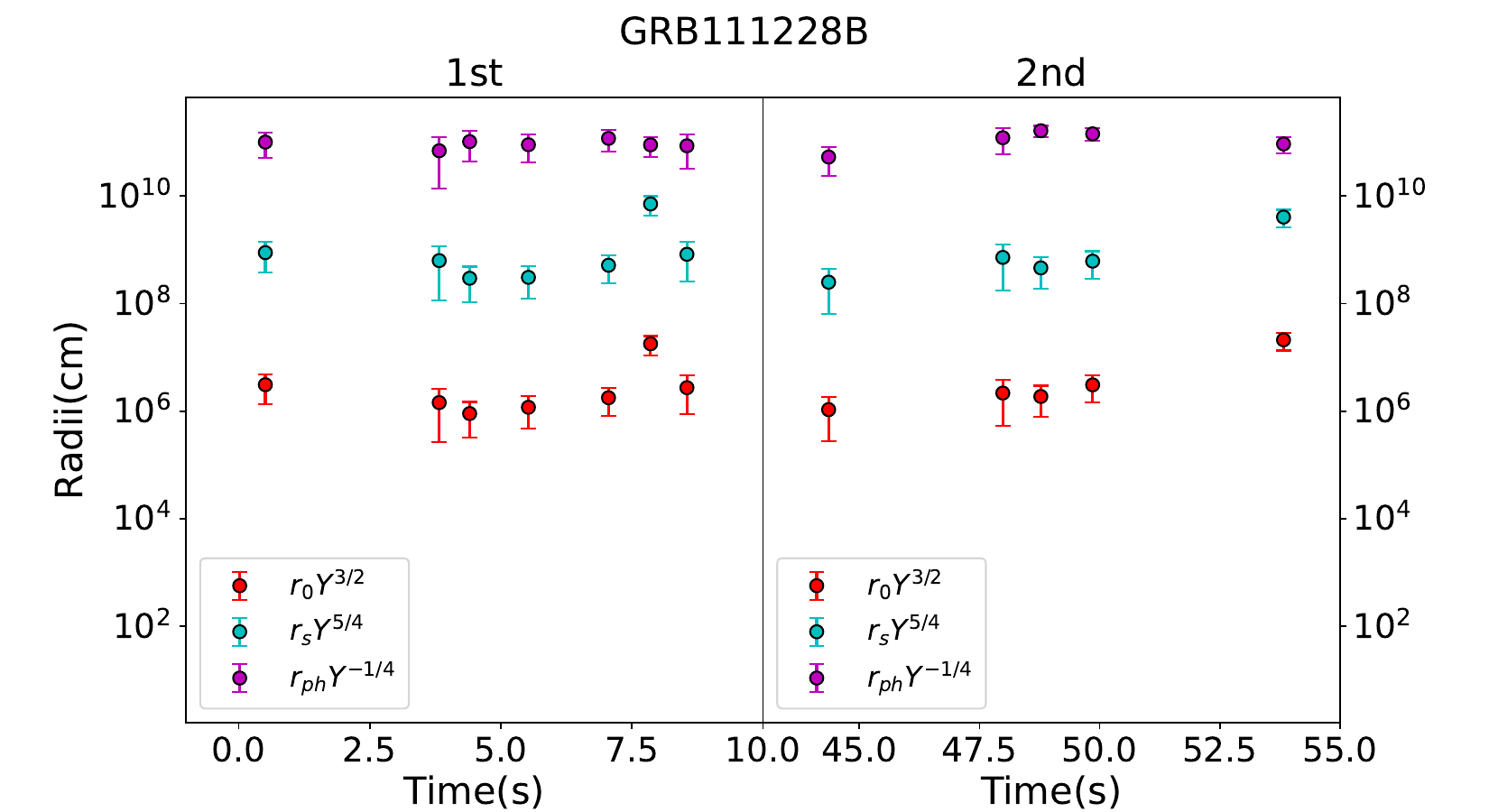}
\includegraphics [width=8cm,height=4cm]{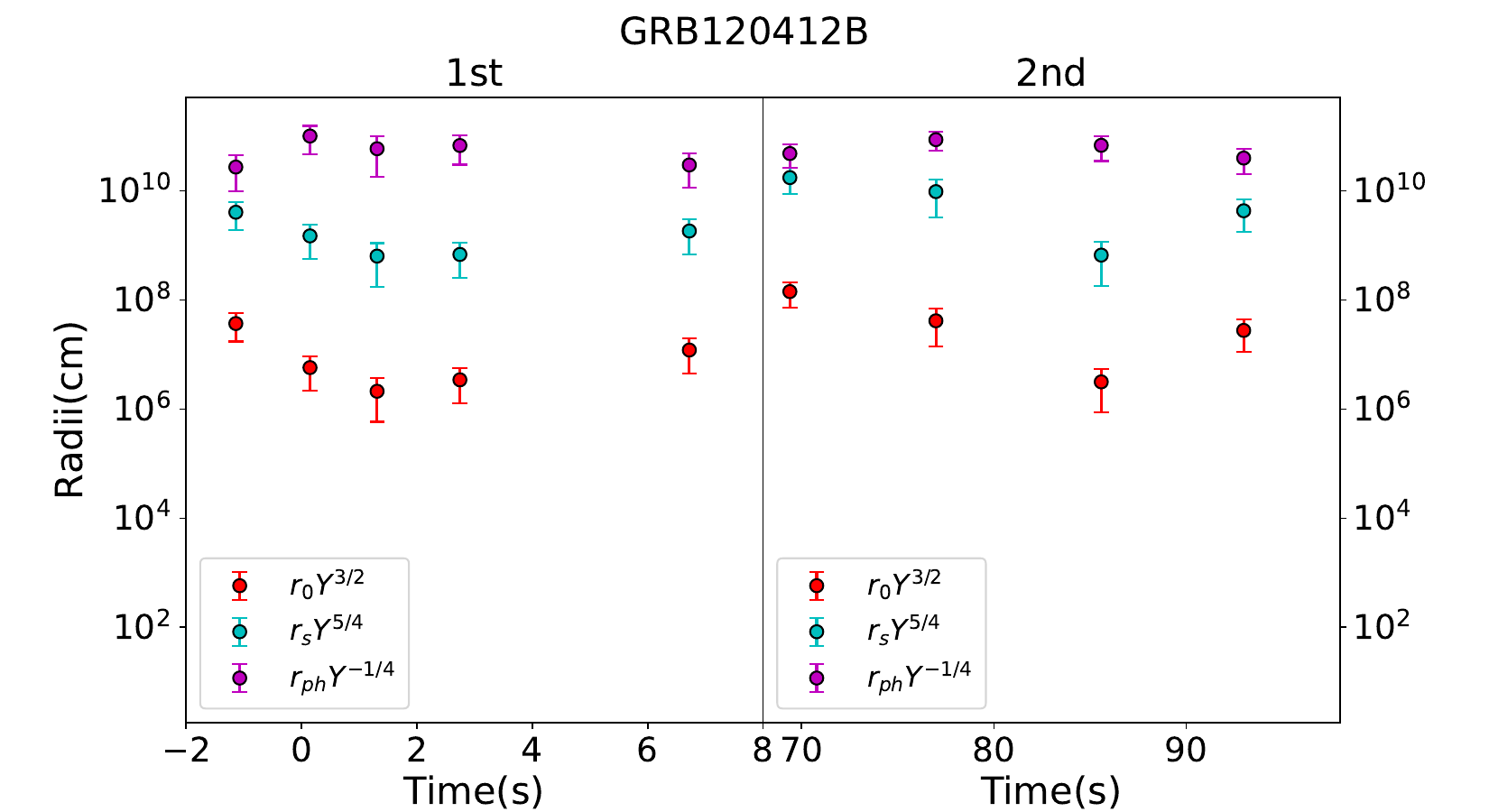}
\includegraphics [width=8cm,height=4cm]{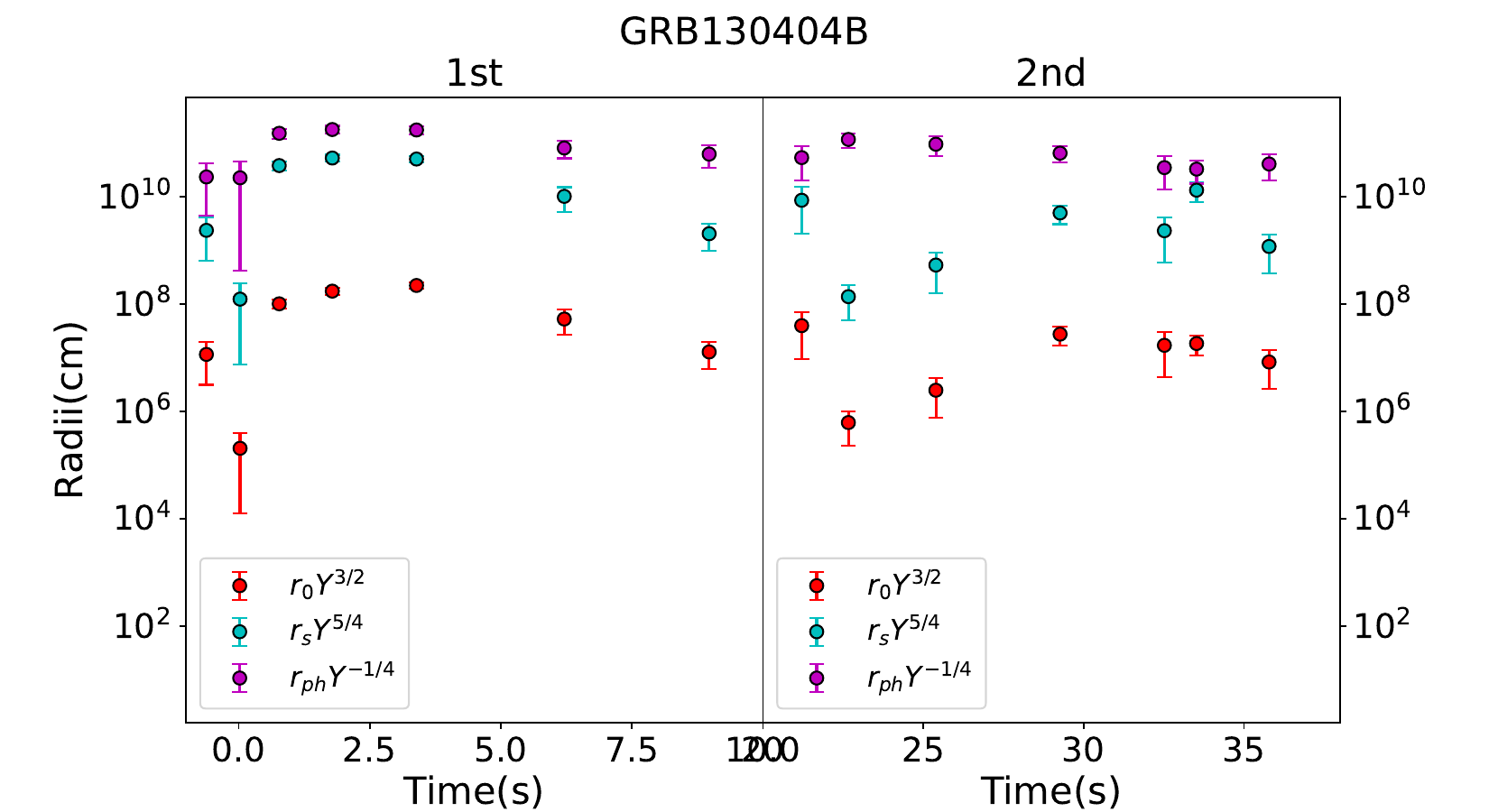}
\includegraphics [width=8cm,height=4cm]{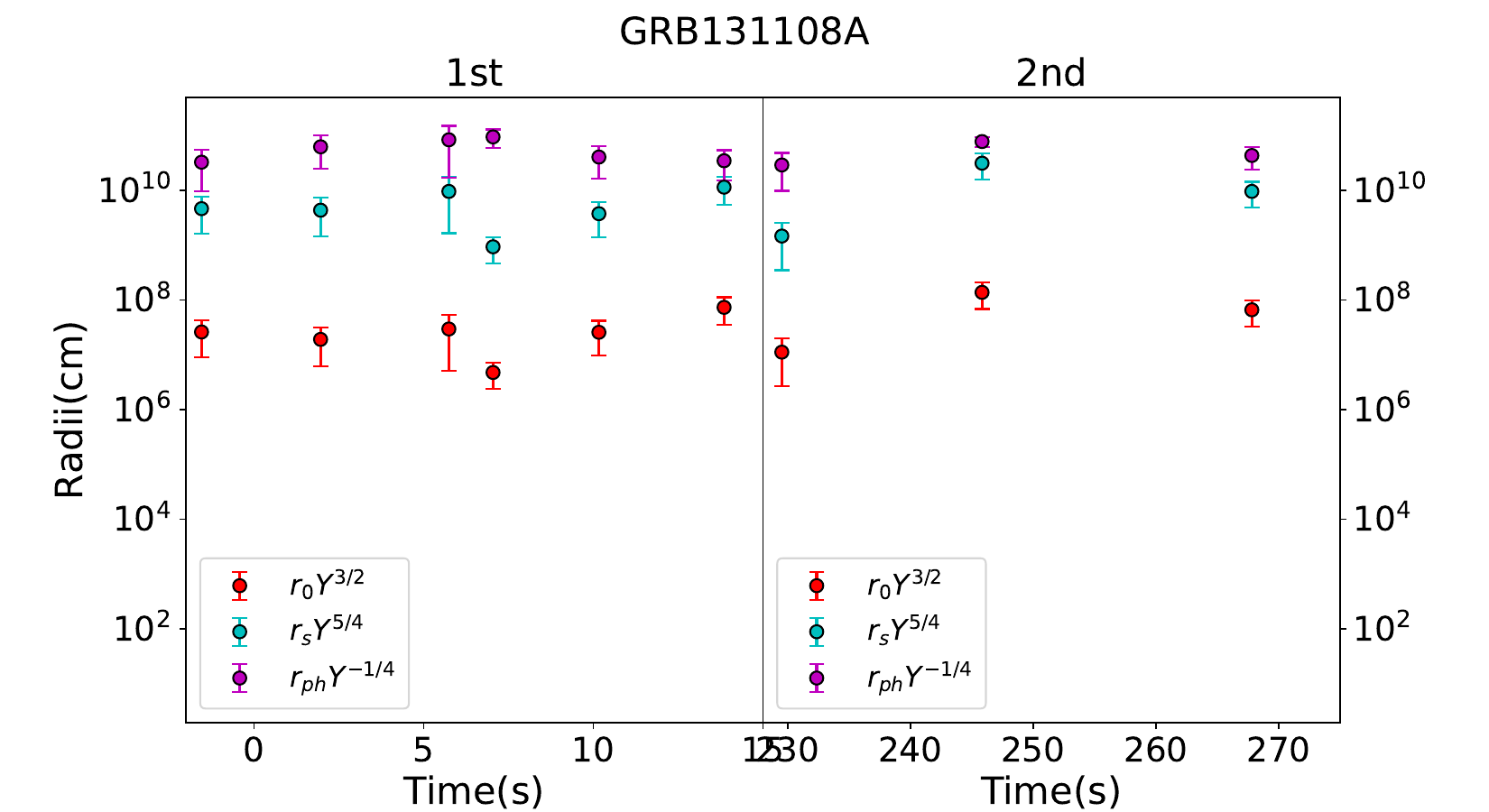}
\includegraphics [width=8cm,height=4cm]{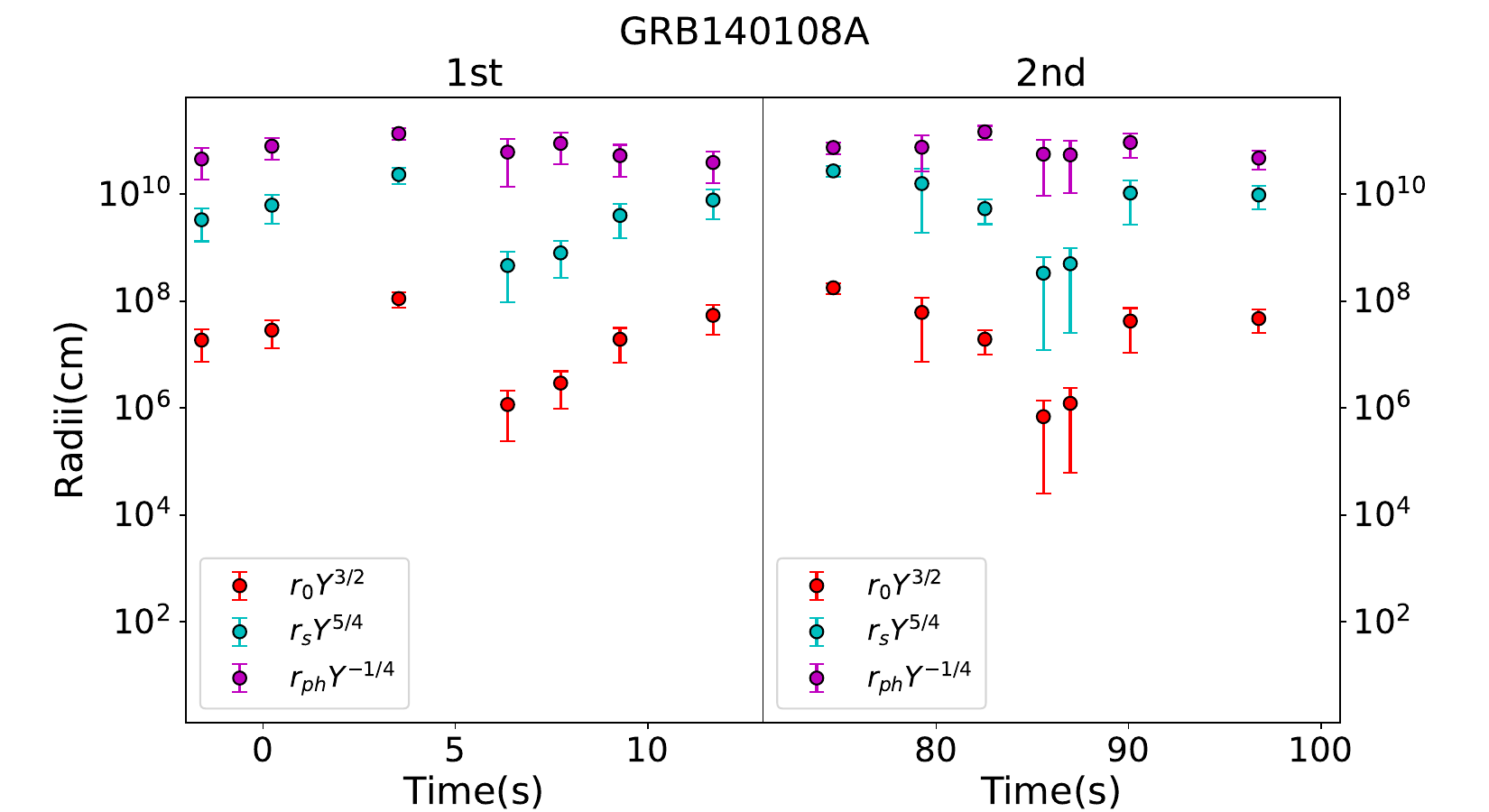}
\includegraphics [width=8cm,height=4cm]{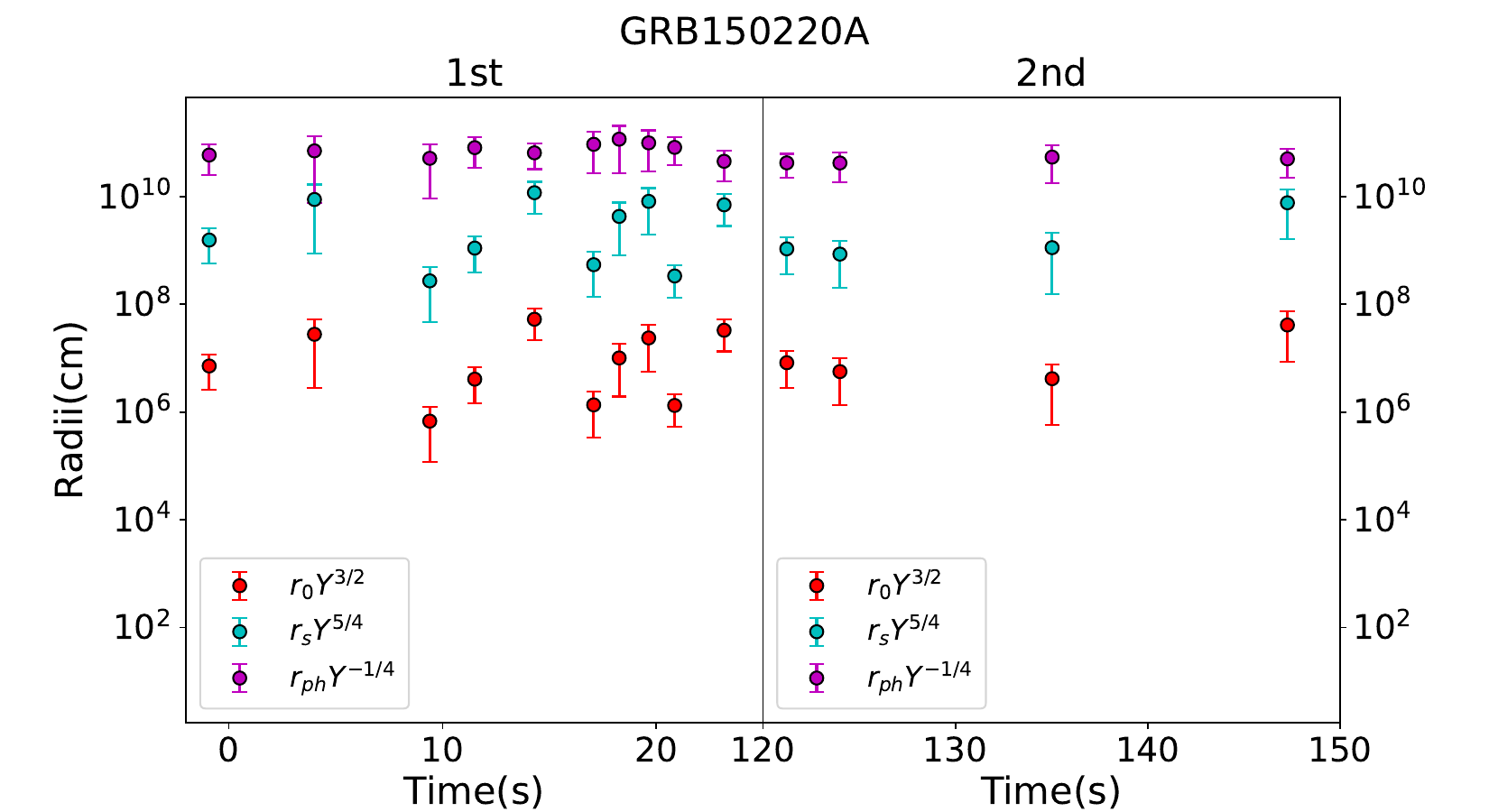}
\includegraphics [width=8cm,height=4cm]{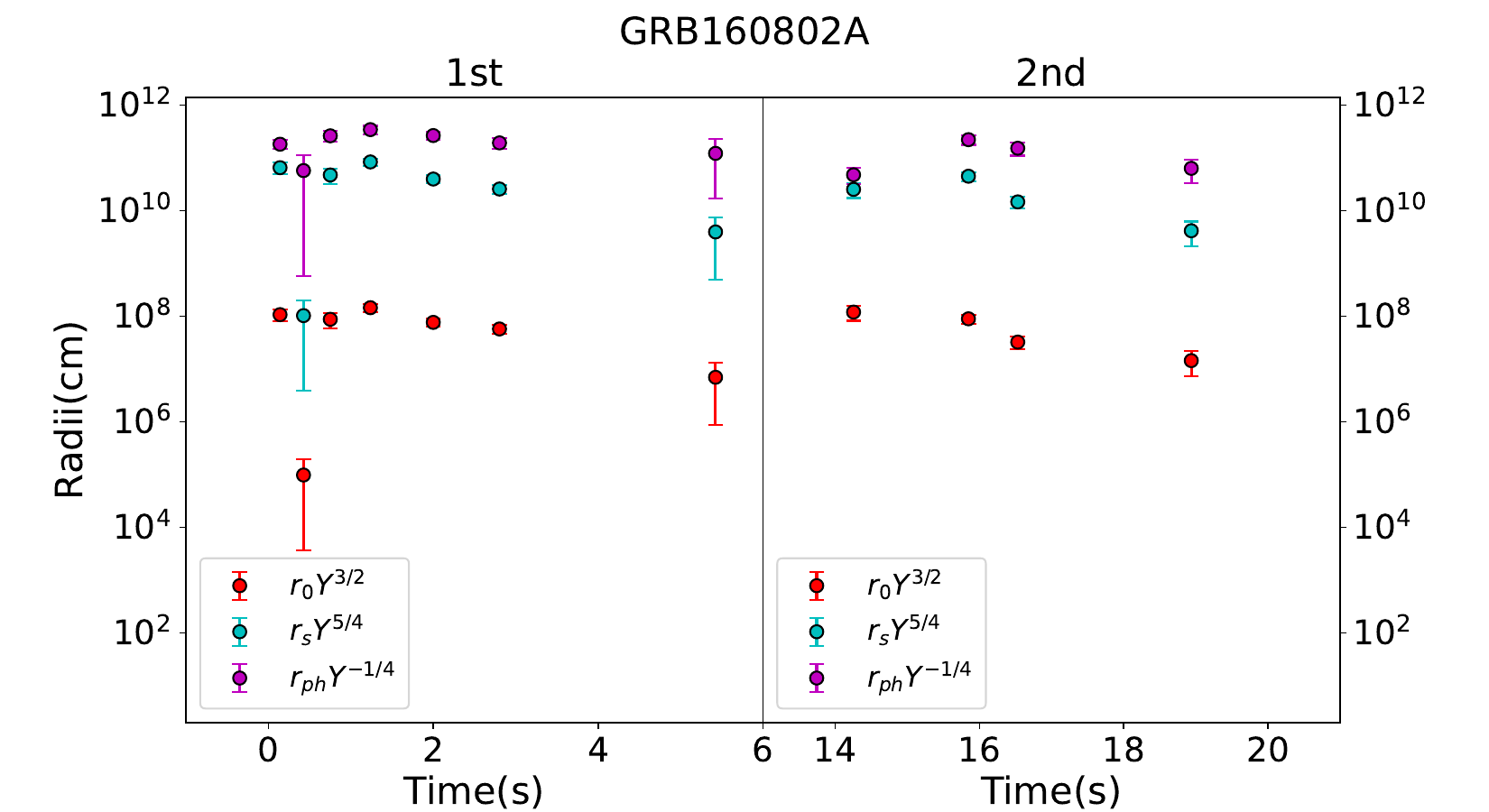}
\includegraphics [width=8cm,height=4cm]{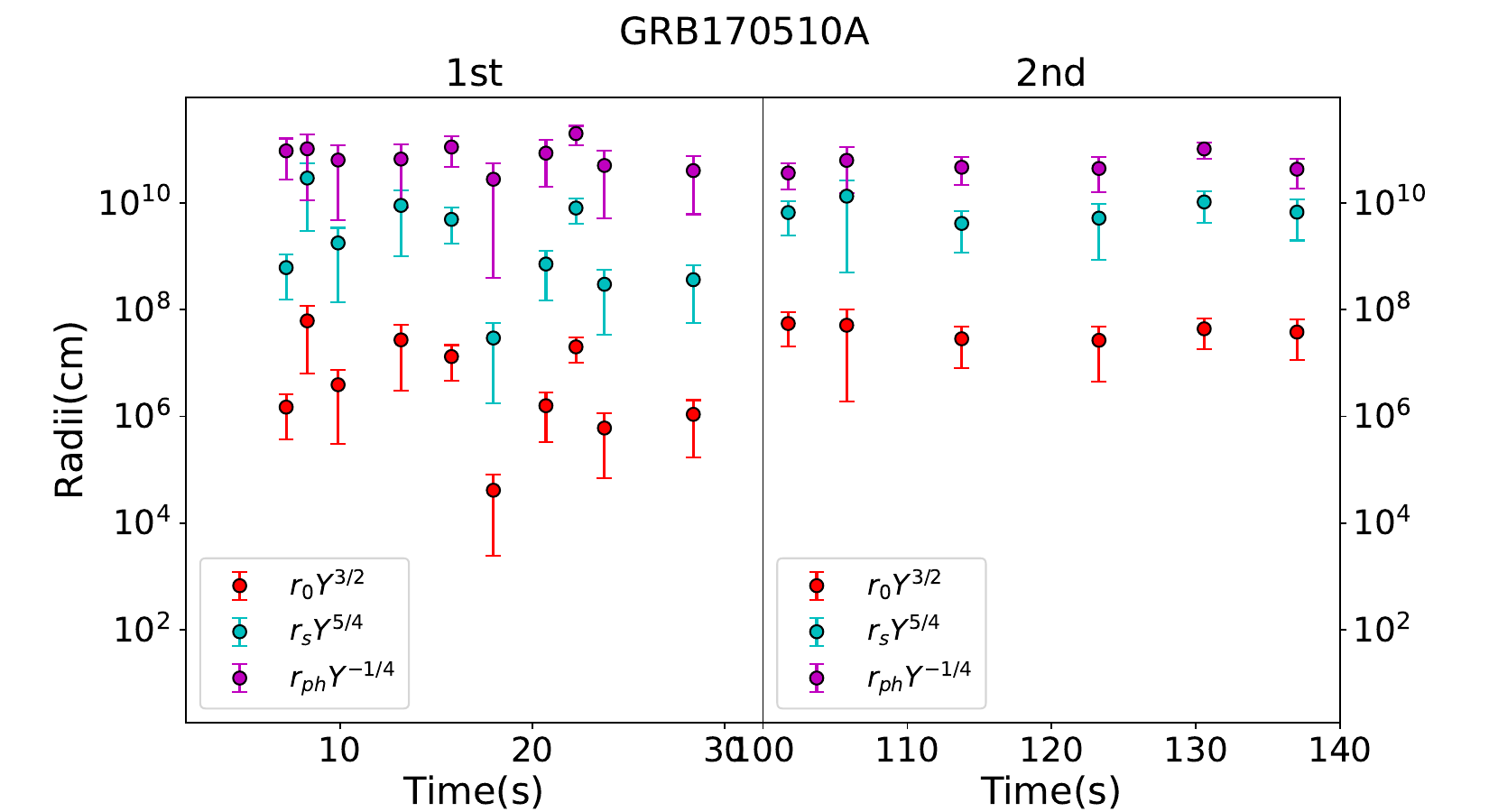}
   \figcaption{Evolution of ${{\rm{r}}_{\rm{0}}}$, ${{\rm{r}}_{\rm{s}}}$, ${{\rm{r}}_{{\rm{ph}}}}$. ``1st'' represents the main burst, and ``2nd'' represents the second burst. \label{fig 17}}
      
\end{figure}

\setcounter{figure}{16}  
\begin{figure}[H]

\centering
\includegraphics [width=8cm,height=4cm]{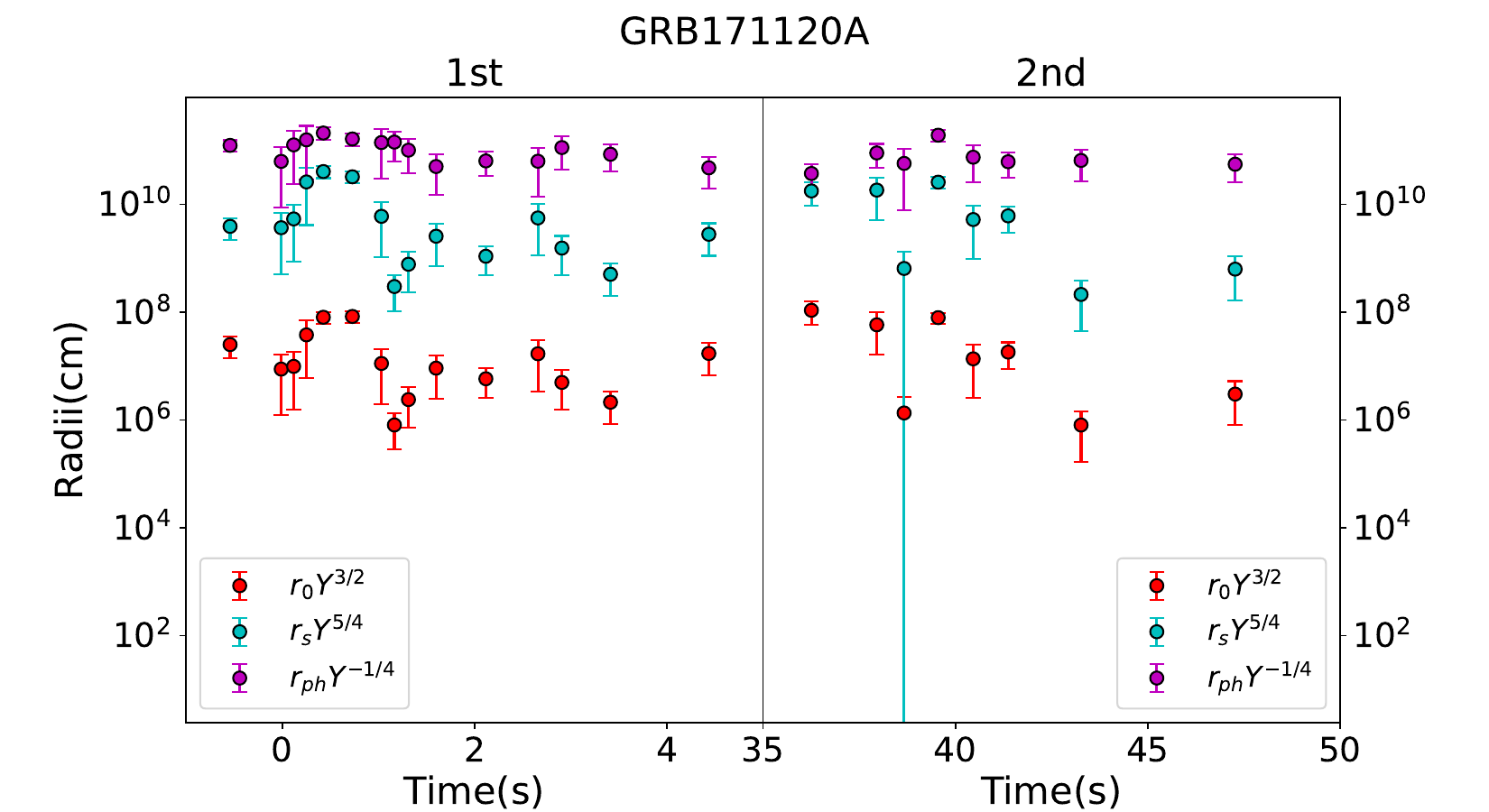}
\includegraphics [width=8cm,height=4cm]{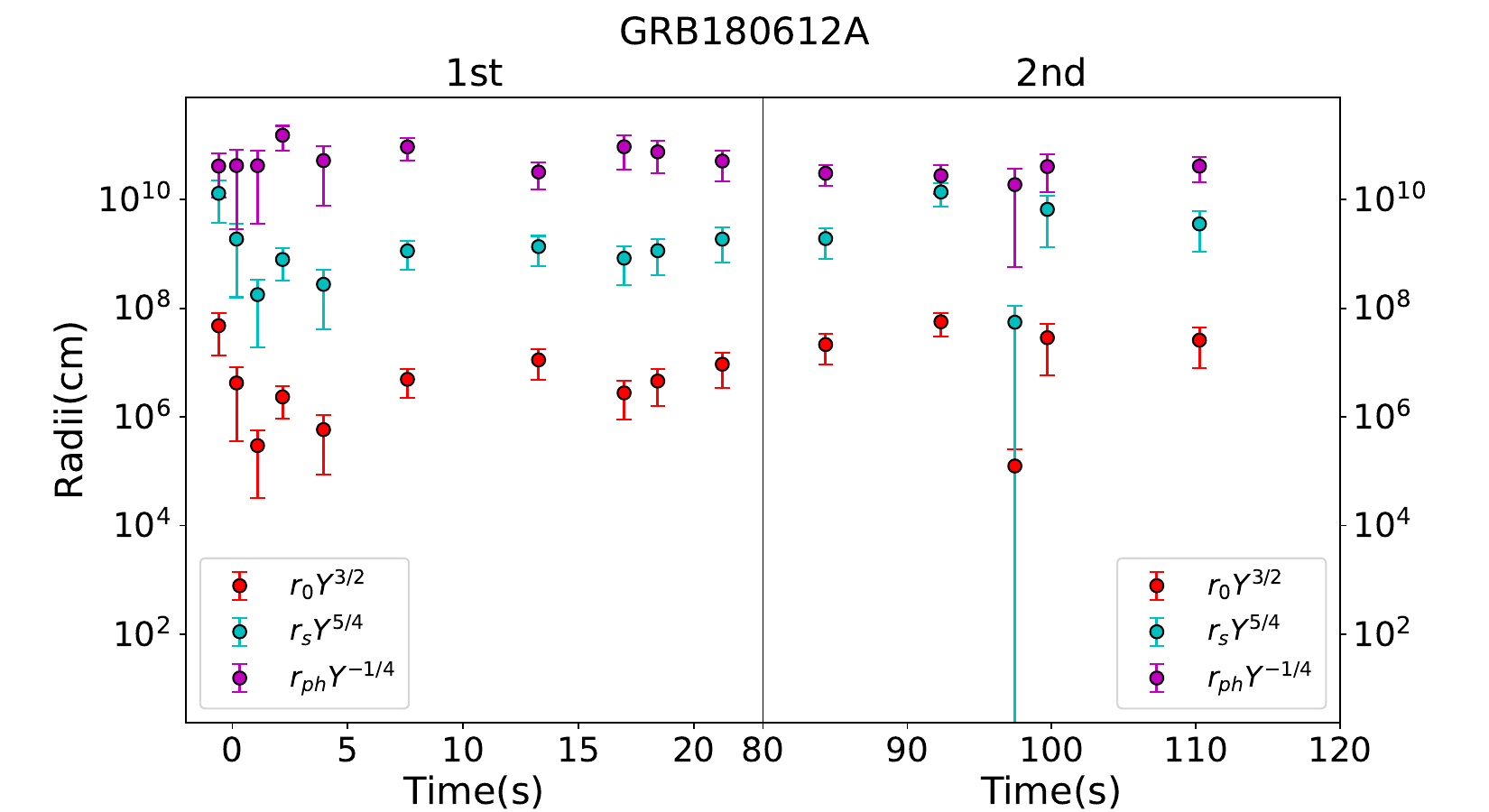}
\includegraphics [width=8cm,height=4cm]{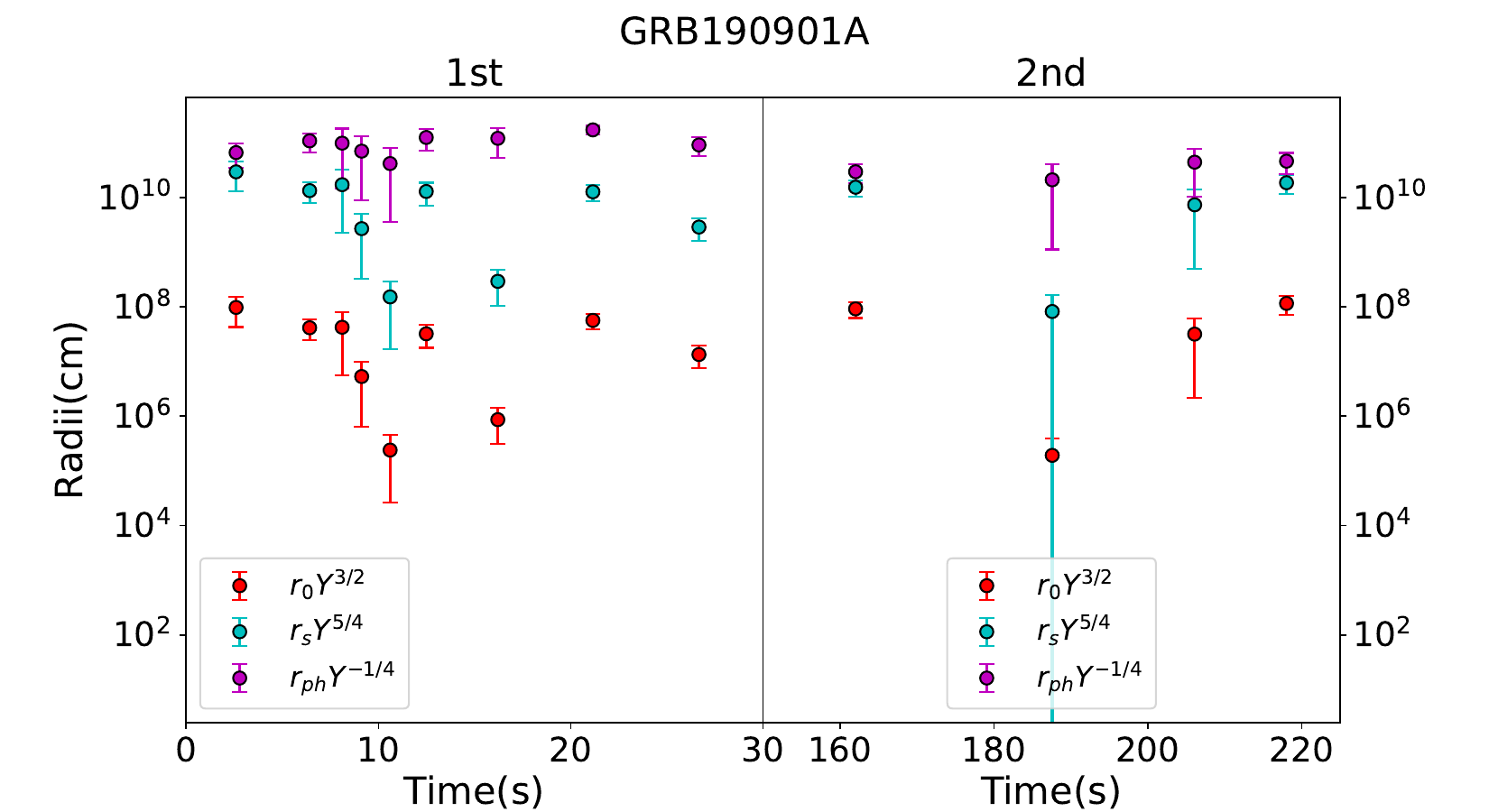}
\includegraphics [width=8cm,height=4cm]{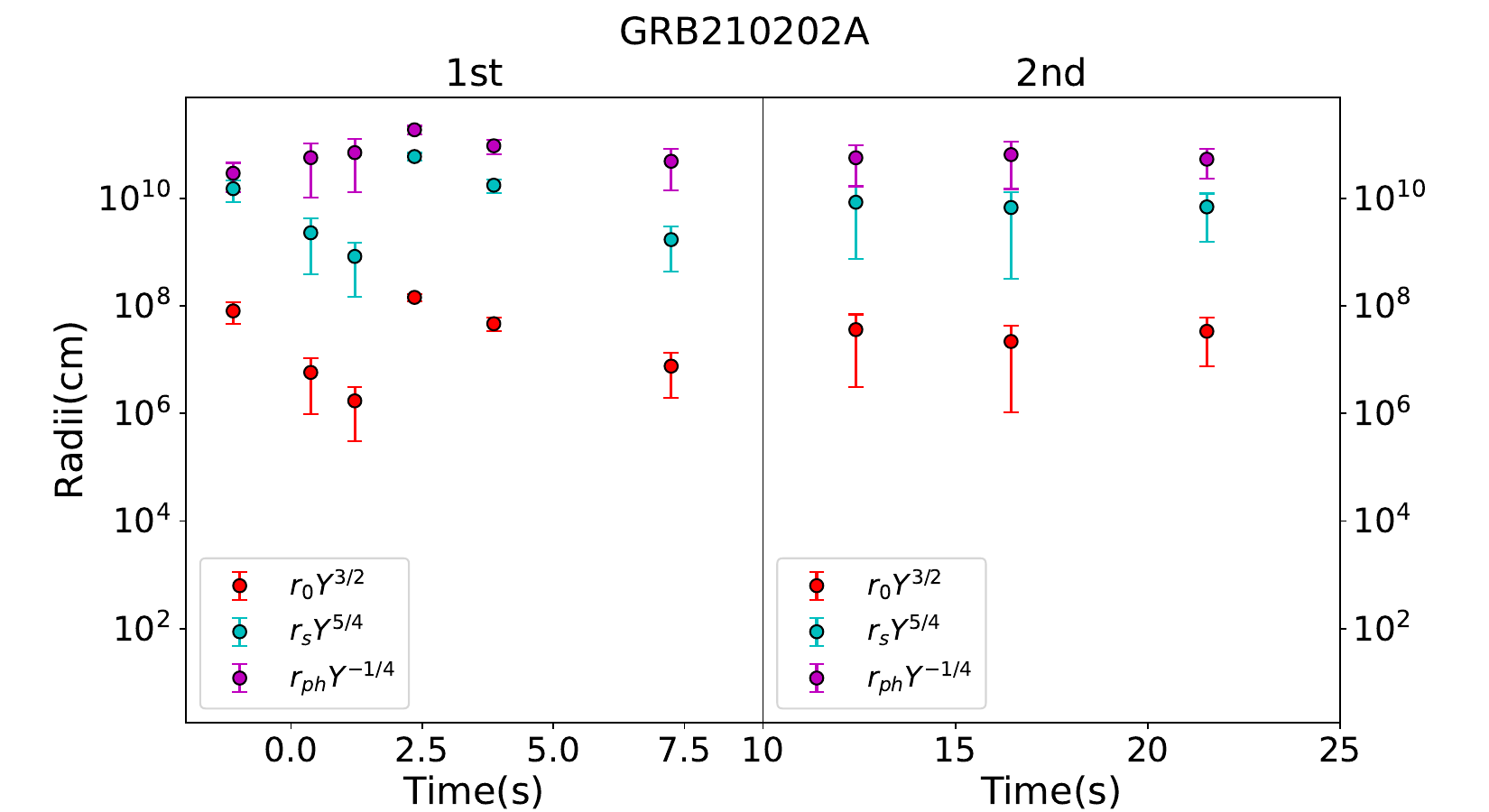}
\includegraphics [width=8cm,height=4cm]{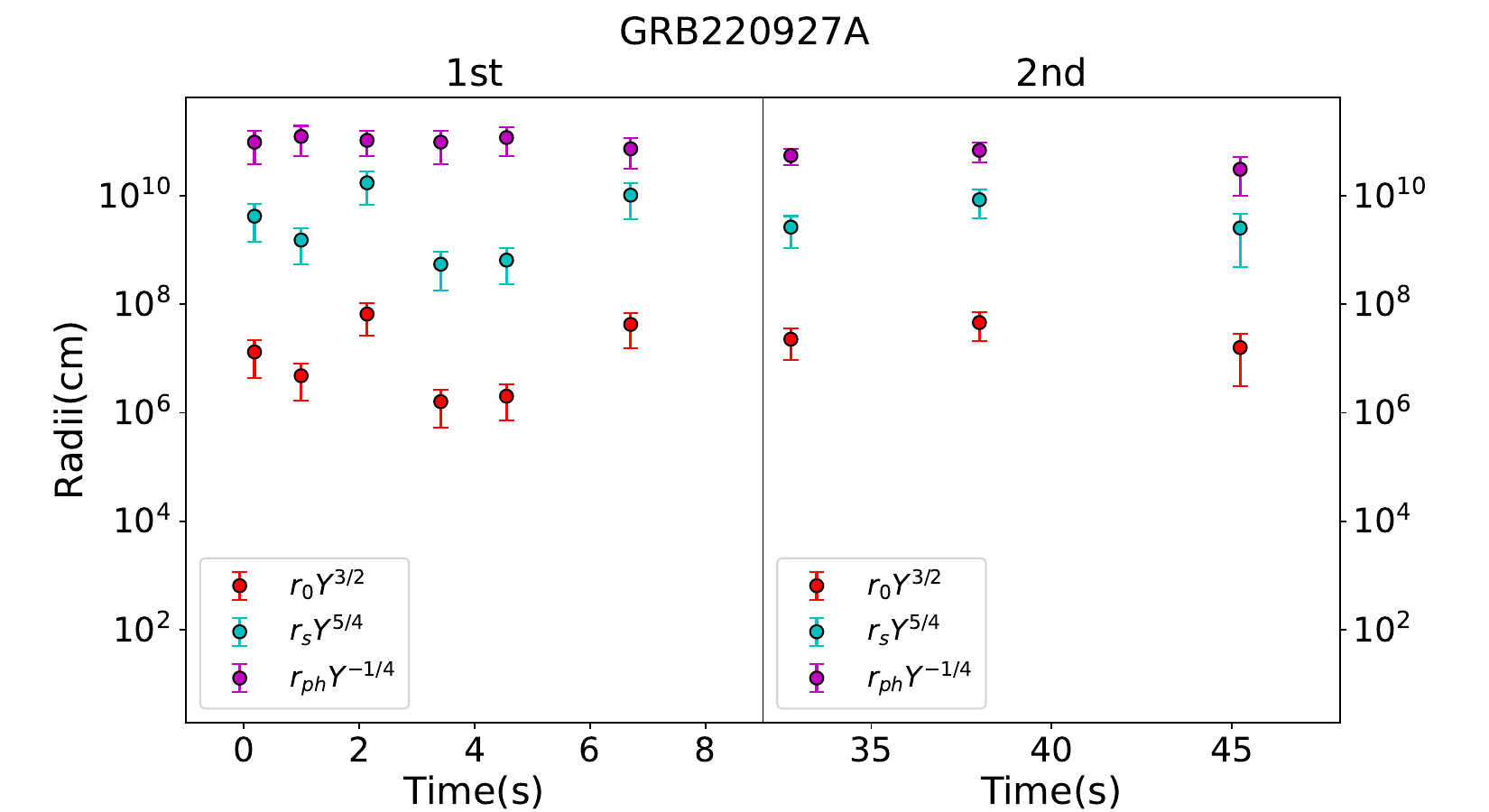}
\includegraphics [width=8cm,height=4cm]{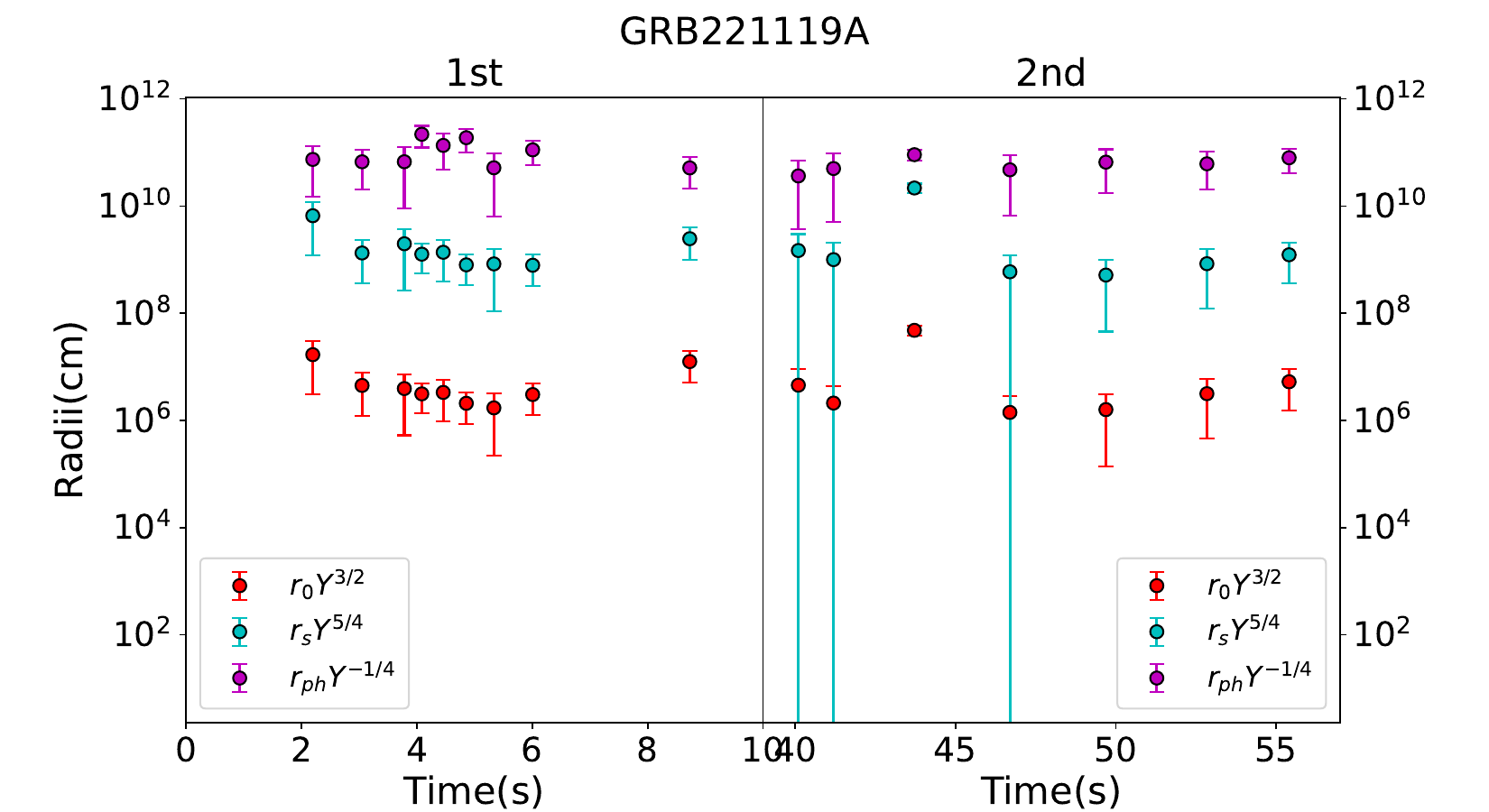}
\includegraphics [width=8cm,height=4cm]{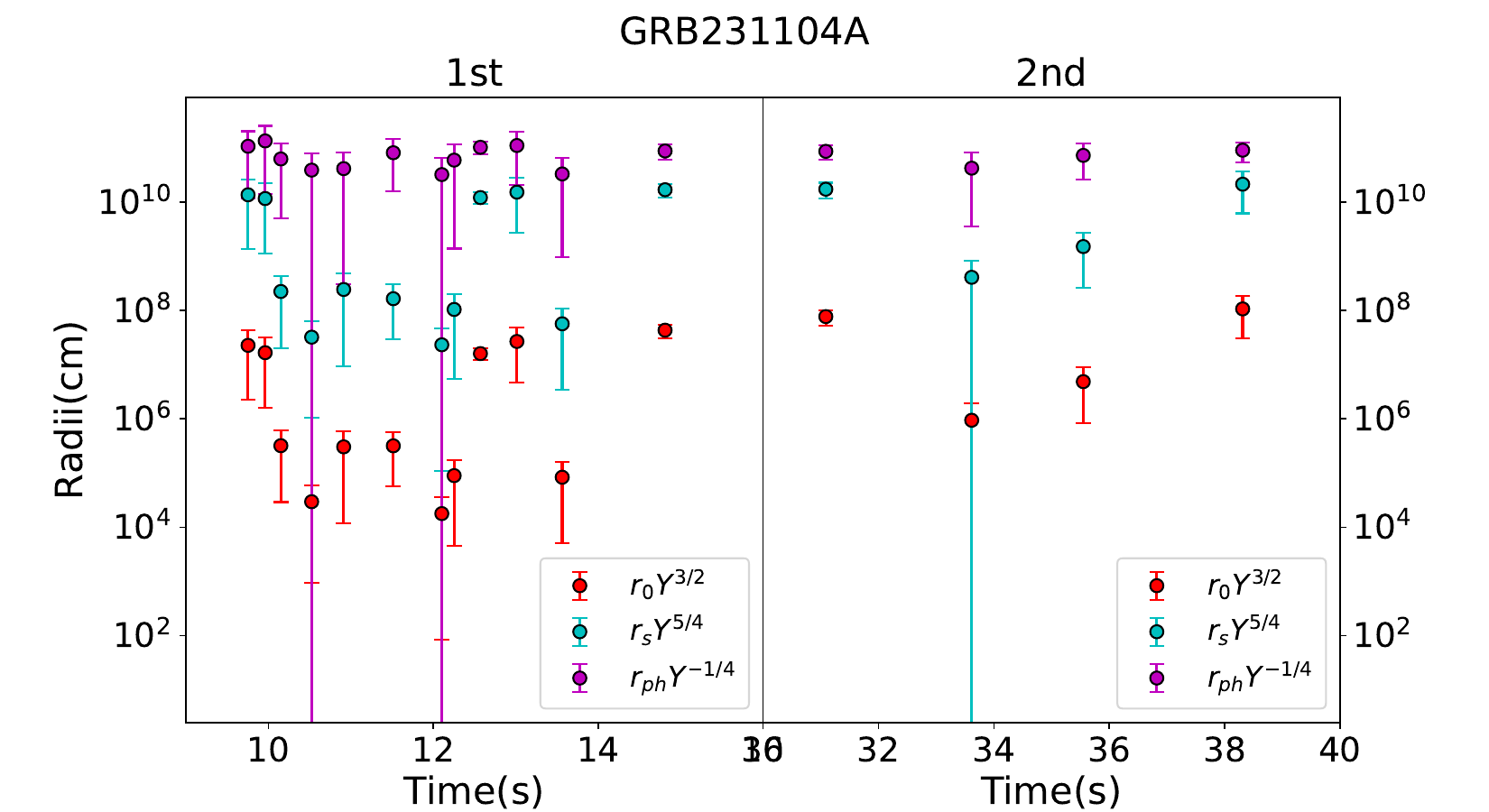}
\includegraphics [width=8cm,height=4cm]{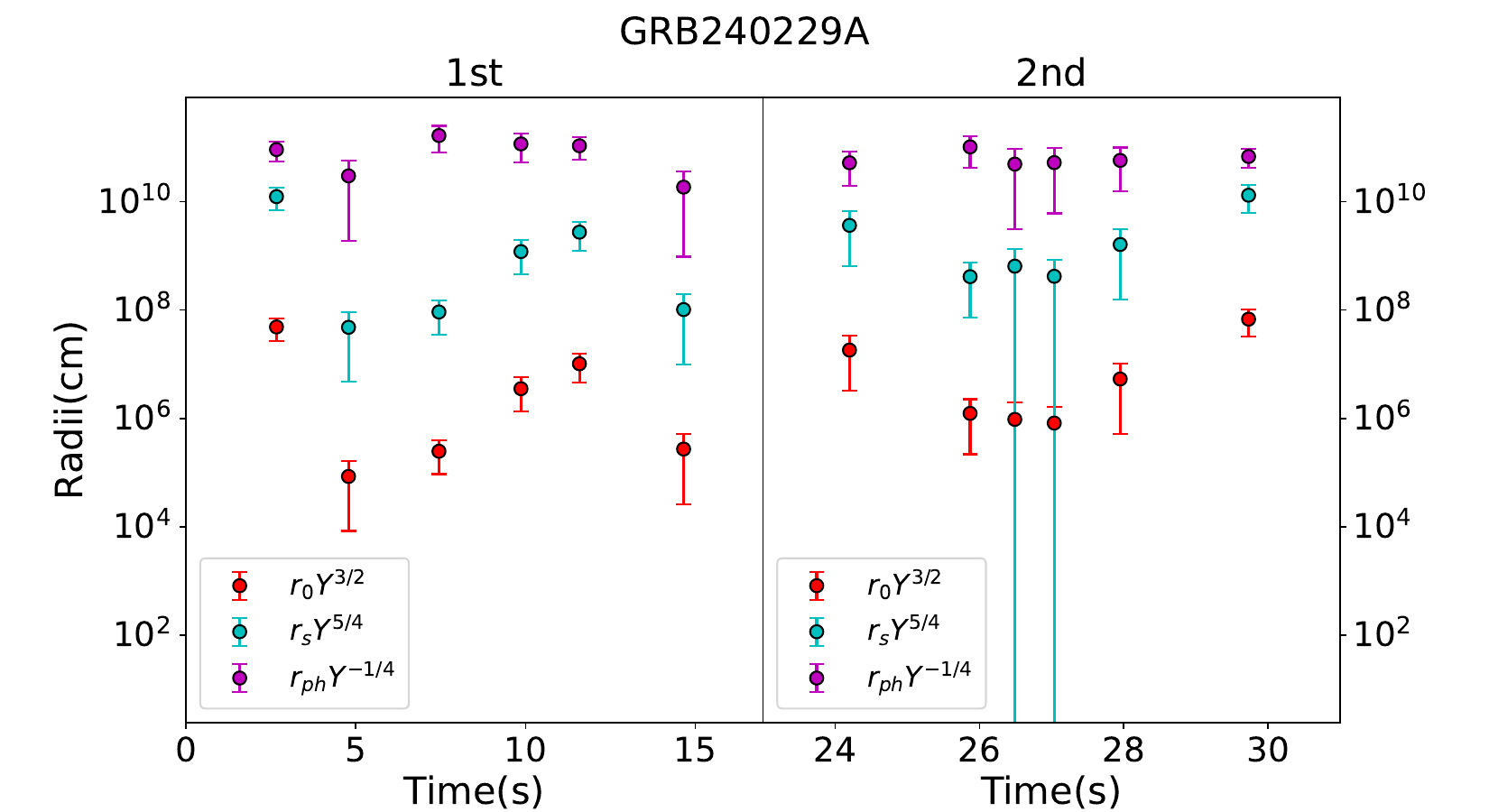}
   \figcaption{(Continued.) \label{fig 17}}
      
\end{figure}

\bibliography{reference}{}
\bibliographystyle{aasjournal}

\end{document}